\documentclass[a4paper,11pt]{article}
\pdfoutput=1 

\usepackage{jcappub} 
\usepackage[toc,page]{appendix}
\usepackage{subcaption}
\usepackage{amsmath}

\DeclareMathOperator{\csch}{csch}
\usepackage{graphicx}                
\usepackage{multirow}
\usepackage[T1]{fontenc} 
\usepackage[table]{xcolor}
\usepackage{float}
\def\noi{{\noindent}}

\definecolor{verde}{rgb}{0,0.5,0}

\def\be{\begin{equation}}
\def\ee{\end{equation}}
\def\bea{\begin{eqnarray}}
\def\eea{\end{eqnarray}}
\def\be{\begin{equation}}
\def\ee{\end{equation}}
\def\ba{\begin{eqnarray}}
\def\ea{\end{eqnarray}}

\hypersetup{pdftitle={Revisiting small-scale fluctuations in alpha-attractor models of inflation}
}
\usepackage{bm}

\title{\boldmath Revisiting small-scale fluctuations in $\alpha$--attractor models of inflation}

\author[a]{Laura Iacconi,}
\author[a]{Hooshyar Assadullahi,} 
\author[a,b]{Matteo Fasiello,}
\author[a]{and David Wands}
\affiliation{$^{a}$Institute of Cosmology \& Gravitation, University of Portsmouth, PO1 3FX, UK}
\affiliation{$^{b}$Instituto de Fisica Téorica UAM/CSIC, calle Nicolas Cabrera 13-15, Cantoblanco, 28049, Madrid, Spain}

\emailAdd{laura.iacconi@port.ac.uk}
\emailAdd{hooshyar.assadullahi@port.ac.uk}
\emailAdd{matteo.fasiello@csic.es}
\emailAdd{david.wands@port.ac.uk}

\abstract{Cosmological $\alpha$--attractors stand out as particularly compelling models to describe inflation in the very early universe, naturally meeting tight observational bounds from cosmic microwave background (CMB) experiments. We investigate $\alpha$--attractor potentials in the presence of an inflection point, leading to enhanced curvature perturbations on small scales. We study both single- and multi-field models, driven by scalar fields living on a hyperbolic field space. 
In the single-field case, ultra-slow-roll dynamics at the inflection point is responsible for the growth of the power spectrum, while in the multi-field set-up we study the effect of geometrical destabilisation and non-geodesic motion in field space. The two mechanisms can in principle be distinguished through the spectral shape of the resulting scalar power spectrum on small scales. These enhanced scalar perturbations can lead to primordial black hole (PBH) production and second-order gravitational wave (GW) generation. 
Due to the existence of universal predictions in $\alpha$--attractors, consistency with current CMB constraints on the large-scale spectral tilt implies that PBHs can only be produced with masses smaller than $10^8\,\text{g}$ and are accompanied by ultra-high frequency GWs, with a peak expected to be at frequencies of order $10\,\text{kHz}$ or above.}

\begin{document}
	\maketitle
	\flushbottom
	
	\section{Introduction}
	
\noi Cosmological inflation provides our best current model for describing the very early universe. Not only does it solve the classical problems of the standard hot big bang cosmology, but it also provides the quantum seeds for the large-scale structure of the cosmic web. On large scales, the main constraints on inflation follow from observations of anisotropies in the cosmic microwave background (CMB), pointing to almost scale-invariant and Gaussian primordial scalar fluctuations, which can be naturally explained in the context of the simplest inflationary models of a single canonical scalar field, slowly rolling down its potential. On smaller scales the primordial power spectrum is much less constrained. An intriguing possibility is that the statistics of the curvature perturbation deviates strongly from the large-scale behaviour, for example displaying a significant enhancement in the scalar power spectrum away from CMB scales.  

Inflationary models supporting features in the scalar power spectrum could lead to primordial black hole (PBH) formation, due to the collapse of large amplitude density fluctuations after horizon entry following inflation~\cite{10.1093/mnras/168.2.399} (see the review~\cite{Sasaki:2018dmp} for other formation mechanisms). PBHs formed in the early universe could potentially explain cold dark matter (or a fraction of it)~\cite{Bird:2016dcv,Bertone:2018krk, Bartolo:2018evs}. A sudden growth of the scalar power spectrum is usually associated with departures from single-field slow-roll inflation~\cite{Motohashi:2017kbs}. In single-field inflation this can be realised by a local feature in the inflaton potential, e.g., an inflection point~\cite{Garcia-Bellido:2017mdw, Germani:2017bcs, Ballesteros:2017fsr, Cicoli:2018asa, Dalianis:2018frf, Passaglia:2018ixg, Bhaumik:2019tvl}. Other mechanisms associated with multi-field models have been proposed, such as a strongly non-geodesic motion~\cite{Fumagalli:2020adf, Palma:2020ejf} and/or a large and negative curvature of the field space~\cite{Braglia:2020eai,Aragam:2021scu} which could cause a transient instability of the isocurvature perturbation, then transferred to the curvature fluctuation, leading to a peak in the scalar power spectrum on small scales.

Primordial density fluctuations induce a stochastic background of primordial gravitational waves (GWs) at second order in perturbation theory~\cite{Ananda:2006af, Baumann:2007zm} (see also the review~\cite{Domenech:2021ztg}), therefore a peak in the scalar power spectrum could lead to a potentially detectable second-order GW signal (see the reviews~\cite{Caprini:2018mtu} for other cosmological sources and~\cite{Regimbau:2011rp} for astrophysical contributions to the stochastic GW background). The detection and characterisation of the GW signal could therefore provide an indirect way of probing the scalar power spectrum on scales much smaller than those where the CMB constraints apply and in turn constrain the physics of inflation. For example, it has been recently shown that in multi-field scenarios characterised by strong and sharp turns in field space, the scalar power spectrum inherits an oscillatory modulation which is then imprinted in the scalar-induced second-order GWs~\cite{Fumagalli:2020nvq, Fumagalli:2021cel, Witkowski:2021raz} (see also~\cite{Braglia:2020taf} for an explicit model). 

In this work we focus on a class of inflationary models which goes by the name of cosmological $\alpha$--attractors~\cite{Kallosh:2013hoa, Kallosh:2013daa, Ferrara:2013rsa, Kallosh:2013pby, Kallosh:2013lkr,Kallosh:2013maa, Kallosh:2013tua, Kallosh:2013yoa}. They stand out as particularly compelling models to describe inflation in the very early universe. On the theoretical side they can be embedded in supergravity theories, while leading to universal predictions for large-scale observables that are independent of the detailed form of the scalar field potential~\cite{Kallosh:2013hoa}, and which at the same time provide an excellent fit to current observational constraints on the primordial power spectra~\cite{Planck:2018jri}. 

Usually $\alpha$--attractors are formulated in terms of a complex field $Z$ belonging to the Poincaré hyperbolic disc~\cite{Kallosh:2015zsa, Carrasco:2015uma}, with potential energy $V(Z,\,\bar{Z})$ which is regular everywhere in the disc. The corresponding kinetic Lagrangian reads
\begin{equation}
\label{kinetic lagr Z}
    \mathcal{L}_\text{kin}=-3\alpha \frac{\partial_\mu Z \partial^\mu\overline{Z}}{(1-Z \overline{Z})^2}\;,
\end{equation}
where the curvature of the hyperbolic field space is constant and negative, $\mathcal{R}_{\text{fs}}=-4/(3\alpha)$. The complex field $Z$ can be parameterized by
\begin{equation}
\label{Z parametrisation}
    Z\equiv r\,e^{i\theta} \,,
\end{equation}
where $r\equiv |Z| <1$, and eq.~\eqref{kinetic lagr Z} can then be rewritten in terms of the fields $r$ and $\theta$ as
\begin{equation}
\label{L kin r theta}
\mathcal{L}_\text{kin}=-\frac{3\alpha}{(1-r^2)^2}\left[(\partial r)^2 + r^2 \, (\partial\theta)^2 \right] \;.  
\end{equation}
As neither of the fields $r$ and $\theta$ are canonically normalised, it is often useful to transform to the canonically normalised radial field $\phi$, defined as 
\begin{equation}
\label{canonical field transformation}
   r\equiv\tanh{\left(\frac{\phi}{\sqrt{6\alpha}}\right)} \;.
\end{equation}
In terms of $\phi$ and $\theta$, the kinetic Lagrangian in eq.~\eqref{L kin r theta} reads
\begin{equation}
\label{alpha attractor metric}
      \mathcal{L}_\text{kin}=-\frac{1}{2}(\partial\phi)^2 - 
      \frac{3\alpha}{4}\sinh^2{\left(\frac{2 \phi}{\sqrt{6\alpha}} \right)} \;(\partial\theta)^2\;.
\end{equation}
Usually it is assumed that the angular field $\theta$ is strongly stabilised during inflation, in which case $\phi$ is the only dynamical field and plays the role of the inflaton~\cite{Carrasco:2015uma}. This leads to an effective single-field description of $\alpha$--attractor models of inflation, characterised by universal predictions for the large-scale cosmological observables which are stable against different choices of the inflaton potential~\cite{Kallosh:2013hoa, Galante:2014ifa, Fumagalli:2016sof}. In particular, the scalar spectral tilt, $n_s-1$, and the tensor-to-scalar ratio, $r_\text{CMB}$, are given at leading order in $(\Delta N_\text{CMB})^{-1}$ as 
\begin{align}
\label{ns prediction alpha attractors}
  n_s&\simeq 1-\frac{2}{\Delta N_\text{CMB}} \;, \\
\label{r prediction alpha attractors}  
  r_\text{CMB}&\simeq 12 \frac{\alpha}{{\Delta N_\text{CMB}}^2} \;,
\end{align}
where $\Delta N_\text{CMB}$ is the number of e-folds that separate the horizon crossing of the CMB comoving scale from the end of inflation. For $50\lesssim\Delta N_\text{CMB}\lesssim 60$ and $\alpha\lesssim \mathcal{O}(1)$ the predictions above sit comfortably within the bounds from the latest CMB observations~\cite{Planck:2018jri, BICEPKeck:2021gln}. 

In some cases both $\phi$ and $\theta$ are light during inflation, implying that the angular field $\theta$ cannot be integrated out and the full multi-field dynamics has to be taken into account. Effects associated with the dynamics of the angular field have been investigated in the context of cosmological inflation~\cite{Brown:2017osf,Mizuno:2017idt,Achucarro:2017ing, Linde:2018hmx,Christodoulidis:2018qdw}\footnote{See~\cite{Krajewski:2018moi,Iarygina:2018kee,Iarygina:2020dwe} for implications of multi-field $\alpha$--attractors for preheating.}. In particular, in~\cite{Achucarro:2017ing} the authors consider a multi-field $\alpha$--attractor model with $\alpha=1/3$ and whose potential depends also on the angular field $\theta$. Under slow-roll and slow-turn approximations, and considering a background evolution close to the boundary of the Poincaré disc, the authors demonstrate that the fields ``roll on the ridge'', evolving almost entirely along the radial direction, and the single-field predictions, eqs.~\eqref{ns prediction alpha attractors} and~\eqref{r prediction alpha attractors}, are stable against the effect of the light angular field. The impact of a strongly-curved hyperbolic field space $(\alpha\ll 1)$ has been investigated in~\cite{Christodoulidis:2018qdw}, showing that for small $\alpha$ the background trajectory could display a phase of angular inflation, a regime in which the fields' evolution is mostly along the angular direction. For the models considered in~\cite{Christodoulidis:2018qdw}, the angular inflation phase shifts the universal predictions \eqref{ns prediction alpha attractors} and~\eqref{r prediction alpha attractors}, whilst it does not lead to an enhancement of the scalar perturbations. 

In this paper we will investigate inflationary models that can support a large enhancement of the scalar power spectrum on small scales and belong to the class of $\alpha$--attractors. Building on the work~\cite{Dalianis:2018frf}, we focus on single-field potentials which feature an inflection point, proposing a potential parametrisation which has a clear physical interpretation. The ultra-slow-roll dynamics associated with non-stationary inflection points can enhance the scalar power spectrum on small scales. We then assess the impact of a light angular direction in the single-field potential, suggesting a simple multi-field extension of the inflection-point model. Within this set-up, the inflationary evolution is realised in two phases, the transition between them being caused by a geometrical destabilisation of the background trajectory and characterised by a deviation from geodesic motion in field space. At the transition the combined effect of a strongly-curved field space and non-geodesic motion could trigger a tachyonic instability in the isocurvature perturbation. The enhanced isocurvature mode couples with and sources the curvature perturbation, delivering a peak in the scalar power spectrum on small scales whose amplitude is set by the curvature of field space and the angular field initial condition. 

Even if the mechanisms enhancing the scalar perturbations differ between the single- and multi-field models, we find that the predicted large-scale observables can be described in both cases by a modified version of the universal predictions for $\alpha$--attractor models, eqs.~\eqref{ns prediction alpha attractors} and~\eqref{r prediction alpha attractors}. Compatibility with the CMB measurements constrains the small-scale phenomenology; in both set-ups the PBHs which can be produced have masses $M_\text{PBH}<10^8\,\text{g}$ and the second-order GW peak at ultra-high frequencies, $f\gtrsim 50\,\text{kHz}$. 

\medskip
This work is organised as follows. We start in section~\ref{sec: single field model} with an analysis of single-field $\alpha$--attractor models featuring an inflection-point potential and discuss the models' predictions for large-scale observables. In section~\ref{sec: phenomenology of the single field model} we discuss the single-field model phenomenology, focusing on PBH production and second-order GW generation. In section~\ref{sec: multi-field extension} we describe the multi-field extension of the single-field inflection-point model, discuss its dynamics, large-scale predictions and small-scale phenomenology. We present our conclusions in section~\ref{sec:conclusions}. For completeness, we provide additional material in a series of appendices. In appendix~\ref{sec: appendix universality class} we review how the universal predictions \eqref{ns prediction alpha attractors} and~\eqref{r prediction alpha attractors} are derived for single-field $\alpha$--attractors. In appendix~\ref{sec:app numerical P_zeta} we illustrate how the numerical computation of the single-field scalar power spectrum is performed. In appendix~\ref{sec: appendix limiting behaviour potential} we study the limiting behaviour of the single-field potential. In appendix \ref{appendix: parameter study multifield potential} we provide a parameter study of the multi-field potential. In appendix~\ref{appendix:compare between polar and planar coordinates} we discuss the two-field model of~\cite{Braglia:2020eai} in terms of polar coordinates mapping of the hyperbolic field space, clarifying its relationship with $\alpha$--attractors models.

\medskip
\textit{Conventions:} Throughout this work, we consider a spatially-flat Friedmann--Lema\^{i}tre--Robertson--Walker universe, with line element $\text{d}s^2=-\text{d}t^2+a^2(t)\delta_{ij}\text{d}x^i\text{d}x^j$, where $t$ denotes cosmic time and $a(t)$ is the scale factor. The Hubble rate is defined as $H\equiv {\dot a}/{a}$,  where a derivative with respect to cosmic time is denoted by $\dot f \equiv {\mathrm{d}f}/{\mathrm{d}t}$. The number of e-folds of expansion is defined as $N\equiv \int\,H(t)\mathrm{d}t$ and $f'\equiv {\mathrm{d}f}/{\mathrm{d}N}$. 
We use natural units and set the reduced Planck mass, $M_\text{Pl}\equiv(8\pi G_N)^{-1/2}$, to unity unless otherwise stated.

\section{Single-field inflection-point model}
\label{sec: single field model}

We will first consider $\alpha$--attractor models where the angular field $\theta$ is stabilised, leading to an effective single-field model. We take the potential to be a non-negative function of the modulus of the original complex field, $f^2(r)$, where $r=|Z|$.
The Lagrangian in terms of the canonically normalised radial field $\phi$, defined in eq.~\eqref{canonical field transformation}, is 
\begin{equation}
\label{effective single field lagrangian}
    \mathcal{L}=\frac{1}{2}R-\frac{1}{2}(\partial\phi)^2-f^2\left(\tanh{\frac{\phi}{\sqrt{6\alpha}}} \right) \;,
\end{equation}
where $f$ is an arbitrary analytic function. 

We will consider models which can successfully support an inflationary stage generating an almost scale-invariant power spectrum of primordial curvature perturbations on large scales, compatible with CMB constraints, and can also amplify scalar curvature fluctuations on smaller scales, potentially producing primordial black holes and/or significant primordial gravitational waves. To do so, the potential $f^2(r)$ must have some characteristics: 

\medskip
(i) at large field values $(\phi\gg1, r\rightarrow 1)$, the potential has to be flat enough to support slow-roll inflation and satisfy the large-scale bounds on the CMB observables. In $\alpha$--attractor models, the flatness of the potential is naturally achieved at the boundary ($r\rightarrow 1$) by the stretching induced by the transformation \eqref{canonical field transformation} so long as $f(r)$ remains finite;

\medskip
(ii) in single-field inflation, a significant amplification of scalar fluctuations on small scales can be achieved by deviations from slow roll~\cite{Motohashi:2017kbs}. In particular, this may be realised with a transient ultra-slow-roll phase~\cite{Kinney:2005vj,Dimopoulos:2017ged,Pattison:2018bct}, where the gradient of the potential becomes extremely small, at intermediate field values. This can be implemented by having an almost stationary inflection point in the potential~\cite{Garcia-Bellido:2017mdw, Germani:2017bcs, Ballesteros:2017fsr, Cicoli:2018asa, Dalianis:2018frf, Passaglia:2018ixg}\footnote{For other mechanisms see, e.g., \cite{Kawai:2021bye,Kawai:2021edk}, where ultra-slow-roll inflation is realised in models comprising a scalar field coupled to the Gauss-Bonnet term.};

\medskip
(iii) at the end of inflation, the condition $V(\phi_\text{end})=0$ ensures that inflation can end without giving rise to a cosmological constant at late times. 

\medskip
In the following, we outline a procedure to fix the potential profile in a way that addresses all the requirements listed above. The potential is constructed in a way similar to~\cite{Dalianis:2018frf}, but our analysis differs in that we present a simplified potential, with a reduced number of parameters and we give a clear dynamical interpretation of each parameter. Furthermore, while in~\cite{Dalianis:2018frf} cases with $\alpha=\mathcal{O}(1)$ have been studied extensively, we will consider configurations with $\alpha<1$, which will enhance the role of the hyperbolic geometry in the model's multi-field extension. 

\subsection{Parameterising the inflection-point potential}

Given the single-field Lagrangian \eqref{effective single field lagrangian}, 
the easiest way to implement an almost stationary inflection point in the potential, $V(\phi)$, is to consider a function $f(r)$ which itself has an almost stationary inflection point. The inflection-point structure of $f(r)$ is then transmitted to the potential
\begin{equation}
\label{V(r)}
    V(\phi)=
f^2\left(r(\phi)\right) \;.
\end{equation}

For a single inflection point it is sufficient to consider a simple cubic polynomial 
\begin{equation}
\label{f function}
    f(r)=f_0 + f_1 r+ f_2 r^2+ f_3 r^3 \;.
\end{equation}
From condition (iii) above we require $V(\phi_\text{end})=0$ at the end of inflation. Here, for simplicity, we set $\phi_\text{end}=0$, which together with \eqref{V(r)} and~\eqref{f function} implies 
\begin{equation}
    \label{cond1}
     f_0=0\;.
\end{equation}
We require $f_1\neq0$ so that $\phi=0$ is a simple minimum with $V''(0)>0$, and given that the potential \eqref{V(r)} is symmetric under $f\to-f$ we then pick $f_1>0$ without loss of generality.

An inflection point in $f(r)$ at $r=r_\text{infl}$, where $0<r_\text{infl}<1$, is defined by the condition $f''(r_\text{infl})=0$. For the function in eq.~\eqref{f function}, this translates into the condition
\begin{equation}
\label{cond2}
    f_3=-\frac{f_2}{3r_\text{infl}} \;,
\end{equation}
where the positivity of $r_\text{infl}$ implies that $f_2$ and $f_3$ have opposite signs. 

The first derivative of the function \eqref{f function} calculated at the inflection point is then
\begin{equation}
\label{first derivative}
    f'(r_\text{infl})=f_1+f_2r_\text{infl} \;.
\end{equation}
In order for $r_\text{infl}$ to be a stationary ($f'(r_\text{infl})=0$) or almost stationary ($f'(r_\text{infl})\simeq0$) inflection point, we require $f_2<0$, which follows from the positivity of $f_1$ and $r_\text{infl}$. From \eqref{cond2}, this implies that $f_3>0$.

%
In order to achieve a significant amplification of the scalar power spectrum on small scales, we will consider models with an approximately stationary inflection point where the first derivative at $r_\text{infl}$ is slightly negative, $f'(r_\text{infl})<0$. As the inflaton rolls from $r>r_\text{infl}$ down towards $r=0$ this acts to further slow the inflaton as it passes through the inflection point, realising an ultra-slow-roll phase.
In this case the inflection point is then preceded by a local minimum (for $r>r_\text{infl}$) and followed by a local maximum (for $r<r_\text{infl}$). 
Using \eqref{first derivative}, both stationary and almost stationary configurations can be described by the condition
\begin{equation}
\label{cond3}
   f_1=-f_2 \left( r_\text{infl}-\xi \right) \;,
\end{equation}
where $\xi=0$ corresponds to the case of a stationary inflection point and an approximate stationary inflection point is realised if $0<\xi\ll r_\text{infl}$.
%

Finally, by substituting \eqref{f function} into \eqref{V(r)} subject to the conditions \eqref{cond1}, \eqref{cond2} and \eqref{cond3}, and transforming to the canonical field $\phi$ defined in eq.~\eqref{canonical field transformation}, the potential can be written as 
\begin{equation}
\label{potential}
V(\phi)
=V_0 \left\{\left(r_\text{infl} -\xi\right) \tanh{\left(\frac{\phi}{\sqrt{6\alpha}}\right)}-\tanh^2{\left(\frac{\phi}{\sqrt{6\alpha}}\right)} +\frac{1}{3r_\text{infl}} \tanh^3{\left(\frac{\phi}{\sqrt{6\alpha}}\right)} \right\}^2         \;,
\end{equation}
where 
we have defined $V_0\equiv {f_2}^2$.
For $\xi=0$ we have a stationary inflection point at $\phi=\phi_\text{infl}$, where we define $\tanh(\phi_\text{infl}/\sqrt{6\alpha})\equiv r_\text{infl}$. More generally we have an approximately-stationary inflation point, with $V'(\phi_\text{infl})={\cal O}\left(\xi/r_\text{infl}\right)$ and $V''(\phi_\text{infl})={\cal O}\left(\xi/r_\text{infl}\right)$ for $0<\xi\ll r_\text{infl}$.

\begin{figure}
\centering
\includegraphics[scale=0.55]{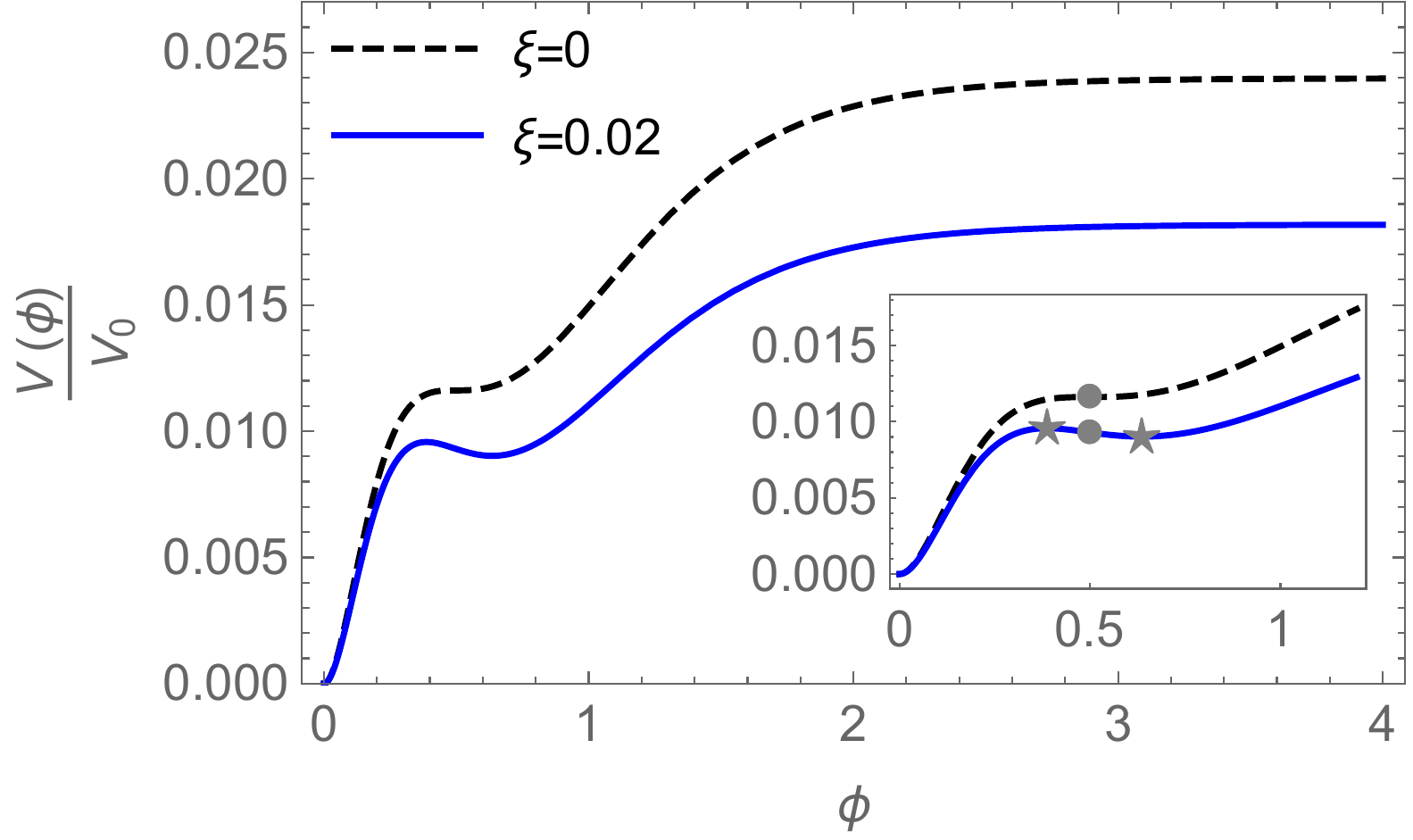}
\caption{Structure of the potential \eqref{potential} with $\alpha=0.1$ and $\phi_\text{infl}=0.5$. The inset zooms around the inflection point. If $\xi=0$ the inflection point is stationary. The case $\xi\neq0$ corresponds to an approximate stationary inflection point, where $\phi_\text{infl}$ (grey dot) is accompanied by a local minimum and a local maximum (grey stars).}
\label{fig: structure around inflection point}
\end{figure}

Starting from an initial set of free parameters $\{
f_0,\, f_1,\, f_2,\, f_3\}$ for a fixed value of $\alpha$, we have reduced it to the set $\{V_0,\, r_\text{infl},\, \xi\}$. 
The normalisation of the potential $V_0$ is fixed at CMB scales in order to reproduce the right amplitude of the scalar fluctuations, leaving only two free parameters to describe the shape of the potential, $\{r_\text{infl},\,\xi\}$, for a given $\alpha$. 
In figure \ref{fig: structure around inflection point}, two configurations of $V(\phi)$ are shown in order to illustrate a stationary inflection point ($\xi=0$) and an approximately stationary inflection point ($0<\xi\ll r_\text{infl}$).

\subsection{Background evolution}

The equations of motion for the homogenous field $\phi(t)$ in an FLRW cosmology are given by the Klein--Gordon and evolution equations
\begin{eqnarray}
\label{single field eq of motion phi}
 \ddot\phi &=& - 3H\dot\phi - V_\phi \,, \\
 \dot{H} &=& - \frac12 \dot\phi^2 \,,
\end{eqnarray}
where $V_\phi\equiv{\mathrm{d}V}/{\mathrm{d}\phi}$, subject to the Friedmann constraint
\begin{equation}
    H^2 =\frac{1}{3}\left(  V+ \frac12\dot\phi^2 \right)\,.
\end{equation}

When studying the evolution of the fields during inflation we will often show this with respect to the integrated expansion or e-folds
\begin{equation}
    N\equiv\int H(t)\,\mathrm{d}t \,.
\end{equation}
%
%
In particular the Hubble slow-roll parameters describe the evolution of $H$ and its derivatives with respect to $N$\footnote{The Hubble slow-roll parameters \eqref{epsilon H}--\eqref{xi H} can be related to the Hubble-flow parameters~\cite{Schwarz:2001vv} $\epsilon_{i+1}=\epsilon_i'/\epsilon_i$, where $\epsilon_{0}=H^{-1}$. In particular we find $\epsilon_1=\epsilon_H$ and $\epsilon_2=2\epsilon_H-2\eta_H$.}
\begin{align}
\label{epsilon H}
    \epsilon_H &\equiv -\frac{H'}{H}
    \;,\\
\label{eta H}    
    \eta_H &\equiv \epsilon_H-\frac{1}{2} 
    \frac{\epsilon_H'}{\epsilon_H}
    \;, \\
\label{xi H}
\xi_H&\equiv \epsilon_H \eta_H -
\eta_H'
\;.
\end{align}
These dimensionless parameters play an important role in the evolution of the scalar perturbations during inflation, as described in appendix~\ref{sec:app numerical P_zeta}.
Inflation is characterised by $\epsilon_H<1$.
\begin{figure}
\centering
\captionsetup[subfigure]{justification=centering}
   \begin{subfigure}[b]{0.48\textwidth}
    \includegraphics[width=\textwidth]{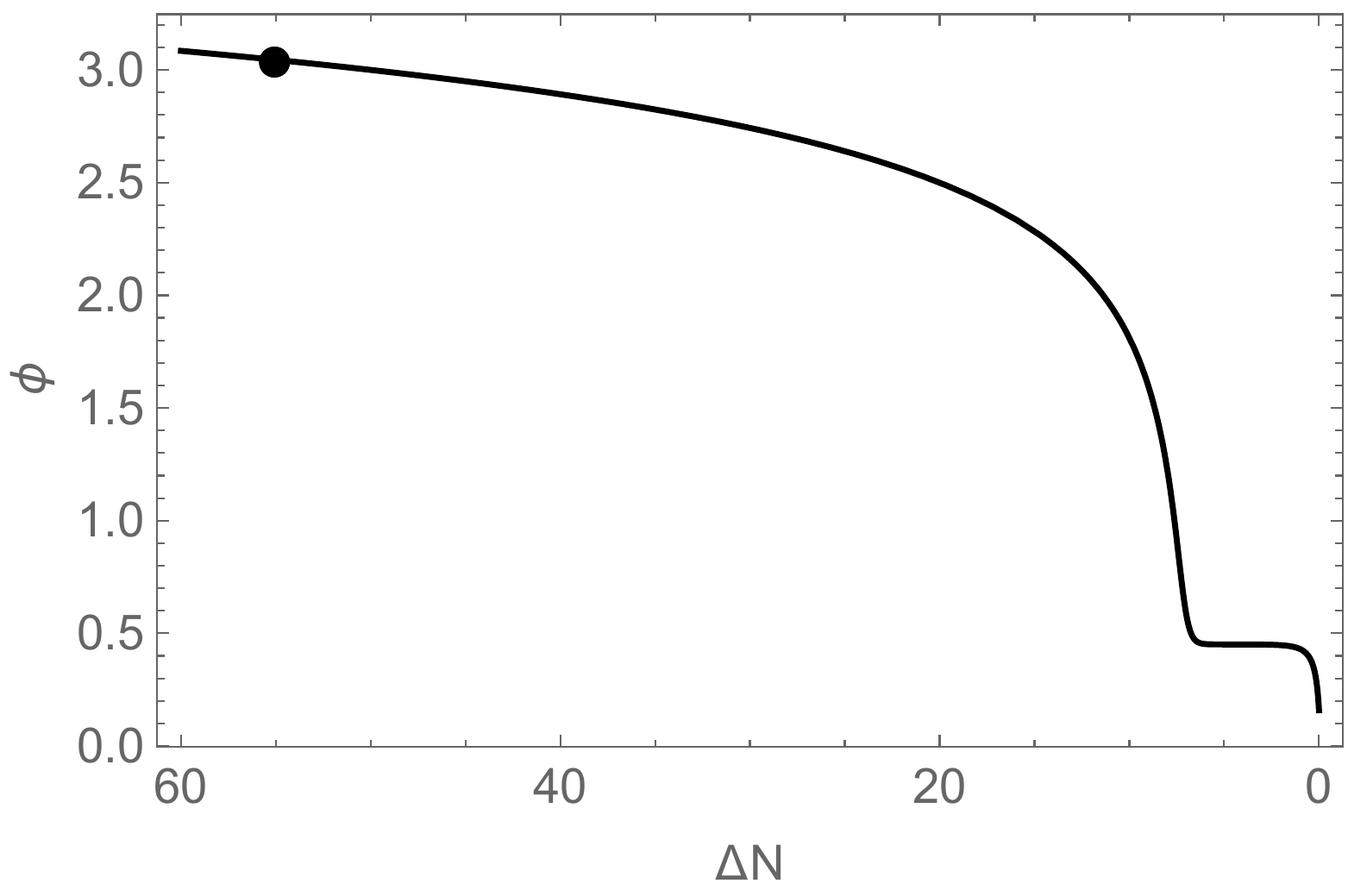}
  \end{subfigure}
  \begin{subfigure}[b]{0.48\textwidth}
    \includegraphics[width=\textwidth]{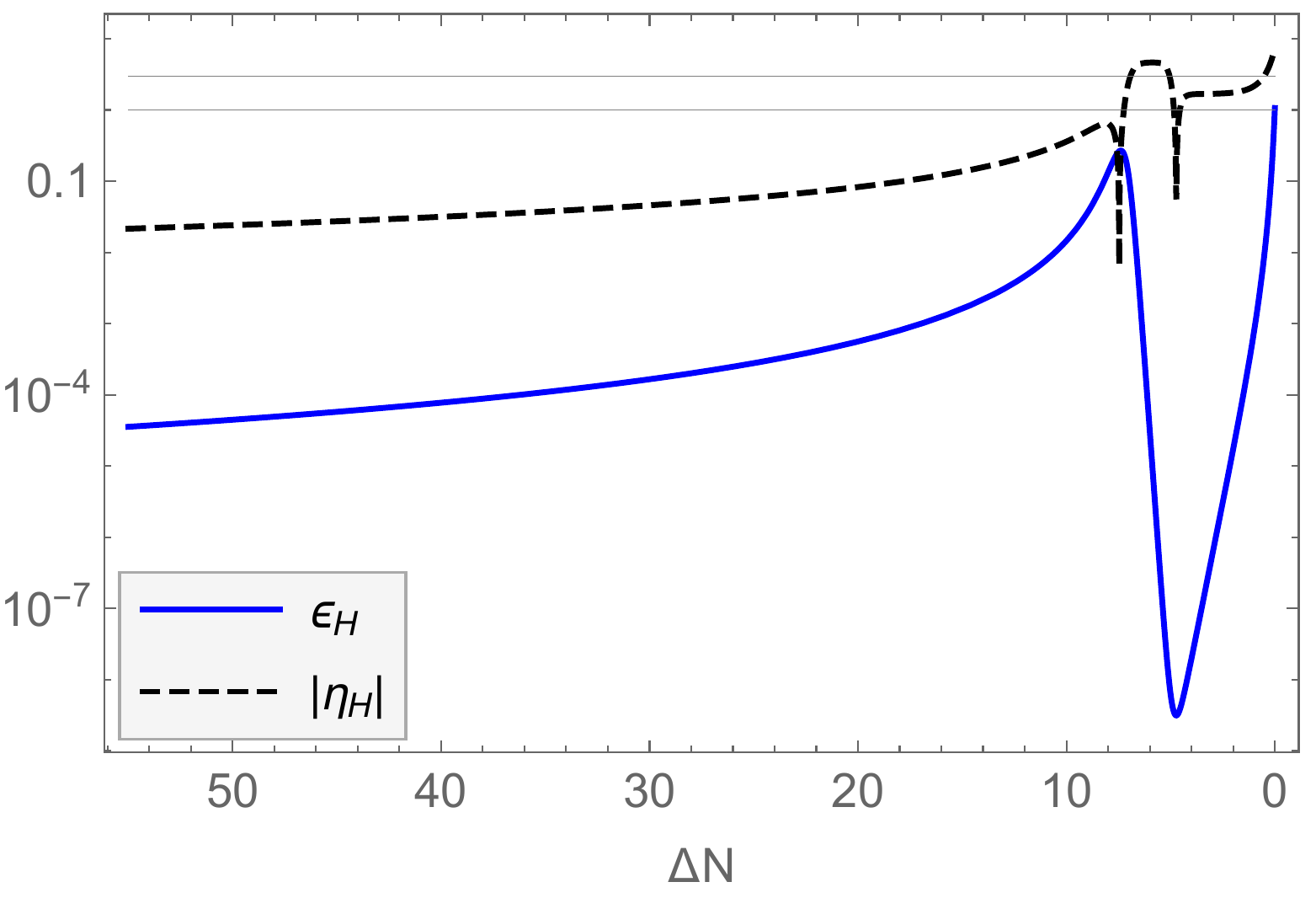}
  \end{subfigure}
 \caption{\textit{Left panel:} the inflaton evolution $\phi(N)$, for a single-field $\alpha$--attractor model with potential $V(\phi)$ given in \eqref{potential}, where we set $\alpha=0.1$, $\phi_\text{infl}=0.5$ and $\xi=0.0035108$. The black dot marks the field value when the CMB scale, $k_\text{CMB}=0.05\,\text{Mpc}^{-1}$, exits the horizon, taking $\Delta N_\text{CMB}=55$. \textit{Right panel:} the corresponding evolution of the first two Hubble slow-roll parameters $\epsilon_H$ and $|\eta_H|$. The two horizontal grey lines correspond to 1 and 3 respectively. Inflation ends when $\epsilon_H=1$ and $\eta_H\approx 3$ signals the ultra-slow-roll regime.}
  \label{fig:background evo}
\end{figure}

As an example, in figure~\ref{fig:background evo} the evolution of the scalar field, $\phi$, and the first two slow-roll parameters, $\epsilon_H$ and $\eta_H$, are displayed in terms of the number of e-folds to the end of inflation, $\Delta N\equiv N_\text{end}-N$, for the case of a single-field $\alpha$--attractor potential, eq.~\eqref{potential}, with $\alpha=0.1$ and an almost stationary inflection point, given by $\{\phi_\text{infl}=0.5,\, \xi=0.0035108\}$. 
The early evolution corresponds to a typical $\alpha$--attractor slow-roll phase with $\epsilon_H\ll |\eta_H| \ll 1$. The inflaton slows down as it approaches the inflection point and enters an ultra-slow-roll regime with $\epsilon_H$ small and rapidly decreasing, such that\footnote{In terms of the Hubble-flow parameter $\epsilon_2$, the ultra-slow-roll regime is described by $\epsilon_2\lesssim-6$. Given that $\epsilon_2=2\epsilon_H-2\eta_H$, the latter becomes $\eta_H\gtrsim3$ in the limit $\epsilon_H\ll|\eta_H|$~\cite{Dimopoulos:2017ged}.} $\eta_H \gtrsim 3$, almost coming to a stop momentarily. After it passes the potential barrier, caused by the local maximum of $V(\phi)$ following the inflection point at $\phi<\phi_\text{infl}$, the inflaton rolls towards the minimum of the potential at $\phi=0$ and inflation ends when $\epsilon_H=1$. 

 
\subsection{CMB constraints} 
\label{sec:connecting with observables}

When studying the phenomenology of an inflationary potential, it will be of key importance to calculate the number of e-folds before the end of inflation when the CMB scale, defined by the comoving wavenumber  $k_\text{CMB}=0.05\,\text{Mpc}^{-1}$, crossed the horizon ($k=aH$) during inflation~\cite{Planck:2018jri}
\begin{equation}
\begin{split}
\label{Nstar}
    \Delta N_\text{CMB} 
    &\equiv N_\text{end}-N_\text{CMB} \\
    &\simeq 67-\ln{\left(\frac{k_\text{CMB}}{a_0 H_0} \right)}+\frac{1}{4}\ln{\left(\frac{V_\text{CMB}^2}{ \rho_\text{end}} \right) }+\frac{1-3w}{12(1+w)}\ln{\left(\frac{\rho_\text{th}}{\rho_{\text{end}}} \right)}-\frac{1}{12}\ln{(g_\text{th})} \,.  
\end{split}
\end{equation}
In this expression $a_0H_0$
is the present comoving Hubble rate, $\rho_\text{end}$ is the energy density at the end of inflation, $V_\text{CMB}$ is the value of the potential when the comoving wavenumber $k_\text{CMB}$ crossed the horizon during inflation, $w$ is the equation of state parameter describing reheating, $\rho_\text{th}$ is the energy scale and $g_\text{th}$ is the number of effective bosonic degrees of freedom at the completion of reheating. Following~\cite{Planck:2018jri}, we fix $g_\text{th}=10^3$.

\begin{figure}
\centering
\includegraphics[scale=0.2]{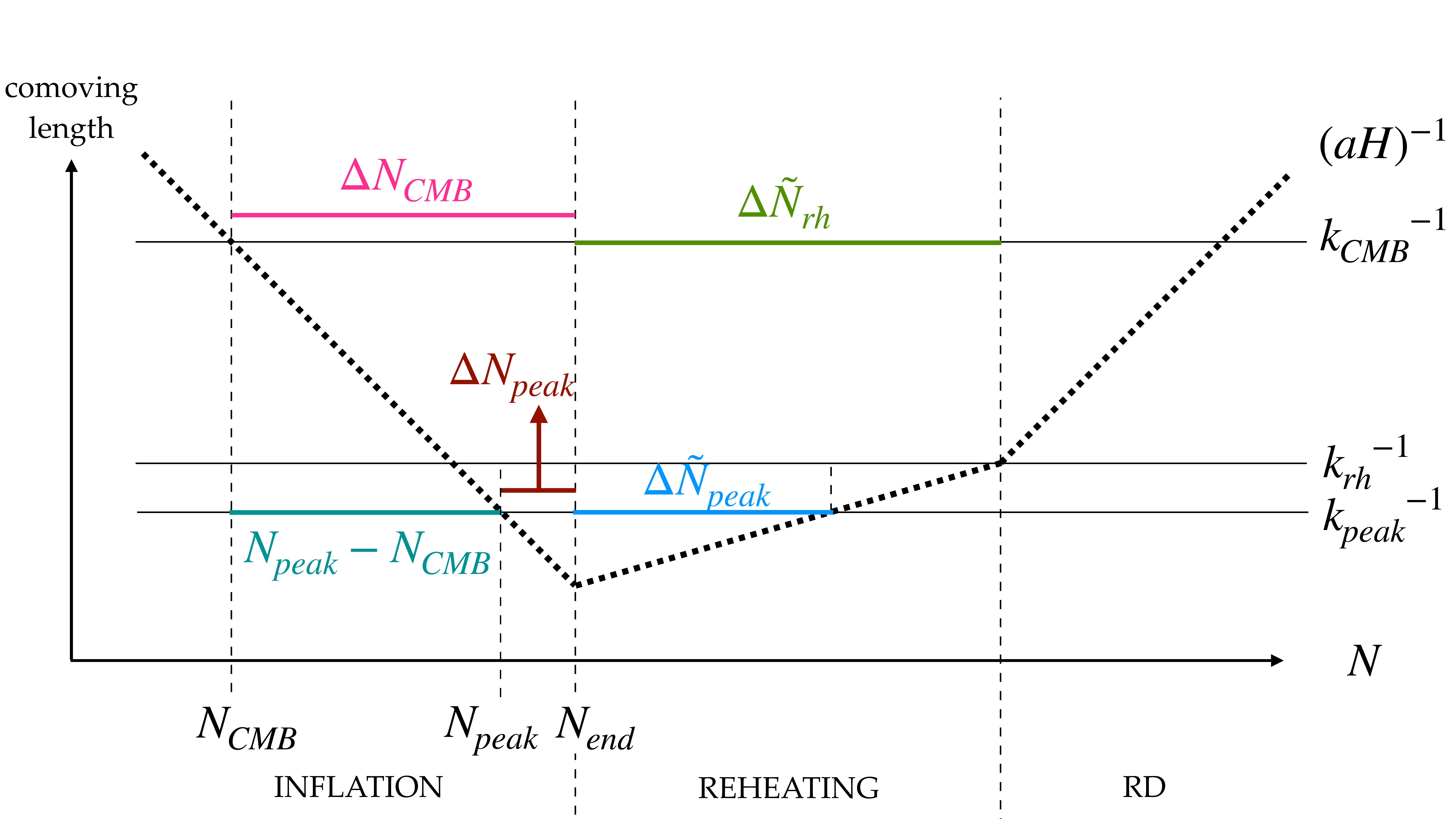}
\caption{Schematic representation showing the horizon crossing of modes with comoving wavenumber $k$ during and after inflation. We use the expression $\Delta N\equiv N_\text{end}-N$ when referring to e-folds elapsed during inflation, and $\Delta \tilde N\equiv N-N_\text{end}$ when referring to e-folds elapsed after inflation. RD stands for radiation domination. }
\label{fig:diagram scales}
\end{figure}

The precise value of $\Delta N_\text{CMB}$ depends on the inflationary potential and the details of reheating~\cite{Martin:2014nya}, as illustrated in figure~\ref{fig:diagram scales}. By assuming instant reheating, $\rho_\text{th}=\rho_\text{end}$, one can obtain the maximum value which $\Delta N_\text{CMB}$ can take (assuming the reheating equation of state $-1<w<1/3$). For $\alpha$--attractor potentials of the type considered here we obtain $\Delta N_\text{CMB, max}\approx 55$ by iteratively solving \eqref{Nstar} for values of $V_0$ compatible with CMB observations. In the following sections, \ref{sec: xi 0 case}--\ref{sec:change alpha}, we will present results assuming that reheating is instantaneous, bearing in mind that in order to describe a complete inflationary scenario it is necessary to include the details and duration of the reheating phase and understand how it impacts the predictions for observable quantities. We will address this topic in section~\ref{sec:reheating}.

Once $\Delta N_\text{CMB}$ is fixed, it is possible to derive the model's predictions for the CMB observables. 
In the slow-roll approximation the scalar power spectrum for primordial density perturbations can be given in terms of the Hubble rate, $H$, and first slow-roll parameter, $\epsilon_H$, evaluated when a given comoving scale, $k$, exits the horizon~\cite{ParticleDataGroup:2020ssz},
\begin{equation}
\label{slow roll power spectrum}
   P_\zeta(k) = \left. \frac{H^2}{8\pi^2\epsilon_H} \right|_{k=aH} \;.
\end{equation}
We parametrise the scalar power spectrum on large scales, which leads to temperature and polarisation anisotropies in the CMB, as~\cite{Planck:2018jri}
\begin{equation}
    P_\zeta(k)=\mathcal{A}_s\left(\frac{k}{k_\text{CMB}} \right)^{(n_s-1)+\frac{\alpha_s}{2} \text{log}(k/k_\text{CMB}) +\cdots} \,,
\end{equation}
where $n_s-1$ is the scalar spectral tilt at $k=k_\text{CMB}$, and $\alpha_s$ its running with scale. The amplitude of the power spectrum of primordial tensor perturbations, arising from quantum vacuum fluctuations of the free gravitational field, is given in terms of the tensor-to-scalar ratio, $r_\text{CMB}\equiv \mathcal{A}_t/\mathcal{A}_s$. 

$n_s$, $\alpha_s$ and $r_\text{CMB}$ can then be calculated in the slow-roll approximation in terms of the slow-roll parameters at horizon exit 
\begin{align}
\label{spectral index}
    n_s&=1-4\epsilon_H+2\eta_H \;, \\
\label{running of spectral index}
    \alpha_s&= -2\xi_H +10 \epsilon_H \eta_H -8\epsilon_H^2 \;, \\
\label{tensor to scalar ratio}
    r_\text{CMB}&=16\epsilon_H \;.
\end{align}
A full numerical computation of the scalar power spectrum (see appendix~\ref{sec:app numerical P_zeta}) confirms that the slow-roll approximation describes well the behaviour on large scales, i.e., far from the inflection point and the end of inflation. In the following we will therefore use eqs.~\eqref{spectral index}--\eqref{tensor to scalar ratio} to calculate the observables quantities relevant to CMB scales for single-field models. 

Model predictions can then be compared with the observational constraints from the latest \textit{Planck} data release~\cite{Planck:2018jri}. In particular, by fitting the \textit{Planck} temperature, polarisation and lensing, plus BICEP2/Keck Array BK15 data with the $\Lambda\text{CDM}+r_\text{CMB}+\alpha_s$ model, the constraints on the tilt, its running and the tensor-scalar ratio, are~\cite{Planck:2018jri} 
\begin{eqnarray}
    \label{bound on ns with alpha s}
    n_s = 0.9639 \pm 0.0044  \;\;\; &(68\,\%\,\text{C.L.})& \;, \\
    \label{bound on alpha s}
    \alpha_s = -0.0069\pm 0.0069\;\;\; &(68\,\%\,\text{C.L.})& \;, \\
    \label{r bound}
    r_{0.002} < 0.065 \;\;\;  &(95\,\%\,\text{C.L.})& \;.
\end{eqnarray}
Here we quote the tensor-to-scalar ratio, $r_{0.002}$, at $k=0.002\,\text{Mpc}^{-1}$, as using the \textit{Planck} plus BK15 data the tensor perturbations are best constrained at $k=0.002\,\text{Mpc}^{-1}$, while the scalar perturbations, and hence the scalar spectral index and its running, are best constrained at $k_\text{CMB}=0.05\,\text{Mpc}^{-1}$~\cite{Planck:2018jri}. 

For the $\alpha$--attractor potentials under consideration, we will show that $\alpha_s$ and $n_s$ are not independent parameters, but rather are related by eq.~\eqref{alphas ns}.  
In particular, the lower observational bound $n_s > 0.9551 \;\;(95\,\%\,\text{C.L.})$ from \eqref{bound on ns with alpha s} implies that $-0.001\lesssim\alpha_s<0$ at $95\,\%$ C.L., about an order of magnitude smaller than the observational uncertainty in eq.~\eqref{bound on alpha s}. 
For these reasons, we neglect the effect of the running when considering bounds on $n_s$ and $r_\text{CMB}$ in the following. We comment further on this topic in section~\ref{sec:a simple explanation}.

Using \textit{Planck}, WMAP and the latest BICEP/Keck data to constrain the tensor-to-scalar ratio at $k_\text{CMB}=0.05\,\text{Mpc}^{-1}$ in the absence of running (i.e., for the $\Lambda\text{CDM}+r_\text{CMB}$ cosmological model) yields the bound~\cite{BICEPKeck:2021gln}
\begin{equation}
\label{r bound new}
    r_\text{CMB}<0.036 \;\;\;  (95\,\%\,\text{C.L.})\;. 
\end{equation}
The predicted value of the tensor-to-scalar ratio changes by about $10\%$ if evaluated at $k_\text{CMB}=0.05\,\text{Mpc}^{-1}$ instead of $k=0.002\,\text{Mpc}^{-1}$. For $\alpha\leq1$, this is irrelevant as the predicted values of the tensor-to-scalar ratio in our model will be at least an order of magnitude below this observational bound. 

For the reasons outlined above, in the following we will impose observational bounds on the scalar spectral index at CMB scales using the baseline $\Lambda\text{CDM}$ cosmology, excluding both $\alpha_s$ and $r_\text{CMB}$. \textit{Planck} temperature, polarisation and lensing data, then require~\cite{Planck:2018jri}
\begin{equation}
\label{CMB 95percent ns bound}
    n_s=0.9649\pm0.0042 \;\;\; (68\, \% \, \text{C.L.}) \;.
\end{equation}  
In particular this  gives a lower bound on the spectral index
\begin{equation}
\label{CMB 95percent ns lower bound}
    n_s > 0.9565 \;\;\; (95\,\%\,\text{C.L.}) \;,
\end{equation}  
which provides the strongest constraint our models, and hence the small-scale phenomenology. 

\subsection[\texorpdfstring{$\xi=0$: stationary inflection point}{xi = 0: stationary inflection point}]{$\bm{\xi=0}$: stationary inflection point}
\label{sec: xi 0 case}
In the case of a stationary inflection point, the only free parameter 
specifying the shape of the function $f(r)$ in the simple cubic polynomial \eqref{f function} is the position of the inflection point $r_\text{infl}$. Along with the hyperbolic curvature parameter, $\alpha$, this then determines the field value at the inflection point, $\phi_\text{infl}$, in the potential, $V(\phi)$ in \eqref{potential}.

\begin{figure}
\centering
\captionsetup[subfigure]{justification=centering}
   \begin{subfigure}[b]{0.48\textwidth}
    \includegraphics[width=\textwidth]{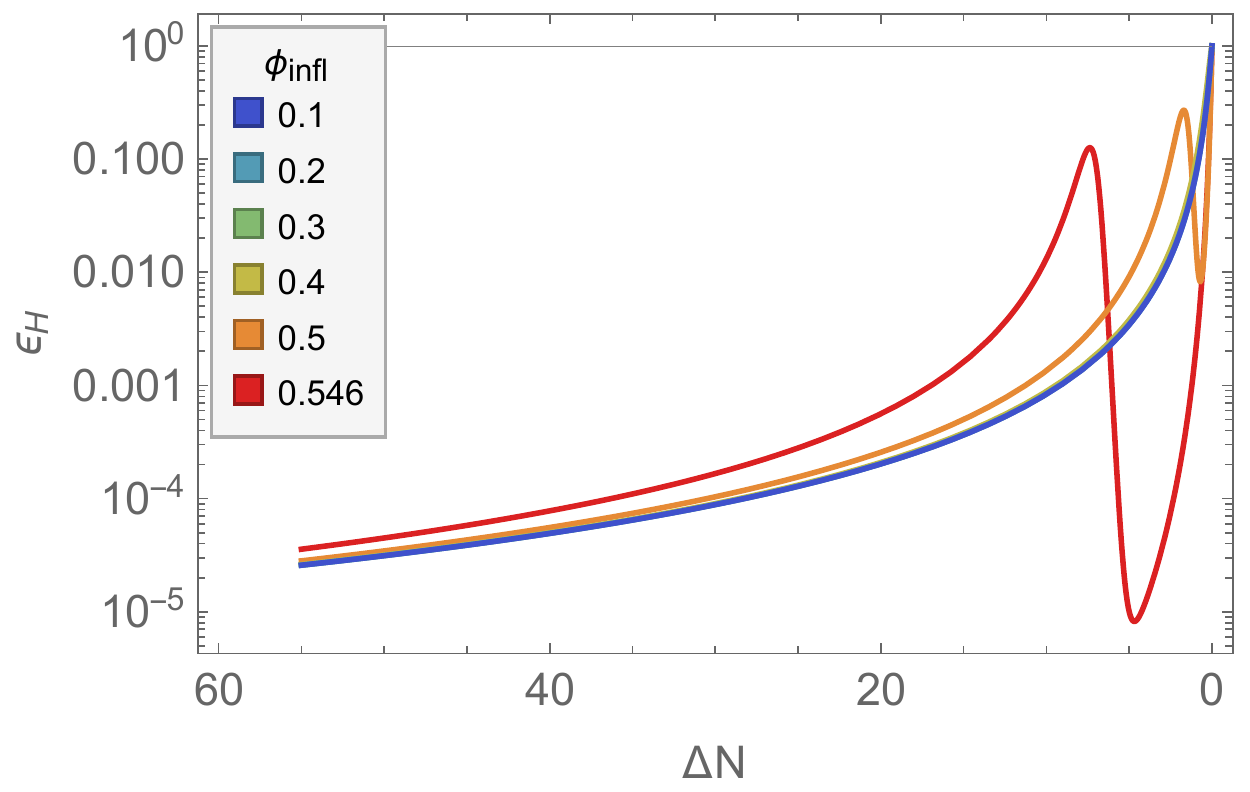}

  \end{subfigure}
  \begin{subfigure}[b]{0.46\textwidth}
    \includegraphics[width=\textwidth]{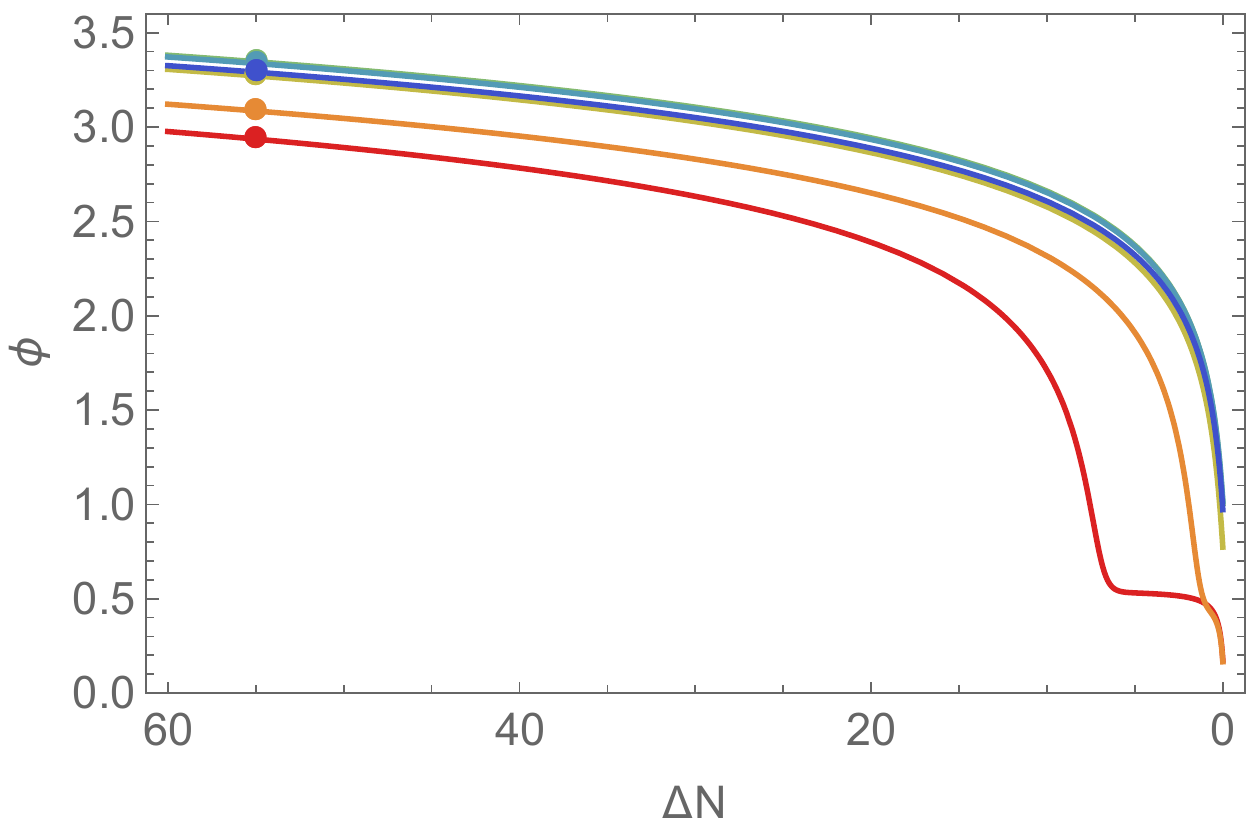}

  \end{subfigure}
 
 \caption{Background evolution of the first slow-roll parameter $\epsilon_H$ (left panel) and of the inflaton field (right panel) for the stationary inflection-point model, $\xi=0$. Different lines correspond to different locations of the inflection point $\phi_\text{infl}$, as displayed in the legend. In the right panel, the points represented on top of $\phi(N)$ signal the field value at which the CMB scale leave the horizon, $\phi_\text{CMB}$. All the configurations represented produce $\Delta N_\text{CMB}\simeq55$.}
  \label{fig:xi=0 background evo}
\end{figure}

For our fiducial value of $\alpha=0.1$,
we find that when $\phi_\text{infl}\geq0.56$ the inflaton, after a brief ultra-slow-roll phase, settles back down into
slow roll towards the inflection point and takes an infinite time to reach it. We therefore exclude that portion of the model's parameter space. We study configurations with $0.1\leq \phi_\text{infl}\leq 0.5465$ and plot the resulting background evolution in figure~\ref{fig:xi=0 background evo}. The limiting behaviour at large or small values of $\phi_\text{infl}$ are explored in appendix~\ref{sec: appendix limiting behaviour potential}.

Let us first discuss the configurations with $0.1\leq\phi_\text{infl}\leq 0.5$. When the inflection point is located at small field values, for $0.1\leq\phi_\text{infl}\leq 0.4$, inflation ends even before the inflaton reaches $\phi_\text{infl}$, making the background evolution effectively indistinguishable between those configurations. The case $\phi_\text{infl}=0.5$ is slightly different, as seen from the corresponding $\epsilon_H$ profile in the left panel of figure \ref{fig:xi=0 background evo}; the inflaton does slow down as it approaches the inflection point and its velocity drops, but only briefly before it passes through the inflection point. 

Using eqs.~\eqref{spectral index}--\eqref{tensor to scalar ratio} we find $0.961\lesssim n_s\lesssim0.963$, $\alpha_s\sim-0.0007$ and $r_{0.002}\sim 4\times 10^{-4}$, for $\{\xi=0,\, 0.1\leq\phi_\text{infl}\leq0.5\}$, showing that this parameter space is compatible with the CMB bounds given in \eqref{r bound new} and~\eqref{CMB 95percent ns bound}. However we find that larger values of $\phi_\text{infl}$, corresponding to a longer permanence of the inflaton around the inflection point (see the left panel of figure~\ref{fig:xi=0 background evo}), lead to smaller values for $n_s$, making the scalar power spectrum redder on CMB scales. This is due to the fact that the large scale CMB measurements test a steeper portion of the inflaton potential as a consequence of the permanence at the inflection point.
We will return this topic in more detail in section \ref{sec:a simple explanation} and give a simple explanation of the connection between the large-scale observations and the inflection-point location. 

\begin{figure}
\centering
\includegraphics[scale=0.6]{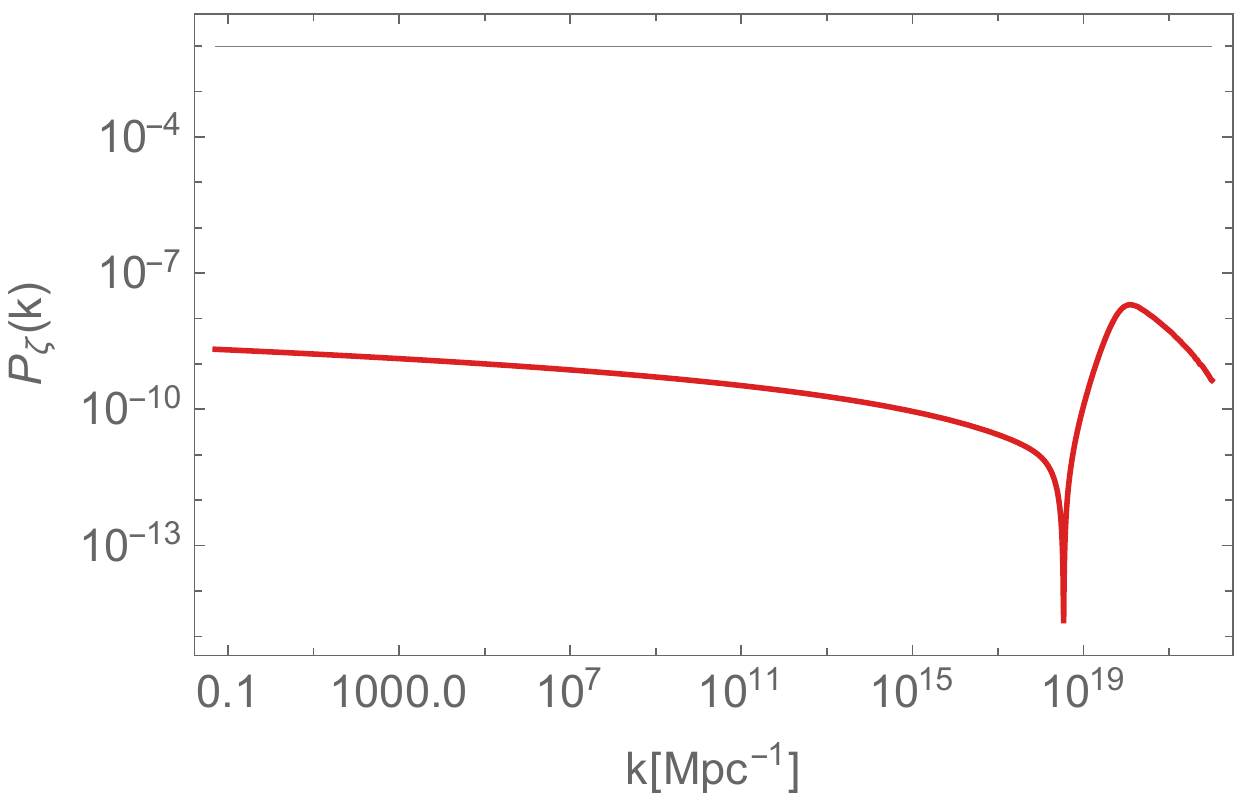}
\caption{Numerical scalar power spectrum for the single-field inflection-point model \eqref{potential} with parameters $\{\alpha=0.1, \, \phi_\text{infl}=0.5465, \, \xi=0\}$, plotted against the comoving scale $k$.}
\label{fig:xi=0 power spectrum limiting case}
\end{figure}

The largest value of $\phi_\text{infl}$ which we find is compatible with the lower limit of the observational bound on the scalar spectral tilt, eq.~\eqref{CMB 95percent ns lower bound}, is $\phi_\text{infl}=0.5465$. The corresponding background evolution is displayed in figure~\ref{fig:xi=0 background evo}. The inflection point does slow down the inflaton field, but without realising a sustained ultra-slow-roll phase. We therefore expect only a limited enhancement of the scalar fluctuations on small scales, which is confirmed by an exact computation of the scalar power spectrum (see appendix \ref{sec:app numerical P_zeta} for a detailed  description of the computation strategy). In figure \ref{fig:xi=0 power spectrum limiting case}, we display $P_\zeta(k)$ obtained numerically for this configuration. 
The power spectrum does exhibit a peak located at $k_\text{peak}=1.3\times10^{20}\,\text{Mpc}^{-1}$, whose amplitude is only one order of magnitude larger with respect to the large-scale power spectrum,  $P_\zeta(k_\text{peak})=2\times10^{-8}$. It is useful to characterise the position of the inflection point through the parametrisation
\begin{equation}
\label{Npeak parametrisation}
   \Delta N_\text{CMB}\equiv\left(N_\text{peak}-N_\text{CMB} \right)+\Delta N_\text{peak} \;,
\end{equation}
which implies that the number of e-folds elapsed between the horizon crossing of the CMB scale and the moment in which $k_\text{peak}$ left the horizon can be expressed as $\Delta N_\text{CMB}-\Delta N_\text{peak}$, see figure \ref{fig:diagram scales}. For the configuration plotted in figure~\ref{fig:xi=0 power spectrum limiting case} its value is $\Delta N_\text{CMB}-\Delta N_\text{peak} \simeq 49.5$.

Surveying the parameter space with $\xi=0$ shows that potentials with a stationary inflection point do not produce a large enhancement of the scalar fluctuations on small scales. In order for inflection-point $\alpha$--attractor models to display an interesting phenomenology on small scales, such as primordial black hole formation and/or significant production of gravitational waves induced at second order, it is necessary to turn to the approximate inflection-point case, $\xi\neq0$.

\subsection[\texorpdfstring{$\xi\neq0$: approximate stationary inflection point}{xi != 0: approximate stationary inflection point}]{$\bm{\xi\neq0}$: approximate stationary inflection point}
\label{sec: xi non 0 xase}

It is possible to obtain a large enhancement of the scalar power spectrum on small scales, $P_\zeta(k_\text{peak})\simeq10^{-2}$, necessary for PBH production after inflation, in simple cubic-polynomial $\alpha$--attractor models with $\xi\neq0$ in eq.~\eqref{potential}. 

\begin{figure}
\centering
\includegraphics[scale=0.5]{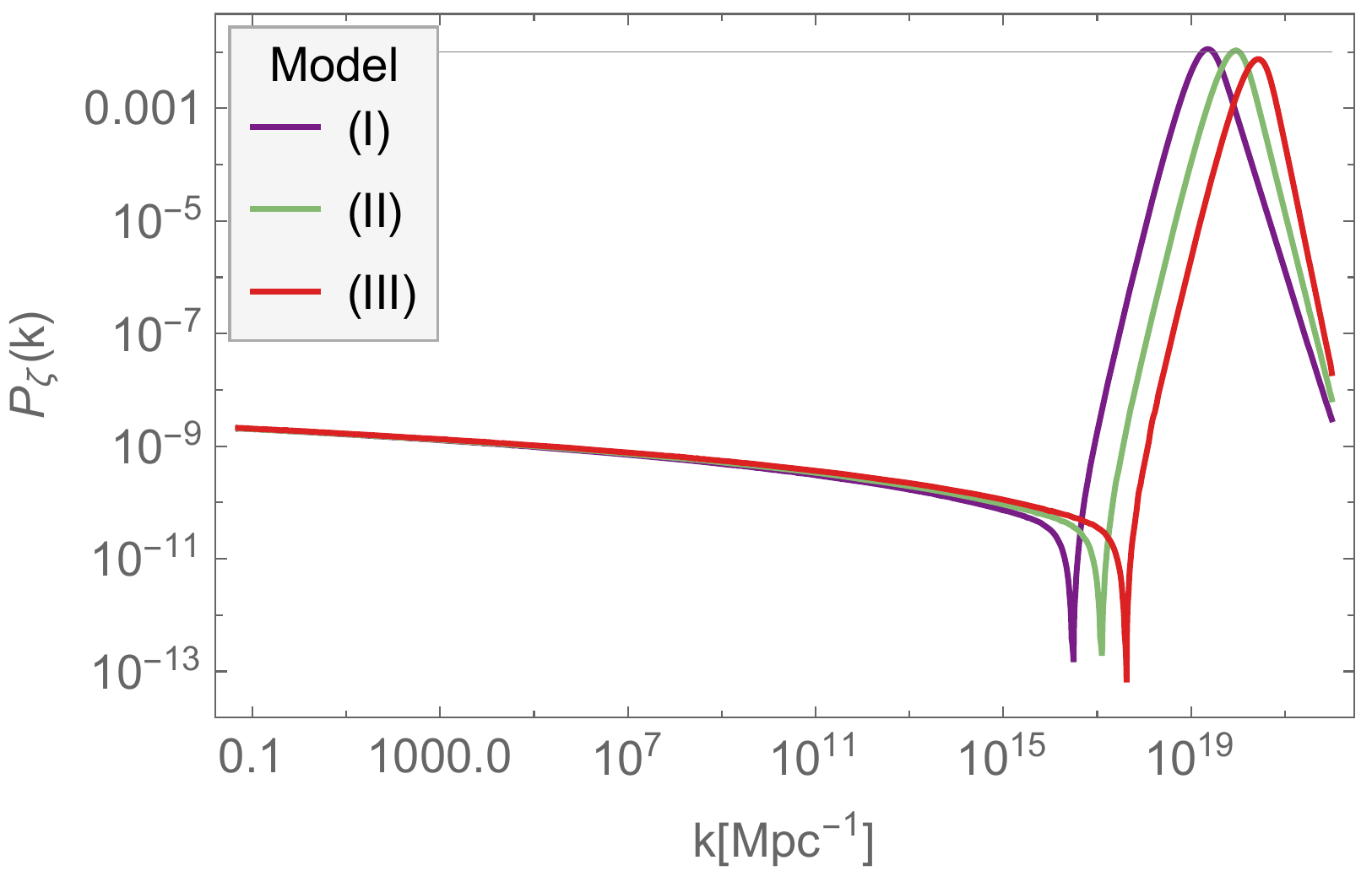}
\caption{Numerical results for the scalar power spectrum $P_\zeta(k)$ for three single-field models with $\alpha=0.1$ and $\xi\neq0$. The values of $\phi_\text{infl}$ and $\xi$ corresponding to each line are listed in table \ref{tab:configs xi non 0}.}
\label{fig:xi non 0 results}
\end{figure}

In table~\ref{tab:configs xi non 0} we display a selection of configurations for our fiducial curvature parameter of $\alpha=0.1$ which produce a peak $P_\zeta(k_\text{peak})\simeq10^{-2}$.
\begin{table}[]
  \centering
   \begin{tabular}{ |c||c|c|c|c|c|c|c|c|c|c|c| }
 \hline
&$\phi_\text{infl}$ & $\xi$  &$\Delta N_\text{CMB}-\Delta N_\text{peak}$ & $k_\text{peak}/\text{Mpc}^{-1}$& $n_s$ & $r_{0.002} $  \\

 \hline
(I) &0.51 & 0.0023495 & 47.8 & $2.2\times10^{19}$ &0.9555& $5.3\times10^{-4}$\\
   \hline
(II) &0.5 & 0.0035108 &  49.3& $9\times10^{19}$  & 0.9569& $4.9\times10^{-4}$\\
  \hline
(III) &0.49 & 0.0049575 & 50.4& $2.7\times10^{20}$ & 0.9579& $4.7\times10^{-4}$ \\
 \hline
\end{tabular}
    \caption{Details of three potentials with $\alpha=0.1$ and approximate stationary inflection points, $\xi\neq0$. The value $\Delta N_\text{CMB}-\Delta N_\text{peak}$ refers to the parametrisation \eqref{Npeak parametrisation}. All the potentials lead to inflation with $\Delta N_\text{CMB}\sim55$, $V_0\sim10^{-10}$ and $\alpha_s\sim-9\times10^{-4}$.}
    \label{tab:configs xi non 0}
\end{table}
We see that the field value at the inflection point, $\phi_\text{infl}$, determines both the location of the peak, $k_\text{peak}$, and the predicted value of the scalar spectral index, $n_s$, on CMB scales. The correspondence between $\phi_\text{infl}$ and $n_s$ holds regardless of the amplitude of the power spectrum peak. In particular, the larger $\phi_\text{infl}$, the smaller $k_\text{peak}$ and $n_s$, as we saw for the case $\xi=0$. For the configurations listed in table~\ref{tab:configs xi non 0}, the inflection-point field value is selected in order to have the power spectrum peak on the largest scale possible, with predicted values for the tilt $n_s$ around the CMB observational lower bound \eqref{CMB 95percent ns lower bound}. The parameter $\xi$ has then been adjusted to obtain $P_\zeta(k_\text{peak})\simeq10^{-2}$. Configuration (I) in table~\ref{tab:configs xi non 0} lies slightly outside the $95\,\%$ C.L. observational bound on $n_s$, while (II) and (III) are within the $95\,\%$ C.L. bound. In figure~\ref{fig:xi non 0 results} numerical results for the power spectra corresponding to these three configurations are displayed.

\subsection[\texorpdfstring{Changing ${\alpha}$}{Changing alpha}]{Changing $\bm{\alpha}$}
\label{sec:change alpha}
In the preceding sections the parameter space $\{\phi_\text{infl},\,\xi\}$ has been studied for a fixed fiducial value of the hyperbolic field-space curvature, corresponding to $\alpha=0.1$. In this section we consider the effect of varying $\alpha$. 

\begin{figure}
\centering
\includegraphics[scale=0.45]{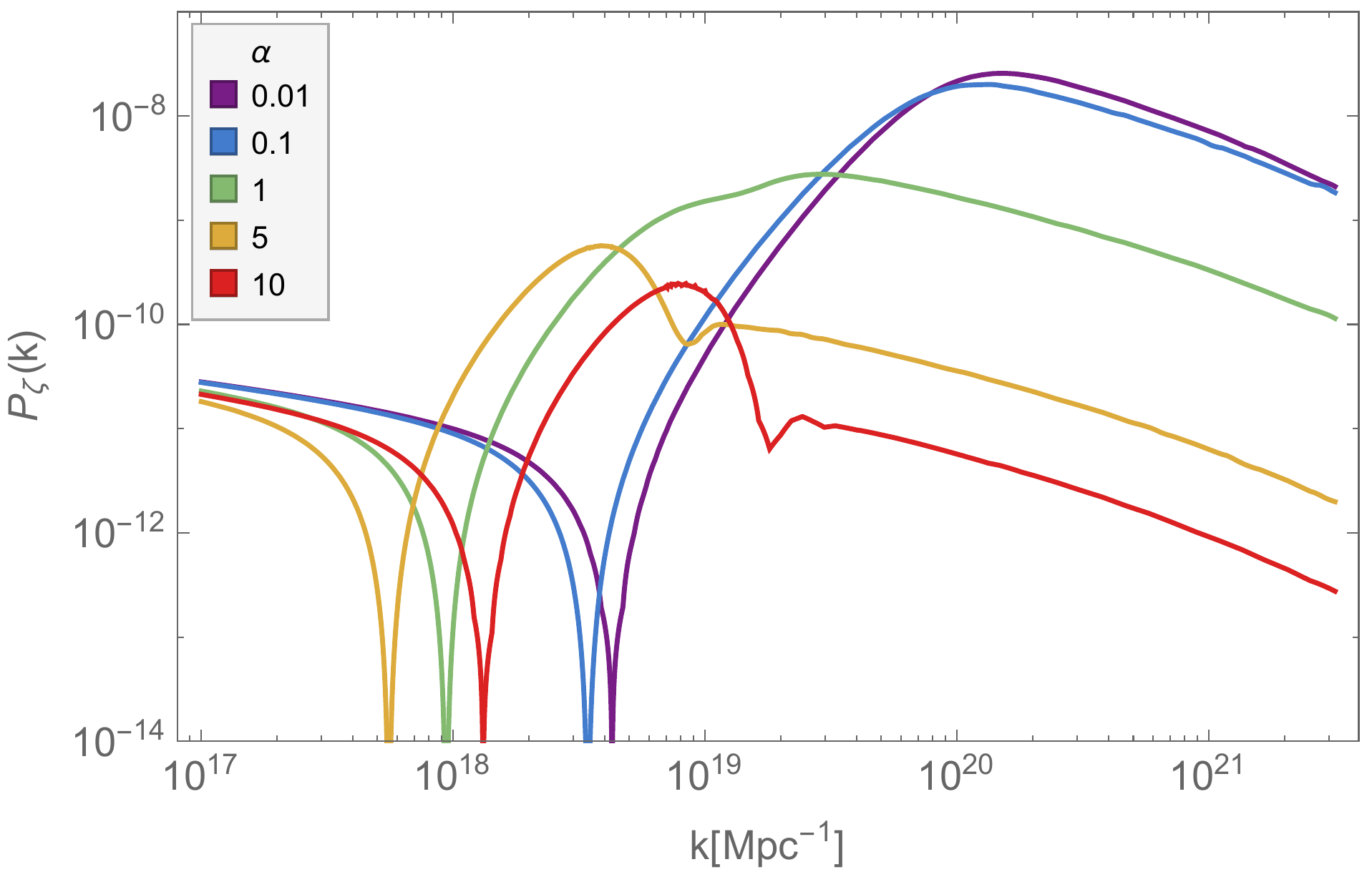}
\caption{Scalar power spectra obtained numerically for the single-field inflection-point models listed in table \ref{tab: change alpha}. Each line corresponds to a different choice of $\alpha$, as detailed in the legend.}
\label{fig:compare different alpha}
\end{figure}

We select five different values of $\alpha\in \{0.01,\,0.1,\,1,\, 5,\, 10\}$, and for simplicity restrict our attention to the case of a stationary inflection point, $\xi=0$. This 
avoids any numerical instabilities, possible when $\alpha>1$ due to fine-tuning of the inflection point when $\xi\neq0$. 
For each case, the value of $\phi_\text{infl}$ is chosen such that the predicted scalar spectral index, $n_s$, is close to the lower observational bound in \eqref{CMB 95percent ns lower bound}. The key parameters for each model are listed in table~\ref{tab: change alpha} and the numerically computed scalar power spectra are displayed in figure~\ref{fig:compare different alpha}. 
\begin{table}[]
  \centering
   \begin{tabular}{ |c||c|c|c|c|c|c|c|c|c|c|c| }
 \hline
$\alpha$ &$\phi_\text{infl}$ &$\Delta N_\text{CMB}$ &$\Delta N_\text{CMB}-\Delta N_\text{peak}$ & $k_\text{peak}/\text{Mpc}^{-1}$& $n_s$ & $r_{0.002} $  \\

\hline
0.01 & 0.255  & 54.3 & 49 & $10^{20}$ &0.9565& $5\times10^{-5}$\\
\hline
0.1 & 0.5465  & 55 & 49.5 & $1.3\times10^{20}$ &0.9565& $5\times10^{-4}$\\
 \hline
1 & 1.009  & 56.3 & 49.4 & $3\times10^{19}$ &0.9565& $4.9\times10^{-3}$\\
   \hline
5 & 1.313 & 57.6 & 49.3 & $4\times10^{18}$  & 0.9565 & $0.0217$\\
  \hline
10 & 1.39 & 58.3 & 48.3 & $8\times10^{18}$ & 0.9565& $0.0385$ \\
 \hline
\end{tabular}
    \caption{Table of parameters for each of the single-field inflection-point models used to generate the scalar power spectra shown in figure~\ref{fig:compare different alpha}.}
    \label{tab: change alpha}
\end{table}

The peak positions for $\alpha<1$ are very close to each other, while for larger $\alpha$ the peak moves, not following a specific trend and always on scales smaller than $10^{18}\,\text{Mpc}^{-1}$. The peak magnitudes vary depending on $\alpha$, whilst being fairly similar for $\alpha<1$. 

The potential normalisation, $V_0$, and hence the values of $r_{0.002}$ differ from each other by roughly one order of magnitude. This is as expected in $\alpha$--attractor models~\cite{Kallosh:2013hoa} where the universal predictions relate the level of primordial gravitational waves at CMB scales to $\alpha$, as shown in eq.~\eqref{r prediction alpha attractors}. Smaller $\alpha$ values are associated with a smaller predicted tensor-to-scalar ratio, as seen in table \ref{tab: change alpha}. Note that the predicted value of $r_{0.002}$ for $\alpha=10$ is in tension with the upper bound \eqref{r bound new}, hence we do not explore $\alpha>10$ (see also~\cite{Kallosh:2021mnu}). 

The fact that the results for $k_\text{peak}$, $P_\zeta(k_\text{peak})$ and $r_{0.002}$ are fairly consistent for small $\alpha$ is consistent with the expected $\alpha$--attractor behaviour. On the other hand the characteristic behaviour of $\alpha$--attractors, formulated on a hyperbolic field space, gets washed away for large $\alpha$, where these models approach the simple chaotic inflation behaviour~\cite{Kallosh:2015zsa}.

\subsection{Modified universal predictions} 
\label{sec:a simple explanation}

The numerical results that we have found for observables on CMB scales from single-field models including an inflection point suggest a simple modification of the $\alpha$--attractors universal predictions for $n_s$ and $r$ given in eqs.~\eqref{ns prediction alpha attractors} and~\eqref{r prediction alpha attractors}, as previously noted in~\cite{Dalianis:2018frf}.
In the presence of an inflection point at smaller field values (after CMB scales exit the horizon), the $\alpha$--attractors universal predictions  still hold if we replace $N_\text{end}$ with $N_\text{peak}$, and hence $\Delta N_\text{CMB}\rightarrow \Delta N_\text{CMB} - \Delta N_\text{peak}$, such that \eqref{ns prediction alpha attractors} and~\eqref{r prediction alpha attractors} are modified for $\Delta N_\text{peak}>0$ to become
\begin{gather}
\label{ns universal prediction inflection point}
  n_s \approx 1-\frac{2}{\Delta N_\text{CMB} - \Delta N_\text{peak}
  } \;, \\
\label{r universal prediction inflection point}
  r_\text{CMB} \approx 12 \frac{\alpha}{\left(\Delta N_\text{CMB} - \Delta N_\text{peak}
\right)^2} \;.
\end{gather}

\begin{figure}
\centering
\captionsetup[subfigure]{justification=centering}
   \begin{subfigure}[b]{0.48\textwidth}
    \includegraphics[width=\textwidth]{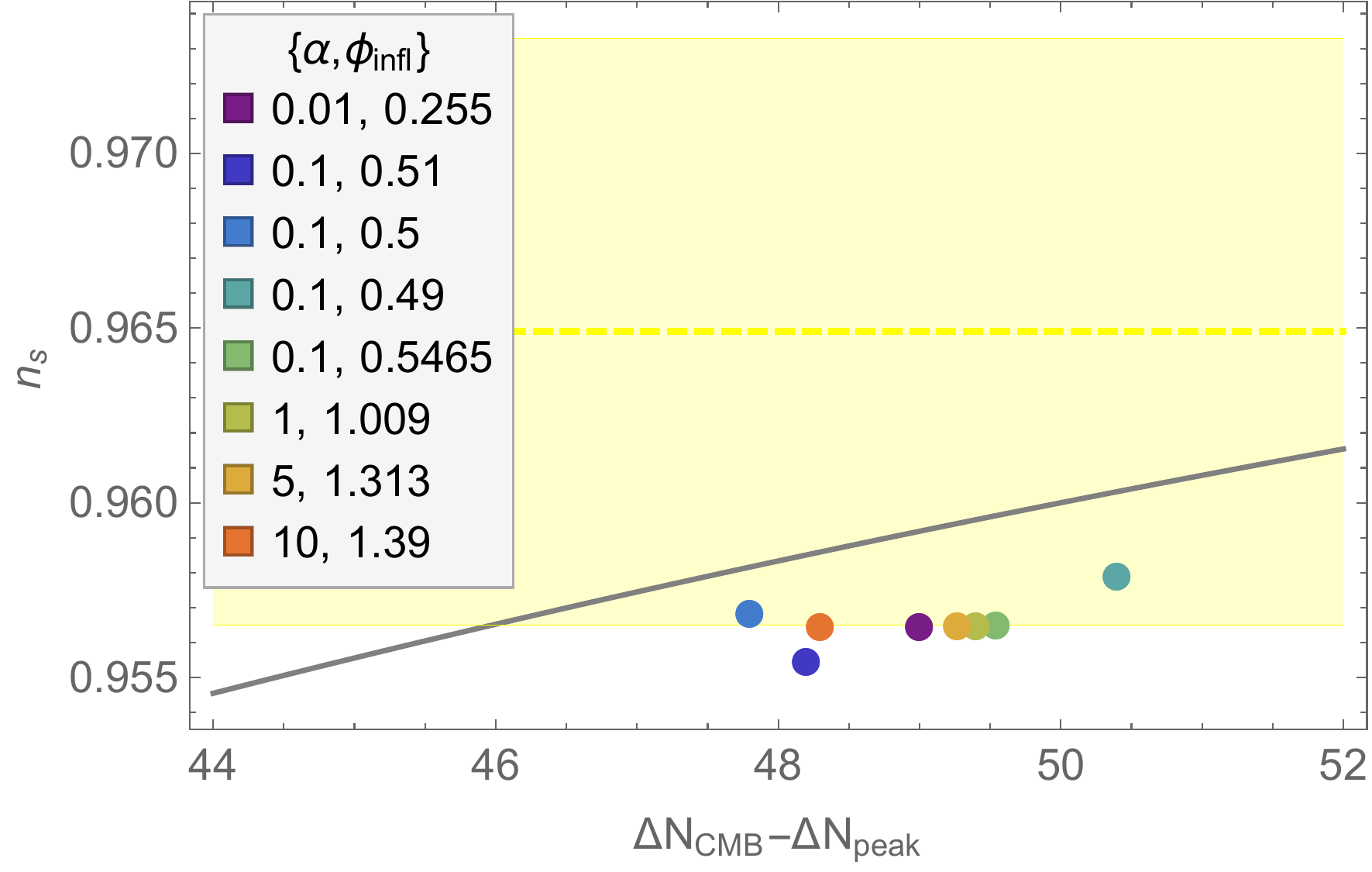}

  \end{subfigure}
  \begin{subfigure}[b]{0.48\textwidth}
    \includegraphics[width=\textwidth]{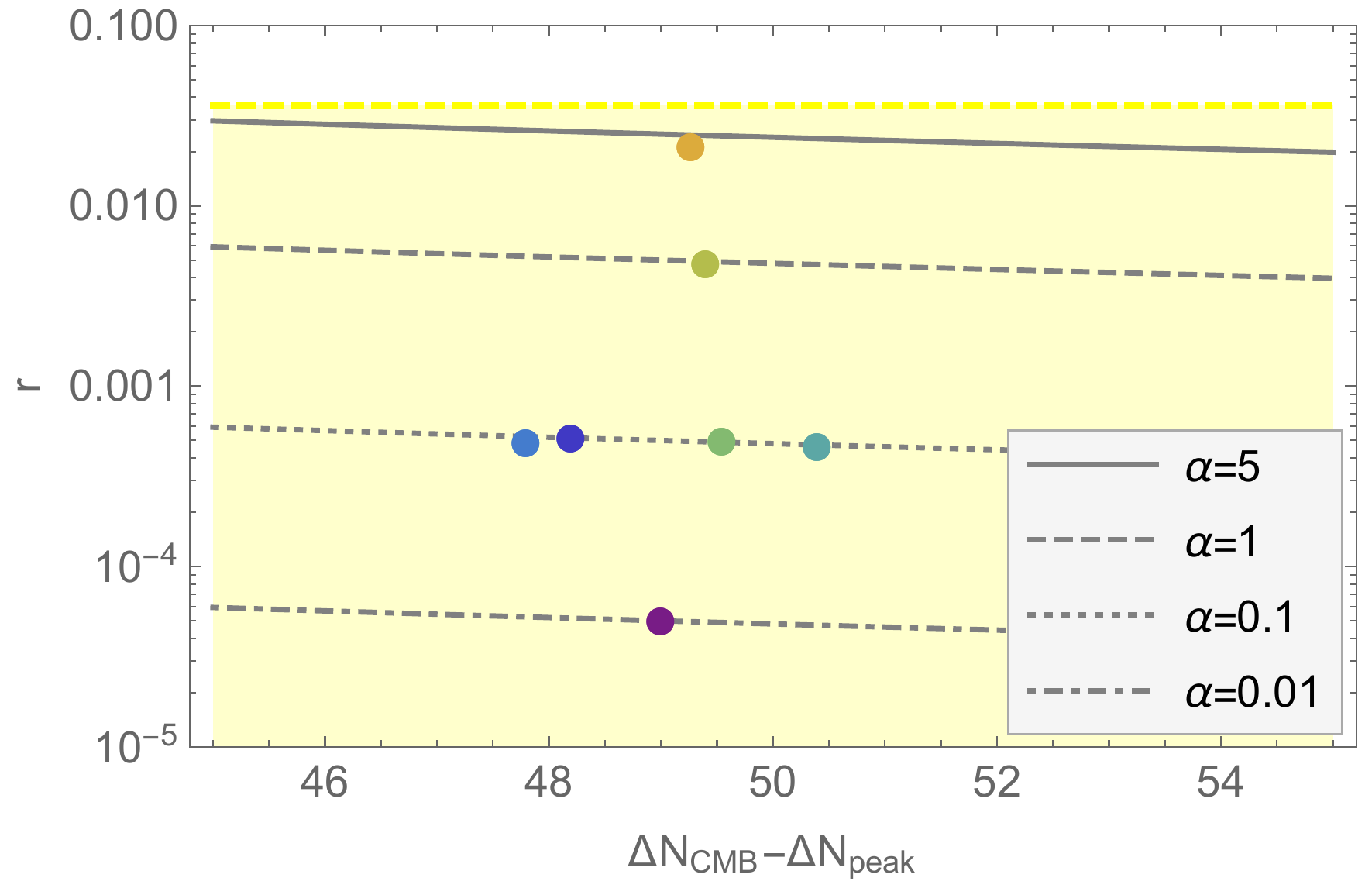}

  \end{subfigure}
 
 \caption{\textit{Left panel:} the approximation \eqref{ns universal prediction inflection point} (grey line) is plotted against numerical results (coloured points) for the scalar spectral index on CMB scale, $n_s$. Each point corresponds to a specific configurations discussed in sections \ref{sec: xi 0 case} and \ref{sec: xi non 0 xase}. The yellow-shaded area highlights the \textit{Planck} $95\,\%$ C.L. region, see \eqref{CMB 95percent ns bound}, with the dashed line representing the central value. \textit{Right panel:} the approximation \eqref{r universal prediction inflection point} is plotted against the numerical results for the tensor-to-scalar ratio, $r_\text{CMB}$.
 Each line corresponds to a different value of $\alpha$. See the left panel for the legend illustrating the coloured points. The yellow-dashed line signals the $95\,\%$ C.L. upper limit \eqref{r bound new}. We do not include the point corresponding to the model with $\alpha=10$, as the predicted value for $r_\text{CMB}$ puts the model in tension with the bound \eqref{r bound new}.}
  \label{fig:ns and r approx}
\end{figure}
In figure~\ref{fig:ns and r approx} we plot the approximations \eqref{ns universal prediction inflection point} and~\eqref{r universal prediction inflection point} together with our numerical results for a number of selected configurations which lie close to the lower bound on $n_s$. The coloured points are centered around values $47\lesssim \Delta N_\text{CMB} - \Delta N_\text{peak}\lesssim 51$ which, while being compatible with CMB measurements, produce a peak in $P_\zeta(k)$ on the largest scales possible. We see that the modified universal predictions describe quite well the numerical points, with a small offset observed in the left panel in figure~\ref{fig:ns and r approx}. We will investigate this further within the multi-field analysis in section \ref{sec:robustness of single-field predictions} and show a simple way of moving the numerical results even closer to the modified universal predictions.

In the following we will use eqs.~\eqref{ns universal prediction inflection point} and~\eqref{r universal prediction inflection point} to explore in a simple and straightforward way the phenomenology of the inflection-point potential \eqref{potential}. Rather than considering all the possibilities, we will focus on configurations that are consistent with the large-scale CMB observational constraints, eqs.~\eqref{r bound new} and~\eqref{CMB 95percent ns bound}. Using eq.~\eqref{ns universal prediction inflection point}, the observational bounds on $n_s$ given in \eqref{CMB 95percent ns bound} translate into
\begin{equation}
\label{Npeak bound}
46 \lesssim \Delta N_\text{CMB} - \Delta N_\text{peak}
\lesssim 75 \;.
\end{equation} 
A lower limit on $\Delta N_\text{CMB} - \Delta N_\text{peak}$ can also be obtained by substituting the upper bound on the tensor-to-scalar ratio \eqref{r bound new} in eq.~\eqref{r universal prediction inflection point}, but for $\alpha\leq1$ it is always weaker than the one given in eq.~\eqref{Npeak bound}. 
The lower bounds become comparable only when $\alpha\gtrsim10$.

During inflation there is a one-to-one correspondence between a scale $k$ and the number of e-folds, $N$, when that scale crosses the horizon, $k=aH$. Calibrating this relation using the values corresponding to the CMB scale yields
\begin{equation}
\label{scale relation}
    k(N)=\frac{a(N)}{a_\text{CMB}}\frac{H(N)}{H_\text{CMB}}\times 0.05\, \text{Mpc}^{-1} \;,
\end{equation}
where $a(N)/a_\text{CMB}=\text{e}^{N-N_\text{CMB}}$. For the scale corresponding to the peak in the scalar power spectrum eq.~\eqref{scale relation} is
\begin{equation}
\label{kpeak Npeak}
k_\text{peak}
\simeq \text{e}^{\Delta N_\text{CMB} - \Delta N_\text{peak}
} \times 0.05\, \text{Mpc}^{-1} \;,
\end{equation}
where we simplify the expression by assuming that the Hubble rate is almost constant during inflation. This equation shows that the largest scale, i.e., the lowest $k_\text{peak}$, corresponds to the lowest allowed value of $\Delta N_\text{CMB} - \Delta N_\text{peak}$. The lower limit in \eqref{Npeak bound} can therefore be used in eq.~\eqref{kpeak Npeak} to derive an estimate of the lowest scale $k_\text{peak}$ for configurations which are not in tension with the CMB observations,
\begin{equation}
\label{k peak bound}
    k_\text{peak}\gtrsim 4.7\times 10^{18}\, \text{Mpc}^{-1}\;,
\end{equation} 
which is valid regardless of the enhancement of the scalar power spectrum, $P_\zeta(k_\text{peak})$. This has important implications for the phenomenology of the model under analysis\footnote{For a counter example see, e.g., \cite{Mishra:2019pzq}, where a localised feature is superimposed on the original global potential.} and is confirmed by the results obtained numerically and presented in tables \ref{tab:configs xi non 0} and \ref{tab: change alpha}. 
\begin{figure}
\centering
\includegraphics[scale=0.45]{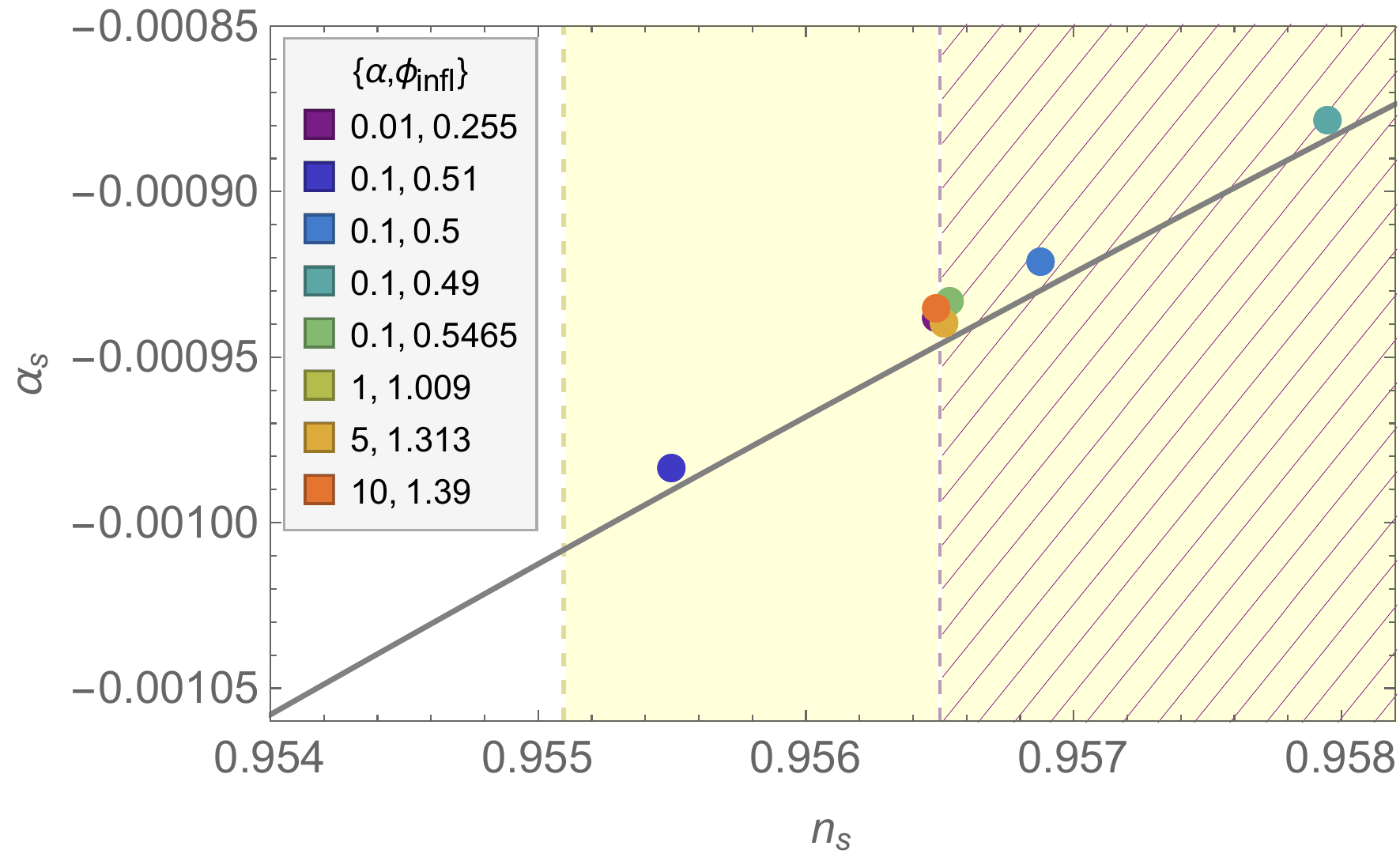}
 \caption{Points representing our numerical results for the spectral index and its running, $(n_s,\,\alpha_s)$. The consistency relation \eqref{alphas ns} is plotted as a solid-grey diagonal line. The yellow area represents part the $95\,\%$ C.L.\ region for $n_s$ when \textit{Planck} data are compared with the $\Lambda\text{CDM}+r_\text{CMB}+\alpha_s$ model,
 \eqref{bound on ns with alpha s}, and the hatch-shaded area to the right represents the $95\,\%$ C.L. region for $n_s$ for the $\Lambda\text{CDM}$ model neglecting running and $r_\text{CMB}$. The range of $\alpha_s$ shown is within the observational bound \eqref{bound on alpha s}.}
  \label{fig:alphas approx}
\end{figure}

Modifying the universal prediction for the running of the tilt, eq.~\eqref{alpha_s universal approximation}, with $\Delta N_\text{CMB}\rightarrow \Delta N_\text{CMB} - \Delta N_\text{peak}$ gives the approximation 
\begin{equation}
\label{alpha s unversal inflection point}
 \alpha_s \approx -\frac{2}{\left(\Delta N_\text{CMB} - \Delta N_\text{peak}
 \right)^2}  \;.
\end{equation}
The numerical results for $\alpha_s$ can be well-approximated by the expression above, with a small offset similar to that seen for $n_s$ in the left panel of figure~\ref{fig:ns and r approx}. We show in appendix~\ref{sec: appendix universality class} that in fact the values of $\alpha_s$ and $n_s$ are well-described the consistency relation 
\begin{equation}
\label{alphas ns}
    \alpha_s\approx -\frac{(n_s-1)^2}{2} \;.
\end{equation}
In figure~\ref{fig:alphas approx} we plot our numerical results for $(n_s,\,\alpha_s)$, and show that they are well-described by the consistency relation \eqref{alphas ns}. In particular, even if we allow for non-zero running, using the lower observational bound on $n_s$ given in eq.~\eqref{bound on ns with alpha s}, the consistency relation \eqref{alphas ns} implies that $\alpha_s>-1.01\times 10^{-3}$ at $95\,\%$ C.L., about an order of magnitude smaller than the observational uncertainty in eq.~\eqref{bound on alpha s}. This justifies what was already anticipated in section~\ref{sec:connecting with observables}, that we can in practice neglect the running when comparing the model predictions with CMB bounds on the tilt, $n_s$.
Thus in the following we will apply the more stringent lower bound on $n_s$, eq.~\eqref{CMB 95percent ns lower bound}, derived for the $\Lambda\text{CDM}$ model without running, in contrast to the approach taken in~\cite{Dalianis:2018frf}.

\section{Extended phenomenology of single-field models}
\label{sec: phenomenology of the single field model}

Building on the numerical results presented in section~\ref{sec: single field model}, we extend here our considerations to the phenomenology of inflection-point models on scales much smaller than those probed by the CMB. 
In section \ref{sec:reheating} we consider the implications of a reheating phase at the end of inflation. In sections \ref{Sec:PBH formation} and \ref{sec:2nd order GW} we review the formation of PBHs and the production of second-order GWs in presence of large scalar perturbations. 
Using the modified universal predictions appropriate for inflection-point models, we restrict our analysis to configurations of the inflection-point potential \eqref{potential} which are not in tension with the large-scale CMB measurements and explore the implications for the masses of the PBHs generated and the wavelengths of the second-order GWs.

\subsection{Reheating}
\label{sec:reheating}

Thus far we have worked under the assumption of instant reheating. Next we will take into account the presence of a reheating stage with finite duration.

At the end of inflation, the inflaton oscillates about the minimum of its potential and its kinetic energy becomes comparable with its potential energy. During this phase, the inflaton, and/or its decay products, must decay into Standard Model particles which rapidly thermalise. The process describing the energy transfer from the inflaton sector to ordinary matter goes by the name of \textit{reheating}. The energy density decreases from $\rho_\text{end}$, at the end of inflation, to $\rho_\text{th}$, when the Standard Model particles are thermalised. 

The duration of reheating measured in terms of e-folds, $\Delta \tilde N_\text{rh}\equiv N_\text{rh}-N_\text{end}$, depends on the effective equation of state parameter, $w$, and the value of $\rho_\text{th}$, and is given by 
\begin{equation}
\label{duration of reheating}
    \Delta \tilde N_\text{rh}\equiv \frac{1}{3(1+w)}\ln{\left(\frac{\rho_\text{end}}{\rho_\text{th}}\right)}  \;.
\end{equation}
We will consider a matter-dominated reheating phase $(w=0)$, as the inflaton behaves as non-relativisitic, pressureless matter when oscillating around a simple quadratic minimum of its 
potential, eq.~\eqref{potential}, for all configurations with $\xi\neq r_\text{infl}$. 

The exact duration of reheating 
depends on the efficiency of the energy transfer process. In order to be as general as possible, we estimate first the maximum duration of reheating and then, within the allowed range, consider the impact of a reheating phase on observable quantities. 

Requiring that reheating is complete before the onset of big bang nucleosynthesis bounds the value of $\rho_\text{th}$ from below. In particular, we follow~\cite{Planck:2018jri} and consider $\rho_\text{th}$ in the range $[(1\,\text{TeV})^4,\rho_\text{end}]$, where the upper limit corresponds to the case of instant reheating. Substituting the lower limit for $\rho_\text{th}$ into eq.~\eqref{duration of reheating} allows us to estimate the maximum duration of reheating as
\begin{equation}
\label{maximum reheating def}
  \Delta \tilde N_\text{rh}\leq\frac{1}{3}\ln{\left(\frac{\rho_\text{end}}{\left(1\,\text{TeV}\right)^4}\right)} \;.
\end{equation}
The inflection-point potential \eqref{potential} predicts $\rho_\text{end}\sim10^{-12}\,{M_\text{Pl}}^4$, with only a weak dependence on $\alpha$, which by means of eq.~\eqref{maximum reheating def} yields
\begin{equation}
\label{actual duration of reheating}
    0\leq\Delta \tilde N_\text{rh}\lesssim38 \;.
\end{equation}
It is instructive to isolate the reheating contribution to the value of $\Delta N_\text{CMB}$ given in eq.~\eqref{Nstar}. For example, for our $\alpha$--attractor models with $\alpha=0.1$ eq.~\eqref{Nstar} gives
\begin{equation}
\label{delta N CMB with reheating explicit}
\Delta N_\text{CMB}\simeq 55-\frac{1}{4}\Delta \tilde N_\text{rh}  \;. 
\end{equation}
Different values of $\Delta \tilde N_\text{rh}$, and hence $\Delta N_\text{CMB}$, can shift the observational predictions for a given inflationary model~\cite{Martin:2014nya}.
CMB constraints, combined with the standard universal predictions for $n_s$ and $r$ in $\alpha$--attractor models, eqs.~\eqref{ns prediction alpha attractors} and~\eqref{r prediction alpha attractors}, already have implications for the duration of reheating in these models. Substituting \eqref{delta N CMB with reheating explicit} in \eqref{ns prediction alpha attractors} and requiring that the duration of reheating does not put the model in tension with the CMB measurement \eqref{CMB 95percent ns bound}, yields the observational bound 
\begin{equation}
\label{new actual duration of reheating}
    0\leq \Delta \tilde N_\text{rh}\lesssim36 \;.
\end{equation}
This restricts the maximum duration of reheating allowed compared with the theoretical range given in eq.~\eqref{actual duration of reheating} and implies $\rho_\text{th}\gtrsim 
(4.5\,\text{TeV})^4$. 

If we now generalise this to include inflection-point $\alpha$--attractor models, giving rise to a peak in the power spectrum on small scales, $k_\text{peak}$ given in eq.~\eqref{kpeak Npeak}, then substituting eq.~\eqref{delta N CMB with reheating explicit} in eq.~\eqref{ns universal prediction inflection point} and imposing the bound on the CMB spectral index \eqref{CMB 95percent ns bound}, yields a stronger bound on the duration of reheating
\begin{equation}
\label{generalised duration of reheating}
    0\leq \Delta \tilde N_\text{rh}\lesssim36-4\Delta N_\text{peak} \;.
\end{equation}
This in turn puts a lower bound on the the thermal energy at the end of reheating
\begin{equation}
    \rho_\text{th}^{1/4}\gtrsim 4.5\,\text{TeV} \times e^{3\Delta N_\text{peak}} \,.
\end{equation}
In practice, 
eq.~\eqref{generalised duration of reheating} will determine the maximum range for the duration of reheating which we consider in the following. 

\subsection{Primordial Black Hole formation}
\label{Sec:PBH formation}
Very large amplitude scalar fluctuations produced during inflation give rise to large density perturbations when they re-enter the horizon after inflation, which can collapse to form primordial black holes~\cite{10.1093/mnras/168.2.399}. Such large fluctuations must be very rare, otherwise the resulting primordial black holes would come to dominate the energy density of the universe, spoiling the successful standard hot big bang cosmology, unless they are so light that they evaporate due to Hawking radiation before the epoch of primordial nucleosynthesis, corresponding to masses $M_\text{PBH}<10^9$~g~\cite{Hooper:2019gtx,Kozaczuk:2021wcl}.

The mass of the PBHs formed is related to the mass contained within the Hubble horizon at the time of formation~\cite{Sasaki:2018dmp}
\begin{equation}
\label{M PBH vs mass hubble horizon}
    M_\text{PBH}\equiv\gamma\, M_\text{H}=\gamma\, \frac{4\pi \rho}{3 H^3} \;,
\end{equation}
where $\rho$ is the energy density at the time of formation and $\gamma$ is a dimensionless coefficient describing the fraction of the Hubble horizon mass which collapses into the PBH. The parameter $\gamma$ depends on details of the gravitational collapse and for illustration we will use the benchmark value $\gamma=0.2$~\cite{1975ApJ...201....1C}. For simplicity, we will assume that there is a one-to-one correspondence between the mass of the PBH formed and the comoving scale of the scalar perturbations which produced it, $M(k)\equiv M_k$. In practice the spectrum of enhanced scalar perturbations will span a range of scales, and the process of critical collapse~\cite{Niemeyer:1997mt} will then lead to a spectrum of PBH masses~\cite{Kalaja:2019uju,Gow:2020bzo}. Nonetheless the Hubble mass \eqref{M PBH vs mass hubble horizon} provides an upper limit on the PBH masses formed.

The PBH formation process (in particular the masses and abundance) differs according to whether the scale corresponding to the peak in the scalar power spectrum re-enters the horizon ($k_\text{peak}=aH$) during reheating or during radiation domination after reheating. 
If $k_\text{peak}$ exits the horizon $\Delta N_\text{peak}$ e-folds before the end of inflation, it re-enters the horizon $\Delta \tilde N_\text{peak}$ e-folds after the end of inflation (see figure~\ref{fig:diagram scales}), where
\begin{equation}
\label{efolds reenter the horizon}
    \Delta \tilde N_{\text{peak}}=\frac{2}{(1+3w)}\Delta N_{\text{peak}} \;.
\end{equation}
In the expression above $w$ is the equation of state parameter describing the background evolution when $k_\text{peak}$ re-enters the horizon. Under the assumption of instant reheating ($\Delta \tilde N_\text{rh}=0$), $k_\text{peak}$ always re-enters the horizon during radiation domination ($w=1/3$), which from eq.~\eqref{efolds reenter the horizon} implies that $\Delta \tilde N_\text{peak}=\Delta N_\text{peak} $. If instead  $\Delta \tilde N_\text{rh}\neq 0$, then $k_\text{peak}$ re-enters the horizon during reheating if $\Delta \tilde N_\text{rh}>\Delta \tilde N_\text{peak}=2\Delta N_\text{peak}  $, where we take $w=0$ in eq.~\eqref{efolds reenter the horizon}.

The mass fraction at formation of PBHs with mass $M_k$ is given by 
\begin{equation}
\label{beta(M) at formation}
    \beta(M_k)\equiv \frac{\rho_\text{PBH}}{\rho_\text{tot}}\Big{|}_\text{at formation}
    \;.
\end{equation}
This is commonly estimated using the Press--Schechter formalism~\cite{1974ApJ...187..425P}, but we note that the peak theory approach~\cite{1986ApJ...304...15B, Young:2014ana, Young:2020xmk} can also be used. 
In the Press-Schechter approach the PBH abundance is determined by the probability that some coarse-grained random field, $\delta$, related to the comoving density perturbation (e.g., the compaction function~\cite{Shibata:1999zs,Musco:2018rwt}) exceeds some critical threshold value, $\delta\geq \delta_c$:
\begin{equation}
\label{beta(M) RD}
    \beta(M_k)
    =2\gamma \int_{\delta_\text{c}}^\infty \, \frac{\mathrm{d}\delta}{\sqrt{2\pi \,\sigma^2(M_k)}} \text{e}^{-\frac{1}{2}\frac{\delta^2}{\sigma^2(M_k)}}
    \;.
\end{equation}
The PBH mass fraction, $\beta(M_k)$, is exponentially sensitive to the variance of the coarse-grained density field, $\sigma^2(M_k)$, and thus to the peak of the primordial power spectrum on small scales~\cite{Dalianis:2018ymb,Sato-Polito:2019hws,Gow:2020bzo}. In eq.~\eqref{beta(M) RD} we assume that the probability distribution of the coarse-grained scalar perturbations, $\delta$, is well-described by a Gaussian distribution, while noting that the abundance of very large density fluctuations could be very sensitive to any non-Gaussian tail of the probability distribution function~\cite{Young:2013oia,Pattison:2017mbe,Ezquiaga:2019ftu,Biagetti:2021eep,Kitajima:2021fpq}.

\subsubsection{PBH formation during radiation domination}
\label{sec:PBH RD}

For modes that re-enter the horizon during the radiation-dominated era after reheating, eq.~\eqref{M PBH vs mass hubble horizon}, assuming conservation of entropy between the epoch of black hole formation and matter-radiation equality, yields~\cite{Wang:2019kaf} 
\begin{equation}
\label{Mofk RD}
    \frac{M(k)}{M_{\odot}}\simeq 
    10^{-16} \left(\frac{\gamma}{0.2} \right) \left(\frac{g(T_k)}{106.75} \right)^{-1/6} \left(\frac{k}{10^{14}\,\text{Mpc}^{-1}} \right)^{-2} \;,
\end{equation}
where $g(T_k)$ is the effective number of degrees of freedom at the time of formation. 
Assuming the Standard Model particle content, we take $g(T_k)=106.75$ and $g(T_\text{eq})=3.38$.

If we consider the non-stationary inflection-point models presented in section~\ref{sec: xi non 0 xase} where we calculated the CMB constraints assuming instant reheating, the PBHs are formed from the collapse of large scalar fluctuations at $k=k_\text{peak}$ which re-enter the horizon during radiation domination. Substituting the numerical values of $k_\text{peak}$ listed in table \ref{tab:configs xi non 0} in eq.~\eqref{Mofk RD} leads to PBH masses $M_\text{PBH}/$g~$\simeq4.2\times 10^{6},\, 2.6\times 10^{5},\, 2.8\times 10^{4}$ for configurations (I), (II) and (III) respectively. 
Thus PBHs resulting from these inflection-point $\alpha$--attractor models would have evaporated before primordial nucleosynthesis.

In section \ref{sec: effect of reheating} we will argue that this is a general result which applies to all $\alpha$--attractor inflection-point models which are not in tension with the CMB measurements on large scales and extends beyond the instant reheating assumption. In particular, the black-dashed line in figure~\ref{fig:PBH masses} shows the range of PBH masses formed when the peak of the power spectrum on small comoving scales re-enters the horizon during the radiation-dominated era after reheating, over the range \eqref{k peak bound} consistent with CMB constraints on large scales.


\subsubsection{PBH formation during matter domination} \label{sec:PBH MD}
As discussed above, it is possible that large scalar perturbations which collapse to form PBHs re-enter the horizon during reheating, corresponding to a transient matter-dominated stage after inflation. The different background evolution during reheating modifies PBH formation; intuitively the collapse is easier in a matter-dominated epoch than in the presence of radiation pressure. 
Another consequence is that the correspondence between the scale of the perturbation that collapses to form the PBH and its mass is modified. In particular, following a procedure similar to the one illustrated for eq.~\eqref{Mofk RD} and taking into account the different background evolution yields~\cite{Dalianis:2018frf}
\begin{equation}
\label{Mofk MD}
    \frac{M(k)}{M_{\odot}}\simeq
    10^{-16} \left(\frac{\gamma}{0.2}\right) \left(\frac{g(T_\text{rh})}{106.75} \right)^{-1/6} \left(\frac{k_\text{rh}}{10^{14}\,\text{Mpc}^{-1}} \right)^{-2} \; \left(\frac{k}{k_\text{rh}} \right)^{-3} \;,
\end{equation}
where the scale 
\begin{equation}
\label{krh}
k_\text{rh}= e^{-3\Delta \tilde N_\text{rh}/4}\times3.8\times10^{22} \,\text{Mpc}^{-1}
\end{equation}
re-enters the horizon at the end of reheating. For perturbations that re-enter the horizon during reheating we have $k>k_\text{rh}$, as sketched in figure \ref{fig:diagram scales}. 
The coloured diagonal lines in figure~\ref{fig:PBH masses} show the range of PBH masses formed when the peak of the power spectrum on small comoving scales re-enters the horizon during reheating for models which are in accordance with CMB constraints on large scales.

\subsubsection{Implications of reheating and modified universal predictions for PBH formation}
\label{sec: effect of reheating}

In the following we examine the implications for the allowed PBHs masses of the modified universal predictions presented in section~\ref{sec:a simple explanation} and the resulting constraints from CMB measurements of the spectral tilt on large scales. 
We consider inflection-point potentials \eqref{potential} with parameters,  $\{\alpha,\,\phi_\text{infl},\,\xi\} $, which generate significant enhancements of the scalar power spectrum on small scales, as we have done for the specific cases discussed in section \ref{sec: single field model}. We take into account the fact that inflation could be followed by a reheating stage, whose duration is bounded by \eqref{generalised duration of reheating} for $\alpha=0.1$. We discuss the effect of varying $\alpha$ at the end of this section.

As already discussed, it is the hierarchy between $k_\text{peak}$ and $k_\text{rh}$ in the presence of reheating that determines the setting for PBH formation, during either radiation or matter domination. Equivalently one can consider the hierarchy between $\Delta N_\text{peak}$ and $\Delta N_\text{rh}= \Delta \tilde N_\text{rh}/2$.
The bound \eqref{generalised duration of reheating} can be written as
\begin{equation}
\label{Npeak bound with reheating}
\Delta N_\text{rh}+2\Delta N_\text{peak} \lesssim 18 
\;. 
\end{equation}
and we discuss here the implications of the expression above for the mass of the PBHs formed within three different scenarios.

\begin{figure}
\centering
\includegraphics[scale=0.45]{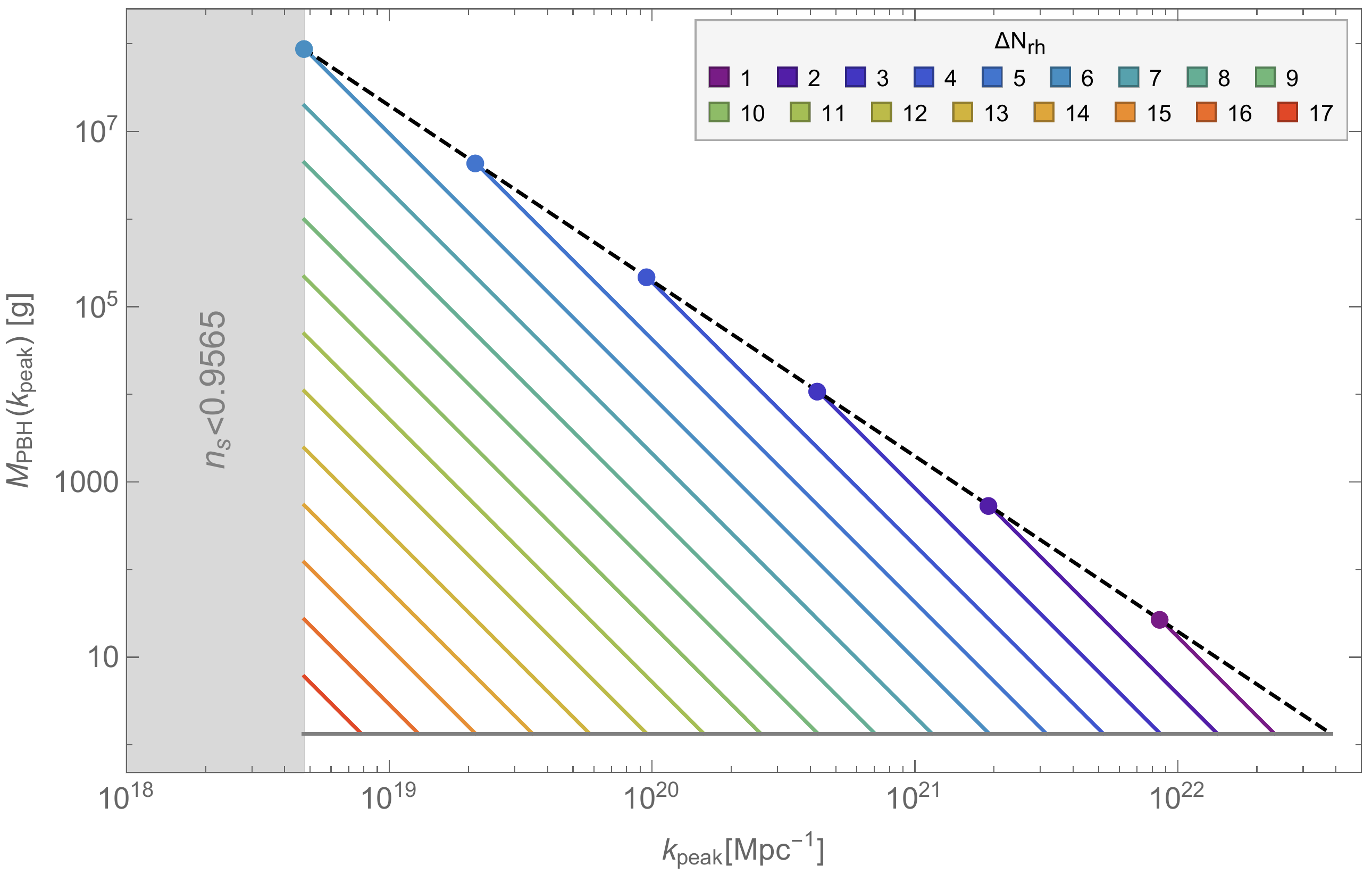}
\caption{Masses of PBHs generated during or after reheating as a function of $k_\text{peak}$ for models with $\alpha=0.1$. Diagonal coloured lines correspond to PBHs produced by modes re-entering the horizon during a period of reheating ($w=0$), where each coloured line corresponds to a given duration of reheating, $\Delta N_\text{rh}$. The region on the left highlighted in grey is excluded by the $95\,\%$ C.L. lower bound on $n_s$, eq.~\eqref{CMB 95percent ns lower bound}. The black-dashed line corresponds to PBHs produced during radiation domination. The lower horizontal grey line corresponds to scales that re-enter the horizon at the start of reheating, immediately after the end of inflation.}
\label{fig:PBH masses}
\end{figure}

\medskip \textbf{(i) Instantaneous reheating ($\bm{\Delta N_\text{rh}=0}$):} In this case $k_\text{peak}$ always re-enters the horizon during radiation domination and it is bounded by \eqref{k peak bound}. 
In figure \ref{fig:PBH masses} the black-dashed line represents $M_\text{PBH}$ against $k_\text{peak}$ over the range $4.7\times10^{18}\,\text{Mpc}^{-1}<k_\text{peak}<k_\text{end}$, compatible with \eqref{k peak bound}, where $k_\text{end}\simeq4\times10^{22}\,\text{Mpc}^{-1}$ for models with $\alpha=0.1$ and instantaneous reheating. The modified universal predictions therefore imply that the mass is maximised for the smallest $k_\text{peak}$ and in general 
\begin{equation}
    M_\text{PBH}< 10^{8}\,\text{g} \;,
\end{equation}
which means that PBHs produced in this case have evaporated before primordial nucleosynthesis and are not a candidate for dark matter. Explicit realisations of this scenario have been discussed in section \ref{sec:PBH RD}.

\medskip \textbf{(ii) PBH formation after reheating is complete ($\bm{\Delta N_\text{peak} > \Delta N_\text{rh}}$):} In this case the PBHs form during radiation domination. The requirement that scales $k_\text{peak}$ re-enter the horizon after reheating
together with \eqref{Npeak bound with reheating} implies that
\begin{equation}
    0<\Delta N_\text{rh} <6 
    \quad \text{and} \quad
    \Delta N_\text{rh}<\Delta N_\text{peak}\lesssim9-\frac{1}{2} \Delta N_\text{rh} \;.
\end{equation}
For fixed $\Delta N_\text{rh}$, using \eqref{scale relation} in the expression above gives a range of possible scales 
\begin{equation}
  4.7 \times 10^{18} \,\text{Mpc}^{-1}\lesssim k_\text{peak}< k_\text{rh}
\;,
\end{equation}
where the reheating scale is given by eq.~\eqref{krh}.

The mass of the PBHs formed is still set by \eqref{Mofk RD}, corresponding to the black-dashed line in figure~\ref{fig:PBH masses} for $M_\text{PBH}(k_\text{peak})$, but in contrast to the case of instant reheating, $k_\text{peak}$ can now only run up to $k_\text{rh}$. This means that only part of the black-dashed line in figure~\ref{fig:PBH masses} for $M_\text{PBH}(k_\text{peak})$ is accessible for a given value of $\Delta N_\text{rh}$. In particular, the coloured points on the black-dashed line signal the largest allowed value of $k_\text{peak}=k_\text{rh}$ for a fixed $\Delta N_\text{rh}<6$. In this case the largest PBH mass produced is again $M_\text{PBH}\sim 10^{8}\,\text{g}$ and it corresponds to $k_\text{peak}=4.7\times 10^{18}\, \text{Mpc}^{-1}$. 

\medskip \textbf{(iii) PBH formation during reheating ($\bm{\Delta N_\text{peak} \leq \Delta N_\text{rh}}$): } 
\label{sec:PBHdom}
In this case the PBHs form before reheating is complete, i.e., during a matter-dominated era. This implies a hierachy, $k_\text{peak}\geq k_\text{rh}$, which together with \eqref{Npeak bound with reheating} results in either
\begin{equation}
   0<\Delta N_\text{peak}\leq \Delta N_\text{rh}\leq6 \;,
\end{equation}
or
\begin{equation}
    6<\Delta N_\text{rh}<18 
    \quad \text{and} \quad
    0<\Delta N_\text{peak}\leq 9-\frac{1}{2} \Delta N_\text{rh} \;. 
\end{equation}
For a given value of $\Delta N_\text{rh}$ and hence a given value of $k_\text{rh}$, see eq.~\eqref{krh}, we have
\begin{align}
    \label{k peak range PBH during reheating 1}
    k_\text{rh} \leq k_\text{peak}< 
    e^{\Delta N_\text{rh}} k_\text{rh}
    \quad \text{if}\quad   
    0<\Delta N_\text{rh}\leq6 
\;, \\
    \label{k peak range PBH during reheating 2}
    4.7 \times 10^{18} \,\text{Mpc}^{-1}\leq 
    k_\text{peak}< 
    e^{\Delta N_\text{rh}} k_\text{rh}
    \quad \text{if}\quad   
    6<\Delta N_\text{rh}<18\;.
\end{align}

The masses of the PBHs produced is set by \eqref{Mofk MD} and it is shown as a function of $k_\text{peak}$ in figure~\ref{fig:PBH masses}.
For a given $k_\text{peak}$, the masses produced during a matter-dominated ($w=0$) reheating stage are all below the corresponding masses produced during radiation domination, because $k_\text{peak}$ re-enters the horizon before the onset of radiation domination and this suppresses the PBH mass by a factor $\left(k_\text{rh}/k_\text{peak} \right)^3$, see eq.~\eqref{Mofk MD}. The PBH masses approach those generated in radiation domination in the limit $\Delta N_\text{peak}\to\Delta N_\text{rh}$. In this case $k_\text{peak}\to k_\text{rh}$ and therefore the formula \eqref{Mofk MD} coincides with \eqref{Mofk RD}. The cases representing $\Delta N_\text{peak}=\Delta N_\text{rh}$ are plotted in figure \ref{fig:PBH masses} with the coloured points, which mark the intersection between the coloured lines and the black-dashed line. The case $\Delta N_\text{rh}=\Delta N_\text{peak}=6$ maximises the PBH mass which could be produced in this scenario, $M_\text{PBH}\sim 10^{8}\,\text{g}$. 

For any duration of reheating, $\Delta N_\text{rh}$, substituting the upper value $k_\text{peak}=e^{\Delta N_\text{rh}} k_\text{rh}$ in \eqref{Mofk MD} 
results in a PBH mass independent of $\Delta N_\text{rh}$, which justifies why all the coloured lines lie above the horizontal grey line in figure \ref{fig:PBH masses} corresponding to $M_\text{PBH}\sim 1$~g. 

The right vertex of allowed values in figure \ref{fig:PBH masses} corresponds to the case $\Delta N_\text{rh}=0$ and $k_\text{peak}=k_\text{end}$. This is the limiting case where the peak is produced at the very end of inflation. While it may be possible to have configurations which produce a peak a few e-folds before the end of inflation, the limited growth of the scalar power spectrum in single-field models~\cite{Byrnes:2018txb} would not allow the 7 orders of magnitude enhancement with respect to the CMB scales which is necessary for significant production of PBHs. 

The analysis above is performed for our fiducial value $\alpha=0.1$. The parameter $\alpha$ sets the maximum value of $\Delta N_\text{CMB}$ (corresponding to $\Delta N_\text{rh}=0$) as illustrated in table \ref{tab: change alpha}. Thus the expression \eqref{delta N CMB with reheating explicit} gets modified for different $\alpha$, which in turns changes the scales involved, see eq.~\eqref{scale relation}. In particular, the lower bound on the PBH mass that can be produced during reheating corresponds to $k_\text{peak}=\text{e}^{\Delta N_\text{CMB, max}-\frac{1}{2}\Delta N_\text{rh}}\times 0.05\,\text{Mpc}^{-1}$, moving the horizontal grey line in figure \ref{fig:PBH masses} up for $\alpha<0.1$ and down for $\alpha>0.1$. On the other hand it is the lower bound on $n_s$ \eqref{CMB 95percent ns lower bound} that bounds $k_\text{peak}$ from below and the modified universal prediction for $n_s$, eq.~\eqref{ns universal prediction inflection point}, does not depend on the parameter $\alpha$. This implies that the largest PBH mass that can be produced is the same for all $\alpha$. 

In summary the maximum PBH mass that can be produced in any of these scenarios is $M_\text{PBH}\simeq  10^{8}\,\text{g}$ which corresponds to a peak on scales $k_\text{peak}=4.7\times 10^{18}\,\text{Mpc}^{-1}$ which re-enter the horizon during radiation domination, after reheating.
PBHs with this mass would have evaporated by today and cannot constitute a candidate for dark matter. This strong constraint on $M_\text{PBH}(k_\text{peak})$ comes from the CMB observational lower bound on $n_s$, eq.~\eqref{CMB 95percent ns lower bound}, in these $\alpha$--attractor models. 

PBHs with masses $M_\text{PBH}\lesssim 10^{8}\,\text{g}$ would have evaporated before the onset of big bang nucleosynthesis and cannot therefore be directly constrained. Nevertheless, it is possible that these ultra-light PBHs are produced with such a large abundance that they come to dominate the cosmological density before they evaporate, giving rise to a period of early black hole domination~\cite{Anantua:2008am, Zagorac:2019ekv, Martin:2019nuw, Inomata:2020lmk, Hooper:2019gtx}. In this scenario, there are various sources of GW production (see e.g., recent work~\cite{Papanikolaou:2020qtd, Domenech:2020ssp, Kozaczuk:2021wcl}), which open up the possibility of constraining ultra-light PBHs using GW observatories, see also the discussion in section \ref{sec:2nd order GW}.

Another possibility is that primordial black holes could leave behind stable relics, instead of evaporating completely, see e.g., the early works~\cite{MacGibbon:1987my, Barrow:1992hq, Carr:1994ar}. Stable PBH relics could constitute the totality of dark matter, a possibility that has been investigated in the context of different inflationary models, see e.g.,~\cite{Dalianis:2019asr} where an $\alpha$--attractor single-field inflationary model is considered. 

We leave for future work the exploration of early PBH domination or stable PBH relics in the context of $\alpha$--attractor models of inflation.

\subsection{Induced gravitational waves at second order}
\label{sec:2nd order GW}

\subsubsection{Induced GWs after reheating}
\label{sec:2nd order GW during RD}
First-order scalar perturbations produced during inflation can source a stochastic background of primordial gravitational waves at second order from density perturbations that re-enter the horizon and oscillate during the radiation-dominated era after reheating~\cite{10.1143/PTP.37.831,Matarrese:1997ay,Ananda:2006af,Baumann:2007zm,Assadullahi:2009jc,Domenech:2021ztg}. In particular, the present-day energy density associated with these second-order GWs can be given as a function of the comoving scale
\begin{multline}
\label{Omega GW}
    \Omega_\text{GW}(k)=\frac{\Omega_\text{r,0}}{36} \int_0^{1/\sqrt{3}}\mathrm{d}d\int_{1/\sqrt{3}}^\infty \mathrm{d}s \; \left[\frac{(d^2-1/3)(s^3-1/3)}{s^2-d^2} \right]^2 \; P_\zeta\left(\frac{k\sqrt{3}}{2}(s+d) \right) \\
    \times P_\zeta\left(\frac{k\sqrt{3}}{2}(s-d) \right) \left[\mathcal{I}_c(d,s)^2+\mathcal{I}_s(d,s)^2 \right]\;,
\end{multline}
where $\Omega_\text{r,0}=8.6\times10^{-5}$ and the functions $\mathcal{I}_c$ and $\mathcal{I}_s$ are defined in eq.~(D.8) in~\cite{Espinosa:2018eve}. The expression above is derived assuming a $\Lambda\text{CDM}$ evolution. For example, a single narrow peak in the scalar power spectrum at the scale $k_\text{peak}$ produces a principal peak in $\Omega_\text{GW}$ from resonant amplification located at $k=2/\sqrt{3}\, k_\text{peak}$~\cite{Ananda:2006af}. 

\begin{figure}
\centering
\includegraphics[scale=0.5]{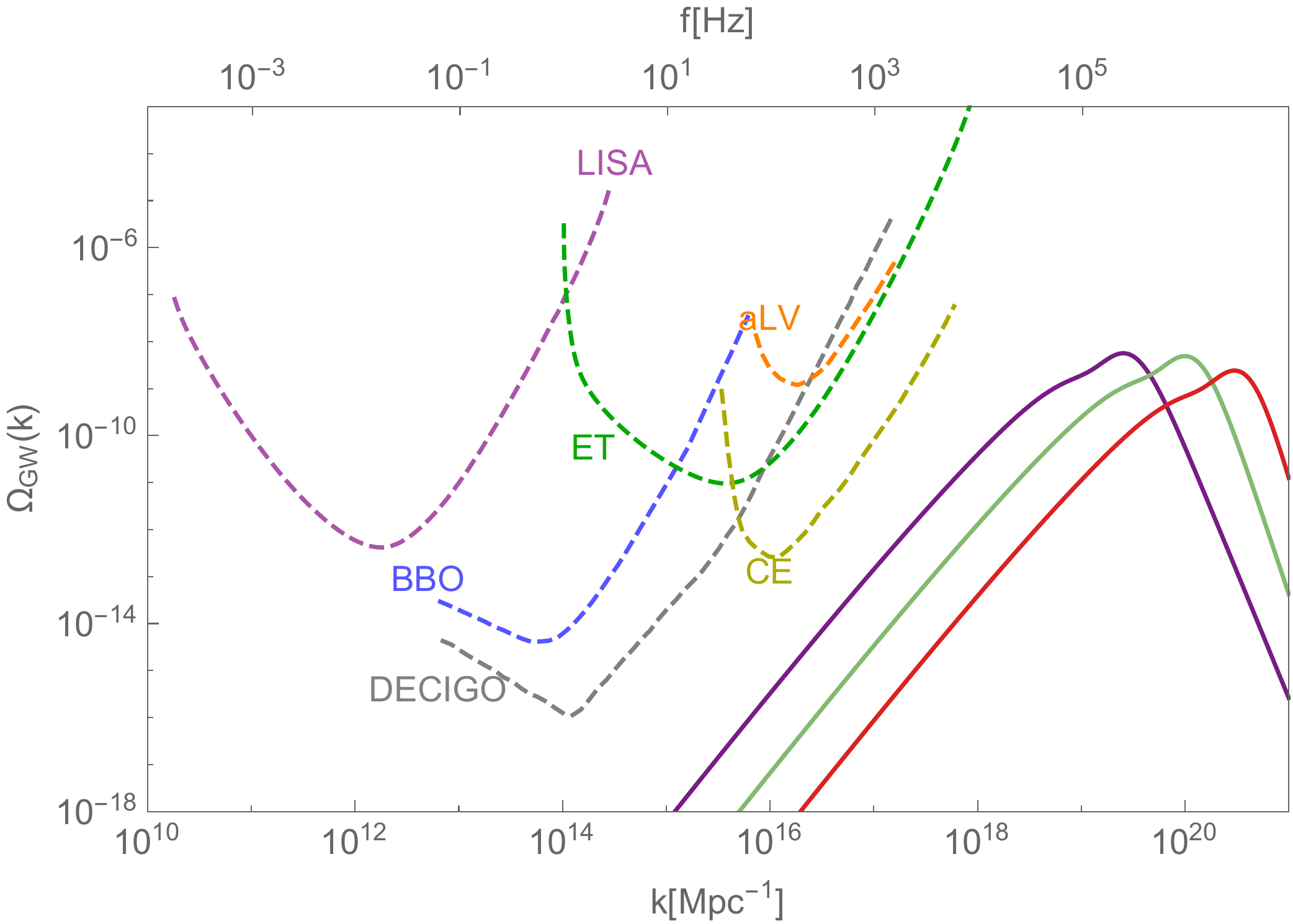}
\caption{GWs produced at second order by the large scalar perturbations generated in single-field inflection-point models with $\xi\neq0$. The legend is the same as in figure \ref{fig:xi non 0 results} and details about the parameters $\{\phi_\text{infl},\, \xi\}$ are listed in table \ref{tab:configs xi non 0}.}
\label{fig:xi non 0 GW}
\end{figure}

We numerically evaluate $\Omega_\text{GW}(k)$ for the gravitational waves induced from the peak in the scalar power spectrum on small scales in the inflection-point models with $\xi\neq0$ discussed in section \ref{sec: xi non 0 xase}. In figure~\ref{fig:xi non 0 GW} the results are represented together with the sensitivity curves of upcoming Earth- and space-based GW observatories, operating up to frequencies in the $\text{kHz}$. 

\begin{figure}
\centering
\captionsetup[subfigure]{justification=centering}
   \begin{subfigure}[b]{0.48\textwidth}
    \includegraphics[width=\textwidth]{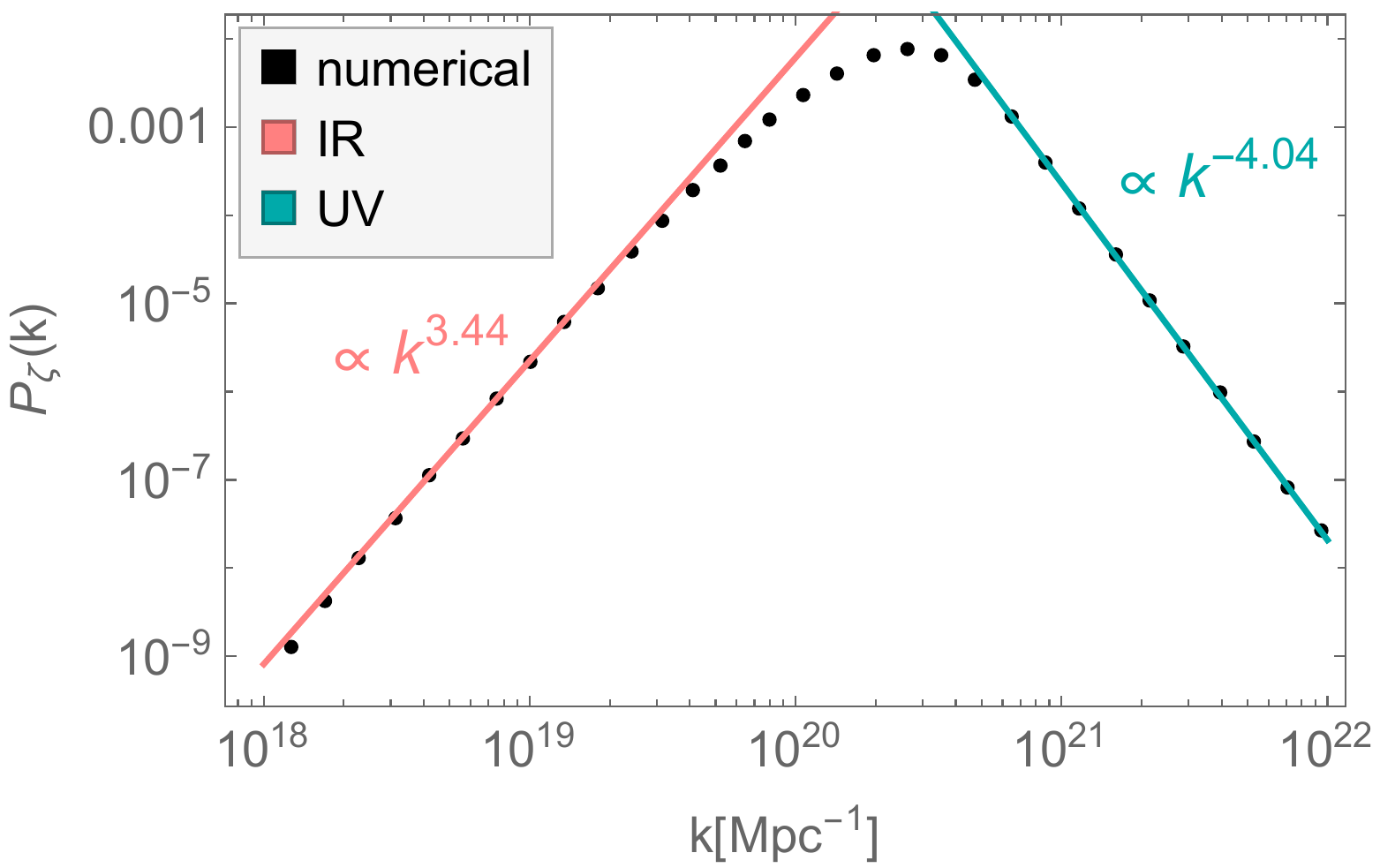}
  \end{subfigure}
   \begin{subfigure}[b]{0.48\textwidth}
    \includegraphics[width=\textwidth]{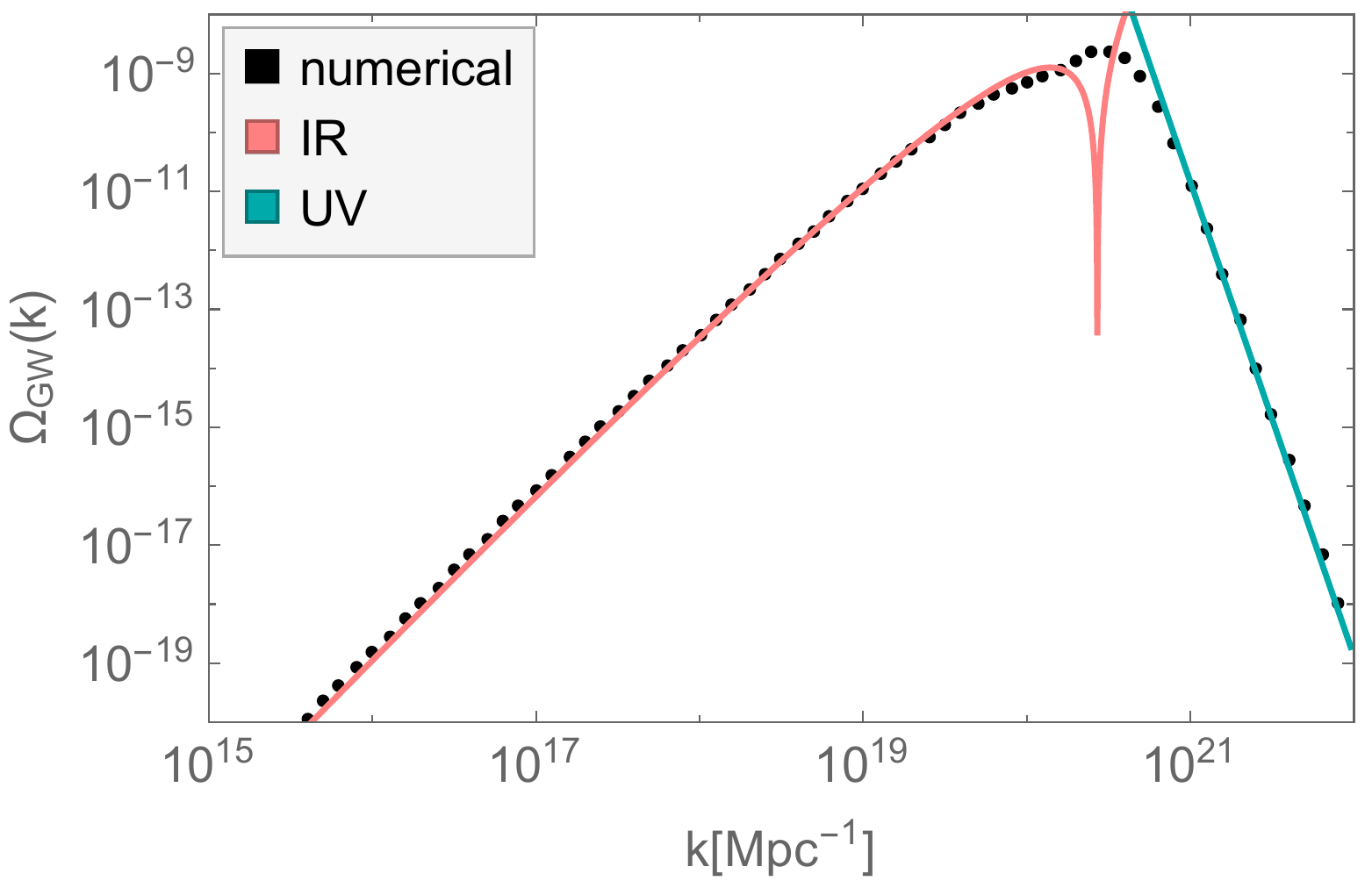}
  \end{subfigure}
\caption{Spectral shape of the scalar power spectrum (left) and second-order GWs (right) for a non-stationary inflection-point model with $\{\alpha=0.1,\, \phi_\text{infl}=0.49,\, \xi=0.0049575\}$. The scalar power spectrum is well approximated by a broken power-law and the IR and UV scaling of $P_\zeta(k)$ explain the IR and UV tails of the GW numerical results. In both plots, the black dots represent numerical results.} 
\label{fig:IR and UV behaviour singlefield}
\end{figure}

The spectral shape of the GW signal for the non-stationary inflection-point models can be understood in terms of the infrared ($k\ll k_\text{peak}$) and ultraviolet ($k\gg k_\text{peak}$) tilt of the peak in $P_\zeta(k)$~\cite{Atal:2021jyo, Domenech:2021ztg}. To demonstrate this, we select configuration (III), see table \ref{tab:configs xi non 0} for the model's parameters, and represent in the left panel of figure \ref{fig:IR and UV behaviour singlefield} the approximate IR and UV scaling of $P_\zeta(k)$ around the peak on top of the numerical results (black dots). We note that the IR tilt is in accordance with the estimate of the maximum growth of the scalar perturbations for single-field inflationary models, $n_\text{IR}\leq 4$~\cite{Byrnes:2018txb}. The IR and UV scaling of $P_\zeta(k)$ determine the IR and UV tails of the second-order GWs, see eqs. (5.16) and (5.20) in~\cite{Domenech:2021ztg}. In the right panel of figure \ref{fig:IR and UV behaviour singlefield}, we represent the numerical results for $\Omega_\text{GW}(k)$ together with the IR and UV approximations aforementioned, which well describe the numerical IR and UV tails.

The principal peak of $\Omega_\text{GW}(k)$ is located at very small scales, as a consequence of the position of the peak in the scalar power spectrum. In particular, the lower bound \eqref{k peak bound} on $k_\text{peak}$ implies that the GWs produced at second order exhibit a principal peak at $k\gtrsim 6\times10^{18}\,\text{Mpc}^{-1}$.
This equivalently implies that the GW signal peaks at frequencies $f\gtrsim10^5 \,\text{Hz}$, as confirmed by the numerical results plotted in figure \ref{fig:xi non 0 GW}. Configurations which are in accordance with CMB measurements on large scales cannot be probed on small scales by currently planned GW observatories.

\subsubsection{Induced GWs during reheating}
\label{sec:2nd order GW during reheating}
Second-order GWs resulting from first-order scalar perturbations that re-enter the horizon during reheating are in general suppressed~\cite{Inomata:2019zqy,Domenech:2021ztg}. 
First-order scalar metric perturbations, in the longitudinal gauge for example, on sub-Hubble scales during a matter-dominated era, remain constant rather than oscillating as they do in a radiation-dominated universe. While these scalar perturbations support second-order tensor metric perturbations in the longitudinal gauge during the matter era~\cite{Matarrese:1997ay,Baumann:2007zm,Assadullahi:2009nf}, these tensor perturbations are not freely-propagating gravitational waves and indeed they are gauge-dependent~\cite{Hwang:2017oxa,Tomikawa:2019tvi}. At the end of the reheating epoch, when the Hubble rate drops below the decay rate of the inflaton ($\Gamma\approx H$), the scalar metric perturbations decay slowly with respect to the oscillation time for sub-horizon GWs ($k/a\gg \Gamma$). Thus the tensor metric perturbations that they support also decay adiabatically on sub-horizon scales. The resulting power spectrum for freely propagating second-order GWs in the subsequent radiation-dominated era is therefore strongly suppressed on scales that re-enter the horizon during reheating. This gives an upper bound on the comoving wavenumber of any second-order GWs produced by modes re-entering the horizon after inflation, $k\lesssim k_\text{rh}$.

The only exception could be if there is a sudden transition from matter domination to radiation domination (rapid with respect the oscillation time, $a/k$)~\cite{Inomata:2019zqy,Inomata:2019ivs}. 
This could indeed occur in an early pressureless era dominated by light PBHs which decay and reheat the universe before primordial nucleosynthesis, as mentioned in section~\ref{sec: effect of reheating}. For a sufficiently narrow range of PBH masses and therefore lifetimes, the final evaporation of PBHs would be an explosive event and could lead to a sudden transition from an early PBH-dominated era after inflation to the conventional radiation-dominated era, leading to an enhancement of the spectrum of induced GWs from first-order scalar perturbations on sub-horizon scales at the transition~\cite{Inomata:2020lmk}. We leave the study of GWs from a possible early PBH-dominated era for future work.

\section{Multi-field extension}
\label{sec: multi-field extension}

Cosmological $\alpha$--attractor models are naturally formulated in terms of two fields living in a hyperbolic field space, therefore we explore here the consequences of embedding in a multi-field setting the single-field inflection-point model studied in the preceding sections. Our aim is to establish whether the single-field predictions are robust against multi-field effects and under which conditions it may be possible to enhance the scalar power spectrum through inherently multi-field effects. 

\subsection{Multi-field dynamics}
\label{sec: multi-field dynamics}
When considering the extension from single-field inflation into a multi-field scenario, there are two novel ingredients which enter the inflationary evolution; the field-space geometry and the multi-field potential. The action of the multi-field model can be written as
\begin{equation}
    \mathcal{S}=\int \mathrm{d}^4x \sqrt{-g} \left[-\frac{1}{2} \mathcal{G}_{IJ}\left(\phi^K \right)\partial_\mu \phi^I \partial^\mu \phi^J -U\left(\phi^K \right)\right] \;,
\end{equation}
where $\mathcal{G}_{IJ}\left(\phi^K\right)$ is the metric on the field space and $U\left(\phi^K\right)$ is the multi-field potential.
For simplicity, from now on we focus on the case of two-field models (equivalent to a single complex field) in a hyperbolic field space. In a FLRW universe, the equations of motion for the evolution of the background fields read
\begin{gather}
\label{H eq multi field t}
3 H^2=\frac{1}{2} \dot\sigma^2 +U \;,\\
\label{Hdot eq multi field t}
\dot{H}=-\frac{1}{2}  \dot{\sigma}^2\;, \\
\label{fields eq multi field t}
\mathcal{D}_t \dot{\phi^I} +3H \dot{\phi^I} +\mathcal{G}^{IJ} U_{,J}=0 \;,
\end{gather}
where $U_{,J}\equiv {\mathrm{d}U}/{\mathrm{d}\phi^J}$, $\dot{\sigma}^2 \equiv \mathcal{G}_{IJ}\dot{\phi}^I\dot{\phi}^J$ is the kinetic energy of the fields, $\mathcal{D}_t A^I=\dot{A^I} +\Gamma^I_{JK} \dot\phi^{J} A^K$ and $\Gamma^I_{JK}$ are the Christoffel symbols on the field space. After some manipulation, eq.~\eqref{fields eq multi field t} can be rewritten as
\begin{equation}
\label{eq sigma dot dot}
   \Ddot{\sigma}+3H\dot{\sigma}+U_{,\sigma}=0 \;, 
\end{equation}
where $U_{,\sigma}\equiv {\dot\phi^I U_{,I}}/{\dot{\sigma}}$. 

In order to ensure that the study of scalar field fluctuations relies on quantities which are covariant under field-space transformations, the covariant perturbation in the spatially-flat gauge $\mathcal{Q}^I$ is used~\cite{Gong:2011uw}. The equations of motion for the linear perturbations are then~\cite{Sasaki:1995aw,GrootNibbelink:2001qt,Langlois:2008mn} (see also the review~\cite{Gong:2016qmq})
\begin{equation}
   \mathcal{D}_t\mathcal{D}_t \mathcal{Q}^I+3H \mathcal{D}_t \mathcal{Q}^I+\frac{k^2}{a^2} \mathcal{Q}^I+{\mathcal{M}^I}_J \mathcal{Q}^J=0 \;,    
\end{equation}
where the mass matrix, ${\mathcal{M}^I}_J$, is defined as 
\begin{equation}
    {\mathcal{M}^I}_J\equiv {U_{;}^I}_J-{\mathcal{R}^I}_{KLJ}\dot{\phi}^K \dot{\phi}^L-\frac{1}{a^3}\mathcal{D}_t \left(\frac{a^3}{H}\dot{\phi}^I\dot{\phi}_J \right) \;.
\end{equation}
The first component of ${\mathcal{M}^I}_J$ is the Hessian of the multi-field potential $U_{;IJ}\equiv U_{,IJ}-\Gamma^{K}_{IJ}U_{,K}$, defined by means of a covariant derivative in field space in order to take into account the non-trivial geometry. The second term also depends on the geometry of the field space, whose Riemann tensor is ${\mathcal{R}^I}_{KLJ}$. For a two-dimensional field space, 
the Riemann tensor is $\mathcal{R}_{IJKL}=\frac{1}{2}\mathcal{R}_{\text{fs}}\left(\mathcal{G}_{IK}\mathcal{G}_{JL}-\mathcal{G}_{IL}\mathcal{G}_{JK} \right)$, where $\mathcal{R}_{\text{fs}}$ is the intrinsic scalar curvature of the field space. The third term encodes the gravitational backreaction due to spacetime metric perturbations induced by the field fluctuations at first order. 

When studying the dynamics of the perturbations, instead of directly using the variables $\mathcal{Q}^I$ it is often convenient to project the fluctuations along the instantaneous adiabatic and entropic directions~\cite{Gordon:2000hv,GrootNibbelink:2001qt}. The adiabatic direction follows the background trajectory in field space and the entropic direction is orthogonal to it. More precisely, the new basis is described by the unit vectors $(\hat{\sigma}^I,\, \hat{s}^I)$, where
\begin{gather}
    \hat{\sigma}^I \equiv \frac{\dot{\phi}^I}{\dot{\sigma}} \;, \\
    \hat{s}^I\equiv \frac{\omega^I}{\omega} \text{ where } \omega^I\equiv \mathcal{D}_t\hat{\sigma}^I \;,\\
    \hat{\sigma}^I \hat{s}_I=0\;, \;\;\; \hat{\sigma}^I \hat{\sigma}_I=\hat{s}^I \hat{s}_I=1 \;.
\end{gather}
Usually $\omega$ is referred to as the turn rate in field space, while the dimensionless bending parameter 
\begin{equation}
    \label{eta perp}
    \eta_\perp\equiv \frac{\omega}{H} 
\end{equation}
measures the deviation of the background trajectory from a geodesic in field space. Using eqs.~\eqref{fields eq multi field t} and~\eqref{eq sigma dot dot}, the components of the turn rate can be expressed as 
\begin{equation}
    \omega^I=-\frac{\mathcal{G}^{IJ}U_{,J}}{\dot{\sigma}} +\frac{\dot{\phi}^I}{\dot{\sigma}^2} U_{,\sigma}
\end{equation}
and 
\begin{equation}
    \omega^2\equiv \mathcal{G}_{IJ}\omega^I\omega^J= \frac{\mathcal{G}^{KM}U_{,K}U_{,M}}{\dot{\sigma}^2}-\frac{(U_{,\sigma})^2}{\dot{\sigma}^2} \;.
\end{equation}

Projecting the perturbations $\mathcal{Q}^I$ in the adiabatic and entropic directions allows us to define the adiabatic and entropic perturbations as $\mathcal{Q}_\sigma\equiv \hat{\sigma}_I \mathcal{Q}^I$ and $\mathcal{Q}_s\equiv \hat{s}_I \mathcal{Q}^I$ respectively. From these, the dimensionless comoving curvature and isocurvature perturbations are given by
\begin{equation}
    \zeta\equiv \frac{H}{\dot{\sigma}} \mathcal{Q}_\sigma, \;\;\;\;\; \mathcal{S}\equiv \frac{H}{\dot{\sigma}} \mathcal{Q}_s \;.
\end{equation}
The presence of isocurvature perturbations, $\mathcal{S}$, gives rise to multi-field effects. The equations of motion for $\mathcal{Q}_\sigma$ and $\mathcal{Q}_s$ are~\cite{Sasaki:1995aw,GrootNibbelink:2001qt,Langlois:2008mn}
\begin{align}
\label{adiabatic eq}
    \Ddot{\mathcal{Q}}_\sigma +3H \dot{\mathcal{Q}}_\sigma + \left(\frac{k^2}{a^2} +{m_\sigma}^2 \right)\mathcal{Q}_\sigma &= \left( 2H \eta_\perp \mathcal{Q}_s \right)^{\cdot} -\left(\frac{\dot{H}}{H} +\frac{U_{,\sigma}}{\dot{\sigma}} \right) 2H \eta_\perp \mathcal{Q}_s\;, \\ 
\label{isocurvature eq}    
    \Ddot{\mathcal{Q}}_s +3H \dot{\mathcal{Q}}_s + \left(\frac{k^2}{a^2} +{m_s}^2 \right)\mathcal{Q}_s &= -2\dot{\sigma} \eta_\perp \dot{\zeta} \;.
\end{align}
These equations show that the adiabatic and entropic perturbations are coupled in the presence of a non-zero bending of the trajectory ($\eta_\perp\neq0$), i.e., non-geodesic motion in field space~\cite{Gordon:2000hv}. The squared-masses of the adiabatic and isocurvature fluctuations are ${m_\sigma}^2$ and ${m_s}^2$ respectively. At leading order in slow roll the adiabatic squared-mass is ${m_\sigma}^2=-\frac{3}{2}\epsilon_2 +\mathcal{O}(\epsilon^2)$, while the entropic squared-mass is
\begin{equation}
\label{isocurvature mass}
    \frac{{m_s}^2}{H^2}\equiv \frac{U_{;ss}}{H^2}+\epsilon_1 \mathcal{R}_{\text{fs}}-\eta_\perp^2 \;,  
\end{equation}
where $U_{;ss}\equiv \hat{s}^I \hat{s}^JU_{;IJ}$. 

In the super-horizon regime $(k\ll aH)$ 
the curvature perturbation obeys
\begin{equation}
    \dot{\zeta} \simeq 2\eta_\perp \frac{H^2}{\dot{\sigma}}\mathcal{Q}_s
    \;,
\end{equation}
which demonstrates that in multi-field inflation the curvature perturbation, $\zeta$, is not constant in the super-horizon regime for non-geodesic trajectories. 
Substituting this expression into eq.~\eqref{isocurvature eq} for $\mathcal{Q}_s$ we obtain 
\begin{equation}
    \Ddot{\mathcal{Q}}_s+3H\dot{\mathcal{Q}}_s+{m_{s,\, \text{eff}}}^2 \mathcal{Q}_s \simeq 0 \;,
\end{equation}
where the entropic effective squared-mass in the super-horizon regime is~\cite{Renaux-Petel:2015mga}
\begin{equation}
\label{eff mass isocurvature}
    \frac{ {m_{s,\,\text{eff}}}^2}{H^2}\equiv \frac{U_{;ss}}{H^2}+\epsilon_1 \mathcal{R}_{\text{fs}}+3\eta_\perp^2 \;.
\end{equation}
From the equations above one can identify two important regimes characterising the multi-field dynamics in a hyperbolic field space:

\medskip (i) geometrical destabilisation: the effective squared-mass of the isocurvature perturbation \eqref{eff mass isocurvature} receives a contribution from the curvature of the field space, ${\mathcal R}_\text{fs}$, which on a hyperbolic geometry is negative \cite{Brown:2017osf, Mizuno:2017idt}. If the combination $\epsilon_1\mathcal{R}_\text{fs}$ is large enough to overcome the other contributions in \eqref{eff mass isocurvature}, this can lead to geometrical destabilisation~\cite{Turzynski:2014tza,Renaux-Petel:2015mga}. In this case, the entropic fluctuation is tachyonic and renders the background trajectory unstable. As a consequence, inflation might end prematurely, affecting the inflationary observables~\cite{Renaux-Petel:2017dia}, or the geometrical instability drives the system away from its original trajectory into a new, side-tracked, field-space trajectory~\cite{Garcia-Saenz:2018ifx,Garcia-Saenz:2018vqf, Grocholski:2019mot};

\medskip (ii) strongly non-geodesic motion: a large bending of the background trajectory ($\eta_\perp\gg1$) could drive the entropic squared-mass, ${m_{s}}^2$ in eq.~\eqref{isocurvature mass}, to negative values.  
In this case the entropic fluctuation may undergo a transient instability in the sub-horizon regime where it is exponentially amplified. However, while contributing negatively to the squared-mass on sub-horizon scales, ${m_s}^2$ in eq.~\eqref{isocurvature mass}, a large bend in the trajectory contributes positively to the effective squared-mass on super-horizon scales, ${m_{s,\, \text{eff}}}^2$ in eq.~\eqref{eff mass isocurvature}, therefore keeping the background trajectory stable. In the case of hyperbolic field-space geometry and strongly non-geodesic regime, the bispectrum is enhanced in the flattened configuration~\cite{Fumagalli:2019noh}. Moreover, as a consequence of the transfer between the entropic and adiabatic modes (whose efficiency is set by $\eta_\perp$), the exponentially-enhanced isocurvature fluctuations can source curvature perturbations~\cite{Fumagalli:2020adf, Palma:2020ejf, Braglia:2020taf}. In this case, the scalar power spectrum can grow faster than would be allowed in single-field ultra-slow-roll inflation~\cite{Byrnes:2018txb}. Depending on the duration of the turn in field space, it can be classified as broad (taking several e-folds) or sharp (less than one e-fold), as will be discussed later after eq.~\eqref{eta perp gaussian profile}. In the case of sharp turns $P_\zeta(k)$ exhibits characteristic oscillatory patterns~\cite{Palma:2020ejf, Fumagalli:2020adf,Braglia:2020taf, Fumagalli:2020nvq}, see also~\cite{Achucarro:2010da, Chen:2010xka} for earlier works on features in $P_\zeta(k)$ produced by sudden turns of the inflationary trajectory.

\medskip In summary, multi-field dynamics in a hyperbolic field-space geometry can lead to a very rich phenomenology, because of geometrical effects and non-geodesic motion. This has been studied in the context of the generation of features in the primordial power spectrum on large scales\footnote{ For other multi-field effects arising from the direct coupling of the inflaton to oscillating `clock' fields see, e.g.,~\cite{Chen:2014cwa, Braglia:2021ckn, Braglia:2021sun, Braglia:2021rej}. }~\cite{Braglia:2020fms}, PBH production~\cite{Palma:2020ejf,Fumagalli:2020adf,Braglia:2020eai}, and second-order GW generation~\cite{Braglia:2020taf,Fumagalli:2020nvq, Fumagalli:2021cel,Witkowski:2021raz, Fumagalli:2021mpc}. 

\medskip In the following, we consider the multi-field set-up of $\alpha$--attractor models, with $\phi^I=\{\phi,\,\theta\}$. The geometry of field space is hyperbolic, with curvature $\mathcal{R}_\text{fs}=-4/(3\alpha)$. The kinetic Lagrangian for the fields $\phi$ and $\theta$ is given in eq.~\eqref{alpha attractor metric}. The Christoffel symbols associated with the hyperbolic metric are
\begin{equation}
    \Gamma^{\phi}_{\theta\theta}=-\frac{1}{2}\sqrt{\frac{3}{2}\alpha}\; \sinh{\left(2\sqrt{\frac{2}{3\alpha}}\phi \right)}\, , \;\;\;
    \Gamma^\theta_{\phi \theta}=\frac{2}{\sqrt{6\alpha}}\tanh^{-1}{\left(\sqrt{\frac{2}{3\alpha}}\phi\right)} \;.
\end{equation} 
In this way the equations of motion for the background evolution \eqref{Hdot eq multi field t}--\eqref{fields eq multi field t} can be written explicitly for the fields $\phi$ and $\theta$ as
\begin{gather}
\label{EoM multifield 1}
    -\frac{H'}{H}=\frac{1}{2}\left(\phi'^2+\frac{3\alpha}{2}\sinh^2{\left(\sqrt{\frac{2}{3\alpha}}\phi \right)}\theta'^2\right) \;, \\
\label{EoM multifield 2}
    H^2\phi''+HH'\phi'+3H^2\phi'+\Gamma^{\phi}_{\theta\theta}H^2\theta'^2 +U_{,\phi}=0 \;, \\
\label{EoM multifield 3}
    H^2\theta''+HH'\theta'+3H^2\theta'+2 \Gamma^{\theta}_{\phi\theta}H^2 \theta'\phi' +\left[\frac{3\alpha}{2}\sinh^2{\left(\sqrt{\frac{2}{3\alpha}}\phi \right)}\right]^{-1}U_{,\theta}=0 \;,
\end{gather}
where a prime denotes a derivative with respect to the number of e-folds, $N$.

In section \ref{sec:justified multi-field potential} we illustrate one possible multi-field embedding of the single-field inflection-point potential and discuss its phenomenology in sections \ref{sec: multi field potential exploration} and \ref{sec: change alpha multifield}. In section \ref{sec:robustness of single-field predictions} we establish the robustness of the modified universal predictions given in eqs.~\eqref{ns universal prediction inflection point}--\eqref{r universal prediction inflection point} for single-field models against multi-field effects, and consider the small-scale phenomenology of multi-field models which are compatible with CMB measurements.

\subsection{Multi-field embedding of the single-field inflection-point potential}
\label{sec:justified multi-field potential}
In section \ref{sec: single field model} we outlined the construction of an inflection-point potential in the context of single-field $\alpha$--attractor models, where the building block is the cubic function $f(r)$. This construction can easily be extended to a multi-field set-up. In analogy with the single-field case, let us consider a function $F(r,\,\theta)$ cubic in $r$, in terms of which the multi-field potential is  
\begin{equation}
\label{U as function of F}
    U(\phi,\,\theta)\equiv F^2(r(\phi),\,\theta)\;.
\end{equation}
In constructing a cubic function of $r$, we have at our disposal the complex field $Z$, as defined in \eqref{Z parametrisation}, its complex conjugate $\bar{Z}$ and their combinations
\begin{align}
\label{ingredients}
    Z\Bar{Z}&=r^2 \;,\\
\label{ingredients 2}
    \frac{Z+\Bar{Z}}{2}&=r \cos{\theta}\;.
\end{align}
In particular, the former is symmetric under a phase-shift while the latter depends on $\theta$ explicitly. The general form of $F(r,\,\theta)$ arising from terms proportional to $(Z \bar Z)^{n/2}(Z+\bar Z)^m\propto \cos^m{(\theta)}r^{n+m}$ is 
\begin{equation}
\label{F_mn}
   F(r,\,\theta) = \sum_{m,n} F_{n+m,m}\cos^m{(\theta)}\,  r^{n+m} \;.
\end{equation}
We note that our potential will thus be symmetric under the reflection $\theta\to-\theta$.

As in the single-field case, we set $F_{0,0}=0$ such that the potential \eqref{U as function of F} has a minimum at $U(0,\,\theta)=0$. For $F(r,\,\theta)$ to be a cubic function of $r$, there are potentially nine terms contributing in \eqref{F_mn}. 
For simplicity we select just the 3 remaining phase-independent terms to be non-zero and one $\theta$-dependent term, such that 
\begin{equation}
\label{F stage1}
 F(r,\,\theta) = F_{1,0} r + F_{2,0}(1+\gamma \cos{(\theta)}) r^2 + F_{3,0} r^3 \;,
\end{equation}
where $\gamma\equiv F_{2,1}/F_{2,0}$.

Identifying the potential $U(r,\,0)\equiv F^2(r,\,0)$ along the direction $\theta=0$ with the single-field potential in \eqref{potential}, with an inflection point in the radial direction located at $r=r_\text{infl}$, gives the coefficients
\begin{equation}
\label{F parameters}
 F_{1,0}=r_\text{infl}-\xi\;, \;\;\;   F_{2,0}=-1/(1+\gamma)\;, \;\;\;F_{3,0}= 1/(3r_\text{infl}) \;.  
\end{equation}
Substituting these coefficients into eq.~\eqref{F stage1} yields 
\begin{equation}
    \label{F stage2}
    F(r,\,\theta) = (r_\text{infl}-\xi) \,r - \frac{1+\gamma \cos{(\theta)}}{1+\gamma}\, r^2 + \frac{1}{3r_\text{infl}}\, r^3 \;.
\end{equation}
Away from the particular direction $\theta=0$ the function \eqref{F stage2} has an inflection point in the radial direction at
\begin{equation}
\label{r star multifield}
    \tilde{r}_\text{infl}(\theta) = \left( \frac{1+\gamma \cos{(\theta)}}{1+\gamma} \right) r_\text{infl} \;.
\end{equation}
For $-(1-r_\text{infl}^2)/r_\text{infl}<\xi<r_\text{infl}$ there is a stationary inflection point (where $\partial F/\partial r=0$) when
\begin{equation}
    \cos{(\theta_\text{st})} = \frac{(1+\gamma)(r_\text{infl}-\xi)^{1/2}-r_\text{infl}^{1/2}}{\gamma r_\text{infl}^{1/2}} \;.
\end{equation}
and
\begin{equation}
    \tilde{r}_\text{infl}(\theta_\text{st}) = r_\text{infl}^{1/2}(r_\text{infl}-\xi)^{1/2} \;.
\end{equation}
If $\xi=0$ then $F(r,\,\theta)$ has only one inflection point in the radial direction, located at $r=r_\text{infl}$ along $\theta=0$, and it is stationary. 
This property simplifies the form of the potential and it is for this reason that in the following we consider two-field models with $\xi=0$ and leave the analysis of the non-stationary inflection-point case, or a stationary inflection point away from the symmetric $\theta=0$ direction, to future work. 

Substituting \eqref{F stage2} with $\xi=0$ in \eqref{U as function of F} yields
\begin{equation}
\label{potential multifield}
    U(\phi,\,\theta)=U_0\,\Big\{r_\text{infl}\, \tanh{\left(\frac{\phi}{\sqrt{6\alpha}}\right)}- \frac{1+\gamma \cos{(\theta)}}{1+\gamma} \tanh^2{\left(\frac{\phi}{\sqrt{6\alpha}}\right)} + \frac{1}{3r_\text{infl}}  \tanh^3{\left(\frac{\phi}{\sqrt{6\alpha}}\right)} \Big\}^2    \;, 
\end{equation}
which is written in terms of the canonical field $\phi$, defined in eq.~\eqref{canonical field transformation}.
The profile of the multi-field potential along the direction $\theta=0$ is represented by the black-dashed line in figure~\ref{fig: structure around inflection point} for a configuration with $\{\alpha=0.1,\,\phi_\text{infl}=0.5\}$. 
%


Once the field-space curvature, $\alpha$, and the position of the inflection point along $\theta=0$, $r_\text{infl}$, are fixed, the only remaining free parameter in the potential \eqref{potential multifield} is $\gamma$. 
We impose some simple conditions on $U(\phi,\,\theta)$ to ensure a successful inflationary scenario, which will restrict the allowed range of $\gamma$.
In particular, we require that the potential has a non-negative derivative in the radial direction, a condition which forbids the radial field, $\phi$, from running back towards larger (radial) field values at late times. 
Thus we require 
\begin{equation}
\label{dF/dr}
\frac{\partial F(r,\,\theta)}{\partial r} \geq0 \;\;\;\forall\, r,\theta \;,
\end{equation}
which one can show implies
\begin{equation}
    -1\leq \dfrac{1+\gamma \cos{(\theta)}}{1+\gamma} \leq 1 \;.
\end{equation}
Thus we will restrict our analysis to the case $\gamma>0$ where the condition that the potential has a non-negative derivative in the radial direction holds for any angle $\theta$. In addition, we can see from eq.~\eqref{potential multifield} that the effective squared-mass of the angular field, $\theta$, is non-negative along $\theta=0$ for $\gamma>0$ (see also the discussion in appendix \ref{appendix: parameter study multifield potential}). Thus we expect to recover the single-field behaviour for evolution along the symmetric direction, $\theta=0$, while the potential can exhibit a richer phenomenology in the two-dimensional field space for $\theta\neq0$.

We plot the profile of the multi-field potential \eqref{potential multifield} with $\{\alpha=0.1, \, \phi_\text{infl}=0.542,\,\gamma=10 \}$ in figure~\ref{fig:multifield evo alpha0.1 vary thetain}. The direction $\theta=0$ corresponds to a minimum of the potential in the angular direction, as expected for $\gamma>0$. 

An interesting comparison can be made between multi-field $\alpha$--attractor potentials, which remain non-singular throughout the hyperbolic field space, and other inflation models discussed in the literature which employ a different coordinate chart in the hyperbolic field space. In particular, the two-field model of~\cite{Braglia:2020eai} is formulated in terms of planar coordinates on the hyperbolic field space and supports a strong enhancement of the scalar power spectrum on small scales. We show in appendix~\ref{appendix:compare between polar and planar coordinates} that the multi-field potential in~\cite{Braglia:2020eai} diverges at a point on the boundary of the hyperbolic disc. At this point, the potential shares the same singularity as the kinetic Lagrangian, and initial conditions which support a small-scale peak in the scalar power spectrum are close to the singularity. In this case, the large-scale observables are then sensitive to characteristics of the potential and initial conditions, as already noted in~\cite{Garcia-Saenz:2018ifx} in the context of side-tracked inflation. The model in~\cite{Braglia:2020eai}, while being of interest in its own right, lies outside the class of $\alpha$--attractors that we consider here. 

\subsection{Exploring the multi-field potential: turning trajectories and geometry at play}
\label{sec: multi field potential exploration}
In the following we perform a numerical analysis of the background evolution stemming from the multi-field potential \eqref{potential multifield}. Initially we will explore a range of possibilities which follow from the form of the potential and the consequences of different choices of parameters and initial conditions. Later, in section \ref{sec:robustness of single-field predictions}, we will restrict our attention to configurations which have been specifically selected to be consistent with CMB measurements on large scales and explore the consequences that CMB observations have for this model. 

\begin{figure}
\centering
\captionsetup[subfigure]{justification=centering}
   \begin{subfigure}[b]{0.48\textwidth}
    \includegraphics[width=\textwidth]{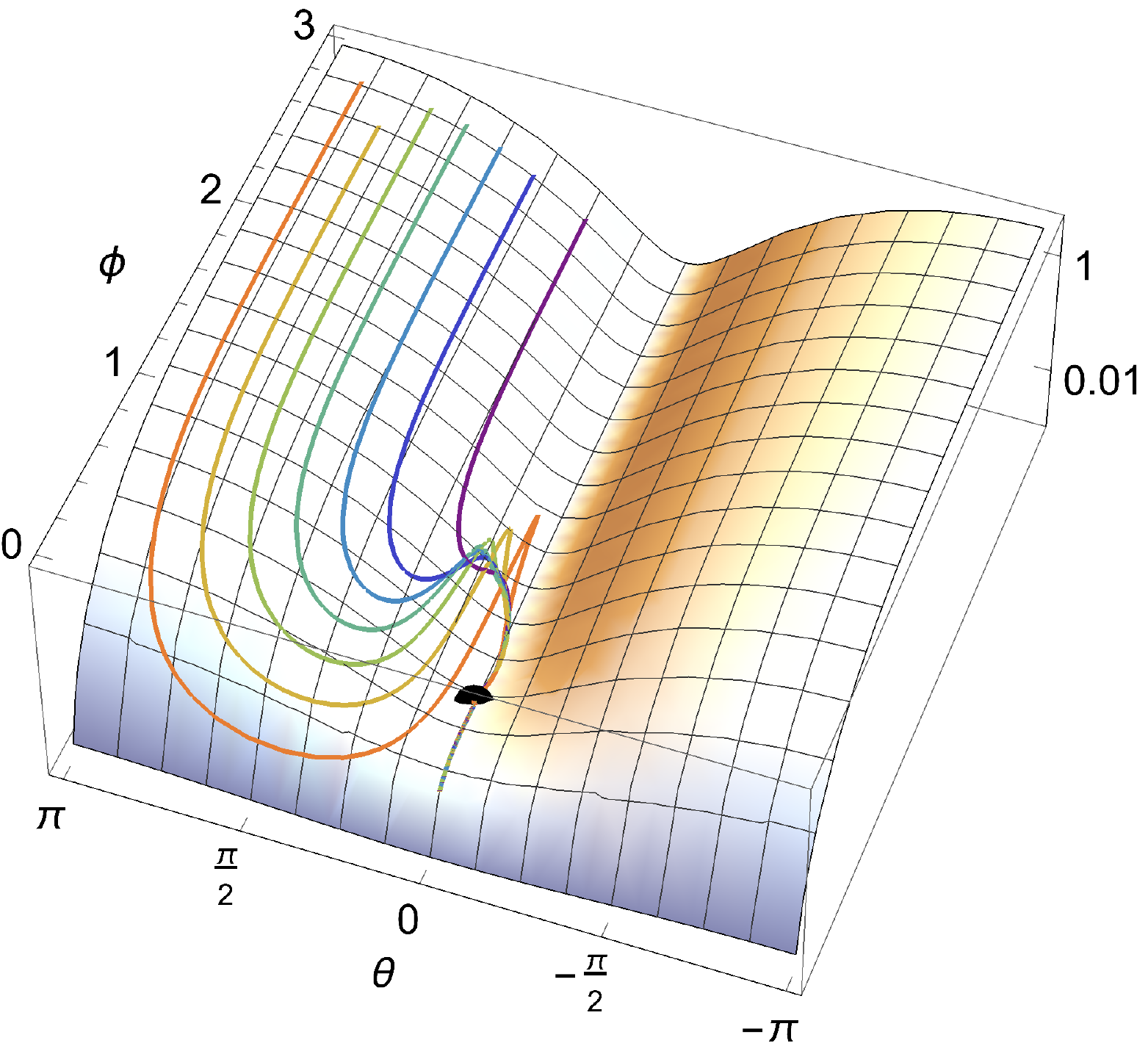}
  \end{subfigure}
  \begin{subfigure}[b]{0.49\textwidth}
    \includegraphics[width=\textwidth]{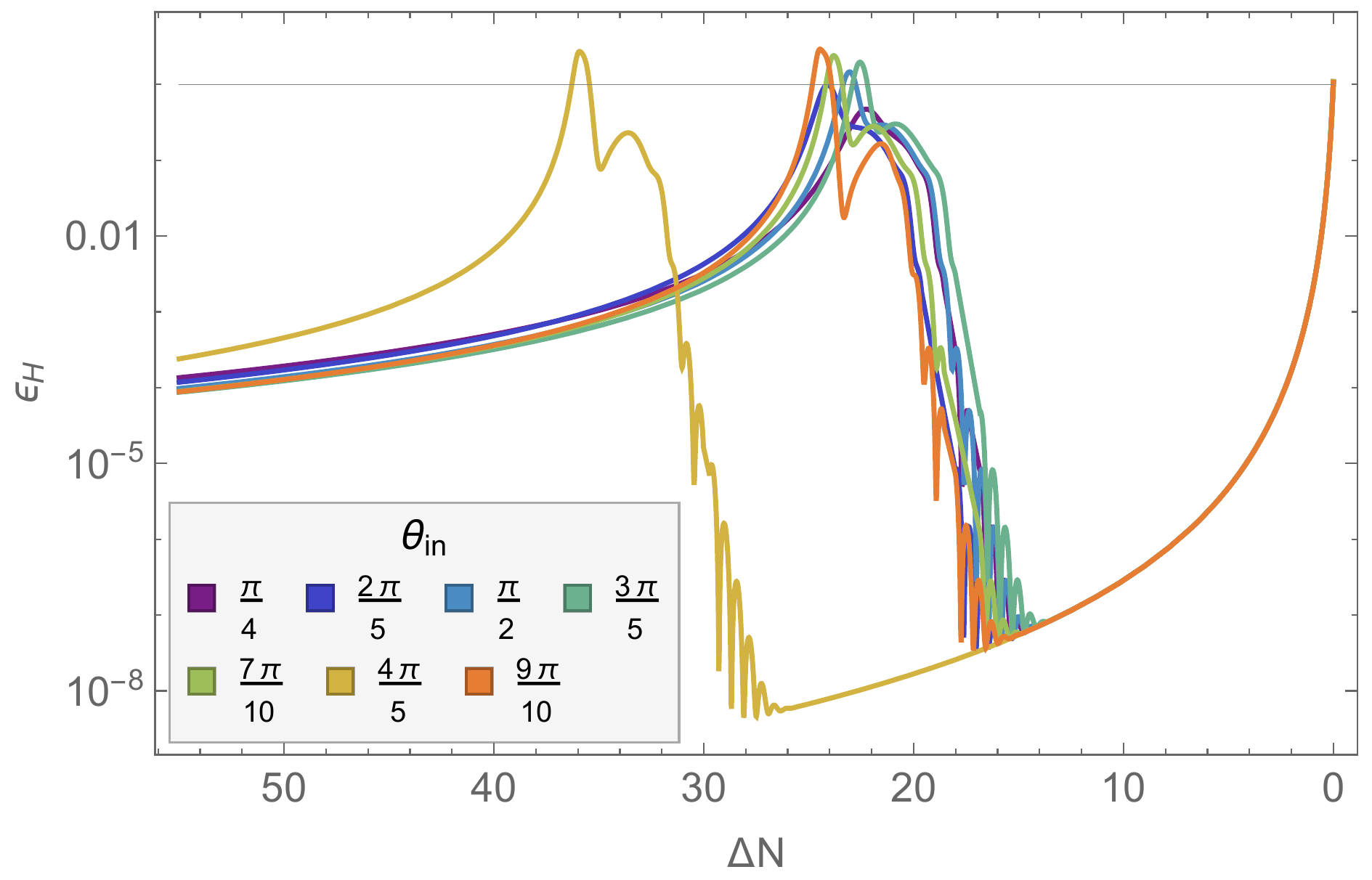}
  \end{subfigure}
  \caption{Background evolution obtained numerically for the potential \eqref{potential multifield} with model parameters $\{\alpha=0.1,\, \phi_\text{infl}=0.542,\, \gamma=10\}$ and different initial conditions $\theta_\text{in}$. On the left the fields trajectories are represented on top of the potential profile, with different colours corresponding to different $\theta_\text{in}$, see the right panel for the legend. The black point located at $\theta=0$ and $\phi=\phi_\text{infl}$ highlights the position of the inflection point. On the right we display the evolution of the slow-roll parameter $\epsilon_H$ against $\Delta N\equiv N_\text{end}-N$. }
  \label{fig:multifield evo alpha0.1 vary thetain}
\end{figure}

The first parameter we fix is $\alpha$, which determines the Ricci curvature of the field space, $\mathcal{R}_\text{fs}=-4/(3\alpha)$. As we did in the single-field case, we start by considering $\alpha=0.1$, which corresponds to $\mathcal{R}_\text{fs}\simeq-13.3$.
The profile of the potential is then parametrised by $\{\phi_\text{infl},\,\gamma\}$, and here we select $\{\phi_\text{infl}=0.542,\, \gamma=10\}$, as shown in figure~\ref{fig:multifield evo alpha0.1 vary thetain}. 
The effect of different choices for $\gamma$ and $\phi_\text{infl}$ is discussed in appendix~\ref{appendix: parameter study multifield potential}. 
The background evolution is derived by numerically solving the differential equations \eqref{EoM multifield 1}--\eqref{EoM multifield 3}. We consider vanishing initial velocities  for the fields, but in practice the fields rapidly settle into single-field, slow-roll attractor  solution at early times. We select $\phi_\text{in}$ such that the model supports at least $55$ e-folds\footnote{This choice is made in analogy with the single-field case, where $\Delta N_\text{CMB}\simeq55$ for models with $\alpha=0.1$ and assuming instant reheating.} before the end of inflation after the background evolution reaches the attractor solution.

In figure~\ref{fig:multifield evo alpha0.1 vary thetain} we show the field evolution (left panel) and first slow-roll parameter (right panel) for several initial conditions for the angular field in the range $\pi/4\leq\theta_\text{in}\leq9\pi/10$. 
All the trajectories share some common features. 
Initially, the angular field, $\theta$, is frozen and only the radial field, $\phi$, is evolving. This is a well known effect in hyperbolic field space, referred to as ``rolling on the ridge''~\cite{Achucarro:2017ing},  where the geometry is responsible for suppressing the potential gradient in the equation of motion for $\theta$, see the term $2/(3\alpha)\, \sinh^{-2}{
\phi/\sqrt{3\alpha/2}}$ multiplying $U_{,\theta}$ in eq.~\eqref{EoM multifield 3}. As long as $\phi\gg\sqrt{3\alpha/2}$, this term is suppressed, effectively freezing $\theta$ at its initial value during the early stages of inflation. 

When $\phi\sim \sqrt{3\alpha/2}$, the angular field $\theta$ starts evolving and there is a turn in the trajectory, which is shallower or sharper depending on $\theta_\text{in}$. During the turn, the field $\phi$ can be driven back towards larger values, this effect being more or less pronounced depending again on $\theta_\text{in}$. The change of sign of $\phi'(N)$ is due to the motion of $\theta$, which switches on the geometrical contribution, $\Gamma^{\phi}_{\theta\theta}H^2\theta'^2$, in the equation of motion for $\phi$,  eq.~\eqref{EoM multifield 2}. This effect also appears in other multi-field $\alpha$--attractor models, e.g., angular inflation~\cite{Christodoulidis:2018qdw}. Once $\theta$ starts oscillating around its minimum, $\theta=0$, the fields cross the radial inflection point and inflation comes to an end soon afterwards. In the right panel of figure~\ref{fig:multifield evo alpha0.1 vary thetain} we display $\epsilon_H\equiv -H'/H$ (see eq.~\eqref{EoM multifield 1}) against $\Delta N\equiv N_\text{end}-N$, where each coloured line corresponds to a different $\theta_\text{in}$. Depending on $\theta_\text{in}$ the profile of $\epsilon_H$ changes, with some trajectories temporarily violating slow roll and ending inflation ($\epsilon_H\gtrsim1$). Despite these differences, all trajectories end up on the same attractor after crossing the inflection point, due to the `levelling' effect of the inflection point, suppressing the inflaton velocity regardless of the preceding dynamics. 

\begin{figure}
\centering
\captionsetup[subfigure]{justification=centering}
\begin{subfigure}[b]{0.8\textwidth}
  \centering
    \includegraphics[scale=0.5]{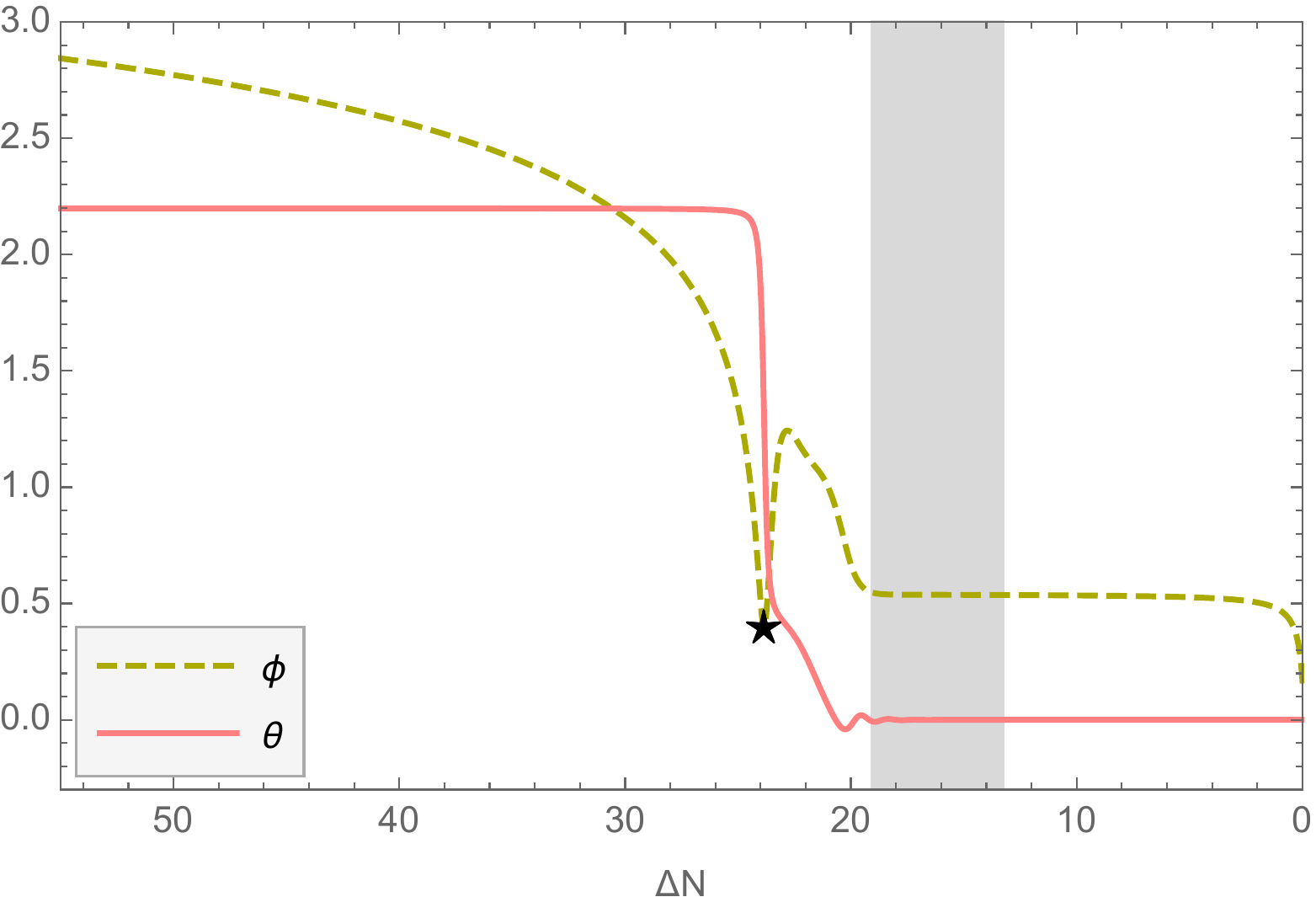}
  \end{subfigure}
  \caption{Numerical evolution of the fields for the multi-field model with parameters $\{\alpha=0.1,\, \phi_\text{infl}=0.542,\, \gamma=10\}$ and  $\theta_\text{in}=7\pi/10$. The black star signals the moment when $\phi=\sqrt{3\alpha/2}$ and the grey area corresponds to the radial field being within 1\% of the inflection point, $|\left(\phi-\phi_\text{infl}\right)/\phi_\text{infl}|\leq 0.01$.}
  \label{fig:multifield evo alpha0.1 one case fields evo}
\end{figure}

To get a better understanding of the background evolution we will focus on a single case. We select $\theta_\text{in}=7\pi/10$ and represent the evolution of $\phi$ and $\theta$ against $\Delta N$ in figure \ref{fig:multifield evo alpha0.1 one case fields evo}. When $\phi$ becomes comparable with the curvature length of the field space, $\sqrt{3\alpha/2}$, signaled by the black star in the plot, the angular field, $\theta$, which was previously frozen, starts evolving. The plot shows the transient change of direction of $\phi$ and its subsequent persistence at the inflection point before finally rolling down to the global minimum, ending inflation. In particular, the grey region highlights the phase of the evolution when the radial field, $\phi$, is within 1\% of the inflection point, $\phi_\text{infl}$.
\begin{figure}
\centering
\captionsetup[subfigure]{justification=centering}
   \begin{subfigure}[b]{0.48\textwidth}
    \includegraphics[width=\textwidth]{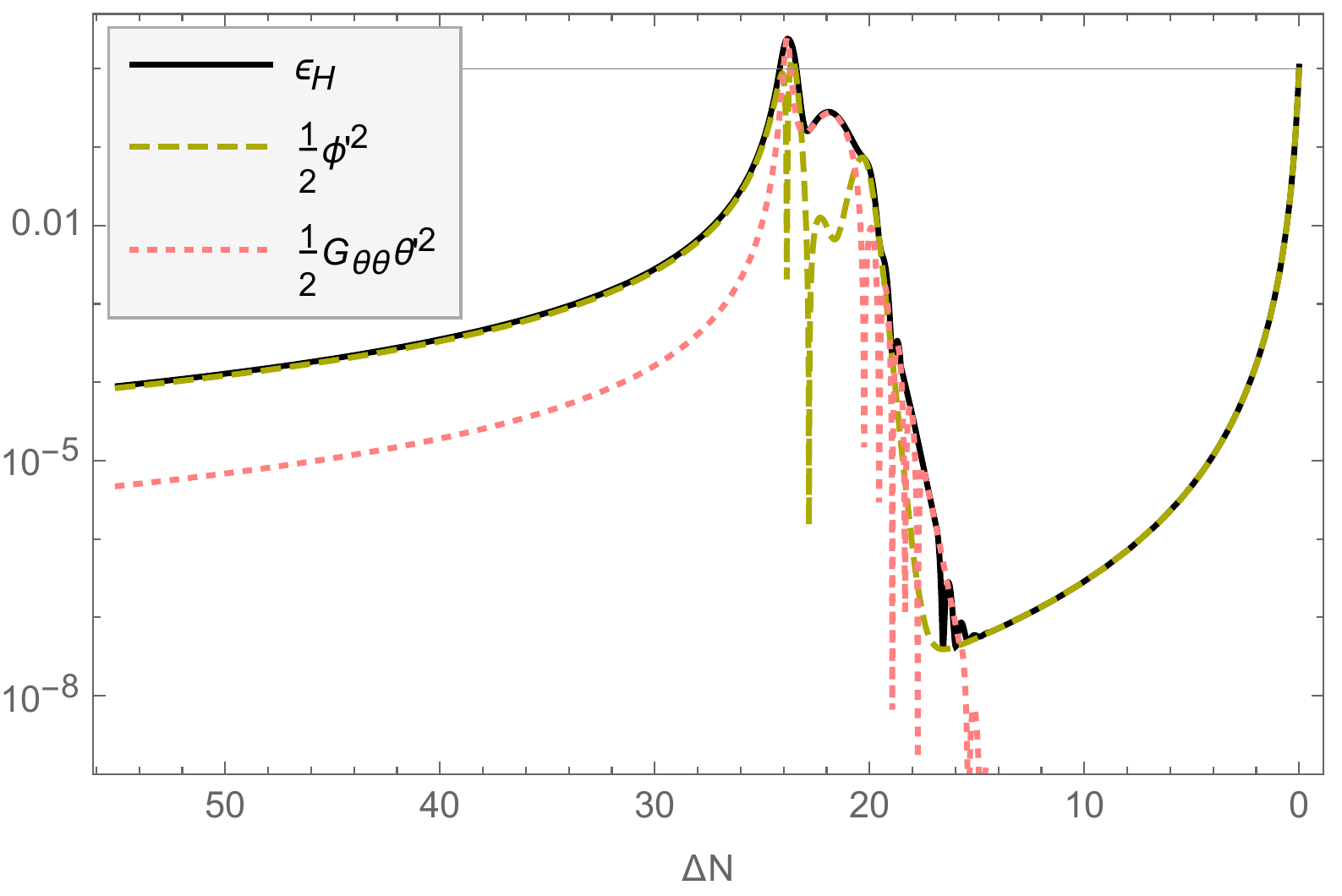}
  \end{subfigure}
   \begin{subfigure}[b]{0.49\textwidth}
    \includegraphics[width=\textwidth]{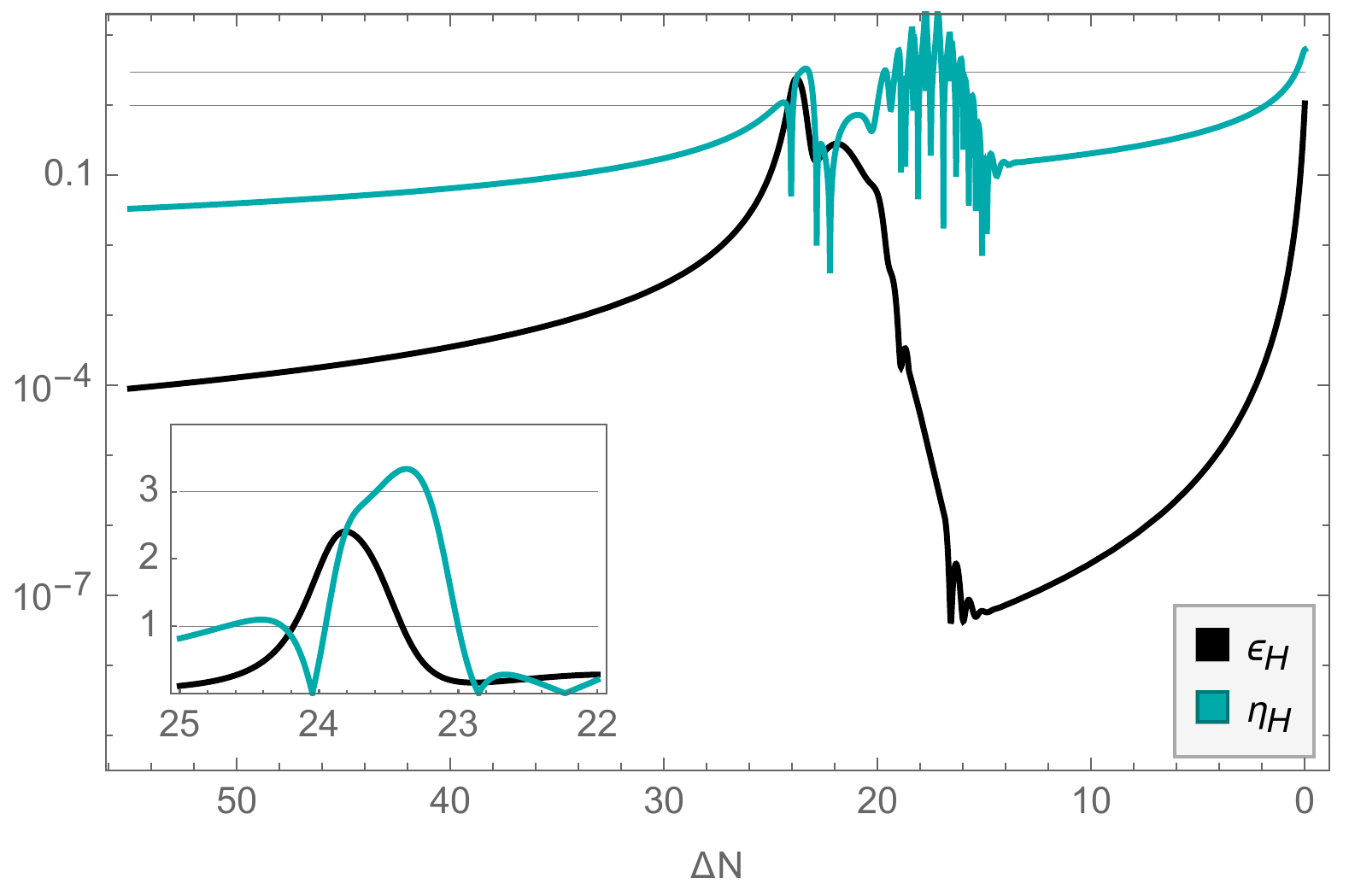}
  \end{subfigure}
\caption{\textit{Left panel:} slow-roll parameter $\epsilon_H$ (black line), decomposed into two parts coming from the kinetic energy of the radial (green-dashed line) and angular (pink-dotted line) fields. The horizontal grey line corresponds to $1$. \textit{Right panel:} $\epsilon_H$ (black line) and $|\eta_H|$ (blue line). The horizontal grey lines highlight the values 1 and 3 respectively.  Both panels show the model $\{\alpha=0.1,\, \phi_\text{infl}=0.542,\, \gamma=10\}$ with $\theta_\text{in}=7\pi/10$.}
  \label{fig:multifield evo alpha0.1 one case epsilon eta}
\end{figure}

In the left panel of figure \ref{fig:multifield evo alpha0.1 one case epsilon eta} $\epsilon_H$ is plotted for this case, together with its two component parts coming from the evolution of $\phi$ (green-dashed) and $\theta$ (pink-dotted), see eq.~\eqref{EoM multifield 1}. At the beginning, $\epsilon_H$ is dominated by the kinetic energy of $\phi$, which is slowly rolling towards smaller values. Then, when $\phi\approx\sqrt{3\alpha/2}$, $\theta$ gets released, its kinetic energy becomes comparable to that of $\phi$, and $\phi$ changes direction. The simple ultra-slow-roll behaviour of $\epsilon_H$ observed in the single-field case (see e.g., figure~\ref{fig:background evo}) is modified due to the change of direction of $\phi$ and the contribution of $\theta$, which oscillates around its minimum. Overall $\epsilon_H$ decreases, until $\phi$ crosses the inflection point and rolls away from it towards the global minimum, bringing inflation to an end. One can see that, similar to the single-field case, inflation is made up of two slow-roll phases driven by $\phi$, separated by an intermediate phase with rapidly decreasing $\epsilon_H$. The transition between the two slow-roll solutions is an effect of the destabilisation induced in the background trajectory by the hyperbolic geometry of field space, see the discussion in section \ref{sec: multi-field dynamics}. 

In the right panel of figure \ref{fig:multifield evo alpha0.1 one case epsilon eta} the second slow-roll parameter, $\eta_H$ defined in \eqref{eta H}, is plotted against $\Delta N$ together with $\epsilon_H$. The first and last phases of inflationary evolution are distinguished by slow roll where $\epsilon_H\ll|\eta_H|$, with an intermediate interval in which slow roll is violated, $|\eta_H|\gtrsim1$. In particular, $|\eta_H|\simeq3$, signals a very brief (less than 1 e-fold) ultra-slow-roll phase, as shown in the inset plot. In this example the first slow-roll parameter, $\epsilon_H$, also briefly exceeds unity, signalling that inflation is interrupted (also for less than one e-fold) about this point, sometimes referred to as ``punctuated'' inflation~\cite{Jain:2008dw,Ragavendra:2020sop}.
\begin{figure}
\centering
\captionsetup[subfigure]{justification=centering}
   \begin{subfigure}[b]{0.55\textwidth}
    \includegraphics[width=\textwidth]{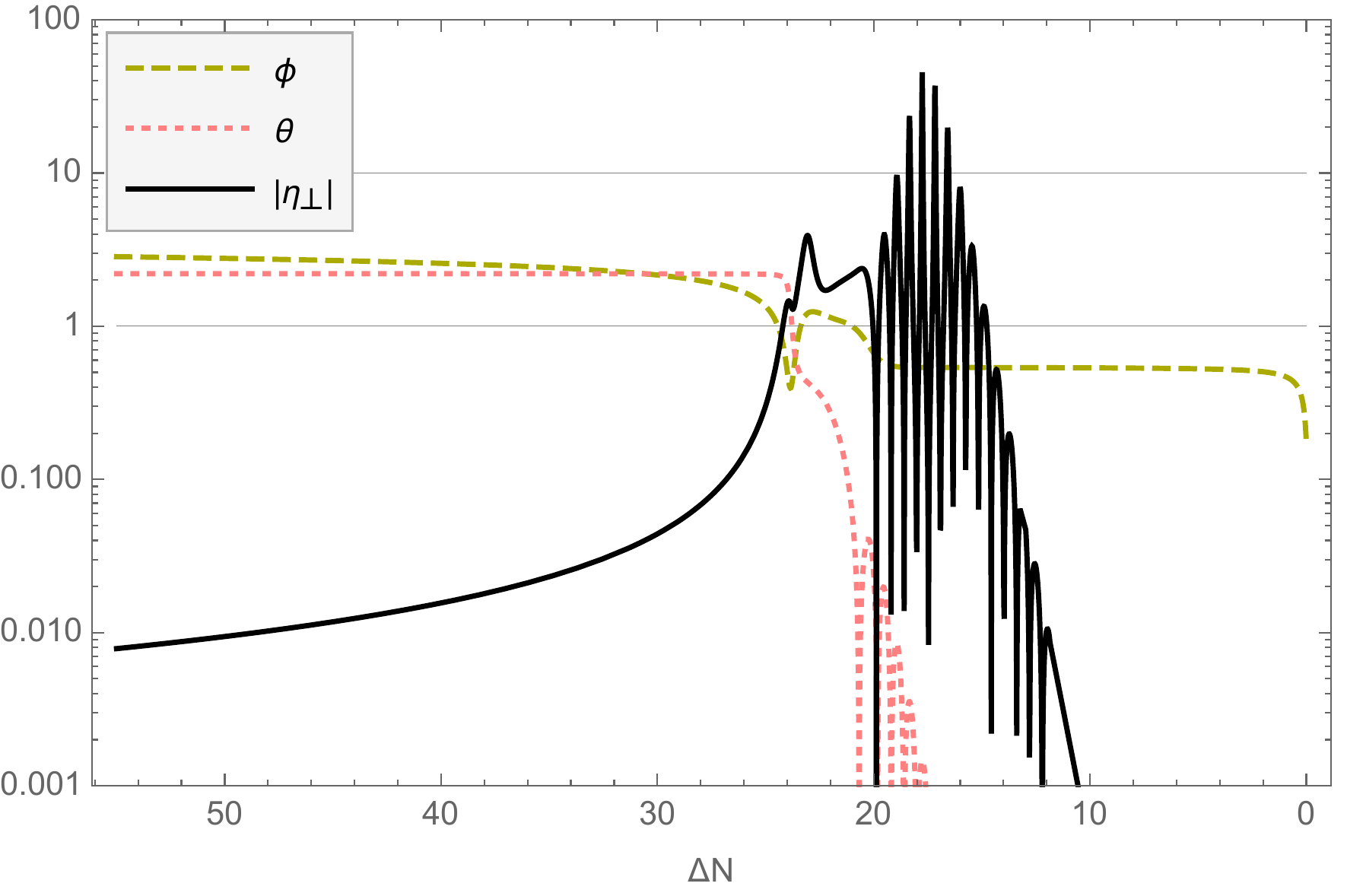}
  \end{subfigure}
   \begin{subfigure}[b]{0.445\textwidth}
    \includegraphics[width=\textwidth]{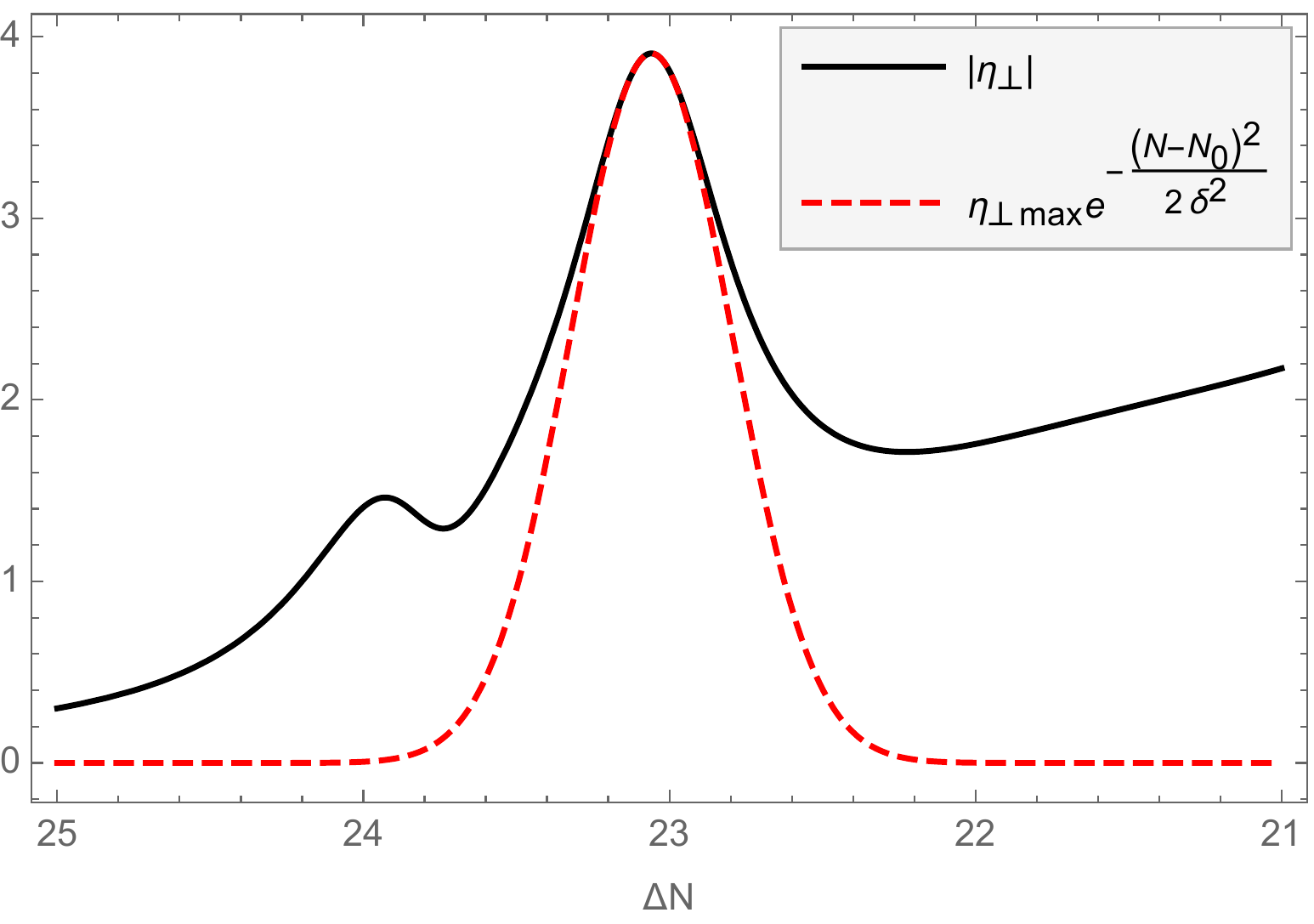}
  \end{subfigure}
  \begin{subfigure}[b]{0.45\textwidth}
    \includegraphics[width=\textwidth]{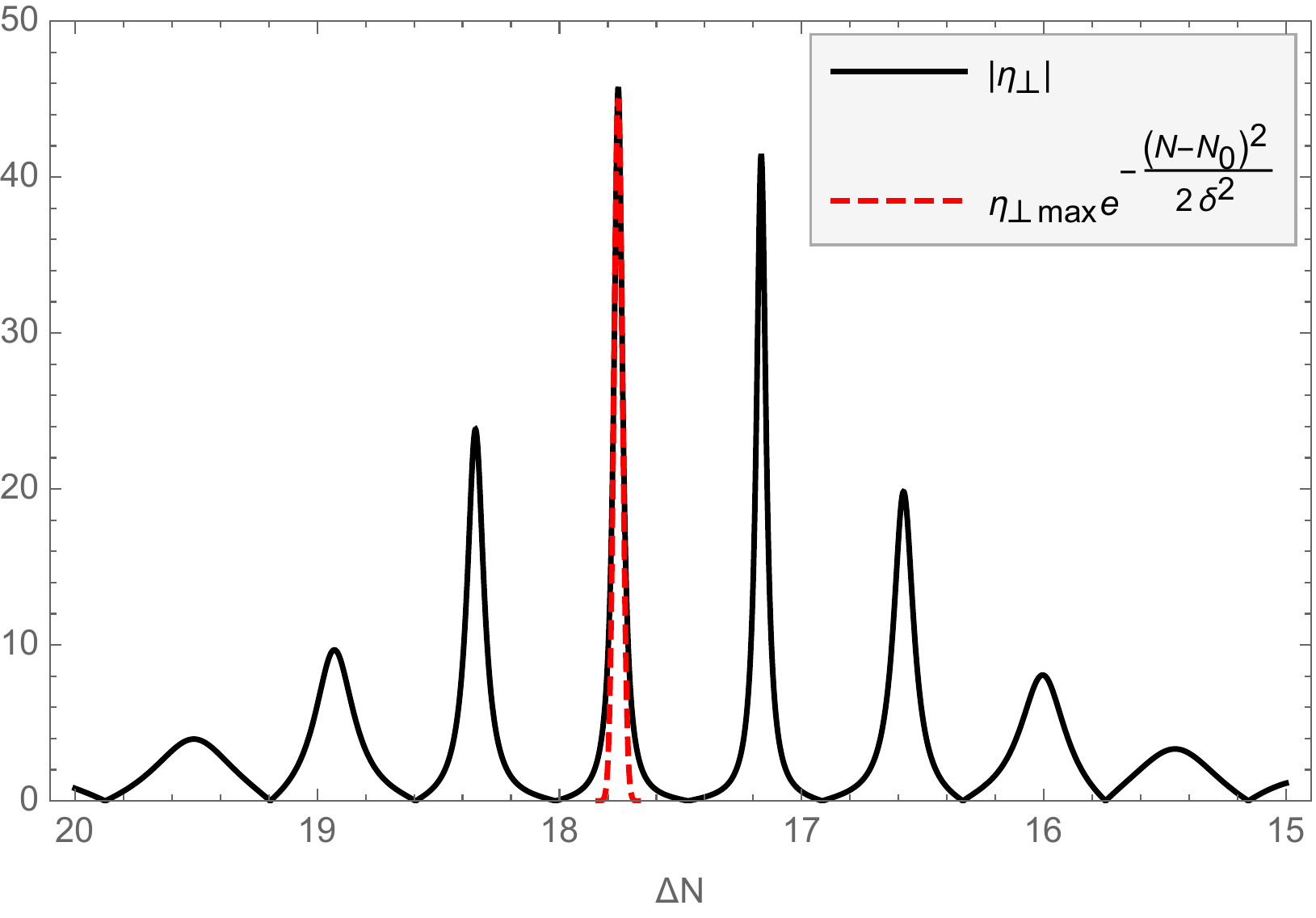}
  \end{subfigure}
 \caption{Evolution of the absolute value of the turn rate, $|\eta_\perp|$, with respect to $\Delta N$ (black line), shown together with the field values, $\phi$ (green-dashed line) and $\theta$ (pink-dotted line), for the model $\{\alpha=0.1,\, \phi_\text{infl}=0.542,\, \gamma=10\}$ and $\theta_\text{in}=7\pi/10$. The two thin horizontal lines highlight the values 1 and 10. The bottom panels show blow-ups of the behaviour of $|\eta_\perp|$ in restricted ranges of $\Delta N$. The red-dashed lines show a Gaussian fit to the evolution about the maxima, see eq.~\eqref{eta perp gaussian profile}.}
  \label{fig:multifield evo alpha0.1 one case eta perp}
\end{figure}

From the results above it is clear that the potential \eqref{potential multifield} can produce a rich background evolution whose properties depend on the initial condition $\theta_\text{in}$. Although we selected $\theta_\text{in}=7\pi/10$ as an example, each case will be different, e.g., not all $\theta_\text{in}$ would produce $|\eta_H|\gtrsim3$. 

As reviewed in section~\ref{sec: multi-field dynamics}, a strong turn in field space ($\eta_\perp\gg1$) and/or a highly curved field space ($\mathcal{R}_\text{fs}\ll-1$) can lead to a situation in which enhanced isocurvature perturbations source the curvature fluctuation, with the coupling between them set by the bending parameter, $\eta_\perp$, see \eqref{eta perp}. In the top panel of figure~\ref{fig:multifield evo alpha0.1 one case eta perp} we represent the evolution of the absolute value of $\eta_\perp$ for the same model considered above, $\{\alpha=0.1,\, \phi_\text{infl}=0.542,\, \gamma=10, \theta_\text{in}=7\pi/10\}$, together with $\phi(N)$ and $\theta(N)$. In the first slow-roll phase, when $\theta$ is effectively frozen, $\eta_\perp\ll1$. When $\theta$ is released and starts evolving, $\eta_\perp$ becomes $\mathcal{O}(1)$, signalling a turning trajectory. In order to compare with the results previously presented, e.g., in~\cite{Fumagalli:2020adf, Fumagalli:2020nvq}, we fit the shape of $\eta_\perp$ around the peak with the Gaussian profile
\begin{equation}
\label{eta perp gaussian profile}
    \eta_\perp(N) = \eta_{\perp,\,\text{max}}\, e^{-\frac{(N-N_0)^2}{2\delta^2}} \;, 
\end{equation}
where $\delta^2\ll1$ signals sharp turns in field space. In the bottom-left panel of figure \ref{fig:multifield evo alpha0.1 one case eta perp} we zoom in on the first localised peak of $\eta_\perp$ and plot it together with the Gaussian profile in \eqref{eta perp gaussian profile} described by $(\eta_{\perp,\,\text{max}}=3.9, \, N_0=23, \delta^2= 0.07)$. The (sharp) bending is not as large as considered, e.g., in~\cite{Fumagalli:2020adf} for producing PBHs. During the subsequent field evolution, the oscillations that the field $\theta$ performs around its minimum are reflected in oscillations of $\eta_\perp$, signalling a series of turns. We zoom into $20\leq \Delta N\leq15$ in the bottom-right panel of figure~\ref{fig:multifield evo alpha0.1 one case eta perp}, where we fit the peak with largest amplitude with the Gaussian profile \eqref{eta perp gaussian profile} and parameters $(\eta_{\perp,\,\text{max}}=45, \, N_0=17.7, \delta^2= 0.0003)$. Again, these turns in field space are strong and sharp.

\begin{figure}
\centering
\captionsetup[subfigure]{justification=centering}
   \begin{subfigure}[b]{0.48\textwidth}
    \includegraphics[width=\textwidth]{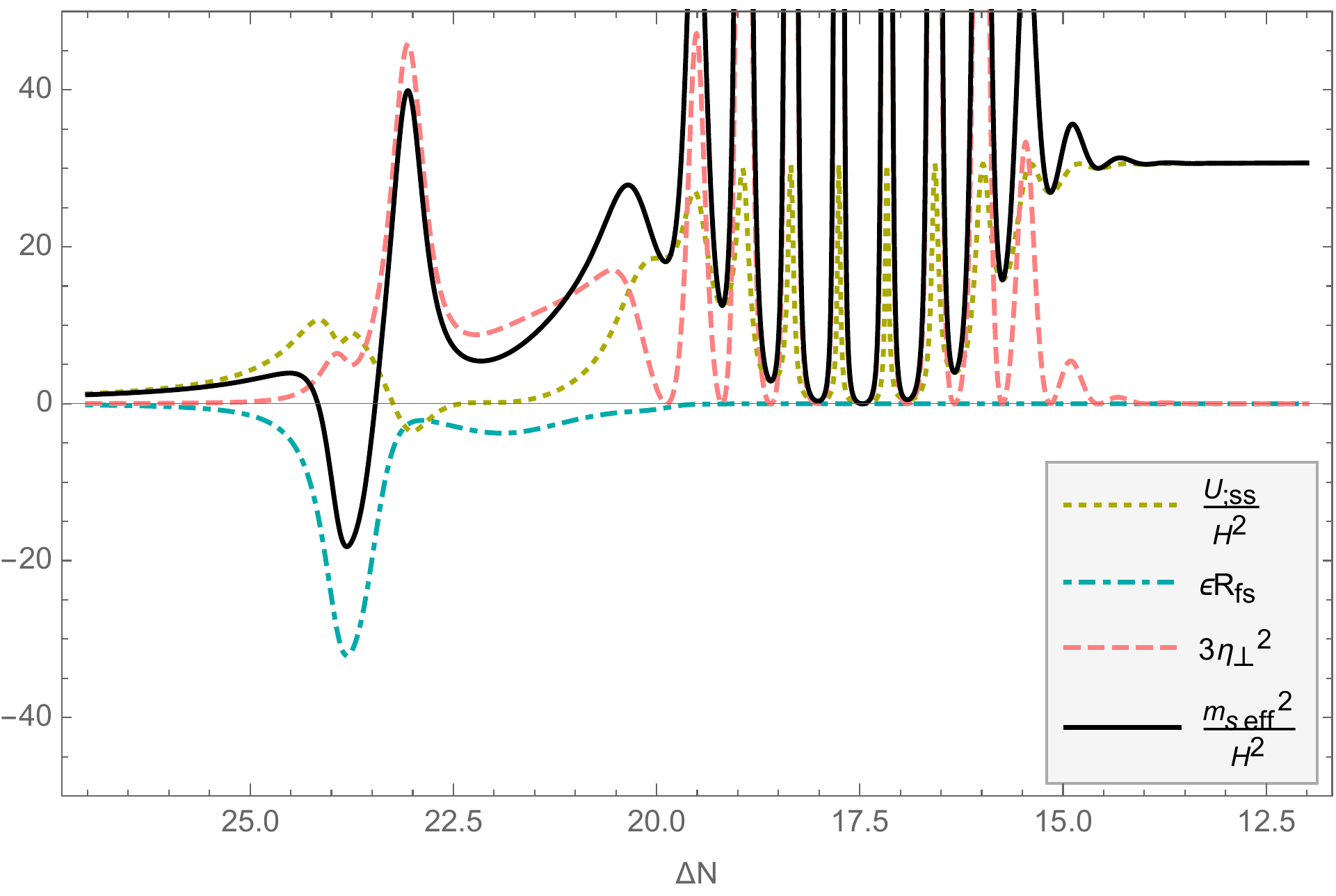}
  \end{subfigure}
   \begin{subfigure}[b]{0.48\textwidth}
    \includegraphics[width=\textwidth]{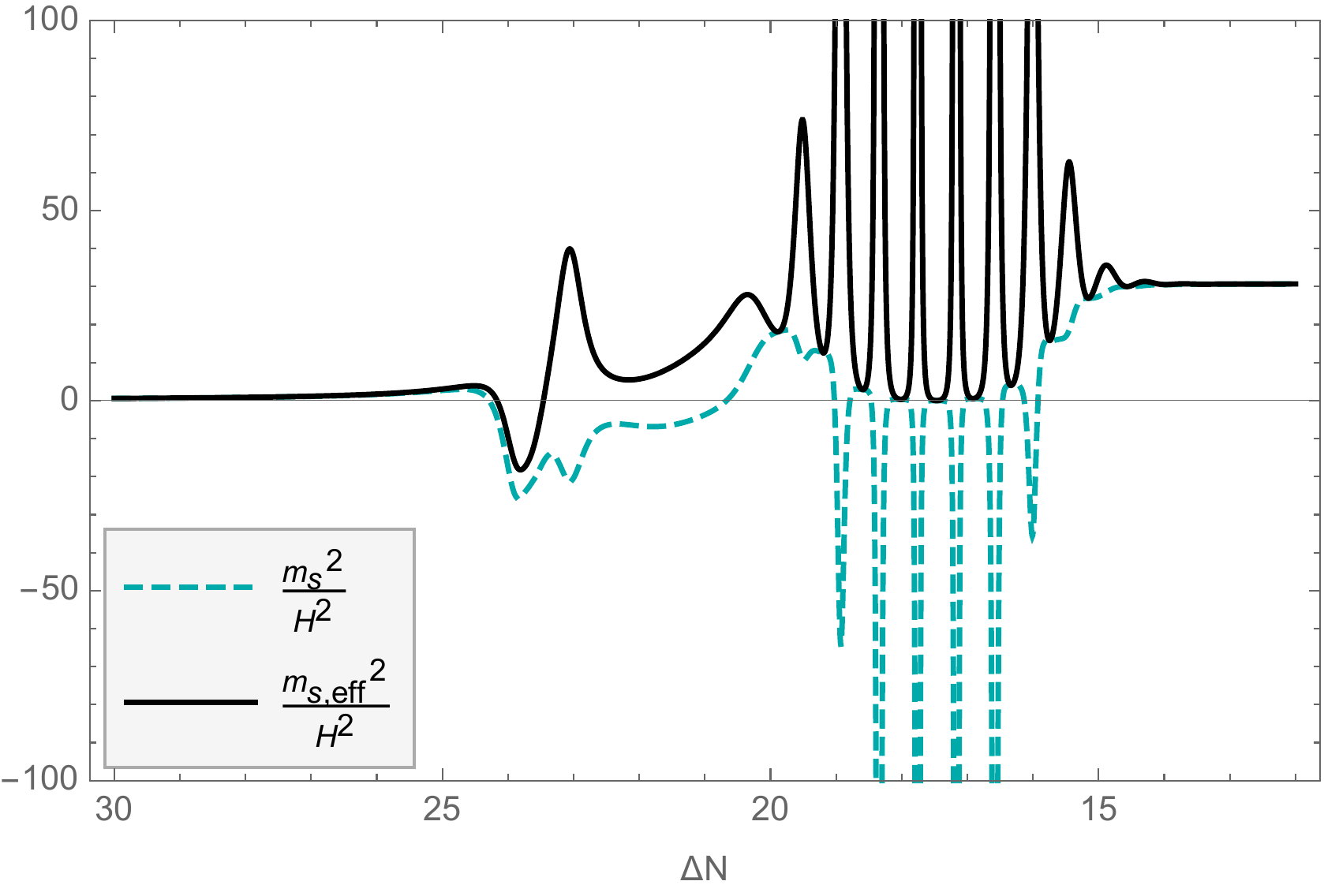}
  \end{subfigure}
\caption{Evolution of the effective squared-mass of the isocuvature perturbation together with its contributions (left) and comparison with the squared-mass (right) for the model $\{\alpha=0.1,\, \phi_\text{infl}=0.542,\, \gamma=10\}$ and $\theta_\text{in}=7\pi/10$.}
  \label{fig:multifield evo alpha0.1 one case masses}
\end{figure}

The behaviour of the isocurvature perturbation is determined by its squared-mass \eqref{isocurvature mass} and its super-horizon effective squared-mass \eqref{eff mass isocurvature}. We display ${m_{s,\, \text{eff}}}^2/H^2$ in the left panel of figure~\ref{fig:multifield evo alpha0.1 one case masses}. Around $24$ e-folds before the end of inflation the super-horizon effective squared-mass turns negative, signalling a destabilisation of the background trajectory, and a transient instability of the isocurvature perturbation for the super-horizon modes. The plot displays several coloured lines accounting for the different components of ${m_{s,\, \text{eff}}}^2/H^2$, see \eqref{eff mass isocurvature}. In particular, it is the geometrical contribution $\epsilon_1 \mathcal{R}_\text{fs}$ that causes the squared-mass to become negative, along the lines of what was investigated in~\cite{Turzynski:2014tza,Renaux-Petel:2015mga, Garcia-Saenz:2018ifx} (see also~\cite{Braglia:2020eai}). In the right panel we plot ${m_{s}}^2/H^2$ and ${m_{s,\, \text{eff}}}^2/H^2$ together. The difference between the squared-mass and the effective squared-mass is due to the contribution from the turn rate, which adds a negative contribution ($-\eta_\perp^2$) to ${m_{s}}^2/H^2$, and a positive contribution ($+3\eta_\perp^2$) to ${m_{s,\, \text{eff}}}^2/H^2$ on super-horizon scales. The negative contributions from the geometry and the strong turn drive ${m_{s}}^2/H^2$ to negative values, signalling a tachyonic growth of the isocurvature perturbations.
\begin{figure}
\centering
\includegraphics[scale=0.6]{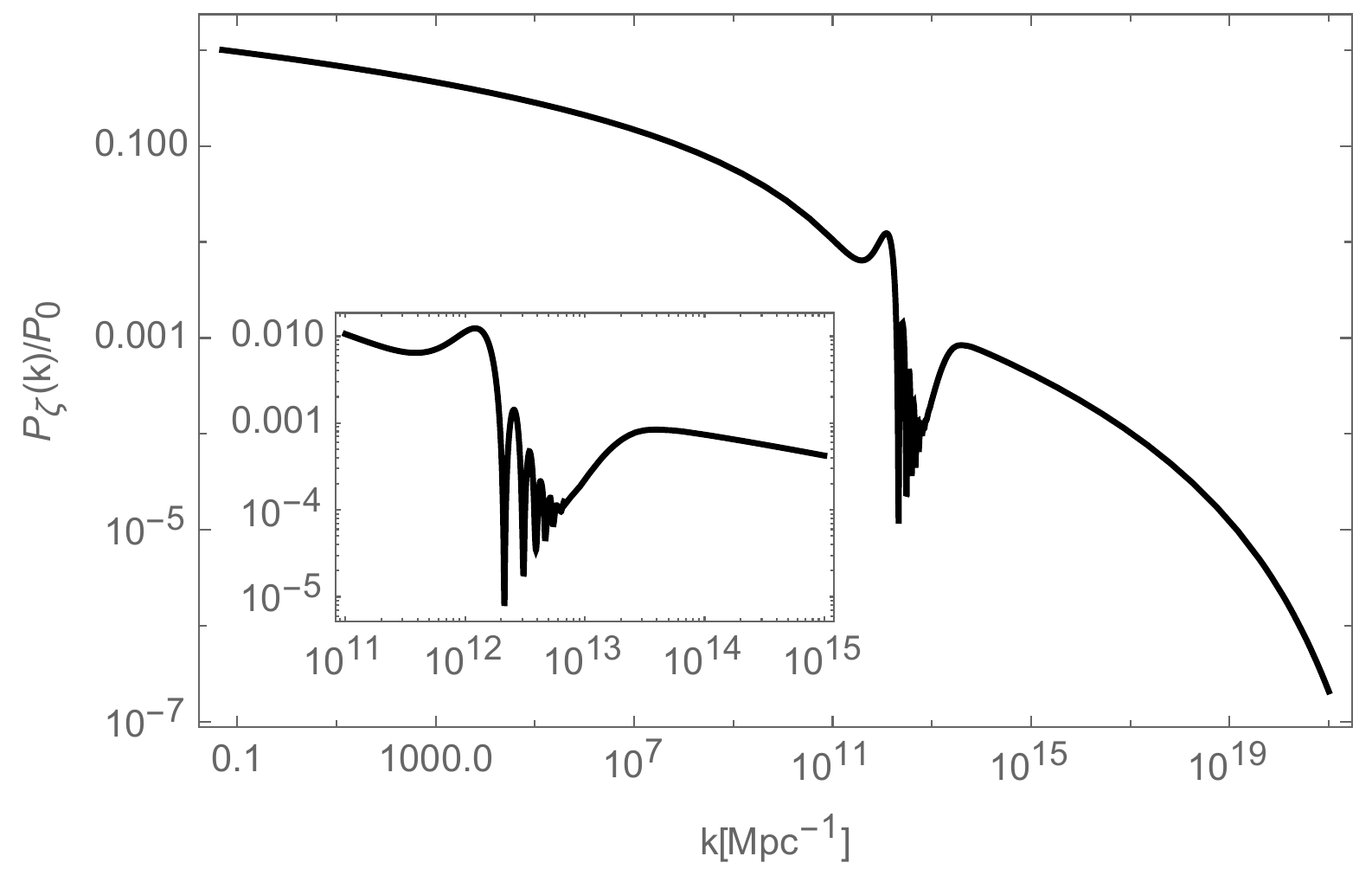}
\caption{Normalised scalar power spectrum, $P(k)/P_0$, for the model described by $\{\alpha=0.1,\, \phi_\text{infl}=0.542,\, \gamma=10\}$ and initial condition $\theta_\text{in}=7\pi/10$. Here $P_0=2.1\times10^{-9}$.}
  \label{fig:multifield one case power spectrum}
\end{figure}

We numerically evaluate the resulting scalar power spectrum, $P_\zeta(k)$, for this model using the \textit{mTransport} Mathematica code provided in~\cite{Dias:2015rca}, with $\Delta N_\text{CMB}=55$. In figure~\ref{fig:multifield one case power spectrum} we represent the power spectrum, $P_\zeta(k)/P_0$, normalised at $k_\text{CMB}=0.05\,\text{Mpc}^{-1}$ where $P_0=2.1\times10^{-9}$. As expected, on small scales the power spectrum grows due to the transient instability of the isocurvature perturbation, displaying a local peak around $10^{12}\,\text{Mpc}^{-1}$. In this example the growth is very limited and it does not lead to an overall enhancement with respect to the power spectrum on CMB scales. In terms of the characteristics of the localised turn in field space, i.e., its maximum amplitude, $\eta_{\perp,\,\text{max}}$, and its duration, $\delta$, the overall amplification of $P_\zeta(k)/P_0$ following a strong turn is roughly given by the factor $e^{\eta_\perp \delta}$~\cite{Fumagalli:2020nvq}. In this case, for the first local peak of the bending parameter this factor is only $\sim 2.8$, which is consistent with the limited growth that we see. 

The sharp turn in the field-space trajectory happening around $\Delta N =23$ (see the bottom-left panel of figure \ref{fig:multifield evo alpha0.1 one case eta perp}) results in an oscillatory pattern in $P_\zeta(k)$ shown in figure~\ref{fig:multifield one case power spectrum}, which is magnified in the inset plot. The decrease in $\epsilon_H(N)$ about the inflection point (see figure~\ref{fig:multifield evo alpha0.1 one case epsilon eta}) explains the subsequent local maximum in $P_\zeta(k)$, around $k=3\times10^{13}\,\text{Mpc}^{-1}$. Although the subsequent evolution displays many sharp turns in field space (as shown in the bottom-right panel in figure~\ref{fig:multifield evo alpha0.1 one case eta perp}) as $\theta$ oscillates about its minimum, the resulting features in the scalar power spectrum are suppressed relative to the first peak. Eventually the evolution returns to slow-roll, as seen on scales $k\gtrsim 10^{14}\,\text{Mpc}^{-1}$, and the power spectrum gradually decreases as $\epsilon_H(N)$ grows. 
\begin{figure}
\centering
\includegraphics[scale=0.45]{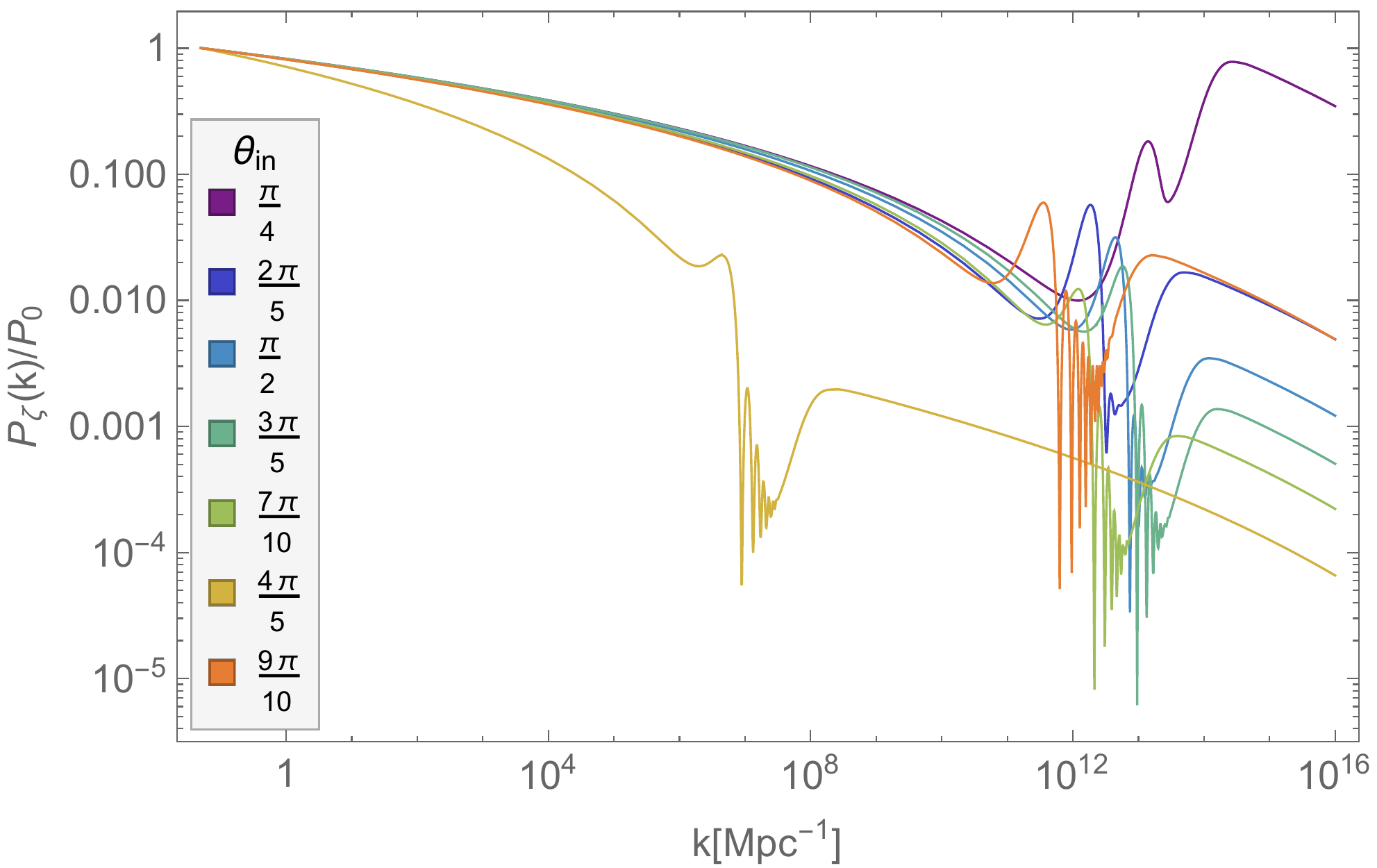}
\caption{Normalised primordial scalar power spectrum for the potential \eqref{potential multifield} with $\{\alpha=0.1, \, \phi_\text{infl}=0.542, \, \gamma=10\}$ and different initial conditions $\theta_\text{in}$. Here $P_0=2.1\times10^{-9}$.}
  \label{fig:multifield alpha 0.1 power spectra}
\end{figure}

In figure~\ref{fig:multifield alpha 0.1 power spectra} we show $P_\zeta(k)/P_0$ resulting from the potential \eqref{potential multifield} with the same model parameters $\{\alpha=0.1, \, \phi_\text{infl}=0.542, \, \gamma=10\}$ but different choices of the initial condition $\theta_\text{in}$.  Each initial condition leads to a different outcome and with this choice of parameters the largest enhancement is produced with $\theta_\text{in}=\pi/4$. Despite the rich and diverse behaviour, one can see that, for the model with $\alpha=0.1$, none of the cases considered here can produce a significant amplification of the scalar power spectrum on small scales above the power on CMB scales. 

\subsection{Changing the hyperbolic field-space curvature}
\label{sec: change alpha multifield}

In the previous section we considered the dynamics for $\alpha=0.1$. In order to investigate the role of the field-space curvature we consider now models with $\alpha=0.01$ and $\alpha=0.005$, which correspond to $\mathcal{R}_\text{fs}\simeq -133.3$ and $\mathcal{R}_\text{fs}\simeq -266.6$ respectively. We study a fixed initial angular direction, $\theta_\text{in}=7\pi/10$, and we also fix $\gamma=10$ to facilitate the comparison. For each $\alpha$, we select the value of $\phi_\text{infl}$ in such a way that the power spectrum starts to grow roughly at the same comoving scale $k$. In particular, we study three different configurations of the potential \eqref{potential multifield}, corresponding to
\begin{align*}
   \text{model}_1 &\rightarrow\{\alpha=0.1,\, \phi_\text{infl}=0.5417\},\\
   \text{model}_2 &\rightarrow\{ \alpha=0.01,\, \phi_\text{infl}=0.19\}, \\
   \text{model}_3
   &\rightarrow\{ \alpha=0.005,\, \phi_\text{infl}=0.103\} \;.
\end{align*}
In the following, we identify each model by the corresponding value of $\alpha$.  

\begin{figure}
\centering
\captionsetup[subfigure]{justification=centering}
   \begin{subfigure}[b]{0.44\textwidth}
    \includegraphics[width=\textwidth]{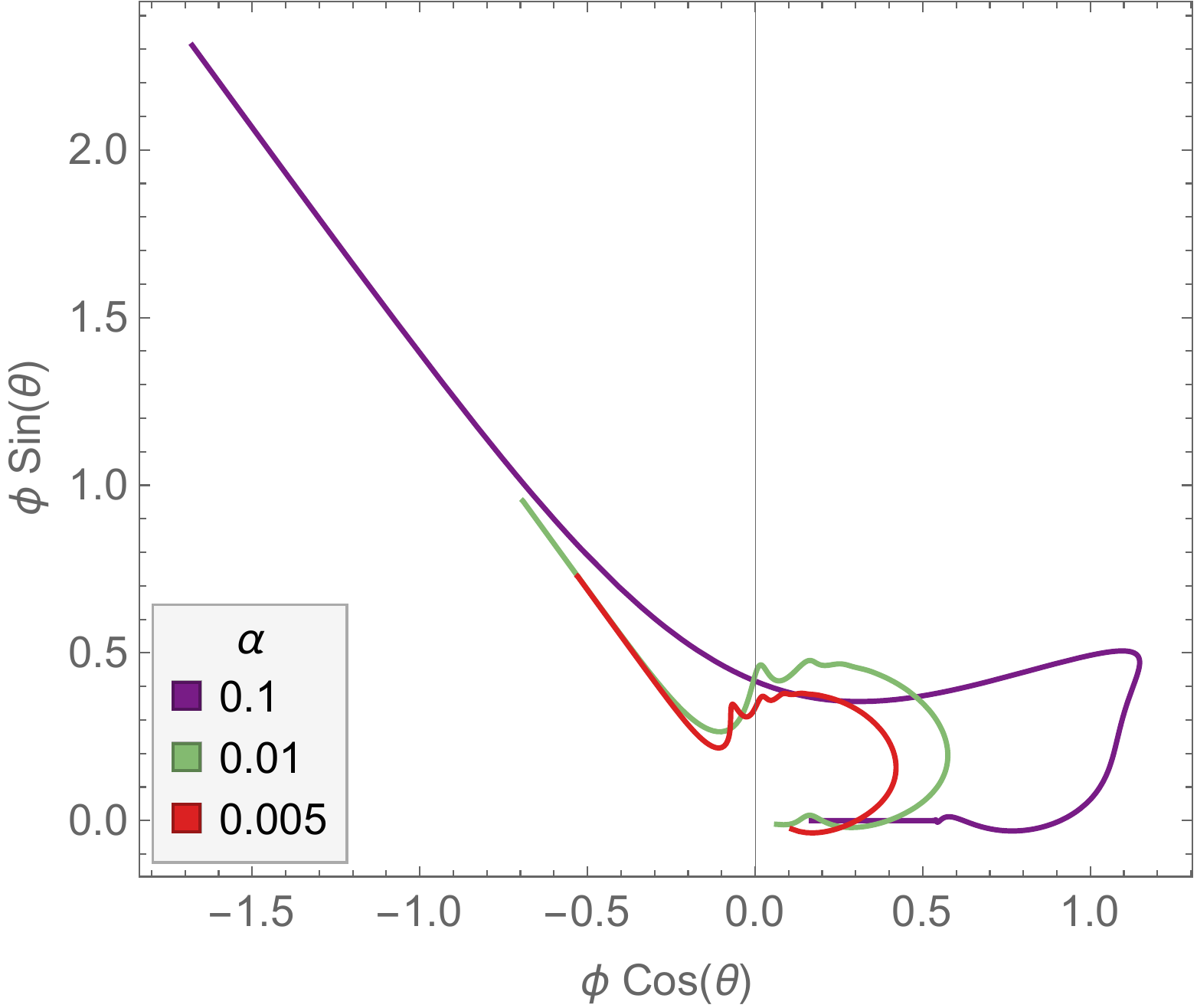}
  \end{subfigure}
   \begin{subfigure}[b]{0.46\textwidth}
    \includegraphics[width=\textwidth]{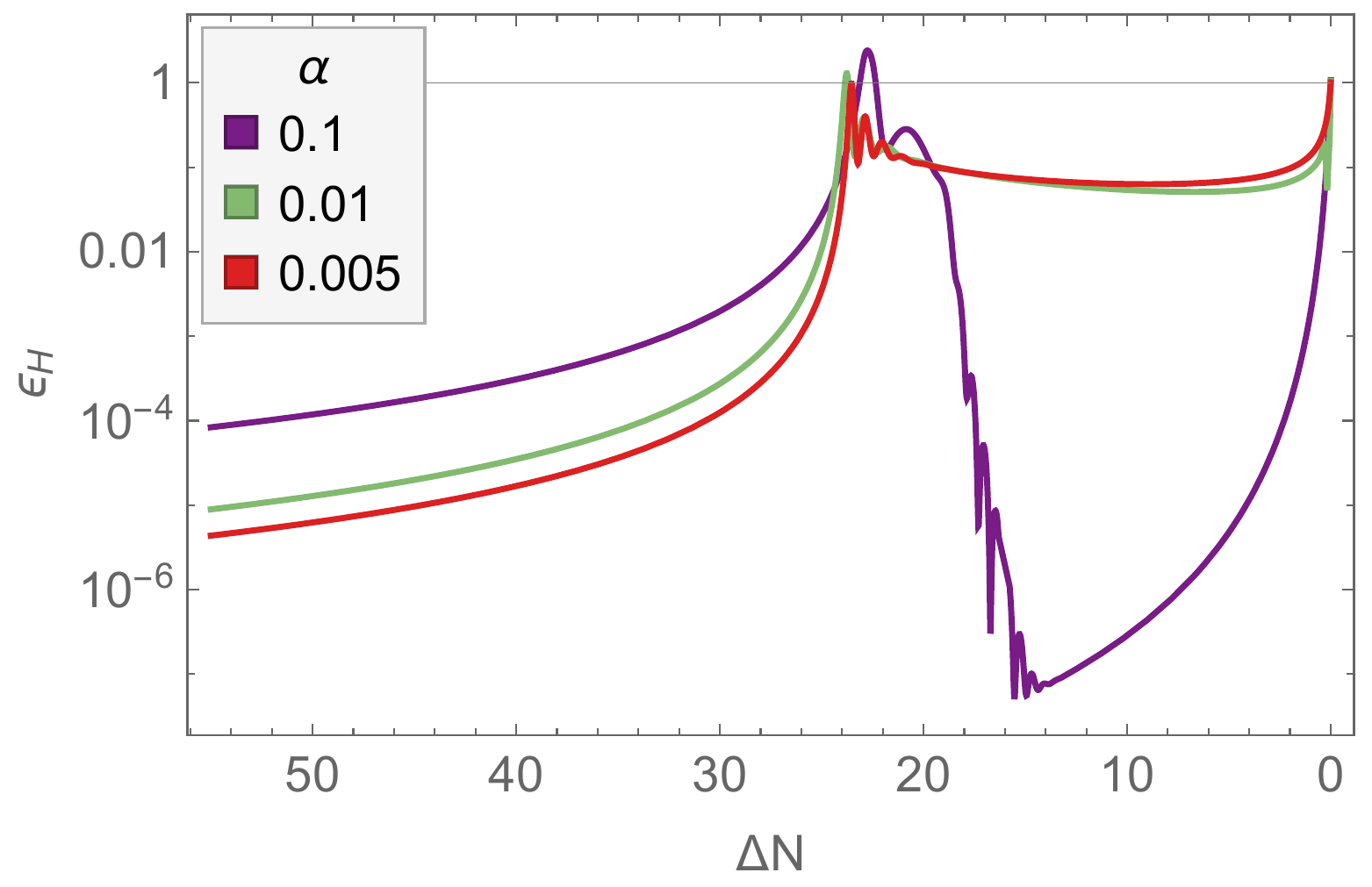}
  \end{subfigure}
\caption{Numerical background evolution for three models with different values for the curvature of the hyperbolic field space, parameterised by $\alpha$. \textit{Left panel}: background trajectories $\{\phi(N) \cos{\theta(N)},\, \phi(N) \sin{\theta(N)}\}$.  \textit{Right panel:} evolution of $\epsilon_H$ with respect to $\Delta N$.}
  \label{fig:multifield change alpha trajetcories}
\end{figure}

We obtain the background trajectories by numerically solving eqs.~\eqref{EoM multifield 1}--\eqref{EoM multifield 3} and represent them in the left panel of figure \ref{fig:multifield change alpha trajetcories}. We parametrise the trajectories in a slightly different fashion with respect to what was done previously, e.g., in figure~\ref{fig:multifield evo alpha0.1 vary thetain}, by plotting $\{\phi(N) \cos{\theta(N)},\, \phi(N) \sin{\theta(N)}\}$ in the last $55$ e-folds of inflation, while in the right panel we display the evolution of $\epsilon_H$ against $\Delta N$. 

Cases with $\alpha=\{0.01,\, 0.005\}$ clearly differ with respect to the evolution for $\alpha=0.1$. The two models with smaller $\alpha$ are characterised by a transient phase of angular inflation, which is defined as a regime in which the field's motion is mostly along the angular direction, with $\phi'(N)$ suppressed~\cite{Christodoulidis:2018qdw}.
\begin{figure}
\centering
\captionsetup[subfigure]{justification=centering}
   \begin{subfigure}[b]{0.45\textwidth}
    \includegraphics[width=\textwidth]{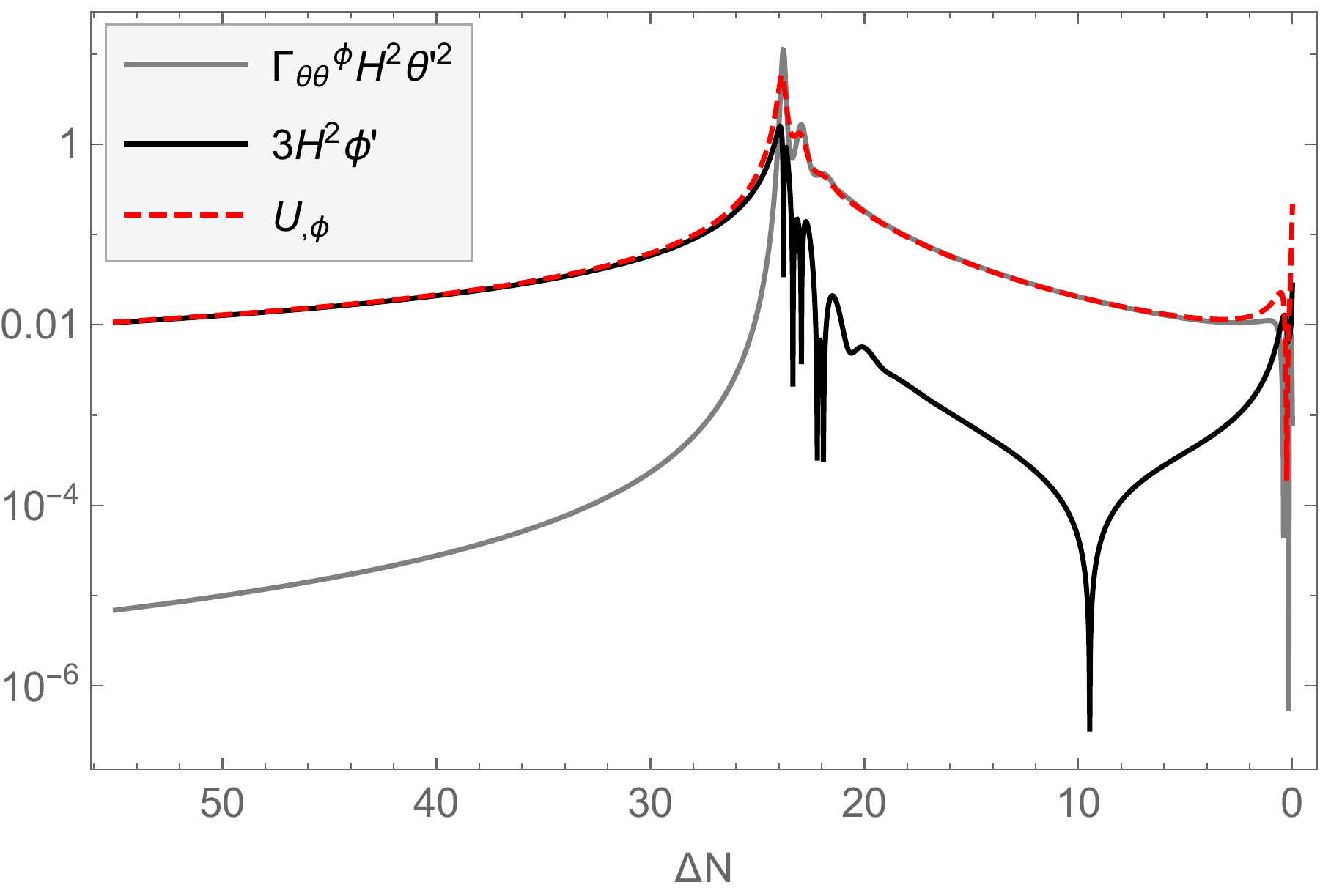}
  \end{subfigure}
   \begin{subfigure}[b]{0.45\textwidth}
    \includegraphics[width=\textwidth]{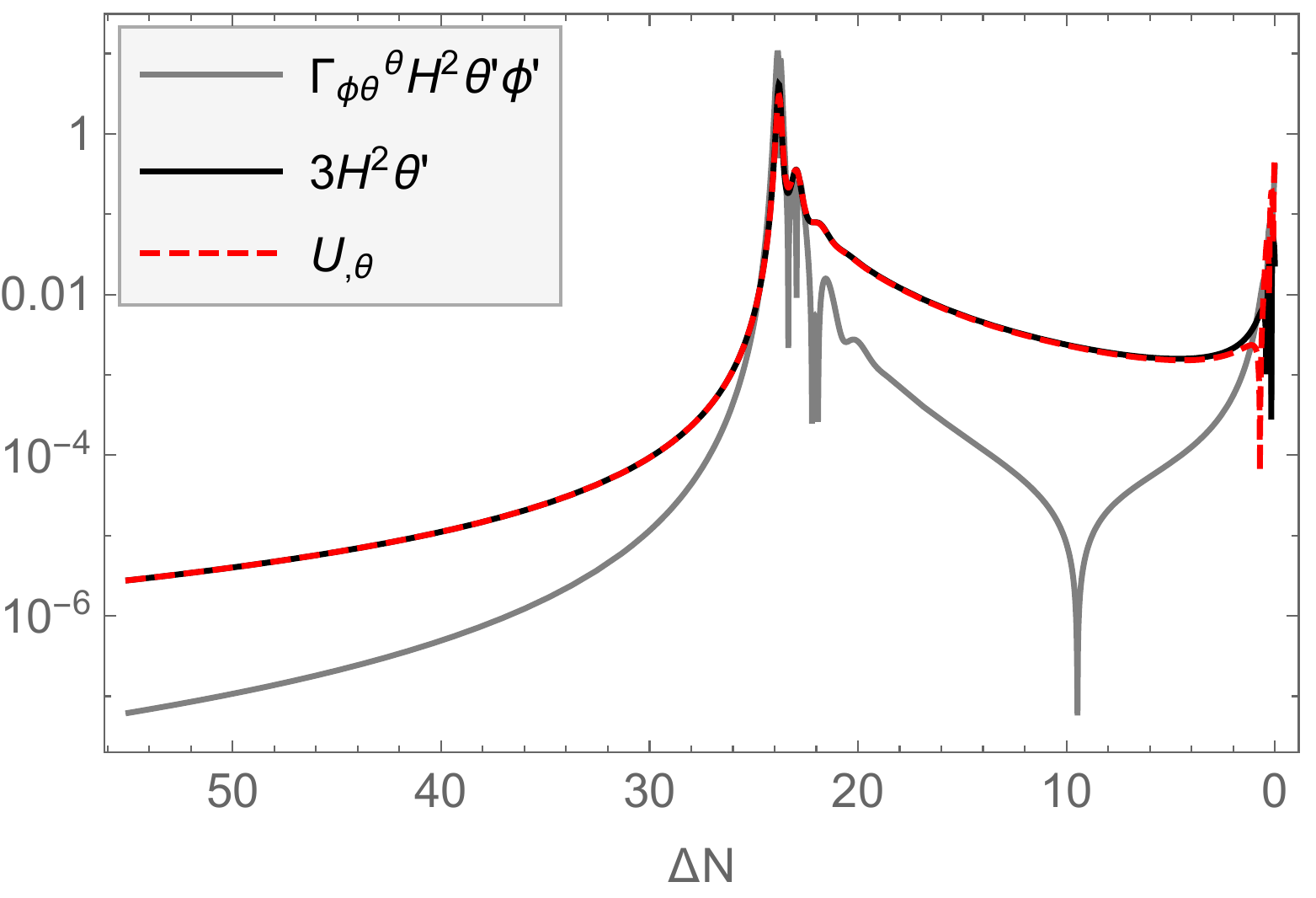}
  \end{subfigure}
\caption{Terms contributing to the equations of motion \eqref{EoM multifield 2} and~\eqref{EoM multifield 3} for the radial field, $\phi$ (left panel), and angular field, $\theta$ (right panel), obtained numerically for the model $\{\alpha=0.01,\, \phi_\text{infl}=0.19,\, \gamma=10\}$ and $\theta_\text{in}=7\pi/10$.}
  \label{fig:multifield change alpha angular inflation regime}
\end{figure}
%
We show in figure \ref{fig:multifield change alpha angular inflation regime} the terms contributing to the equations of motion for $\phi$ and $\theta$, eqs.~\eqref{EoM multifield 2} and~\eqref{EoM multifield 3}, for the model with $\alpha=0.01$. 
Only the models with $\alpha=\{0.01,\,0.005\}$ lead to a phase of angular inflation as these values correspond to a large field-space curvature $\mathcal{R}_\text{fs}$, which destabilises the background trajectory into the new attractor solution. During angular inflation, the geometry of the field space pushes the radial field towards the larger volume in field space at the boundary of the Poincaré disc~\cite{Christodoulidis:2018qdw}. The radial field remains approximately constant while the potential gradient is balanced against the geometrical effect, and the angular field slow rolls, see figure \ref{fig:multifield change alpha angular inflation regime}.
\begin{figure}
\centering
\includegraphics[scale=0.5]{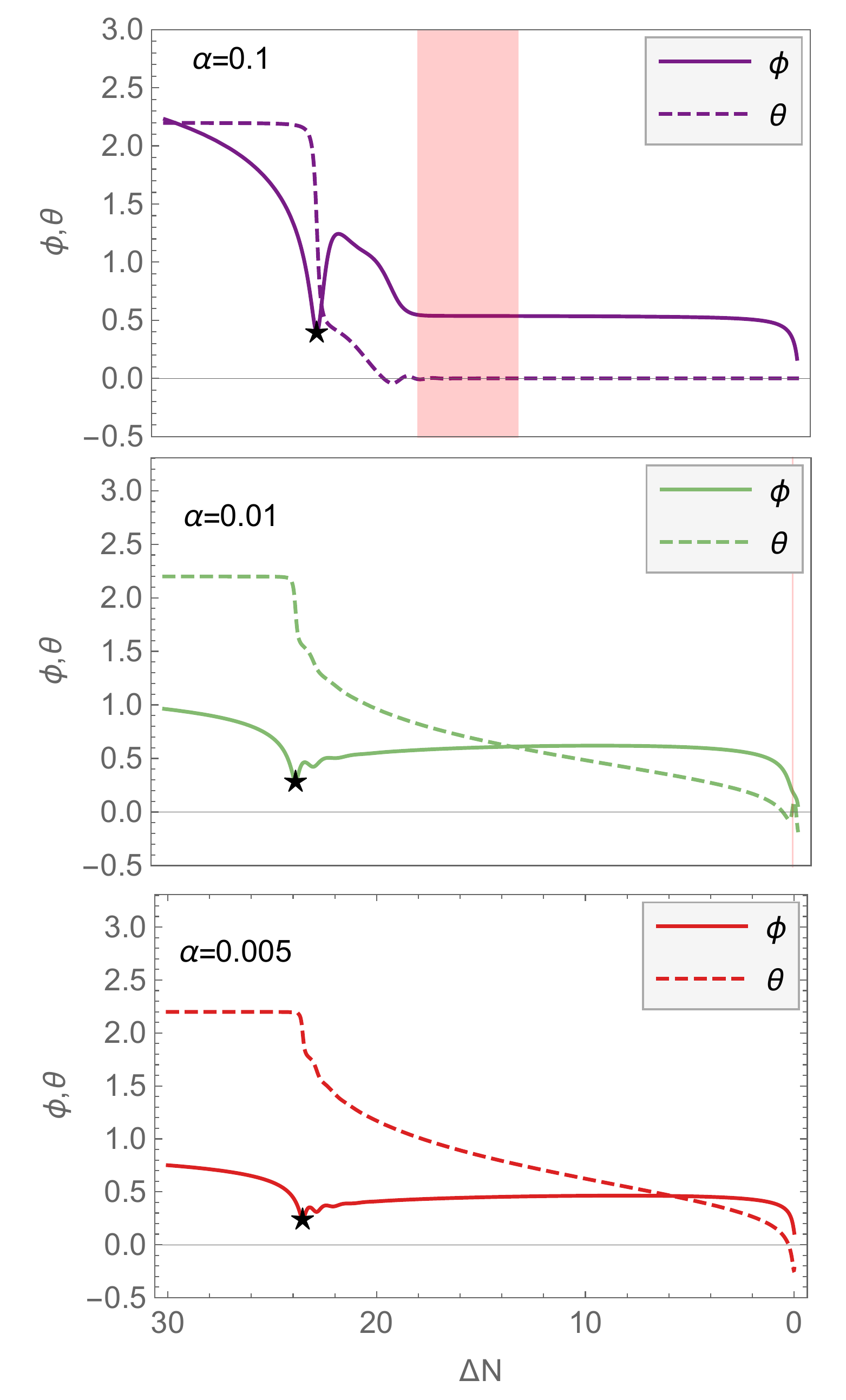}
\caption{Evolution of the radial (continuous line) and angular (dashed line) fields against $\Delta N$ in the last 30 e-folds of inflation, with each panel corresponding to one of the models discussed, see the value of $\alpha$ in the top-left label. The black star signals the moment when the radial field is equal to the field-space curvature length, $\phi=\sqrt{3\alpha/2}$. The red area highlights when the radial field is within 1\% of the inflection point, $|\left(\phi-\phi_\text{infl} \right)/\phi_\text{infl}|\leq0.01$. The permanence at the inflection point is quite extended for the model with $\alpha=0.1$, rather short for $\alpha=0.01$ (see the vertical, thin, red line in the central panel, close to the end of inflation), while inflation ends before $\phi$ reaches the inflection point in the model with $\alpha=0.005$.}
\label{fig:multifield vary alpha effect of the inflection point}
\end{figure}

Another effect of having small $\alpha$ is that the dynamics at the inflection point is changed in the presence of a phase of angular inflation. We clarify this by plotting the fields evolution against $\Delta N$ in the last 30 e-folds of inflation in figure \ref{fig:multifield vary alpha effect of the inflection point}, where each panel corresponds to one of the models discussed. While for $\alpha=0.1$ the radial field gets bounced back only transiently and then is able to settle around the inflection point (see the red area), when $\alpha$ is smaller the fields undergo a phase of angular inflation and $\phi$ is kept away from the inflection point. As displayed in the middle panel, for $\alpha=0.01$ right before the end of inflation $\phi$ crosses the inflection point (see the vertical, thin, red line) and a consequent slight change in its velocity can be seen from the plot. Instead, for the smallest $\alpha$ considered, $\alpha=0.005$, inflation ends before the radial field is able to cross the inflection point. In other words, the effect of the inflection point is washed out from the evolution of $\phi$ for small $\alpha$. Nevertheless, given our parameterisation of the potential, the value of $\phi_\text{infl}$ still has an effect on the large scales observables, as discussed in section \ref{sec:robustness of single-field predictions}, even in cases where inflation ends before the radial field is able to cross the inflection point. This is because $\phi_\text{infl}$, together with the parameter $\gamma$, governs the mass of the angular field $\theta$, as discussed in appendix~\ref{appendix: parameter study multifield potential}. Because of this $\phi_\text{infl}$ determines the position of the transition between the first and second phases of inflationary evolution, which in turn affects the predictions for large-scales observables.
\begin{figure}
\centering
\includegraphics[scale=0.5]{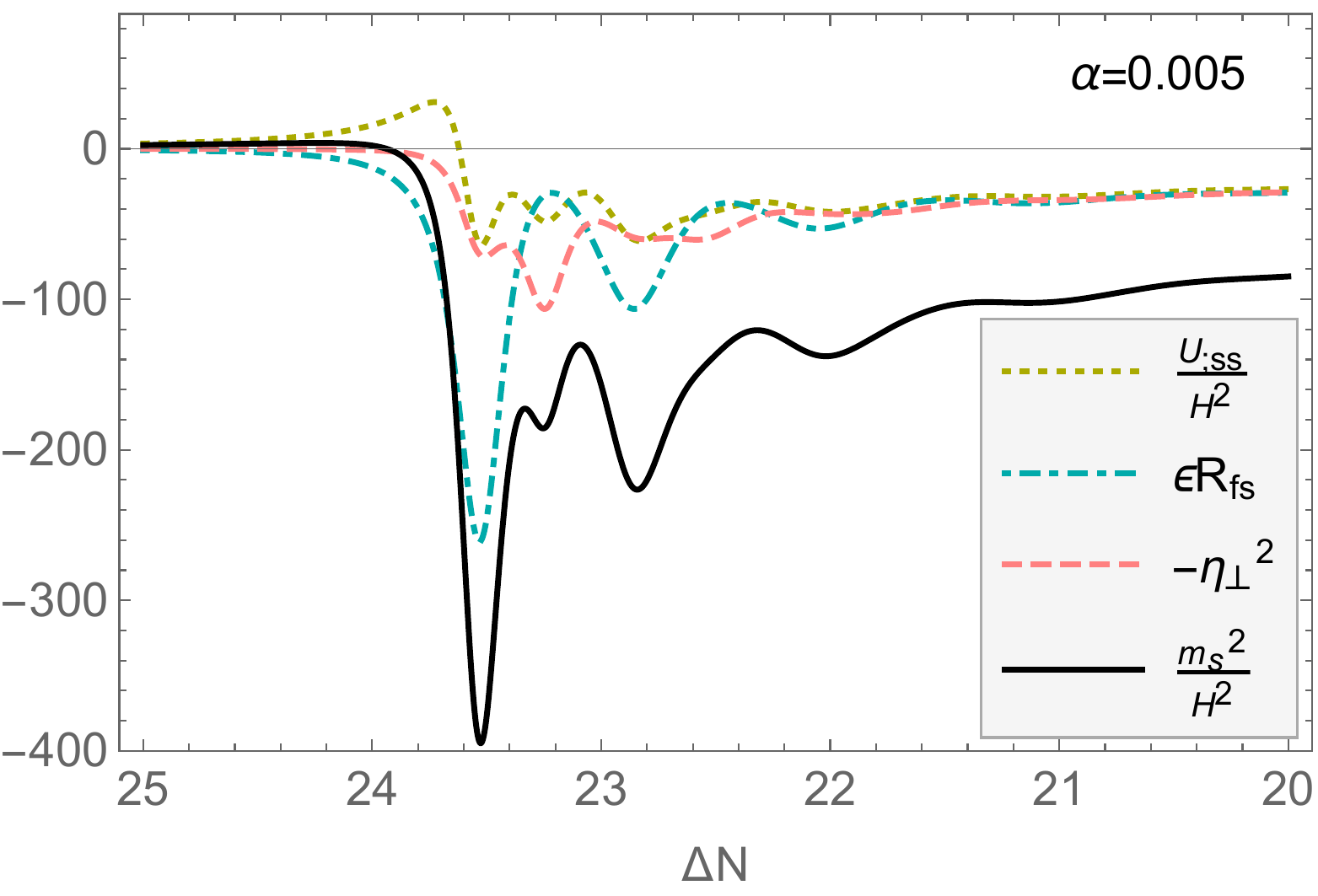}
\caption{Evolution of the squared-mass of the isocurvature perturbation (black line) for the model with $\alpha=0.005$, displayed together with its three components given in eq.~\eqref{isocurvature mass}. }
\label{fig:multifield vary alpha one case ms components}
\end{figure}

In figure \ref{fig:multifield vary alpha one case ms components} the squared-mass of the isocurvature perturbation is displayed for the model with $\alpha=0.005$, together with its three component parts given in eq.~\eqref{isocurvature mass}. The figure shows how the first negative peak in ${m_s}^2/H^2$ is mainly driven by the geometrical component, $\epsilon_1 \mathcal{R}_\text{fs}$. 
\begin{figure}
\centering
\captionsetup[subfigure]{justification=centering}
   \begin{subfigure}[b]{0.45\textwidth}
    \includegraphics[width=\textwidth]{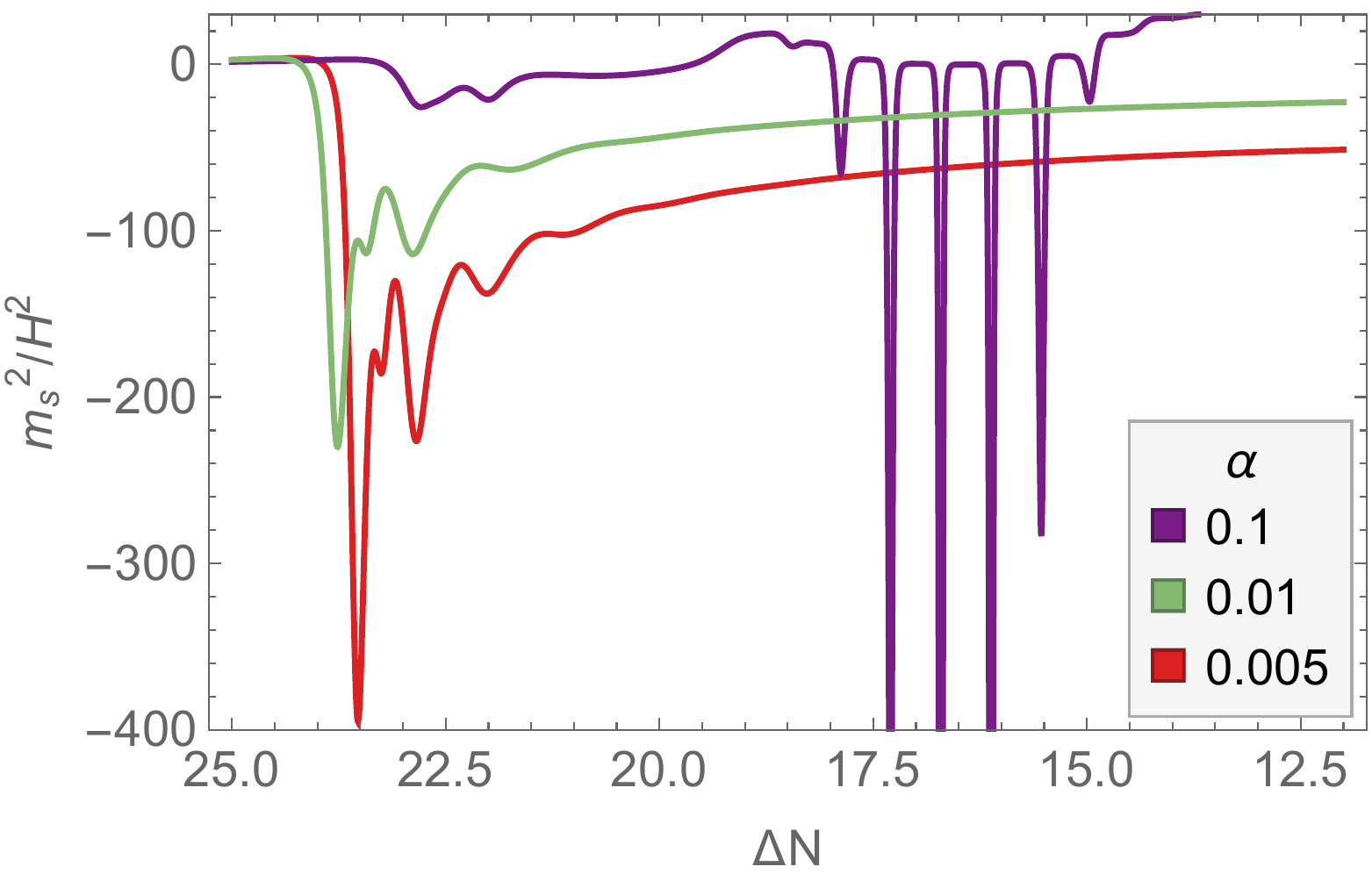}
  \end{subfigure}
   \begin{subfigure}[b]{0.45\textwidth}
    \includegraphics[width=\textwidth]{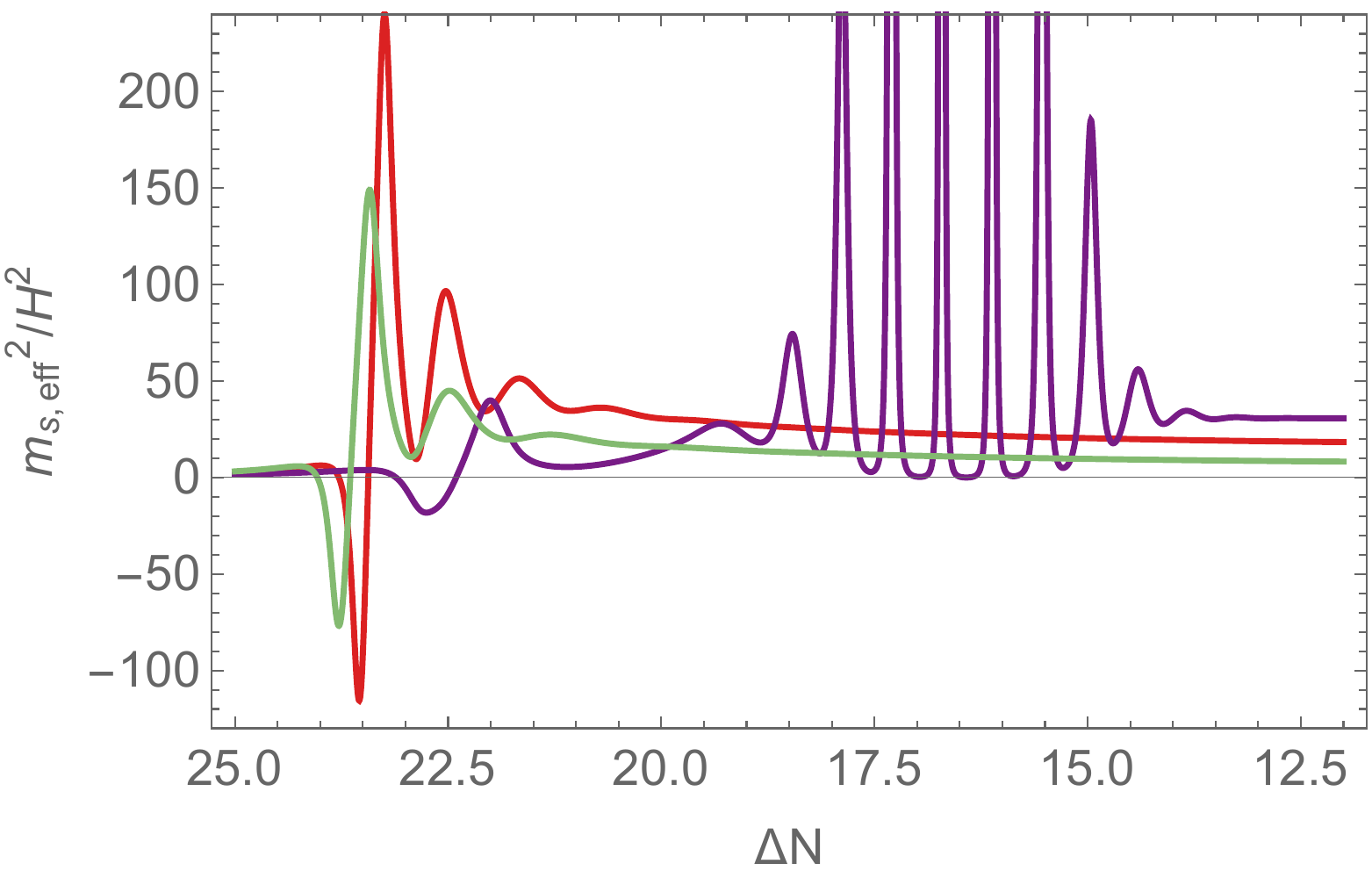}
  \end{subfigure}
   \begin{subfigure}[b]{0.45\textwidth}
    \includegraphics[width=\textwidth]{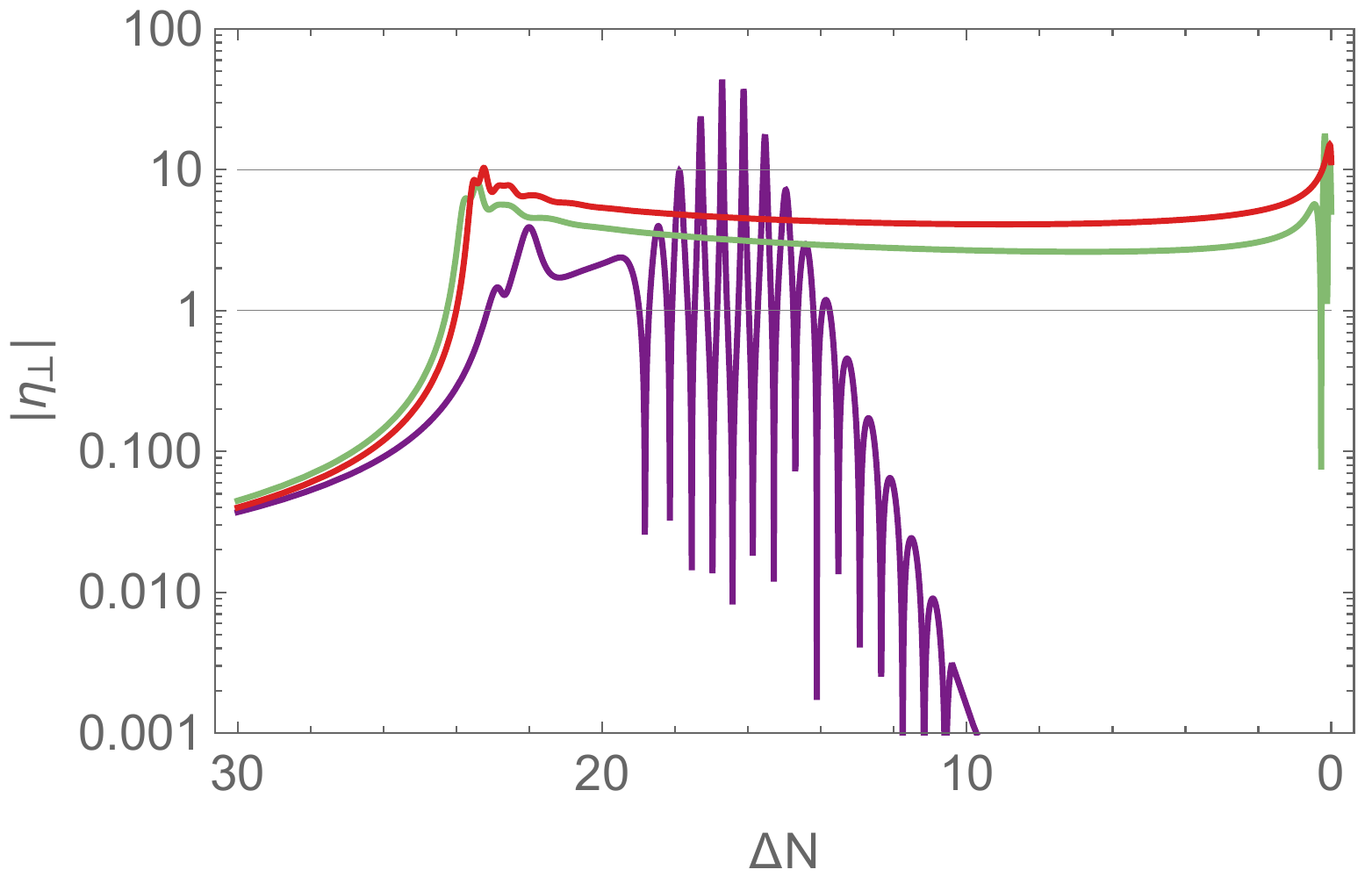}
  \end{subfigure}
   \begin{subfigure}[b]{0.45\textwidth}
    \includegraphics[width=\textwidth]{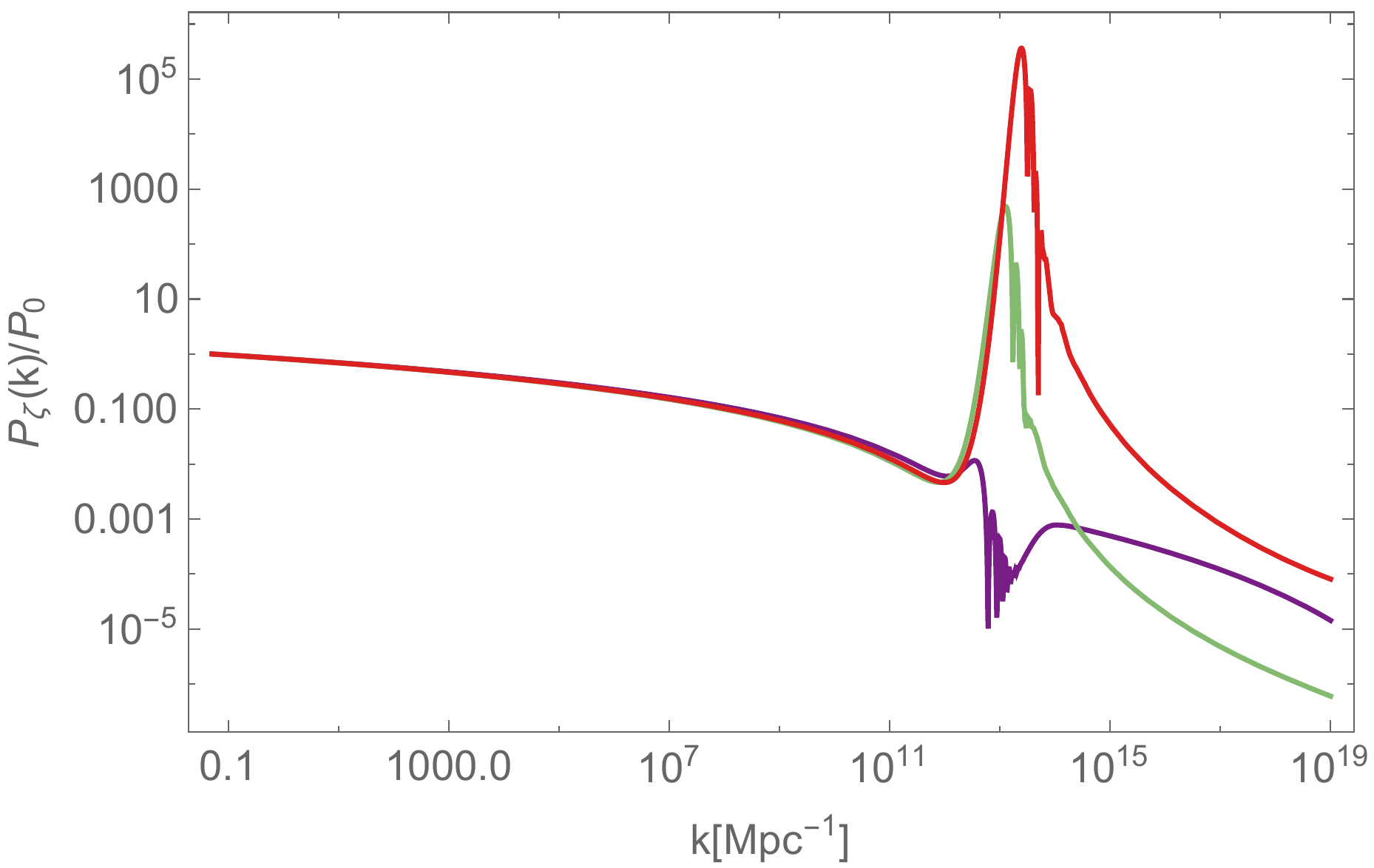}
  \end{subfigure}
\caption{Evolution of the squared-mass \eqref{isocurvature mass} (top-left panel) and super-horizon squared-mass \eqref{eff mass isocurvature} (top-right panel) of the isocurvature perturbation, together with the magnitude of the bending parameter \eqref{eta perp} (bottom-left panel) and numerical scalar power spectrum (bottom-right panel) obtained for the models with different $\alpha$. In the bottom-right panel $P_0=2.1\times10^{-9}$. The legend is the same in each plot, see the top-left panel.}
  \label{fig:multifield change alpha power spectra}
\end{figure}

In the top-left panel of figure~\ref{fig:multifield change alpha power spectra} we show the evolution of the squared-mass of the isocurvature perturbation, eq.~\eqref{isocurvature mass}, for the three different values of $\alpha$. As shown in figure~\ref{fig:multifield vary alpha one case ms components}, the first negative peak in each case is driven by the geometrical contribution $\epsilon_1 \mathcal{R}_\text{fs}$, with $\mathcal{R}_\text{fs}$ inversely proportional to $\alpha$. For smaller $\alpha$ the geometrical contribution to ${m_s}^2/H^2$ is boosted, which explains why ${m_s}^2/H^2$ shows a larger negative profile with decreasing $\alpha$. This has a clear consequence for the curvature perturbations as well; we expect a larger enhancement for smaller $\alpha$, as long as the background trajectory is turning, $\eta_\perp\neq0$, which is the case for the models shown, see the bottom-left panel in the same figure. The top-right panel shows the evolution of the super-horizon effective squared-mass for the isocurvature modes, eq.~\eqref{eff mass isocurvature}, over the same range of $\Delta N$ as in the top-left panel. The bending parameter contributes positively in this case, explaining the difference between the squared-mass and the super-horizon effective squared-mass. Finally, in the bottom-right panel we show the normalised power spectra $P_\zeta(k)/P_0$ for the three cases, obtained numerically with a modified version of the \textit{\textit{mTransport}} code~\cite{Dias:2015rca}, where $\Delta N_\text{CMB}=55$ and $P_0=2.1\times10^{-9}$. The numerical results confirm what was anticipated from the behaviour of ${m_s}^2/H^2$. The field space with the largest curvature considered, corresponding to $\alpha=0.005$, leads to a five orders of magnitude enhancement in the power spectrum with respect to CMB scales. Further decreasing $\alpha$ could lead to even larger enhancements, even the seven orders of magnitude required to possibly produce PBHs. 
\begin{figure}
\centering
\includegraphics[scale=0.5]{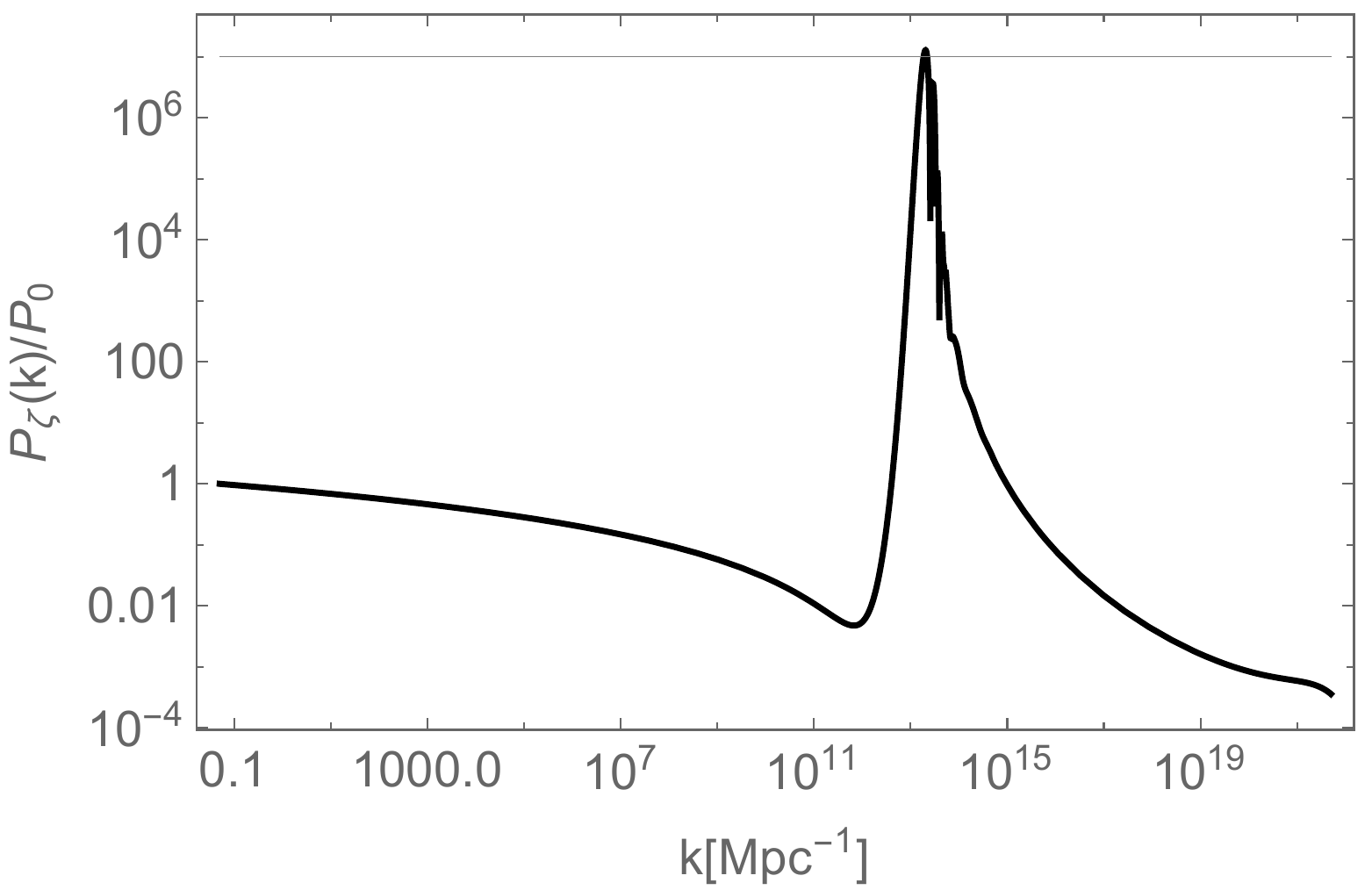}
\caption{Normalised primordial scalar power spectrum for the potential \eqref{potential multifield} with $\{\alpha=0.0035,\,\phi_\text{infl}=0.077,\, \gamma=10,\, \theta_\text{in}=7\pi/10\}$. Here $P_0=2.1\times 10^{-9}$. }
\label{fig:multifield vary alpha seven order of magnitude enhancement}
\end{figure}
As an example, we display in figure~\ref{fig:multifield vary alpha seven order of magnitude enhancement} the normalised scalar power spectrum, $P_\zeta(k)/P_0$, obtained for the potential \eqref{potential multifield} with parameters $\{\alpha=0.0035,\,\phi_\text{infl}=0.077,\, \gamma=10,\, \theta_\text{in}=7\pi/10\}$.  

\subsection{Robustness of single-field predictions}
\label{sec:robustness of single-field predictions}

We have explored multi-field effects in the presence of an inflection point in the context of $\alpha$--attractor models of inflation\footnote{See~\cite{Aldabergenov:2020bpt, Ishikawa:2021xya} for a different multi-field set-up featuring a near-inflection point in the potential.} and seen that these models display a rich phenomenology. Depending on the initial condition, $\theta_\text{in}$, and on the curvature of the hyperbolic field space, the power spectrum, $P_\zeta(k)$, can be significantly enhanced on small scales. 
We now assess what phenomenology is possible in models which are consistent with large-scale CMB observations, specifically of the spectral tilt, $n_s$, and the running of the spectral index, $\alpha_s$.

The multi-field potential \eqref{potential multifield} is parametrised by $\{\alpha,\, \phi_\text{infl},\, \gamma\}$. We will fix $\alpha=0.005$ and $\gamma=10$, and study the effect of varying the position of the inflection point, $\phi_\text{infl}$, which determines the scale at which the peak in the power spectrum is located and therefore also affects the CMB observables on large scales. We also fix the initial condition $\theta_\text{in}=7\pi/10$, which together with $\alpha=0.005$, implies that the scalar power spectrum can be amplified on small scales by roughly five orders of magnitude, as discussed in section~\ref{sec: change alpha multifield}. 

\begin{figure}
\centering
\includegraphics[scale=0.43]{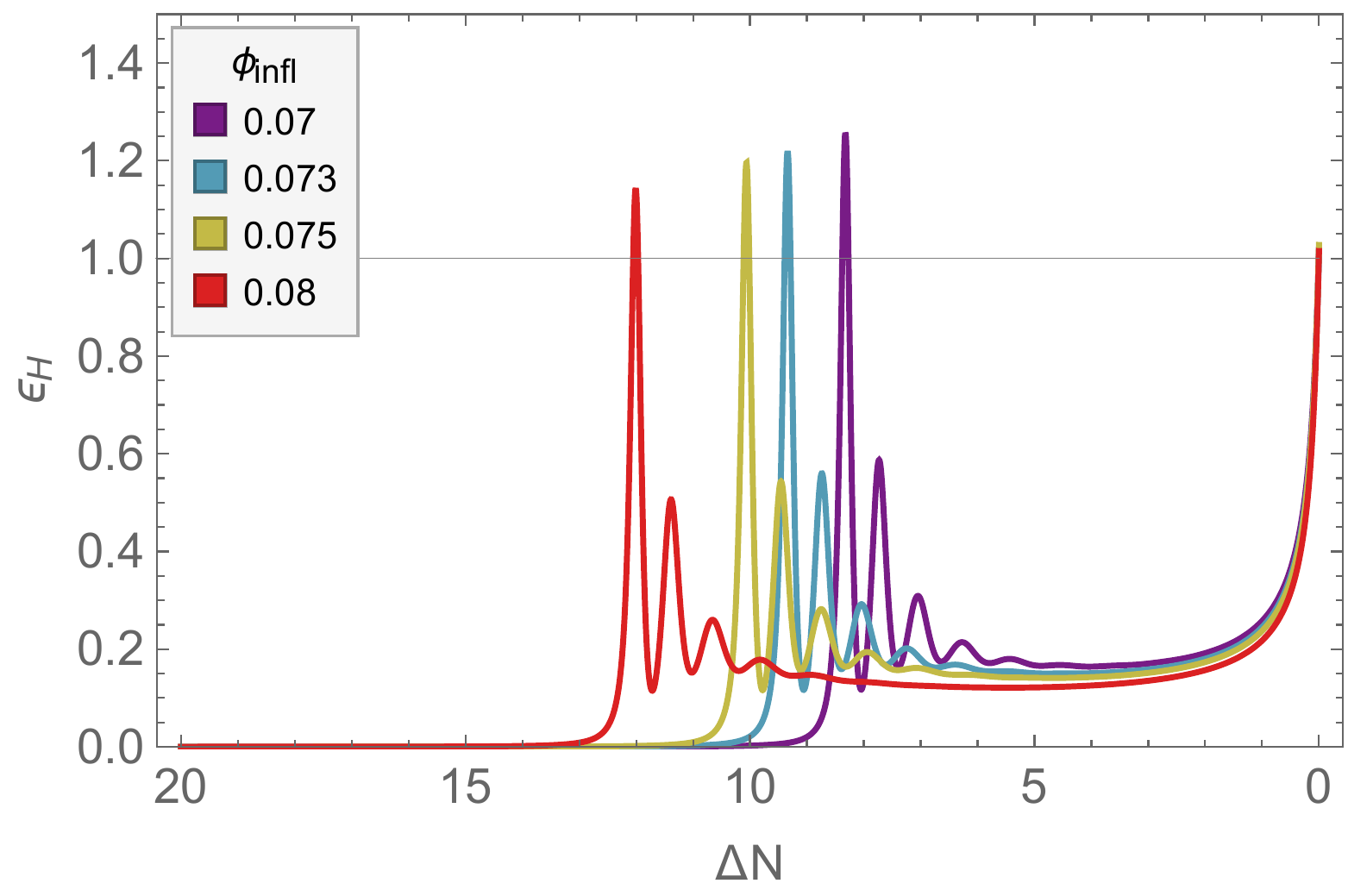}
\caption{Evolution of $\epsilon_H(N)$ plotted against $\Delta N\equiv N_\text{end}-N$ in the last $20$ e-folds of inflation. These results have been obtained numerically for multi-field models with $\{\alpha=0.005,\, \gamma=10\}$ and different positions for the inflection point, $\phi_\text{infl}$. }
\label{fig:multifield CMB compatibility epsilon}
\end{figure}
We show in figure \ref{fig:multifield CMB compatibility epsilon} the evolution of $\epsilon_H(N)$ against $\Delta N$ in the last 20 e-folds of inflationary evolution for  $\phi_\text{infl}=\{0.07,\, 0.073,\, 0.075,\, 0.08\}$. Slow roll is violated close to the end of inflation in each model and this transition moves closer and closer to the end of inflation for smaller $\phi_\text{infl}$. Also, after the angular field starts evolving and $\epsilon_H(N)$ peaks, each model displays a transient phase of angular inflation~\cite{Christodoulidis:2018qdw}, as expected given the small value of $\alpha$ (large field-space curvature) in these cases. 
\begin{figure}
\centering
\captionsetup[subfigure]{justification=centering}
   \begin{subfigure}[b]{0.46\textwidth}
    \includegraphics[width=\textwidth]{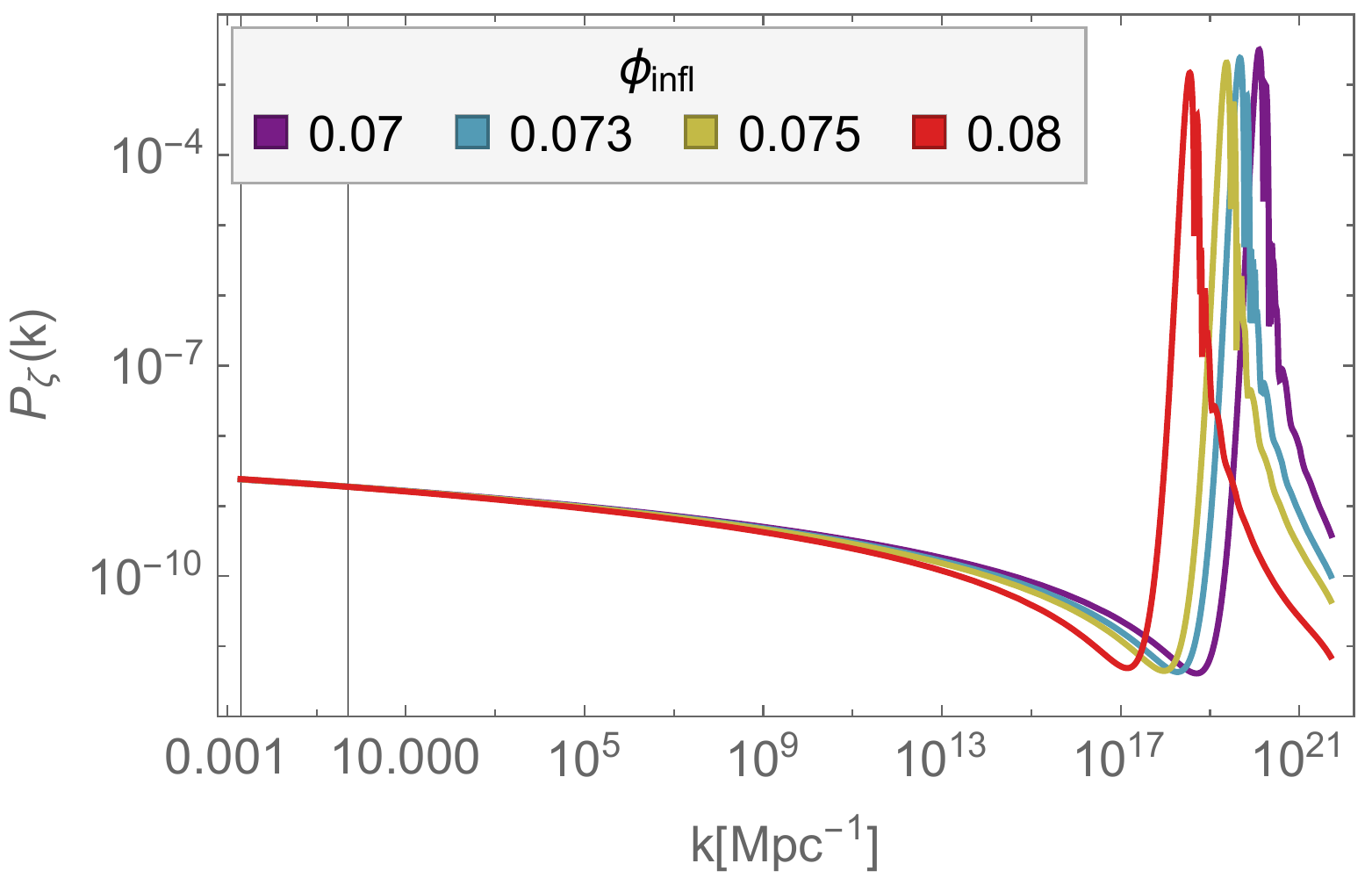}
  \end{subfigure}
   \begin{subfigure}[b]{0.48\textwidth}
    \includegraphics[width=\textwidth]{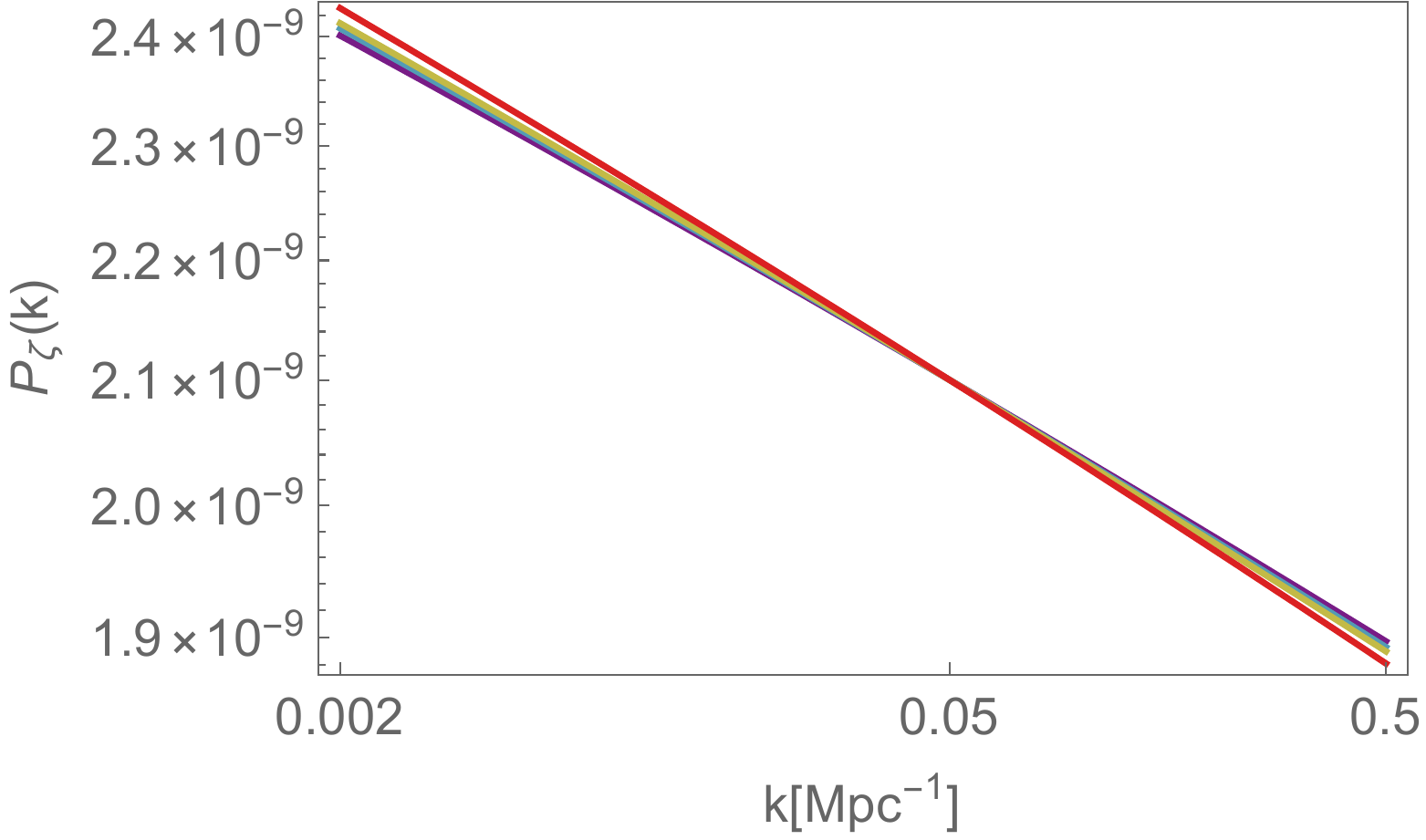}
  \end{subfigure}
\caption{\textit{Left panel:} numerically determined power spectrum, $P_\zeta(k)$, for multi-field models with $\{\alpha=0.005,\, \gamma=10\}$ and different $\phi_\text{infl}$. The two vertical lines correspond to $k=\{0.002\,\text{Mpc}^{-1},\, 0.5\,\text{Mpc}^{-1}\}$ and highlight the CMB scales. \textit{Right panel:} zoomed-in plot of the power spectrum on CMB scales. The spectral tilt of $P_\zeta(k)$ is slightly different for each model. }
  \label{fig:multifield CMB compatibility power}
\end{figure}

We numerically evaluate $P_\zeta(k)$ with a modified version of \textit{\textit{mTransport}}~\cite{Dias:2015rca} and represent the results in the left panel of figure \ref{fig:multifield CMB compatibility power}. Equation \eqref{Nstar}\footnote{For simplicity we assume here instant reheating, $\rho_\text{th}=\rho_\text{end}$.} allows us to estimate $\Delta N_\text{CMB}$. On large scales the power spectrum is almost scale-invariant, while at higher frequencies it is enhanced due to multi-field effects and $k_\text{peak}$ varies depending on the value of $\phi_\text{infl}$, moving towards smaller scales with decreasing $\phi_\text{infl}$. The amplitude of the peak is slightly different among the models, which can be explained in light of the fact that the first negative peak in the squared-mass of the iscocurvature perturbation is driven by the geometrical contribution $\epsilon_1 \mathcal{R}_\text{fs}$, see eq.~\eqref{isocurvature mass}, as shown in figure~\ref{fig:multifield vary alpha one case ms components} for a model with $\{\alpha=0.005, \, \gamma=10, \, \phi_\text{infl}=0.103,\, \theta_\text{in}=7\pi/10\}$. As seen in figure \ref{fig:multifield CMB compatibility epsilon}, $\epsilon_H$ peaks at higher values for decreasing $\phi_\text{infl}$, which means in turn that the negative peak in ${m_s}^2/H^2$ is larger in magnitude for smaller $\phi_\text{infl}$ (for fixed field-space curvature), explaining why the highest peak in $P_\zeta$ is reached for the smallest $\phi_\text{infl}$. 

In the right panel of figure~\ref{fig:multifield CMB compatibility power}, we zoom in on the large-scale behaviour of $P_\zeta(k)$, which shows how the presence of a peak on small scales affects the large-scale observables, in particular the tilt of the power spectrum. 
\begin{table}[]
  \centering
   \begin{tabular}{ |c||c|c|c|c|c|c|c|c|c|c| }
\hline
$\phi_\text{infl}$ &$\Delta N_\text{CMB}$ &$\Delta N_\text{CMB}-\Delta N_\text{peak}$ & $k_\text{peak}/\text{Mpc}^{-1}$& $n_s$& $r_{0.002}$ & $\alpha_s$  \\

\hline
0.07 & 55.1927  & 50.32  & $1.25\times10^{20}$ &0.9569& $2.6\times 10^{-5}$ & -0.00092 \\
\hline
0.073 & 55.2318  & 49.28 & $4.6\times10^{19}$ &0.9560& $2.7\times 10^{-5}$ &-0.00096 \\
 \hline
0.075 & 55.2593  & 48.56 &  $2.3\times10^{19}$ &0.9554& $2.8\times 10^{-5}$ &-0.00099 \\
   \hline
0.08 & 55.3201 & 46.59 & $3.4\times10^{18}$  & 0.9534 & $3\times 10^{-5}$ &-0.0011 \\
  \hline
\end{tabular}
    \caption{Details of multi-field models with $\{\alpha=0.005,\, \gamma=10\}$ and different $\phi_\text{infl}$. For all models the fitted cubic and quartic coefficients are $\mathcal{O}(10^{-5})$ and $\mathcal{O}(10^{-6})$ respectively.}
    \label{tab: multifiled CMB compatibility}
\end{table}
We list in table \ref{tab: multifiled CMB compatibility} key quantities obtained for each of the models considered. We show the predicted values for the spectral tilt, $n_s$, and running, $\alpha_s$, obtained at $k_\text{CMB}=0.05\,\text{Mpc}^{-1}$ by fitting the numerical results for $\ln(P_\zeta(k))$ with a quartic function\footnote{We find that residual noise is minimised when we fit a polynomial that is quartic in $\ln(k/k_\text{CMB})$.} of $\ln(k/k_\text{CMB})$ on scales $0.002  \,\text{Mpc}^{-1}<k<0.5\,\text{Mpc}^{-1}$. The predicted values for the tensor-to-scalar ratio are calculated at $k=0.002\,\text{Mpc}^{-1}$  using the single-field slow-roll approximation on large scales, \eqref{tensor to scalar ratio}. For all models $r_{0.002}$ is well below the current upper bound \eqref{r bound new}. 

\begin{figure}
\centering
\captionsetup[subfigure]{justification=centering}
   \begin{subfigure}[b]{0.48\textwidth}
    \includegraphics[width=\textwidth]{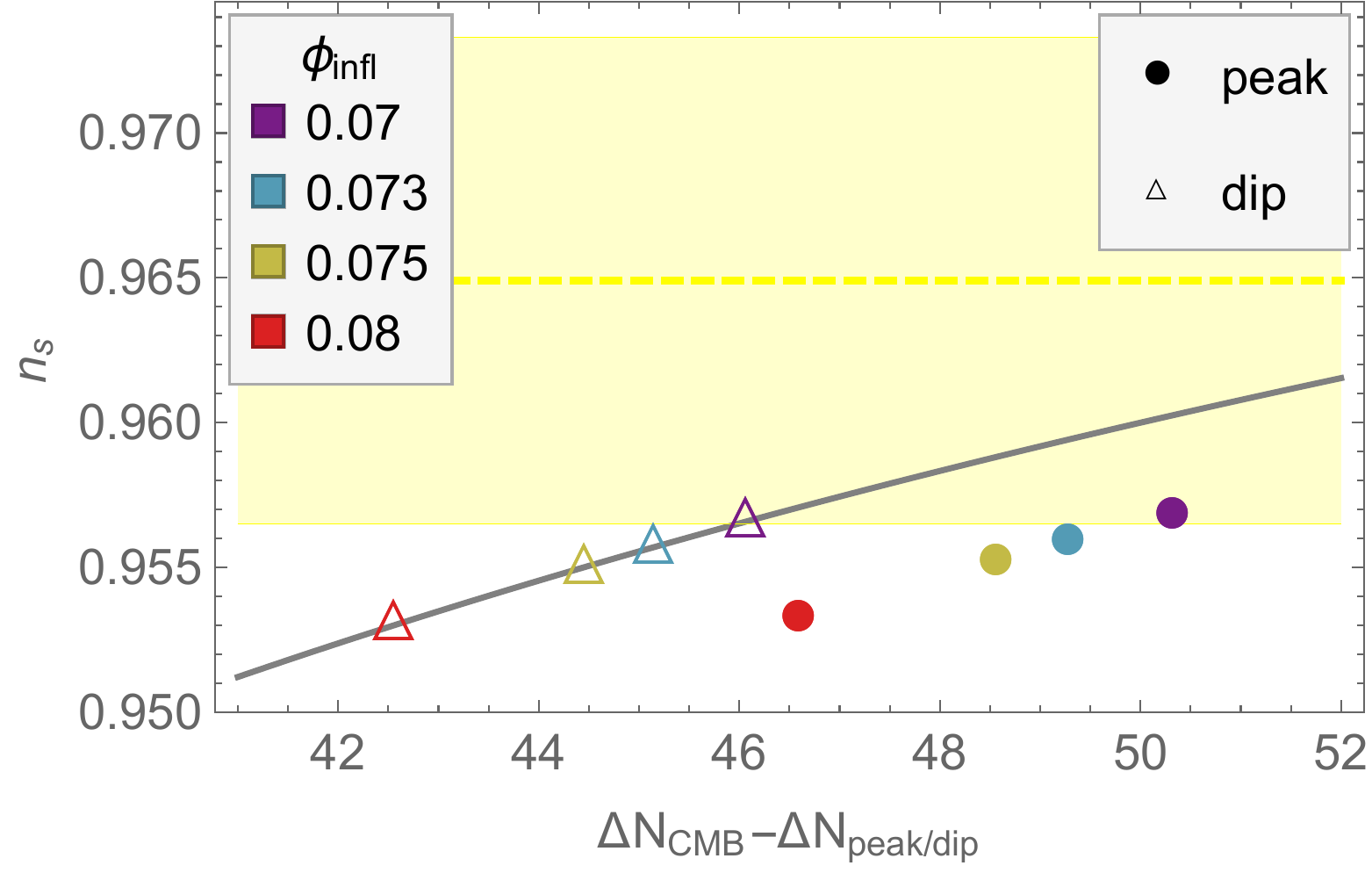}
  \end{subfigure}
   \begin{subfigure}[b]{0.48\textwidth}
    \includegraphics[width=\textwidth]{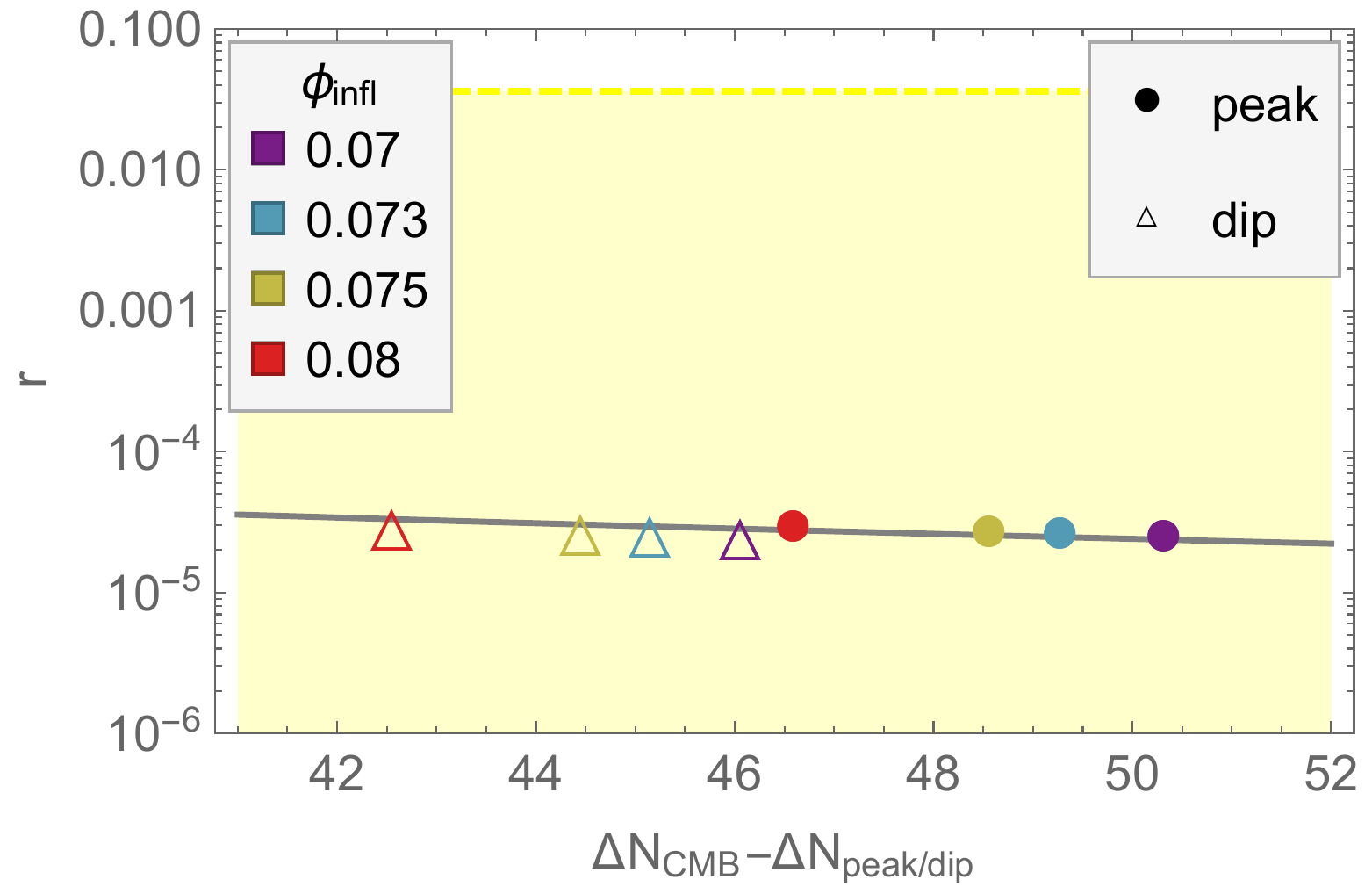}
  \end{subfigure}
\caption{\textit{Left panel:} the modified universal prediction for $n_s$ \eqref{ns universal prediction inflection point} is plotted in grey together with numerical results obtained from multi-field models with $\{\alpha=0.005,\, \gamma=10\}$ and different $\phi_\text{infl}$, as shown in the legend. In particular, coloured points represent numerical results for $n_s$ against $\Delta N_\text{CMB}-\Delta N_\text{peak}$. The empty triangles represent instead the numerical tilt $n_s$ against $\Delta N_\text{CMB}-\Delta N_\text{dip}$, i.e., the scale of the dip in $P_\zeta(k)$ is taken as a reference instead of $k_\text{peak}$. The yellow-shaded area highlights the \textit{Planck} $95\,\%$ C.L. region, see \eqref{CMB 95percent ns bound}. \textit{Right panel:} the approximation \eqref{r universal prediction inflection point} with $\alpha=0.005$ is plotted in grey together with the numerical results for the tensor-to-scalar ratio against $\Delta N_\text{CMB}-\Delta N_\text{peak}$ (coloured points) and $\Delta N_\text{CMB}-\Delta N_\text{dip}$ (empty triangles). Each colour is associated with a specific $\phi_\text{infl}$, as detailed in the legend. The yellow-shaded area represents the $95\,\%$ C.L. region from the bound on $r_\text{CMB}$ \eqref{r bound new}.} \label{fig:multifield CMB compatibility ns and r}
\end{figure}

\begin{figure}
\centering
\includegraphics[scale=0.55]{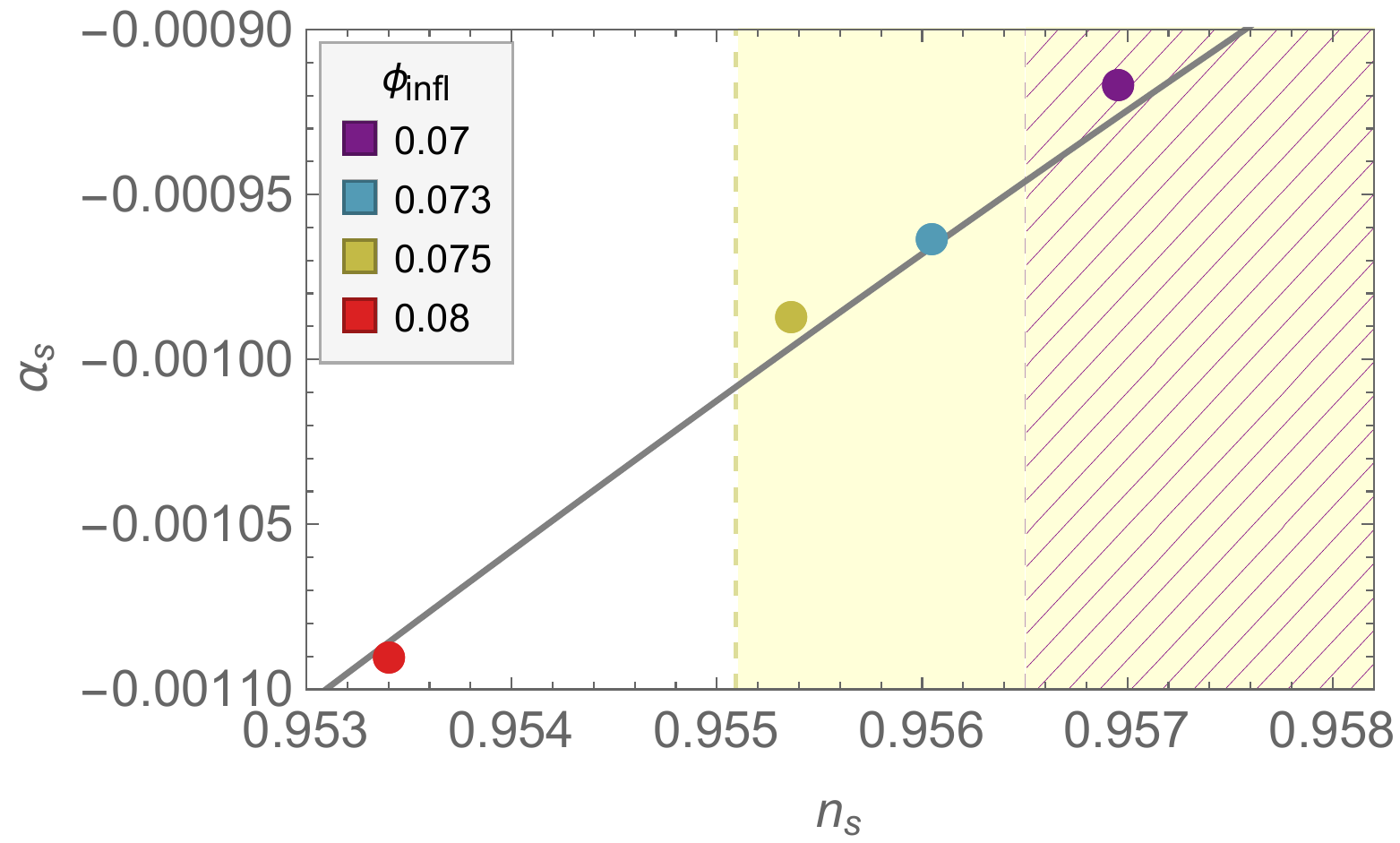}
\caption{The $\alpha$--attractors consistency relation \eqref{alphas ns} is plotted in grey together with the numerical results for the tilt and its running for multi-field models with $\{\alpha=0.005,\, \gamma=10\}$ and different position of the inflection point $\phi_\text{infl}$, as detailed in the legend. The yellow region highlights part of the $95\,\%$ C.L. region when \textit{Planck} data are compared with the $\Lambda\text{CDM}+r_\text{CMB}+\alpha_s$ model, \eqref{bound on ns with alpha s}, while the purple hatch-shaded area represents the lower $95\,\%$ C.L. region of $n_s$ for the $\Lambda\text{CDM}$ model instead. The range of $\alpha_s$ shown is within the observational bound \eqref{bound on alpha s}. }
\label{fig:multifield CMB compatibilty alphas consistency}
\end{figure}

In analogy with the analysis performed for the single-field inflection-point model, we can compare the CMB observables listed in table~\ref{tab: multifiled CMB compatibility} with the modified universal predictions for $n_s$ \eqref{ns universal prediction inflection point}, $r_\text{CMB}$ \eqref{r universal prediction inflection point} 
and $\alpha_s$ versus $n_s$ \eqref{alphas ns}. In the left panel of figure \ref{fig:multifield CMB compatibility ns and r} we plot eq.~\eqref{ns universal prediction inflection point} together with coloured points representing numerical results $\left( \Delta N_\text{CMB}-\Delta N_\text{peak},\, n_s\right)$ for each model considered. Similarly, the right panel shows eq.~\eqref{r universal prediction inflection point} with $\alpha=0.005$ together with coloured points representing the numerical results $\left(\Delta N_\text{CMB}-\Delta N_\text{peak},\, r_\text{CMB}\right)$. 
We do not explicitly show the results for $\alpha_s$ versus $\Delta N_\text{CMB}-\Delta N_\text{peak}$ as the comparison between the universal prediction \eqref{alpha s unversal inflection point} and the numerical results is qualitatively the same as for the tilt $n_s$. Instead, we show $\alpha_s$ versus $n_s$ in figure~\ref{fig:multifield CMB compatibilty alphas consistency}, alongside the $\alpha$--attractors consistency relation \eqref{alphas ns}.

As for the single-field case, the modified universal predictions describe well the numerical results, with a small offset for the case of the tilt $n_s$ and its running $\alpha_s$. The match between the modified universal predictions and the numerical results can be further improved by substituting $\Delta N_\text{CMB}-\Delta N_\text{peak}\rightarrow \Delta N_\text{CMB}-\Delta N_\text{dip}$ or, in other words, by taking as a reference the scale associated with the local minimum in the power spectrum, $k_\text{dip}$, instead of its local maximum, $k_\text{peak}$. We demonstrate this in figure \ref{fig:multifield CMB compatibility ns and r} by including the numerical results for $n_s$ and $r_\text{CMB}$ represented with empty triangles against $\Delta N_\text{CMB}-\Delta N_\text{dip}$. By comparing triangles and circles (of the same colour) with the grey lines, one can see that the CMB observables (especially $n_s$) can be described even better by the position of the dip in $P_\zeta(k)$. However, it is the peak in $P_\zeta(k)$ that could have potentially observable consequences, such as PBH production and second-order GW generation. Therefore we focus on the consequences of the modified universal predictions for $k_\text{peak}>k_\text{dip}$
when it comes to exploring the phenomenology.  
The numerical results for the tilt and its running are well described by the $\alpha$--attractors consistency relation \eqref{alphas ns}, as shown in figure \ref{fig:multifield CMB compatibilty alphas consistency}. As for the single-field case, models which are compatible with the bound on the spectral index \eqref{bound on ns with alpha s} predict $\alpha_s$ about one order of magnitude smaller than the current observational uncertainty in \eqref{bound on alpha s}. We therefore compare the model predictions with the CMB constraints for the $\Lambda\text{CDM}$ model, in particular with the lower bound on $n_s$ given in eq.~\eqref{CMB 95percent ns lower bound} in the absence of running. Given the numerical results for $n_s$ this implies that only configurations with $\phi_\text{infl}\lesssim0.7$ are consistent with the CMB observations and the peak is located on scales $k_\text{peak}\gtrsim10^{20}\,\text{Mpc}^{-1}$.  

\begin{figure}
\centering
\captionsetup[subfigure]{justification=centering}
   \begin{subfigure}[b]{0.48\textwidth}
    \includegraphics[width=\textwidth]{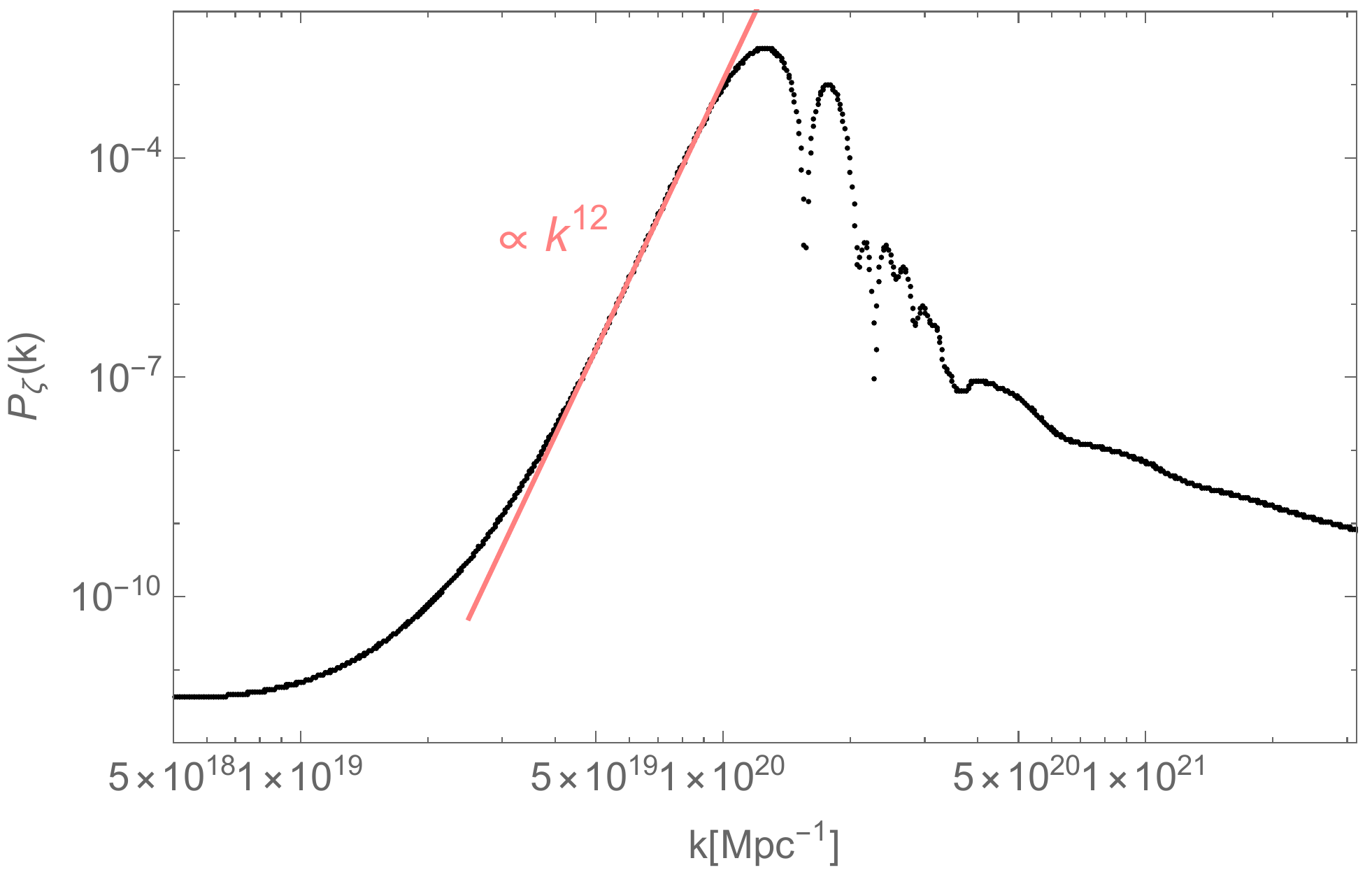}
  \end{subfigure}
   \begin{subfigure}[b]{0.48\textwidth}
    \includegraphics[width=\textwidth]{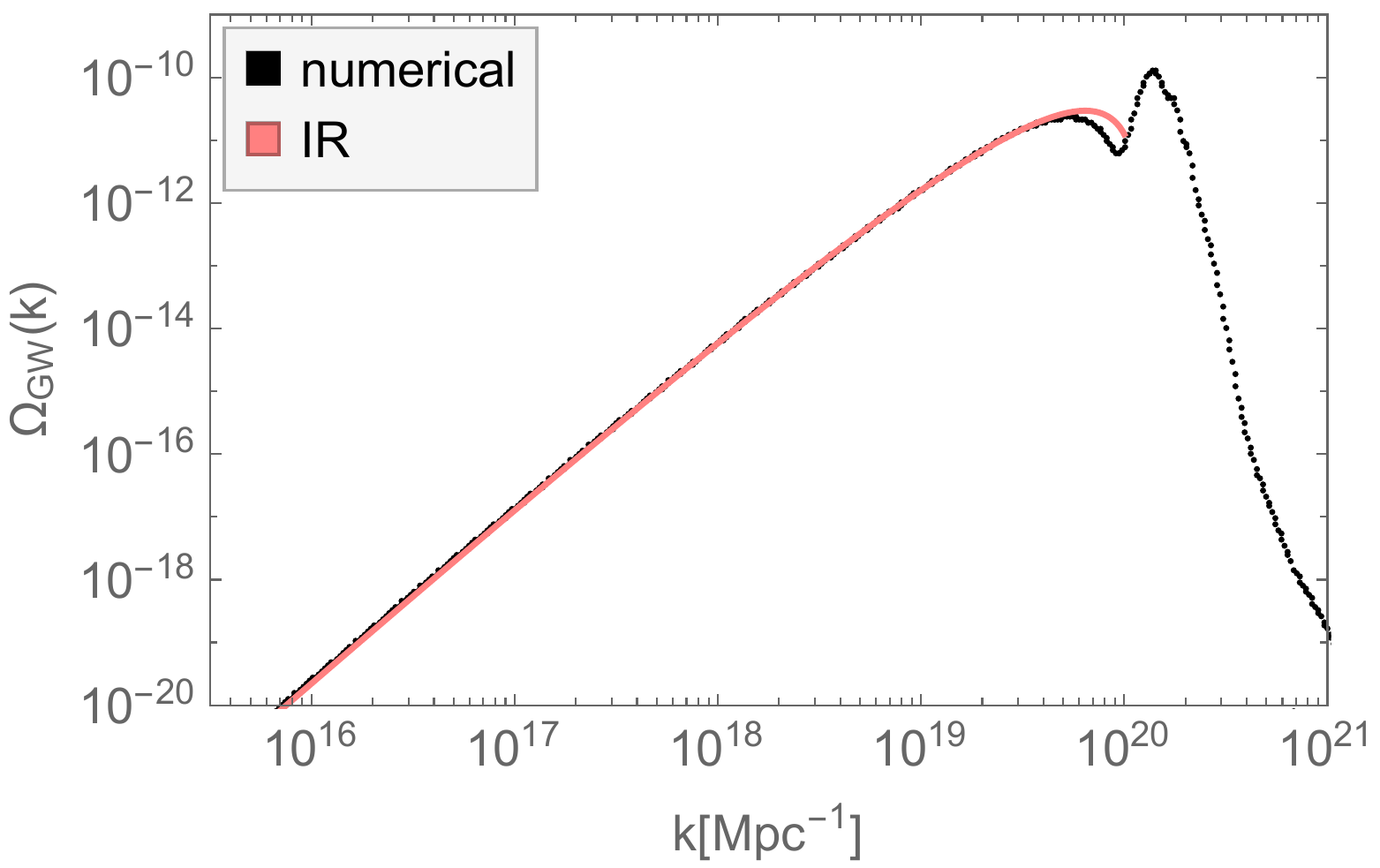}
  \end{subfigure}
\caption{\textit{Left panel:} numerical results of $P_\zeta(k)$ for the multi-field model with parameters $\{\alpha=0.005,\, \gamma=10, \phi_\text{infl}=0.07\}$ and initial condition $\theta_\text{in}=7\pi/10$. We display the results in the region of the peak, with the pink line representing the scaling of the infrared tail of it. \textit{Right panel:} numerical results for the second-order GWs produced during radiation domination by the enhanced scalar perturbations whose power spectrum is displayed in the left panel. The pink line is obtained using eq.(5.16) of~\cite{Domenech:2021ztg} with $n_\text{IR}=12$ and $n_\text{UV}\rightarrow\infty$ and well approximates the spectral shape of the infrared tail of the numerical $\Omega_\text{GW}(k)$.}
\label{fig:IR and UV behaviour multifield}
\end{figure}
In the left panel of figure \ref{fig:IR and UV behaviour multifield} we display the numerical results for the scalar power spectrum in the region around the peak, for a model which is compatible with large-scale (CMB) measurements of $n_s$, $\phi_\text{infl}=0.07$. The pink line shows the scaling of the IR tail of the peak, $n_\text{IR}\simeq12$, which by far exceeds the limits on the growth of the power spectrum possible in single-field models~\cite{Byrnes:2018txb, Carrilho:2019oqg, Ozsoy:2019lyy}. In~\cite{Palma:2020ejf,Fumagalli:2020adf, Braglia:2020taf} it has been shown that multi-field models with strong turns in field space evade the single-field bound on the growth of $P_\zeta$, and we demonstrate here that the same holds when the growth of the curvature perturbation is also due to the strong curvature of the hyperbolic field space. 

The peak in the scalar power spectrum displays a series of peaks which is due to the strong and sharp turn in field space characterising the background evolution, see e.g., the bottom-left panel of figure \ref{fig:multifield evo alpha0.1 one case eta perp} for a model with $\{\alpha=0.1,\, \gamma=10,\,  \phi_\text{infl}=0.542\}$. In~\cite{Fumagalli:2020nvq} it is shown that a strong and sharp turn can lead to an exponentially enhanced amplitude of $P_\zeta$, with an oscillatory modulation. In particular, in~\cite{Fumagalli:2020adf,Fumagalli:2020nvq} the bending parameter $\eta_\perp$ is modelled with a Gaussian profile, see eq.~\eqref{eta perp gaussian profile}, and the Hubble rate is assumed to be smooth and slowly-varying. The multi-field models discussed here are characterised by a more complicated background evolution, see e.g., $\epsilon_H(N)$ in figure \ref{fig:multifield CMB compatibility epsilon}, and a profile for the bending parameter which can only be partially described by a Gaussian function, see e.g., figure~\ref{fig:multifield evo alpha0.1 one case eta perp}. It is therefore not surprising that the oscillations in $P_\zeta$ displayed in figure \ref{fig:IR and UV behaviour multifield} cannot simply be identified as either sharp or resonant features~\cite{Chen:2008wn, Chen:2010xka, Chluba:2015bqa, Slosar:2019gvt}, but are instead more of a combination of the two, similar to models discussed e.g., in~\cite{Braglia:2020taf}.

In the right panel of figure \ref{fig:IR and UV behaviour multifield} we represent the numerical results for the second-order GWs produced by the enhanced scalar fluctuations after horizon re-entry during radiation domination. In particular, we display $\Omega_\text{GW}(k)$ as numerically calculated using eq.~\eqref{Omega GW} and the scalar power spectrum displayed in the left panel of the same figure. The sharpness of the peak of $P_\zeta$ (see the purple line in the left panel of figure \ref{fig:multifield CMB compatibility power}) results in a two-peak structure for $\Omega_\text{GW}(k)$, with a broader and smaller peak followed by a dip and a narrower principal peak located approximately at the scale $2/\sqrt{3}\,k_\text{peak}$. The oscillatory modulation of the scalar power spectrum due to the sharp turn in the background trajectory is imprinted in $\Omega_\text{GW}(k)$ as an oscillatory pattern modulating the principal peak. The IR tail of the signal can be understood in terms of the IR and UV scaling of $P_\zeta$ around the first peak; substituting the values $n_\text{IR}=12$ and $n_\text{UV}\rightarrow\infty$ into eq.(5.12) of~\cite{Domenech:2021ztg}, we get the pink line shown on top of the numerical results. The UV tail of $\Omega_\text{GW}(k)$ (and $P_\zeta$) displays a more complicated scaling, which cannot be understood in terms of such a simple approximation. 
\begin{figure}
\centering
\includegraphics[scale=0.65]{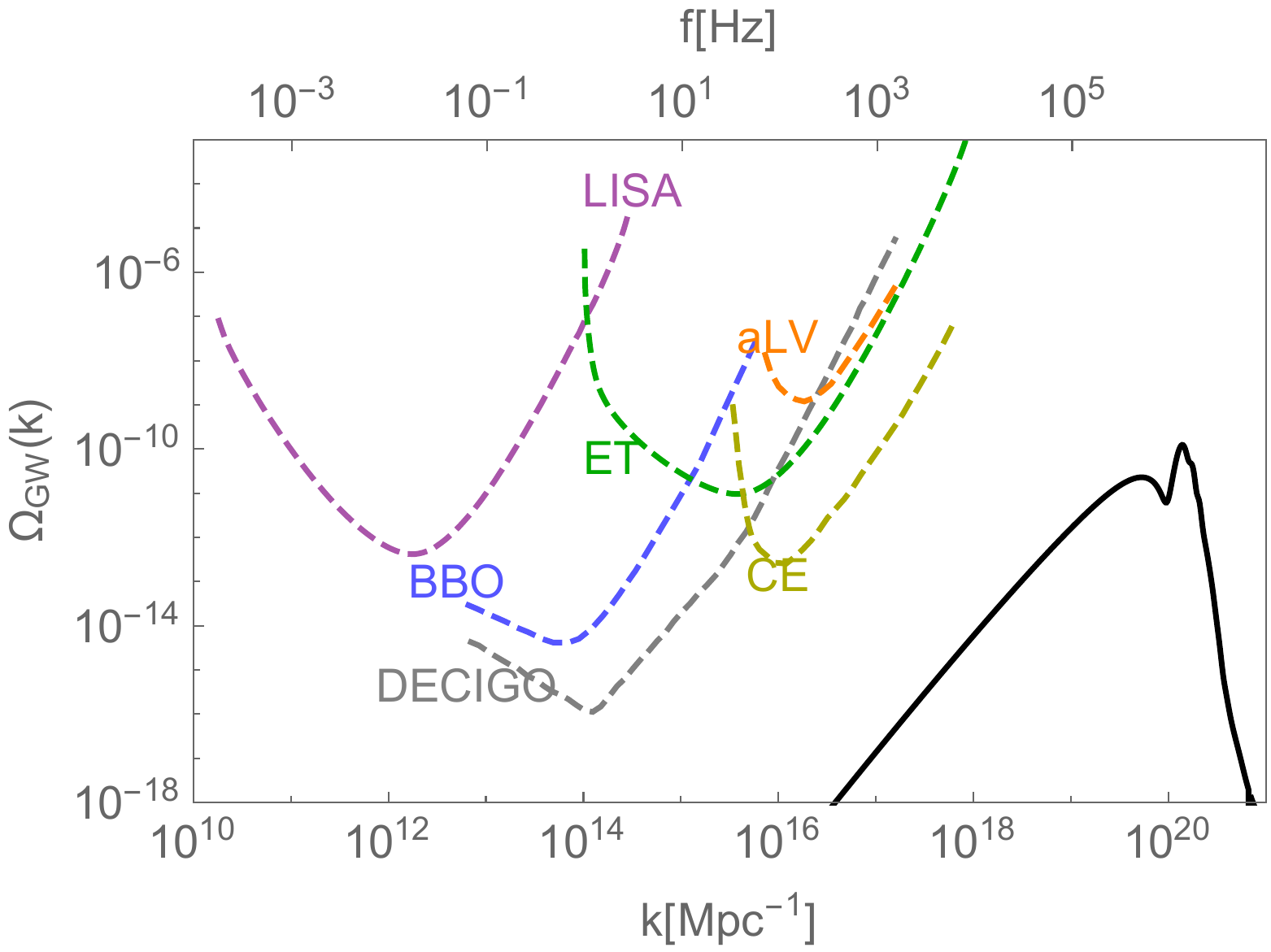}
\caption{Numerical results for the second-order GWs produced within the multi-field model with parameters $\{\alpha=0.005,\, \gamma=10,\, \phi_\text{infl}=0.07\}$ and initial condition $\theta_\text{in}=7\pi/10$.}
\label{fig:multifield GW}
\end{figure}

In figure \ref{fig:multifield GW} the numerical results for $\Omega_\text{GW}(k)$ are plotted together with the sensitivity curves of upcoming space- and Earth-based GW observatories. 
The amplitude of the GW signal is determined by the amplitude of the peak of the scalar power spectrum, which in turn mainly depends on the curvature of the hyperbolic field space, set by the parameter $\alpha$, and the initial condition for the angular field, $\theta_\text{in}$. In particular, after fixing all the other model parameters, reducing $\alpha$ enhances $P_\zeta(k_\text{peak})$ and therefore the amplitude of the principal peak of $\Omega_\text{GW}(k)$. Our numerical results show that the multi-field model described by $\{\alpha=0.005, \, \gamma=10, \, \phi_\text{infl}=0.07, \, \theta_\text{in}=7\pi/10\}$ produces $\Omega_\text{GW}\simeq 10^{-10}$ at its peak. 

The position of the principal peak in $\Omega_\text{GW}(k)$ is set by the position of the largest peak in $P_\zeta$, which is the first peak displayed in the left panel of figure \ref{fig:IR and UV behaviour multifield}. For models which are not in tension with the large-scale measurements of $n_s$, the second-order GW principal peak is located at $f_\text{peak}\gtrsim 50\,\text{kHz}$, which is obtained by substituting the lower bound \eqref{k peak bound} in the position of the principal peak, $2/\sqrt{3}\,k_\text{peak}$, and is consistent with our numerical results. 
Given that the modified universal predictions (in particular, the one for $n_s$ \eqref{ns universal prediction inflection point}) hold also for the multi-field inflection-point potential, the considerations for PBH production discussed in section \ref{Sec:PBH formation} for the single-field case apply also for the multi-field extension. 

In summary, compatibility with the CMB observations of $n_s$ place the peak of the second-order GWs beyond the reach of current and upcoming GW observatories both for the single- and multi-field inflection-point potentials considered in this work, as well as bounding the mass of the PBHs that could possibly be produced to values $M_\text{PBH}<10^8\,\text{g}$, which make the PBHs too light to constitute candidates for dark matter in the Universe today. 

\section{Discussion}
\label{sec:conclusions}

In this work we explore the phenomenology of cosmological $\alpha$--attractor models featuring an inflection point in the potential for scalar fields evolving in a hyperbolic field space.
We consider both single-field examples and multi-field dynamics. 

In all cases we show that the primordial perturbations generated on large (CMB) scales can be described by a simple modification of the universal predictions of $\alpha$--attractors for the scalar spectral index, eq.~\eqref{ns universal prediction inflection point}, and the tensor-to-scalar ratio, eq.~\eqref{r universal prediction inflection point}.
A shift in the universal predictions was previously noted by~\cite{Dalianis:2018frf} for the single-field inflection-point $\alpha$--attractor potential, and also in a different context by~\cite{Linde:2018hmx}.
This universal behaviour leads to a consistency condition relating the scalar spectral tilt and its running, eq.~\eqref{alphas ns}.
A consequence of the tight bounds on the scalar spectral index from CMB observations is that the running of the spectral index must be small, and any deviations from the standard single-field $\alpha$-attractor dynamics is constrained to lie close to the end of inflation. Hence any enhancement in the primordial power spectrum, $P_\zeta$, is only allowed on small comoving scales, $k_\text{peak}\gtrsim5\times 10^{18}\,\text{Mpc}^{-1}$.
By adopting the tight observational bounds on the spectral index obtained from CMB observations, 
in the absence of any running of the spectral index,
eq.~\eqref{CMB 95percent ns lower bound}, which we emphasise is constrained by the consistency relation~\eqref{alphas ns}, we obtain a stronger constraints on the scale of the peak of $P_\zeta$ in comparison to the previous work of Dalianis \textit{et al.}~\cite{Dalianis:2018frf}. 

The lower bound on the comoving wavenumber, $k_\text{peak}$, implies that any primordial black holes resulting from enhanced density perturbations on small scales can only be produced with masses $M_\text{PBH}\lesssim10^8\,\text{g}$. These PBHs have long since evaporated by the present time so do not constitute a candidate for dark matter. 
%
Nonetheless they could yet leave interesting signatures if stable Planck mass relics are left behind~\cite{MacGibbon:1987my, Barrow:1992hq, Carr:1994ar}, or resulting from an early black-hole-dominated era~\cite{Anantua:2008am, Zagorac:2019ekv, Martin:2019nuw, Inomata:2020lmk, Hooper:2019gtx}.

Similarly the lower bound on $k_\text{peak}$ also has implications for the induced gravitational waves produced at second order in perturbation theory after inflation, see eq.~\eqref{Omega GW}. The peak of the GW signal is constrained to be at very high frequencies, $f_\text{peak}\gtrsim 50\,\text{kHz}$, well beyond the reach of current Earth-based or future space-based GW detectors. Rather they could provide a target for future ultra-high frequency detectors discussed in~\cite{Aggarwal:2020olq}.

A key parameter for the inflationary dynamics is the curvature of the hyperbolic field space, $\mathcal{R}_\text{fs}\equiv -4/(3\alpha)$. 
In single-field models the small-scale $P_\zeta$ is enhanced by the presence of the inflection point in the potential, leading to a phase of ultra-slow-roll dynamics. 
The effect of $\alpha$ is primarily to set the relative amplitude of the tensor modes with respect to  scalar perturbations at CMB scales, see eq.~\eqref{r universal prediction inflection point}.
In multi-field embeddings a far richer dynamical behaviour is possible. We have seen that a large and negative curvature, $\alpha\ll1$, can cause a geometrical instability in the inflationary trajectory, as previously studied in~\cite{Renaux-Petel:2015mga, Garcia-Saenz:2018ifx}. 
A strongly-curved field space could also be accompanied by strongly non-geodesic motion and a tachyonic instability in the isocurvature perturbation, which gets transferred to the curvature fluctuation, potentially leading to a small-scale growth of the scalar power spectrum~\cite{Fumagalli:2020adf, Braglia:2020eai, Aragam:2021scu}. 
For these reasons we choose to focus on models characterised by $\alpha\ll1$, see also \cite{Christodoulidis:2018qdw}. 
The behaviour we see in this case differs from that seen in other multi-field $\alpha$-attractor models studied, for example, in~\cite{Achucarro:2017ing} where the authors consider a potential monotonic in the radial direction, with field-space curvature $\alpha=1/3$. 
In that case 
the single-field predictions, eqs.~\eqref{ns prediction alpha attractors} and~\eqref{r prediction alpha attractors}, were found to be stable even in presence of a light angular direction.  
Here we introduce both a feature in the radial potential (the inflection point) and consider larger curvature ($\alpha\ll1$), both of which amplify the geometrical destabilisation of the background trajectory, breaking the slow-roll and slow-turn approximations. While the radial field at early times follows the standard evolution close to the boundary of the Poincaré disc ($r\rightarrow1$), the dynamics of the angular field then leads to a second distinct phase of inflation at late times. Nonetheless on large scales CMB observables can still be explained by means of a simple modification of the standard universal predictions of single-field $\alpha$--attractors, see eqs.~\eqref{ns universal prediction inflection point} and~\eqref{r universal prediction inflection point}.
 
%
%
%

\begin{figure}
\centering
\captionsetup[subfigure]{justification=centering}
   \begin{subfigure}[b]{0.48\textwidth}
    \includegraphics[width=\textwidth]{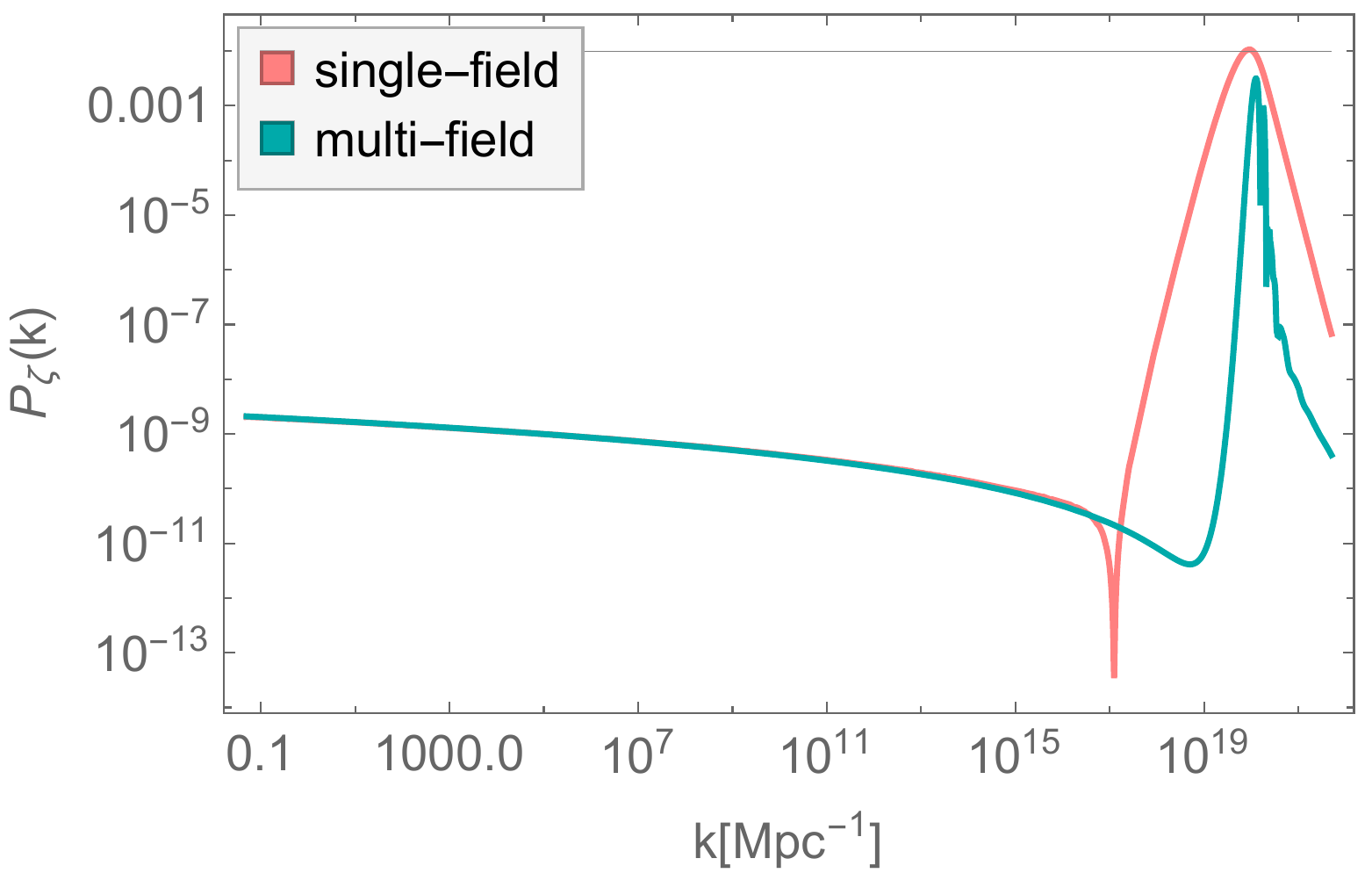}
  \end{subfigure}
   \begin{subfigure}[b]{0.48\textwidth}
    \includegraphics[width=\textwidth]{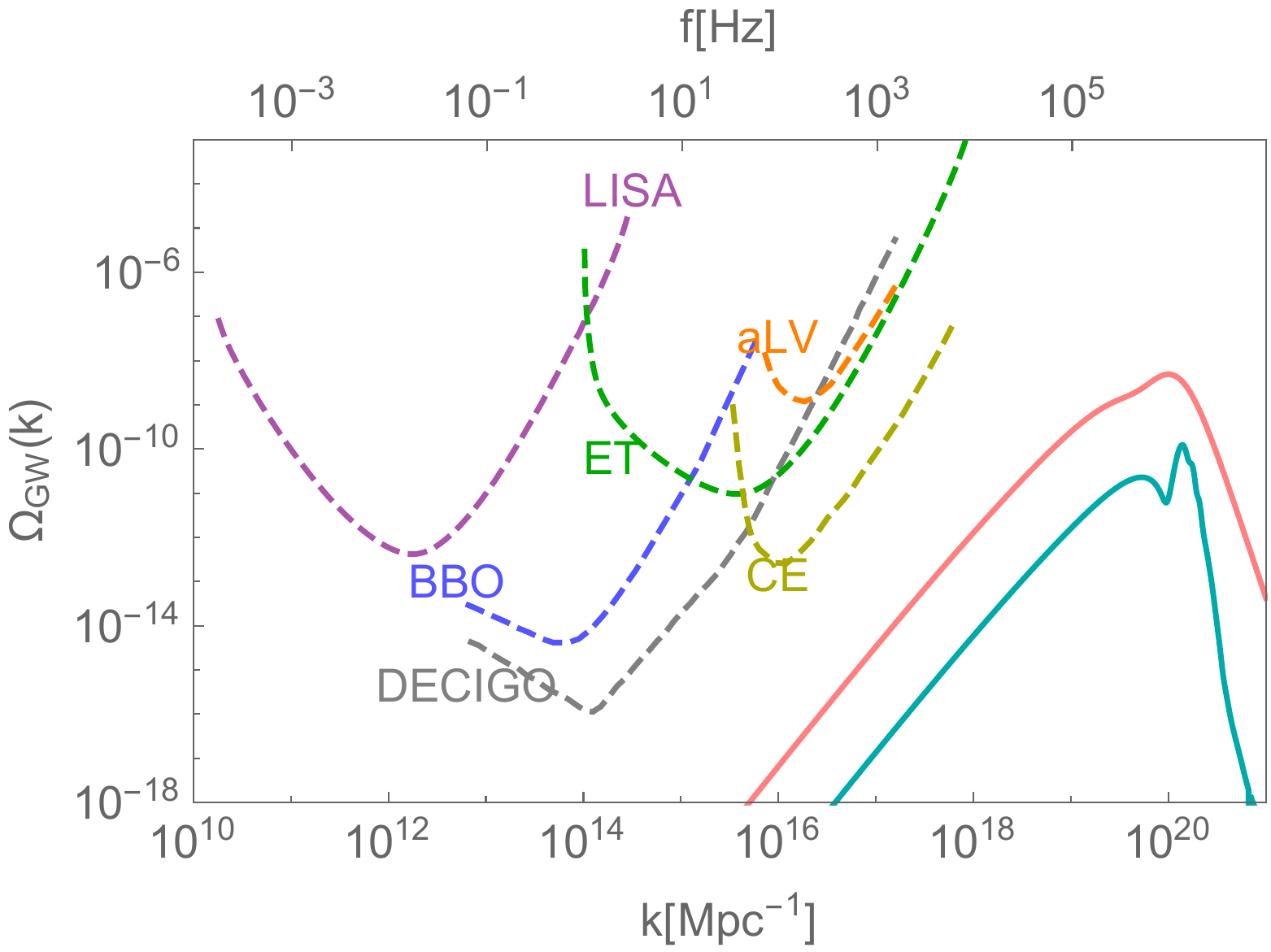}
  \end{subfigure}
\caption{\textit{Left panel:} numerical results of $P_\zeta(k)$ for a single-field inflection-point model, eq.~\eqref{potential}, with parameters $\{\alpha=0.1, \, \phi_\text{infl}=0.5,\, \xi=0.0035108\}$ (pink line) and for a multi-field model, eq.~\eqref{potential multifield}, with parameters $\{\alpha=0.005,\, \gamma=10, \phi_\text{infl}=0.07\}$ and initial condition $\theta_\text{in}=7\pi/10$ (green line). These models predict $n_s=0.9568$ and $n_s=0.9569$ respectively and are both compatible with the CMB lower bound on $n_s$, \eqref{CMB 95percent ns lower bound}. \textit{Right panel:} numerical results for the second-order GWs produced during radiation domination by the enhanced scalar perturbations whose power spectra are displayed in the left panel, together with the sensitivity curves of current and upcoming GW observatories. The colour legend is the same as on the left.}
\label{fig:summary single and multi field}
\end{figure}

We find marked differences in the spectral shape of the scalar power spectrum found in the single- and multi-field cases. 
In the single-field case the peak in $P_\zeta$ is broad ($n_\text{IR}\simeq 3.4,\, n_\text{UV}\simeq -4$), while in the multi-field case it is narrower, with a much steeper infrared growth ($n_\text{IR}\simeq12$), and oscillations following the principal peak. Figure~\ref{fig:summary single and multi field} shows a comparison of the scalar power spectra obtained in a single- and a multi-field inflection-point model. 
Multi-field effects can explain the oscillatory behaviour of the green line after the peak, caused by a sharp and strong turn in field space, providing an explicit realisation of the mechanism discussed in~\cite{Fumagalli:2020nvq, Fumagalli:2021cel} (see also~\cite{Braglia:2020taf}).

The differences in the spectral shape of the peak in $P_\zeta$ are reflected in the power spectrum of the induced GWs, as shown in the right panel of figure~\ref{fig:summary single and multi field}. 
In particular, the narrower, oscillatory peak in the multi-field $P_\zeta$ leads to second-order GWs with a two-peak structure and an oscillatory modulation of the second (principal) peak, see the green line in the right panel of figure~\ref{fig:summary single and multi field}. In the single-field set-up, the two-peak structure is almost wiped out as a consequence of the broadness of the scalar power spectrum peak and the oscillations are absent, see the pink line in the right panel of figure \ref{fig:summary single and multi field}.  
%
%

Strong enhancement of the scalar perturbations on small scales might be expected to be associated with significant non-Gaussianity. This can have an important impact on both the production of PBHs~\cite{Bullock:1996at, Saito:2008em, Byrnes:2012yx, Young:2013oia,Bugaev:2013vba, Young:2014oea, Young:2015cyn, Franciolini:2018vbk, Atal:2018neu, DeLuca:2019qsy, Passaglia:2018ixg,Ozsoy:2021qrg, Taoso:2021uvl, Davies:2021loj} and second-order GWs induced~\cite{Domenech:2021ztg, Cai:2018dig, Unal:2018yaa, Yuan:2020iwf, Atal:2021jyo, Adshead:2021hnm, Ragavendra:2020sop}, and it would be interesting to explore non-Gaussianities in these models and their possible effect on the small-scale phenomenology.

\textbf{Note added:} while this paper was under peer review, several related works appeared on the arXiv \cite{Kallosh:2022feu, Kallosh:2022ggf, Kallosh:2022vha, Fumagalli:2021dtd, Dalianis:2021dbs, Geller:2022nkr, Bhattacharya:2022fze}. 

\acknowledgments
The authors are grateful to Matteo Braglia and Andrew Gow for numerous discussions, as well as Dani Figueroa, Jacopo Fumagalli, Juan García-Bellido, Sébastien Renaux-Petel, Angelo Ricciardone and Lukas Witkowski for very useful comments. The work of LI is supported by UK STFC grant ST/R505018/1 and ST/S505651/1. DW and HA are supported by UK STFC grant ST/S000550/1. M.F. would like to acknowledge support from the “Atracci\'{o}n de Talento” grant 2019-T1/TIC15784, his work is partially supported by the Spanish Research Agency (Agencia
Estatal de Investigaci\'{o}n) through the Grant IFT Centro de Excelencia Severo Ochoa No CEX2020-001007-S, funded by MCIN/AEI/10.13039/501100011033. LI acknowledges the kind hospitality of Jean Allison and her family while this work was completed. For the purpose of open access the authors have applied a Creative Commons Attribution (CC-BY) licence to any Author Accepted Manuscript version arising. Supporting research data are available on reasonable request from the corresponding author.

\appendix 
\section[\texorpdfstring{Universality of $\bm{\alpha}$--attractors}{Universality of alpha-attractors}]{Universality of $\bm{\alpha}$--attractors}
\label{sec: appendix universality class}
Cosmological $\alpha$--attractors correspond to a class of inflationary models which provide robust observational predictions despite having apparently different formulations (see~\cite{Kallosh:2013hoa} and references therein). In particular, the large-scale CMB spectral index and tensor-to-scalar ratio are given by eqs.~\eqref{ns prediction alpha attractors} and~\eqref{r prediction alpha attractors} at leading order in the expansion in terms of $1/\Delta N_\text{CMB}$. 
We review here how the $\alpha$--attractor models form a universality class and derive the observables $(n_s,\; r_\text{CMB}, \; \alpha_s)$ for single-field $\alpha$--attractor models given by a monomial potential in terms of the radial distance from the centre of the Poincaré disc \eqref{canonical field transformation} 
\begin{equation}
\label{potential universality class}
    V(\phi)=V_0 \tanh^p{\left(\frac{\phi}{\sqrt{6\alpha}}\right)} \;.
\end{equation}

In canonical single-field slow-roll inflation, $(n_s,\; r, \; \alpha_s)$ can be given in terms of potential slow-roll parameters as~\cite{Planck:2013jfk}
\begin{equation}
\label{ns and r slow roll V defs}
    n_s=1-6\epsilon_V+2\eta_V \;, \;\;\; r_\text{CMB}=16\epsilon_V \;, \;\;\; 
    \alpha_s=16 \epsilon_V\eta_V-24 {\epsilon_V}^2-2{\xi_V}^2 \;,
\end{equation}
where $\epsilon_V$, $\eta_V$ and $\xi_V$ are expressed in terms of derivatives of the inflaton potential,
\begin{equation}
\label{slow roll parameters potential defs}
    \epsilon_V=\frac{1}{2}\left(\frac{V_\phi}{V} \right)^2\;, \;\;\; \eta_V=\frac{V_{\phi\phi}}{V}\;, \;\;\; {\xi_V}^2=\frac{V_\phi V_{\phi\phi\phi}}{V^2} \;. 
\end{equation}
Given the explicit form of the potential in eq.~\eqref{potential universality class}, it is possible to write these potential slow-roll parameters in terms of $\phi$,
\begin{eqnarray}
 \epsilon_V &=& \frac{p^2}{3\alpha} \csch^2{\left(\frac{2\phi}{\sqrt{6\alpha}}\right)}\:, \\
 \eta_V &=& \frac{2p}{3\alpha}\csch^2{\left(\frac{2\phi}{\sqrt{6\alpha}}\right)} \left(p-\cosh{\left(\frac{2\phi}{\sqrt{6\alpha}}\right)} \right)\;, \\
 \xi_V^2 &=& \frac{2p^2}{9\alpha^2} \csch^4{\left(\frac{2\phi}{\sqrt{6\alpha}}\right)} \left(3+2p^2-6p\cosh{\left(\frac{2\phi}{\sqrt{6\alpha}}\right)} +\cosh{\left(\frac{4\phi}{\sqrt{6\alpha}}\right)}  \right) \;.
\end{eqnarray}
Substituting the above into eq.~\eqref{ns and r slow roll V defs} yields the large scales observables in terms of $\phi$. They are evaluated when the CMB scales left the horizon, i.e., at $\phi=\phi_\text{CMB}$. In order to do so, we use the inflaton equation of motion, eq.~\eqref{single field eq of motion phi}, which in the slow-roll approximation can be simplified to give
\begin{equation}
    \frac{\mathrm{d}\phi}{\mathrm{d}N}\simeq-\frac{V_\phi}{V} \;.
\end{equation}
Integrating the equation above yields the number of e-folds elapsed between the two field values $\phi_\text{CMB}$ and $\phi_\text{end}$, 
\begin{equation}
    \Delta N_\text{CMB}
    \simeq \int_{\phi_\text{end}}^{\phi_\text{CMB}} \mathrm{d}\phi\,\frac{V}{V_\phi}  \;.
\end{equation}
Performing the integration above for the potential in eq.~\eqref{potential universality class} yields
\begin{equation}
\label{N universality class}
    \Delta N_\text{CMB} \simeq \frac{3\alpha}{2p} \left[\cosh{\left(\frac{2\phi_\text{CMB}}{\sqrt{6\alpha}}\right)} -\cosh{\left(\frac{2\phi_\text{end}}{\sqrt{6\alpha}}\right)} \right] \;,
\end{equation}
where the value of $\phi_\text{end}$ is fixed by the condition $\epsilon(\phi_\text{end})=1$, corresponding to 
\begin{equation}
\label{phi end}
    \sinh^2{\left(\frac{2\phi_\text{end}}{\sqrt{6\alpha}}\right)} \simeq \frac{p^2}{3\alpha} \;.
\end{equation}
Substituting eq.~\eqref{phi end} into eq.~\eqref{N universality class} and expressing the equation in terms of $\phi_\text{CMB}$ yields
\begin{equation}
    \sinh^2{\left(\frac{\phi_\text{CMB}}{\sqrt{6\alpha}}\right)} \simeq \frac{p \Delta N_\text{CMB}}{3\alpha} +\frac{\sqrt{3\alpha+p^2}}{2\sqrt{3\alpha}}-\frac{1}{2} \;.
\end{equation}
In this way the CMB observables in eq.~\eqref{ns and r slow roll V defs} can be written as 
\begin{align}
    n_s&\simeq 1-\frac{2\Delta N_\text{CMB}+\frac{1}{p}\sqrt{3\alpha(3\alpha+p^2)}+\frac{3\alpha}{2}}{\Delta N_\text{CMB}^2 +\frac{\Delta N_\text{CMB}}{p}\sqrt{3\alpha(3\alpha+p^2)}+\frac{3\alpha}{4}} \;, \\
    r_\text{CMB}&\simeq \frac{12\alpha}{\Delta N_\text{CMB}^2 +\frac{\Delta N_\text{CMB}}{p}\sqrt{3\alpha(3\alpha+p^2)}+\frac{3\alpha}{4}} \;, \\
    \alpha_s&\simeq -8\frac{18\alpha^2+p(3\alpha+4\Delta N_\text{CMB})\sqrt{3\alpha(3\alpha+p^2)} +p^2[4\Delta N_\text{CMB}^2+3\alpha(1+2\Delta N_\text{CMB})]}{(3p\alpha +4 \Delta N_\text{CMB} \sqrt{3\alpha(3\alpha+p^2)} +4 p \Delta N_\text{CMB}^2 )^2} \;.
\end{align}
In the large $\Delta N_\text{CMB}$ expansion, these expressions reduce to eqs.~\eqref{ns prediction alpha attractors} and~\eqref{r prediction alpha attractors} for the observables $n_s$ and $r_\text{CMB}$ respectively on large (CMB) scales, while we can relate the running of the spectral index to the spectral index itself
\begin{equation}
\label{alpha_s universal approximation}
\alpha_s\approx-\frac{2}{\Delta N_\text{CMB}^2}\approx-\frac{(n_s-1)^2}{2} \,, 
\end{equation}
regardless of the parameters of $V_0$ and $p$ appearing in the potential \eqref{potential universality class}. The spectral index $n_s$ and running $\alpha_s$ are dependent only on $\Delta N_\text{CMB}$, while $r_\text{CMB}$ depends only on $\Delta N_\text{CMB}$ and $\alpha$, such that
\begin{equation}
    r_\text{CMB} \approx 3\alpha(n_s-1)^2 \,.
\end{equation}
This is due to the potential \eqref{potential universality class} remaining finite at the boundary of the moduli space ($r\equiv \tanh{\left(\phi/\sqrt{6\alpha}\right)}  \rightarrow 1$), which is a key feature of $\alpha$--attractor models. The transformation to the canonical field $\phi$ \eqref{canonical field transformation}, renders the potential a function of $\tanh{\left(\phi/\sqrt{6\alpha}\right)}$, which ensures the flatness of the potential for large field values ($\phi \rightarrow \infty$) and makes observational predictions on large scales approximately independent of the precise form of the function describing the potential dependence on $\tanh{\left(\phi/\sqrt{6\alpha}\right)}$.

\section{Numerical computation of the single-field scalar power spectrum}
\label{sec:app numerical P_zeta}
In this appendix we present in detail how we compute the scalar power spectrum, $\mathcal{P}_\zeta(k)$, for the single-field inflation models considered in section~\ref{sec: single field model}. Our procedure is closely related to similar strategies described in the literature, see for example~\cite{Ballesteros:2017fsr, Ballesteros:2020qam}. 

The inflaton potential in eq.~\eqref{potential} features an approximate stationary inflection point located at $\phi_\text{infl}$ whose effect is to introduce a transient ultra-slow-roll phase. The slow-roll approximation breaks down and a full numerical analysis of the Mukhanov--Sasaki equation is needed. 
As an example, we refer to the power spectrum resulting from the specific model $\{\alpha=0.1,\; \phi_\text{infl}=0.5,\; \xi=0.0035108\}$, whose background evolution is displayed in figure~\ref{fig:background evo}. 
For illustrative purposes we consider here the case of instant reheating, which implies $\Delta N_\text{CMB}\simeq55$ (see table~\ref{tab:configs xi non 0}). 

In the slow-roll approximation, valid on large scales, far from the inflection point, it is possible to estimate the scalar power spectrum using the expression for $P_\zeta(N)$ given in \eqref{slow roll power spectrum}, by substituting in the background quantities $H(N)$ and $\epsilon_H(N)$. $P_\zeta(N)$ can be transformed into $P_\zeta(k)$ by means of eq.~\eqref{scale relation} for $k(N)$. We normalise the amplitude of the scalar power spectrum at $k_\text{CMB}=0.05\,\text{Mpc}^{-1}$ by using the \textit{Planck} 2018 measurement of $\mathcal{A}_s$~\cite{Planck:2018jri}, which in turns identifies the amplitude of the potential as $V_0=7.7\times 10^{-10}$. Eqs.~\eqref{spectral index}--\eqref{tensor to scalar ratio} predict for this configuration the large-scale observables 
\begin{equation}
\label{CMB predictions appendix case}
    n_s=0.9569 \;,\;\;\;
    \alpha_s=-0.00092 \;,\;\;\;
    r_{0.002}=4.956\times 10^{-4} \;,
\end{equation}
consistent with the latest \textit{Planck} data release~\cite{Planck:2018jri},
where the spectral tilt and its running are evaluated at $k_\text{CMB}$. 

While eq.~\eqref{slow roll power spectrum} is reliable when the evolution is well described by slow roll, a full numerical analysis of the perturbations is necessary in order to compute the exact scalar power spectrum including non-slow-roll evolution, as occurs through the inflection point and approaching the end of inflation. The Mukhanov--Sasaki equation describes the evolution of the scalar curvature perturbation associated with the comoving wavenumber $k$, $\zeta_k$. We define $\zeta_k\equiv v_k/z$, where $z\equiv a\,\phi'$, and a prime denotes derivatives with respect to e-folds, $N$. The mode $v_k$ is given by the solution of the Mukhanov--Sasaki equation~\cite{Bassett:2005xm}
\begin{equation}
\label{MS}
    v_k''+(1-\epsilon_H)v_k'+\left[\frac{k^2}{a^2H^2} +(1+\epsilon_H-\eta_H)(\eta_H-2)-(\epsilon_H-\eta_H)'\right]v_k=0 \;,
\end{equation}
where the Hubble slow-roll parameters, $\epsilon_H$ and $\eta_H$, are defined in eqs.~\eqref{epsilon H} and~\eqref{eta H}. Note however that this equation does not assume slow roll.

We solve eq.~\eqref{MS} for modes ranging from $k_\text{CMB}=0.05\,\text{Mpc}^{-1}$ to $k_\text{end}=2.6\times 10^{22}\,\text{Mpc}^{-1}$ and follow the evolution of $v_k$ for each wavemode from the sub-horizon regime ($k\gg aH$, where canonical quantum commutation relations give the normalisation for the mode function, $|v_k^2|=1/2k$~\cite{Bassett:2005xm}) to super-horizon scales ($k\ll aH$). These solutions then enable us to calculate the scalar power spectrum on super-horizon scales as 
\begin{equation}
\label{power spectrum MS}
   P_\zeta(k)=\frac{k^3}{2\pi^2}\Big|\frac{v_k}{z} \Big|^2 \;,
\end{equation}
where each mode is evaluated well after horizon crossing ($k\ll aH$), when $\zeta_k\equiv v_k/z$ approaches a constant value. 

There are a few practical considerations regarding the strategy used to solve eq.~\eqref{MS}:
\begin{itemize}
    \item when numerically solving eq.~\eqref{MS}, it is useful to express $\epsilon_H$ and $\eta_H$ in terms of derivatives of the field $\phi(N)$, instead of using their definitions in terms of derivatives of $H(N)$. In particular, the relevant terms in eq.~\eqref{MS} reduce to~\cite{Kefala:2020xsx}
\begin{align}
    1-\epsilon_H&=1-\frac{\phi'^2}{2} \;, \\
    (1+\epsilon_H-\eta_H)(\eta_H-2)-(\epsilon_h-\eta_H)'&=-2-3\frac{\phi''}{\phi'}-\frac{\phi'''}{\phi'}+\frac{\phi'^2}{2}+\frac{\phi'\phi''}{2} \;,
\end{align}
where the second and third derivatives of the field are given by the Klein--Gordon equation \eqref{single field eq of motion phi} for the field, re-written in terms of $\phi(N)$,
\begin{equation}
    H^2 \phi''+HH'\phi'+3H^2\phi'+V_{\phi}=0 \,,
\end{equation}
and its derivative;
\item the factor $k/aH$ can be normalised by noting that the scale $k_\text{CMB}=0.05\,\text{Mpc}^{-1}$ crossed the horizon $\Delta N_\text{CMB}$ e-folds before the end of inflation;
\item in order to minimise numerical errors, one should evolve the background solution for a sufficiently long time before the relevant scales cross the horizon. This can be achieved by choosing a large enough initial value of $\phi$;
\item instead of solving directly for the complex perturbation, $v_k$, it is simpler to solve separately for its real and imaginary parts~\cite{Adams:2001vc}. For each mode, the integration of eq.~\eqref{MS} is started 5 e-folds before horizon crossing, where Bunch--Davies initial conditions are applied~\cite{Bassett:2005xm}, and is integrated up until the end of inflation. In terms of the real and imaginary part of $v_k$, the initial conditions are   
\begin{equation}
\label{vkin}
    \text{Re}\{v_k\}=\frac{1}{\sqrt{2k}} \;,\; \text{Re}\{v_k'\}=0 \;,\;
    \text{Im}\{v_k\}=0 \;,\; \text{Im}\{v_k'\}=-i\frac{\sqrt{k}}{\sqrt{2}k_\text{in}} \;,
\end{equation}
where $k_\text{in}$ is the mode that crossed the horizon when the integration is started;
\item the correct normalisation on CMB scales for the power spectrum in eq.~\eqref{power spectrum MS} at $k=k_\text{CMB}$ is set by fixing $V_0$ and hence the Hubble scale when $k_\text{CMB}$ leaves the horizon.
\end{itemize}

\begin{figure}
\centering
\includegraphics[scale=0.5]{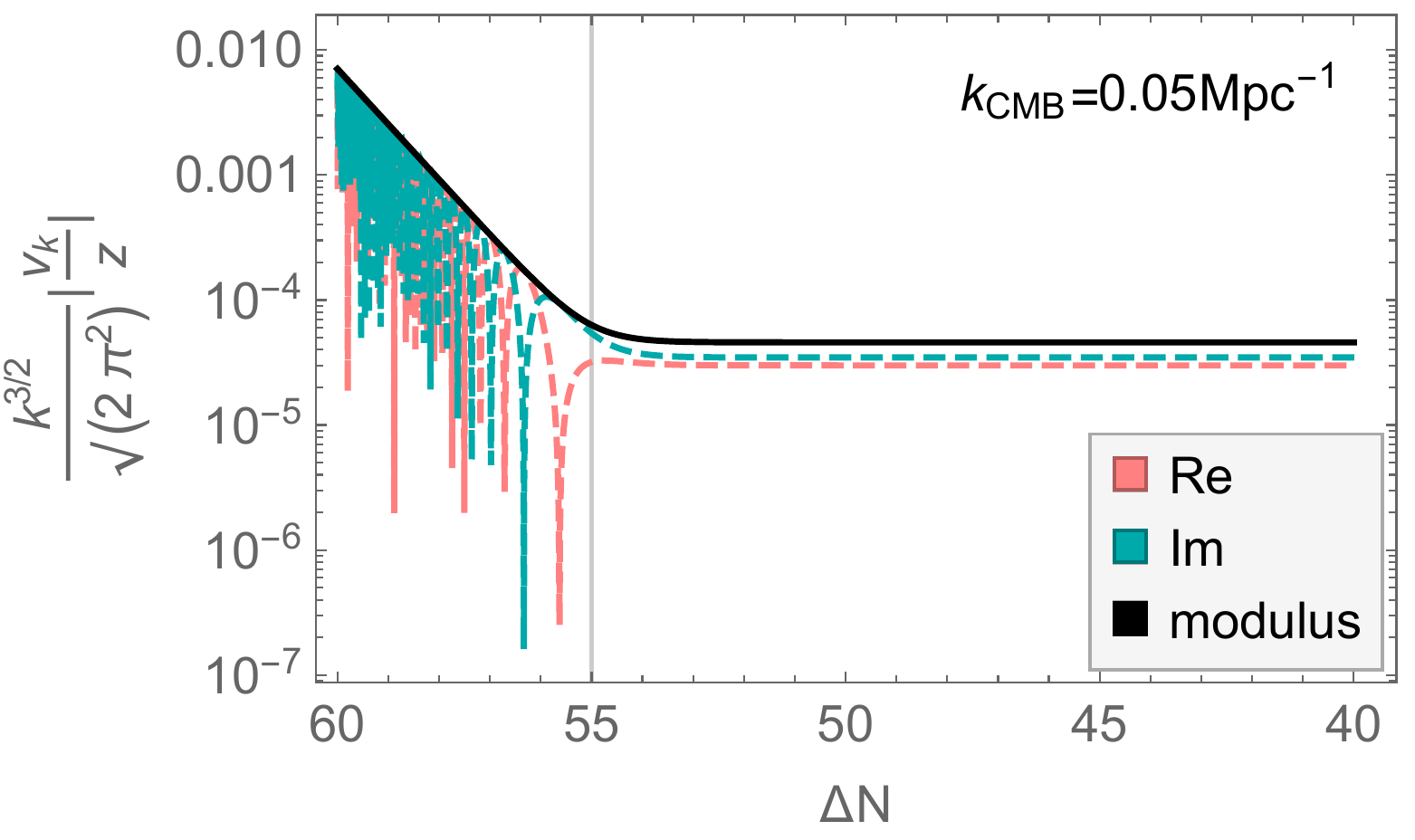}
\caption{Mode evolution, $v_k(N)/z$, for the CMB scale, $k_\text{CMB}$. The real and imaginary parts, and modulus correspond to the pink- and green-dashed lines, and the black-solid line respectively. The vertical thin line signals the moment in which the CMB scale crosses the horizon $\Delta N_\text{CMB}\sim 55$ e-folds before the end of inflation.}
\label{fig:CMB mode evolution}
\end{figure}

As an example, we solve eq.~\eqref{MS} for $k_\text{CMB}$ and plot in figure \ref{fig:CMB mode evolution} the mode evolution, $v_k(N)/z$. The mode starts off in the Bunch--Davies vacuum, oscillates in the sub-horizon regime and freezes to a constant value after crossing the horizon. 

\begin{figure}
\centering
\captionsetup[subfigure]{justification=centering}
   \begin{subfigure}[b]{0.49\textwidth}
    \includegraphics[width=\textwidth]{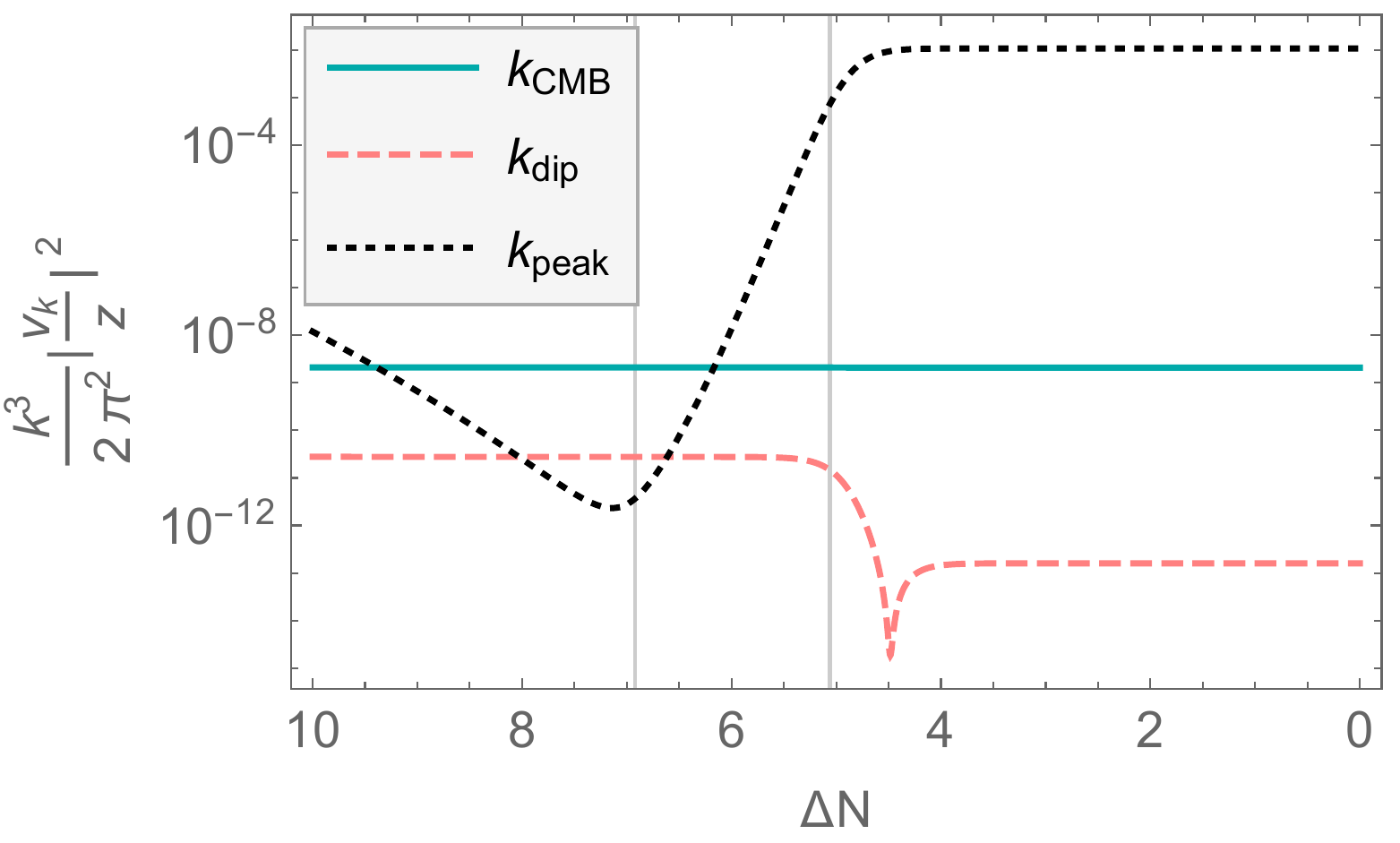}

  \end{subfigure}
  \begin{subfigure}[b]{0.46\textwidth}
    \includegraphics[width=\textwidth]{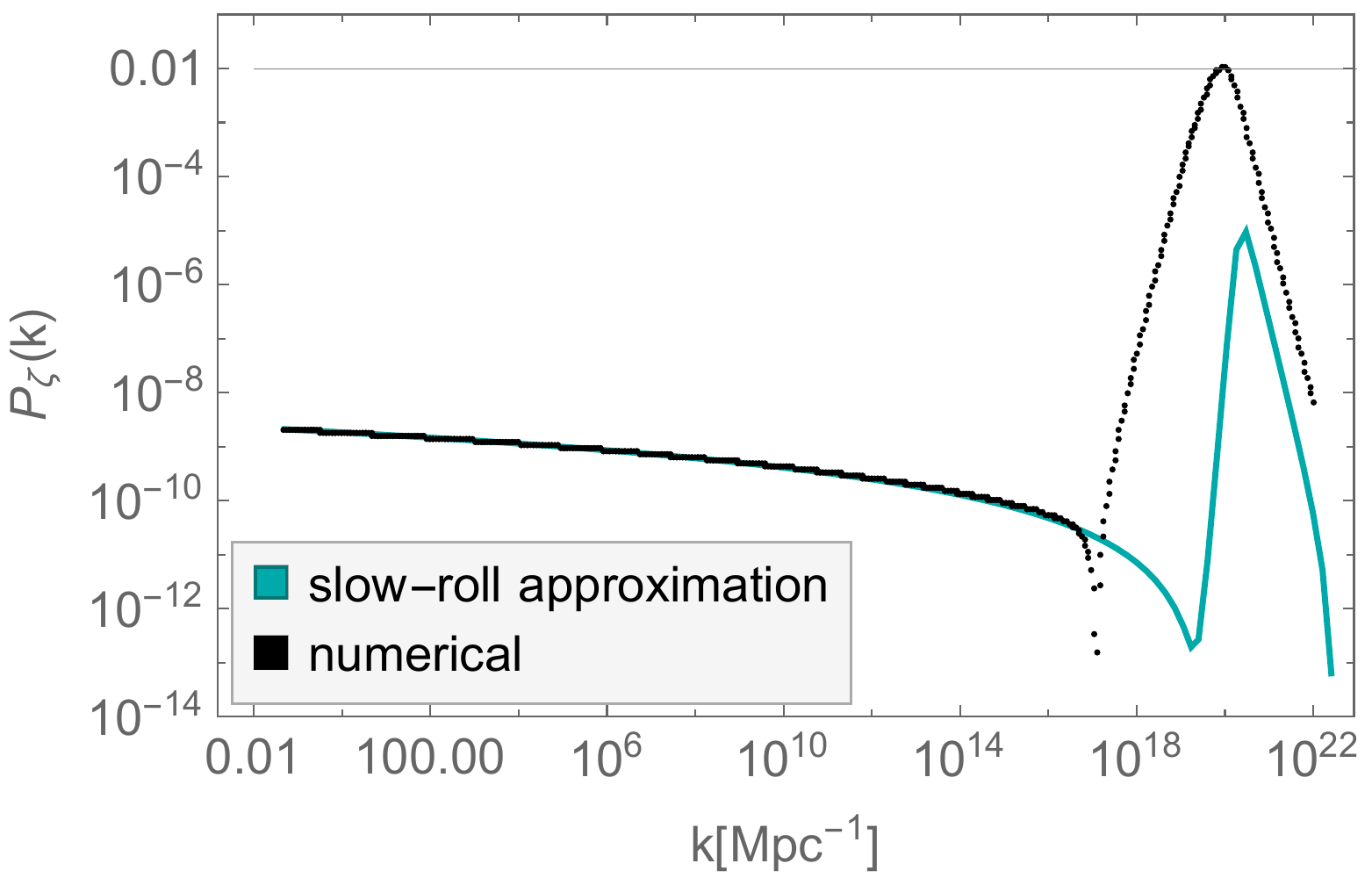}

  \end{subfigure}
 
 \caption{\textit{Left panel:} comparison between different modes' evolution in the last 10 e-folds of inflation. The region between the two thin vertical lines is characterised by $\eta_H>3$. \textit{Right panel:} numerical results for the scalar power spectrum, $P_\zeta(k)$, compared with the slow-roll approximation in eq.~\eqref{slow roll power spectrum}.}
  \label{fig:numerical results}
\end{figure}

In the left panel of figure \ref{fig:numerical results}, the evolution of three different modes is represented for comparison in the last 10 e-folds before the end of inflation. The continuous line is associated with the mode $k_\text{CMB}$, which for the range of e-folds represented is well outside of the horizon and frozen at a constant value. The dashed line describes the scale $k_\text{dip}=10^{17}\,\text{Mpc}^{-1}$, which corresponds to the dip in the scalar power spectrum. Finally, the mode corresponding to the peak in the scalar power spectrum, $k_\text{peak}=9\times 10^{19}\,\text{Mpc}^{-1}$, is plotted with the dotted line. The mode associated to $k_\text{peak}$ experiences the largest growth as it crosses the horizon close to the onset of the ultra-slow-roll phase. See~\cite{Ballesteros:2020qam} for a detailed discussion of the mechanism of growth (suppression) which shapes the modes' evolution and the scalar power spectrum. 

In the right panel of figure \ref{fig:numerical results} we show a comparison of the numerical results for $P_\zeta(k)$ compared with the slow-roll approximation in eq.~\eqref{slow roll power spectrum}. On large scales the numerical results agree very well with the slow-roll approximation, showing that the slow-roll CMB predictions in \eqref{CMB predictions appendix case} are reliable on these scales. On the other hand, on small scales the exact power spectrum differs substantially from the slow-roll approximation, both in terms of the position and the height of the peak. For the configuration under analysis, the power spectrum features a peak of $\mathcal{O}(0.01)$ at the comoving scale $k_\text{peak}$. 

\section{Limiting behaviour of the single-field potential}
\label{sec: appendix limiting behaviour potential}

In sections~\ref{sec: single field model} and \ref{sec: phenomenology of the single field model} we considered the phenomenology of the inflection-point potential in eq.~\eqref{potential} with the aim of realising an ultra-slow-roll phase and enhancing the scalar perturbations on small scales. For completeness we discuss here the limiting behaviour of the same potential when the inflection point is located at small or large $\phi$ values. For simplicity we restrict our discussion to the case $\xi=0$ and choose $\alpha=0.1$. 

In the limit $\phi_\text{infl}\rightarrow 0$, the dominant contribution to the potential in eq.~\eqref{potential} comes from the term proportional to $\tanh^3{(\phi/\sqrt{6\alpha})}$,
\begin{equation}
    \lim_{\phi_\text{infl}\rightarrow0} V(\phi)\simeq V_0 \left[\frac{1}{3} \coth{\left(\frac{\phi_\text{infl}}{\sqrt{6\alpha}}\right)} \right]^2 \, \tanh^6{\left(\frac{\phi}{\sqrt{6\alpha}}\right)} \;.
\end{equation}
It is therefore interesting to analyse the inflationary predictions of the inflection-point potential with $\phi_\text{infl}$ small and compare them with those obtained from the $\alpha$--attractor T-model potential~\cite{Kallosh:2015zsa}
\begin{equation}
\label{potential for comparison}
    U(\phi)=U_0 \tanh^6{\left(\frac{\phi}{\sqrt{6\alpha}}\right)} \;.
\end{equation}

\begin{figure}
\centering
\includegraphics[scale=0.45]{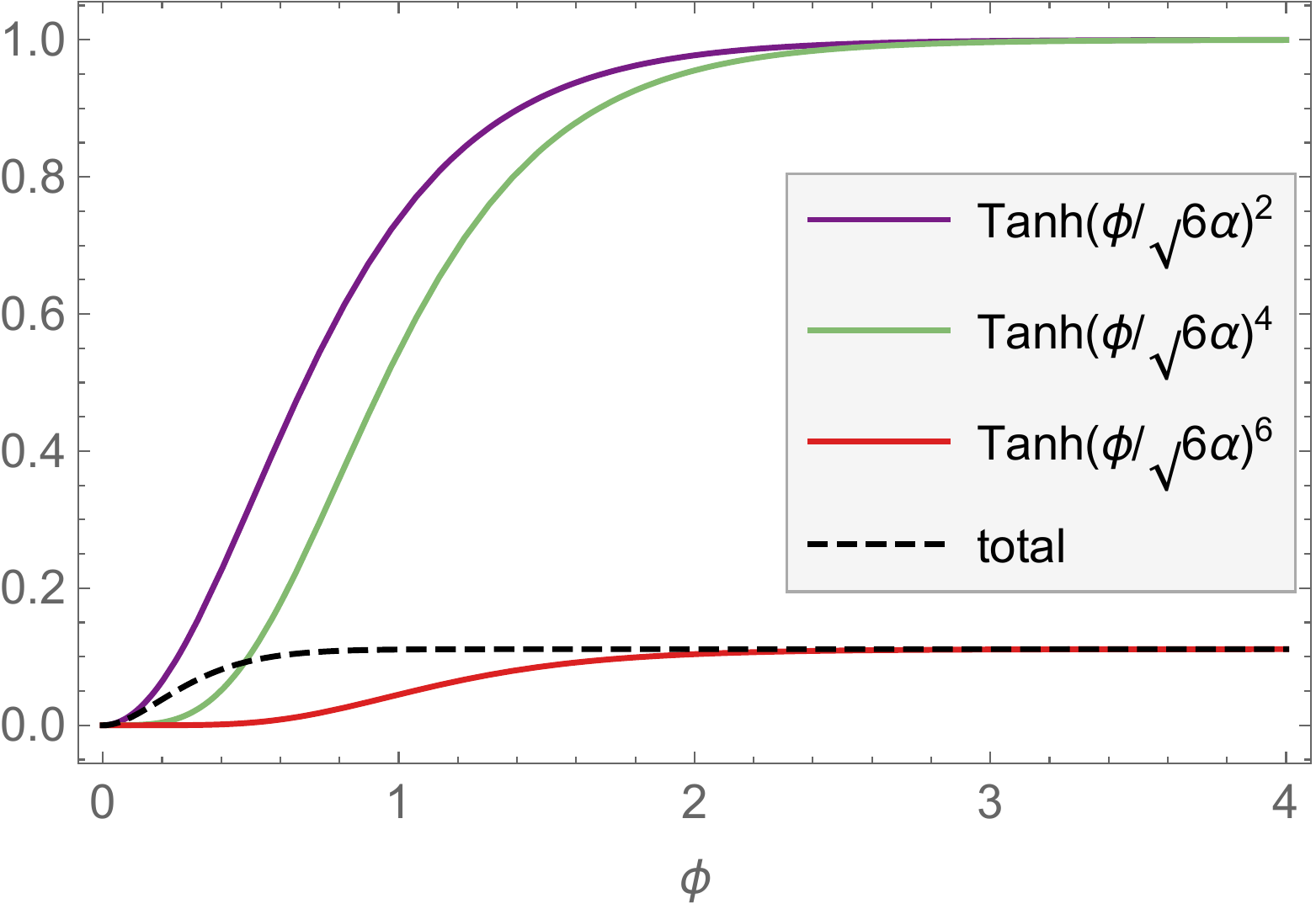}
\caption{Behaviour of the inflection-point potential in the large $\phi_\text{infl}$ limit. The plot is produced with the parameters $\{\alpha=0.1,\,\phi_\text{infl}=10,\,\xi=0\}$. The potential $V(\phi)/V_0$ in eq.~\eqref{potential} (black-dashed line) is plotted together with the contributions coming from the single terms.}
\label{fig:large phi infl limit}
\end{figure}

On the other hand, when $\phi_\text{infl}$ is large $\tanh{(\phi_\text{infl}/\sqrt{6\alpha})}\to1$ and there is no simple limiting behaviour for the inflection-point potential, as illustrated in figure~\ref{fig:large phi infl limit}. For large $\phi$ values ($\phi\gtrsim2$ for the configuration plotted in figure~\ref{fig:large phi infl limit}) the dominant contribution comes from the $\tanh^3{(\phi/\sqrt{6\alpha})}$ term (since $f_1\approx-f_2$ in \eqref{cond3} and the $\tanh{(\phi/\sqrt{6\alpha})}$ and $\tanh^2{(\phi/\sqrt{6\alpha})}$ terms approximately cancel), but for smaller $\phi$ the potential receives contributions from all the terms. 

\begin{figure}
\centering
\captionsetup[subfigure]{justification=centering}
   \begin{subfigure}[b]{0.46\textwidth}
    \includegraphics[width=\textwidth]{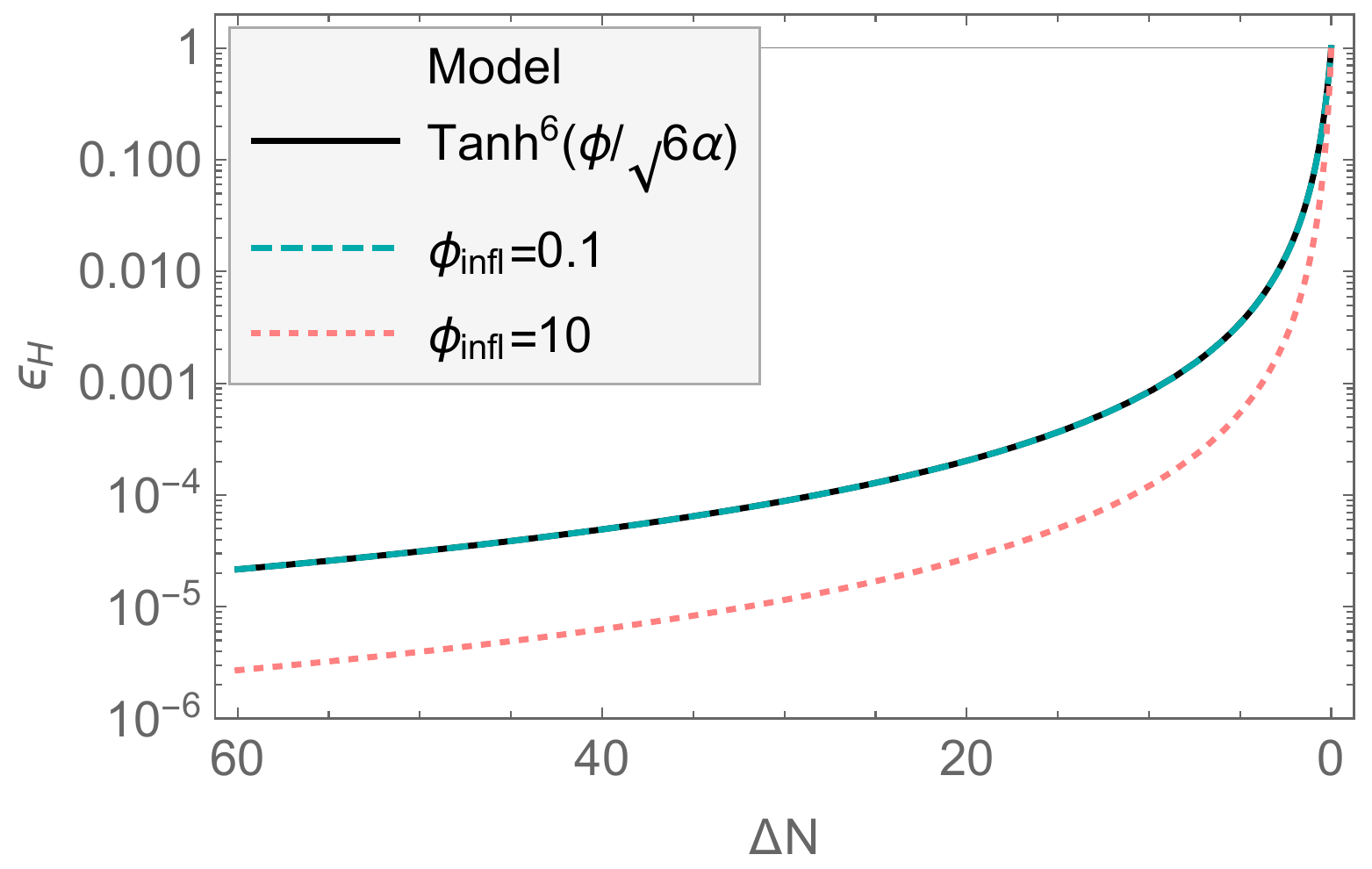}

  \end{subfigure}
  \begin{subfigure}[b]{0.44\textwidth}
    \includegraphics[width=\textwidth]{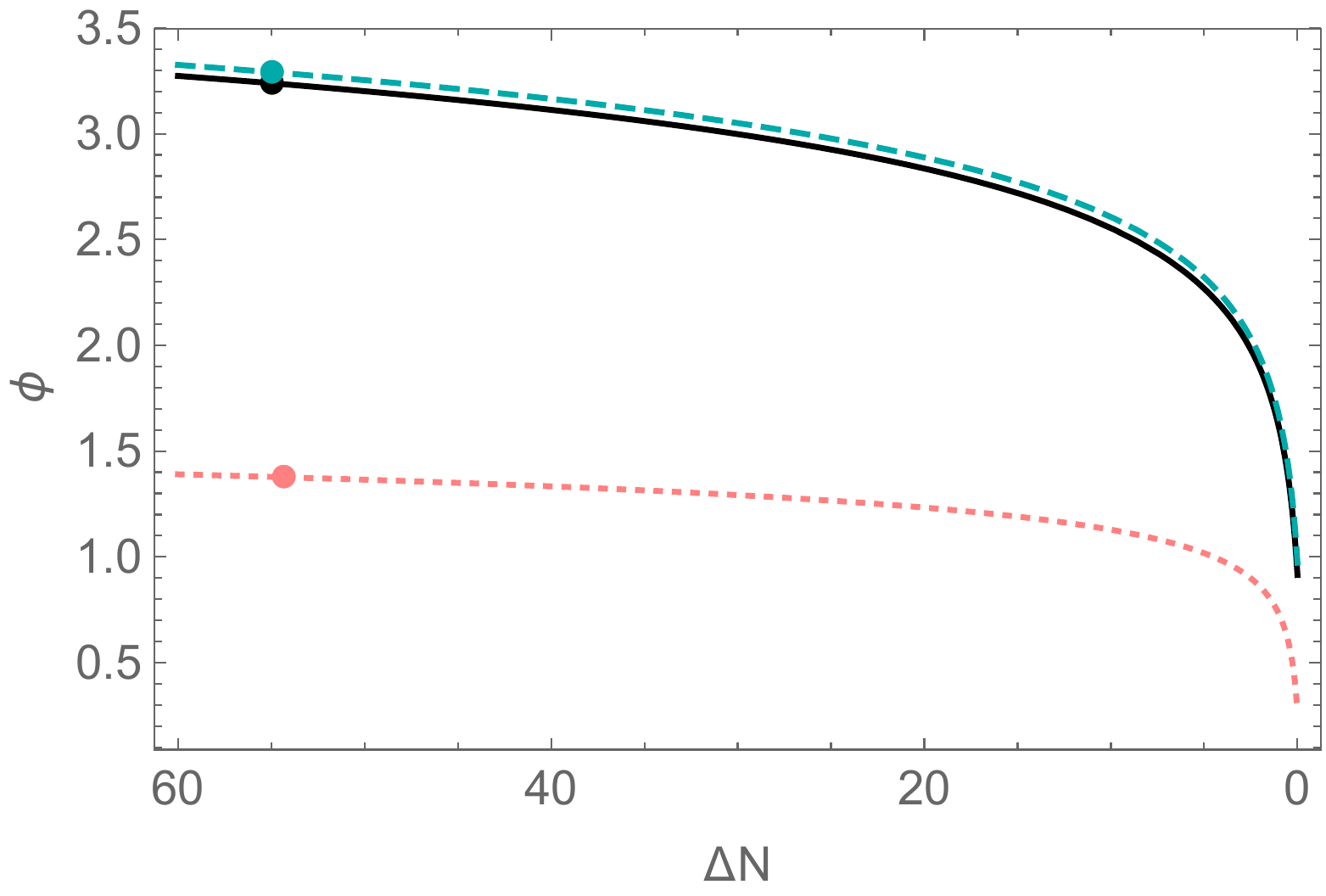}

  \end{subfigure}
   \begin{subfigure}[b]{0.44\textwidth}
    \includegraphics[width=\textwidth]{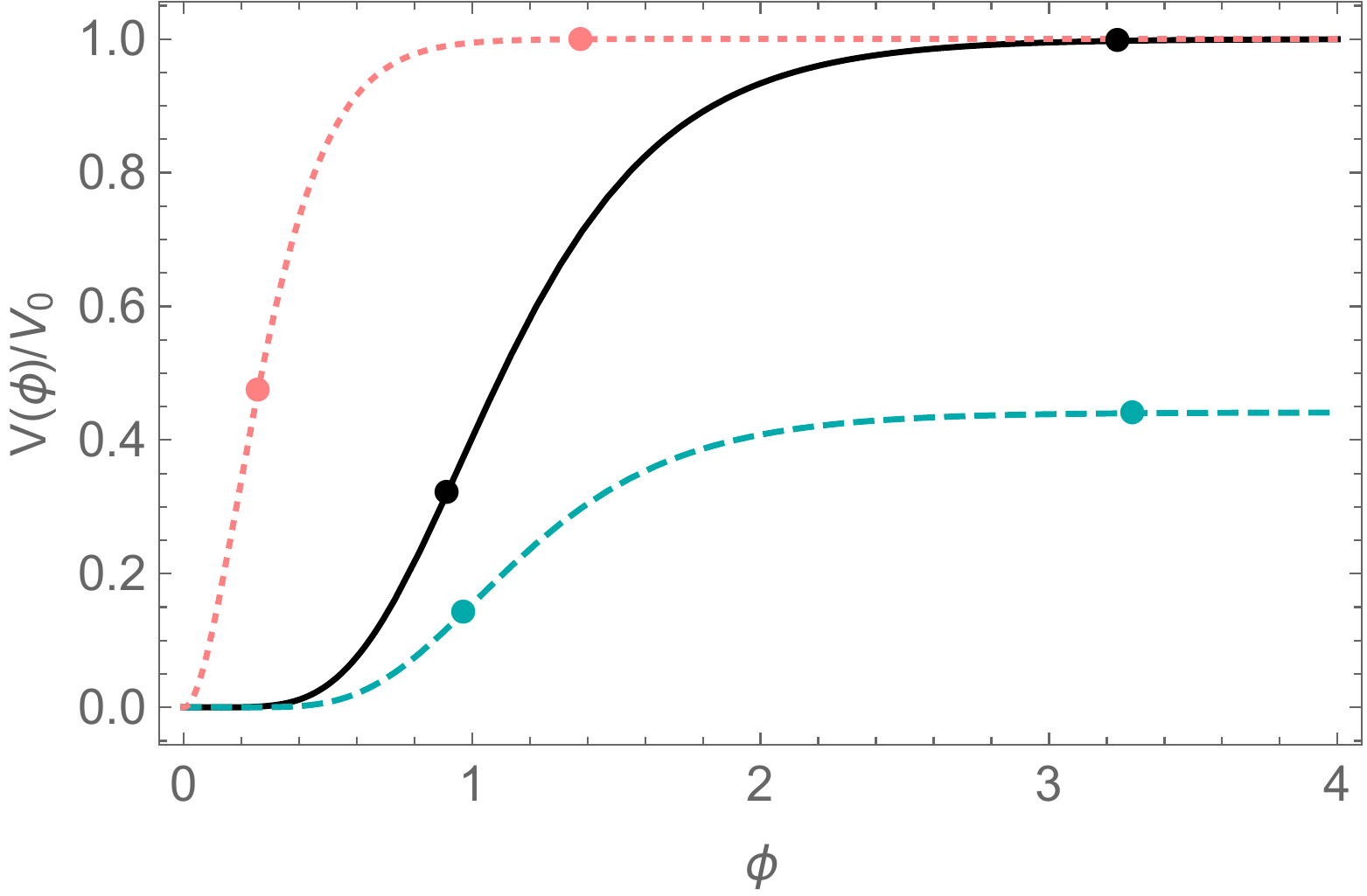}

  \end{subfigure}
  \begin{subfigure}[b]{0.46\textwidth}
    \includegraphics[width=\textwidth]{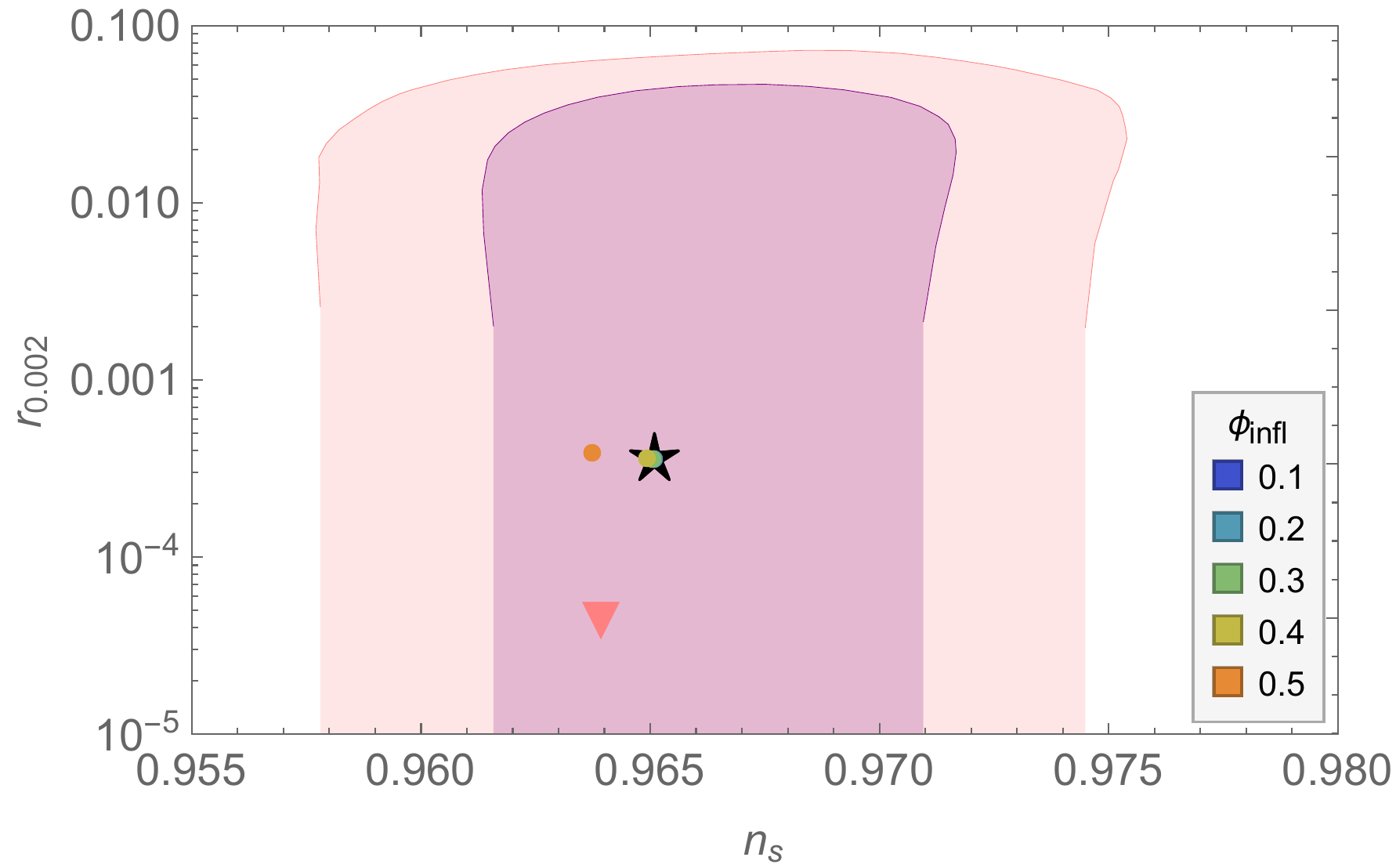}

  \end{subfigure}
 
 \caption{\textit{Top Row:} Comparison between the background evolution of the inflection-point model with $\phi_\text{infl}=\{0.1,\,10\}$ and the $\alpha$--attractor T-model in eq.~\eqref{potential for comparison}. The bullet points in the right-hand plot indicate the field values when the CMB scale crossed the horizon. \textit{Bottom Row:} In the left panel the potentials are plotted together with bullet points indicating the field values corresponding to the CMB scale (right bullet) and the end of inflation (left bullet). The legend identifying each line is the same as in the top-left panel. In the right panel the predictions of the CMB observables are represented together with the marginalised joint $68\,\%$ and $95\,\%$ C.L. regions in the $(n_s,\,r)$ plane at $k=0.002\,\text{Mpc}^{-1}$ as obtained from $\text{\textit{Planck}}+\text{BK}15+\text{BAO}$ data assuming the $\Lambda\text{CDM}+r_\text{CMB}$ cosmological model~\cite{Planck:2018jri}. The black star corresponds to the $\alpha$--attractor T-model potential \eqref{potential for comparison}, the pink triangle corresponds to the inflection-point potential with $\phi_\text{infl}=10$ and the remaining coloured circles correspond to different choices of small $\phi_\text{infl}$ indicated in the legend.}
  \label{fig:limiting behaviour potential}
\end{figure}

We consider two benchmark values, $\phi_\text{infl}=\{0.1,\,10\}$ in eq.~\eqref{potential}, in order to study their background evolution and compare it with that obtained from the potential in eq.~\eqref{potential for comparison}. The evolution of the Hubble slow-roll parameter, $\epsilon_H$, and the inflaton field, $\phi$, are plotted in the top row of figure \ref{fig:limiting behaviour potential} against the number of e-folds to the end of inflation, $\Delta N\equiv N_\text{end}-N$.
As expected, the background evolution for the inflection-point potential with $\phi_\text{infl}=0.1$ is almost identical to that produced by the $\alpha$--attractor T-model potential \eqref{potential for comparison}. The evolution corresponding to $\phi_\text{infl}=10$ is instead quite different and the reason why this is the case is clear from the bottom-left panel of figure~\ref{fig:limiting behaviour potential}, where the potentials and the field values corresponding to scales observed in the CMB are shown.
For the model with $\phi_\text{infl}=10$, the inflationary evolution observable in the CMB is located at $\phi<2$ where the potential is not well-approximated by the function $\tanh^6{(\phi/\sqrt{6\alpha})}$, as discussed above and illustrated in figure~\ref{fig:large phi infl limit}. 
The bottom-right panel in figure~\ref{fig:limiting behaviour potential} shows the CMB observables $(n_s,\,r_\text{CMB})$ at the scale $0.002 \,\text{Mpc}^{-1}$ predicted by each potential. One sees that the predictions obtained with $\phi_\text{infl}=0.1$ are not distinguishable from those produced by the T-model for $\phi_\text{infl}\leq 0.4$. The slightly lower value of $n_s$ predicted by $\phi_\text{infl}=0.5$ could in future allow us to distinguish it form the $\alpha$--attractor T-model. As expected, the predictions of the model with $\phi_\text{infl}=10$ differ from the T-model ones and the two could potentially be distinguished by the predicted value of $r_{0.002}$, lower for the inflection-point potential with $\phi_\text{infl}=10$.

\section{Parameter study of the multi-field potential}
\label{appendix: parameter study multifield potential}
The multi-field potential \eqref{potential multifield} is parametrised by $\{\alpha,\, \phi_\text{infl},\, \gamma\}$. While in the main text we discuss the effect of varying $\alpha$, we consider here the dependence on $\phi_\text{infl}$ and $\gamma$, for trajectories with a fixed initial value $\theta_\text{in}=7\pi/10$. The models discussed in this appendix are not necessarily compatible with the CMB measurements on large scales, but instead they are selected because they demonstrate the impact of changing $\phi_\text{infl}$ and $\gamma$. 
\begin{figure}
\centering
\captionsetup[subfigure]{justification=centering}
   \begin{subfigure}[b]{0.48\textwidth}
    \includegraphics[width=\textwidth]{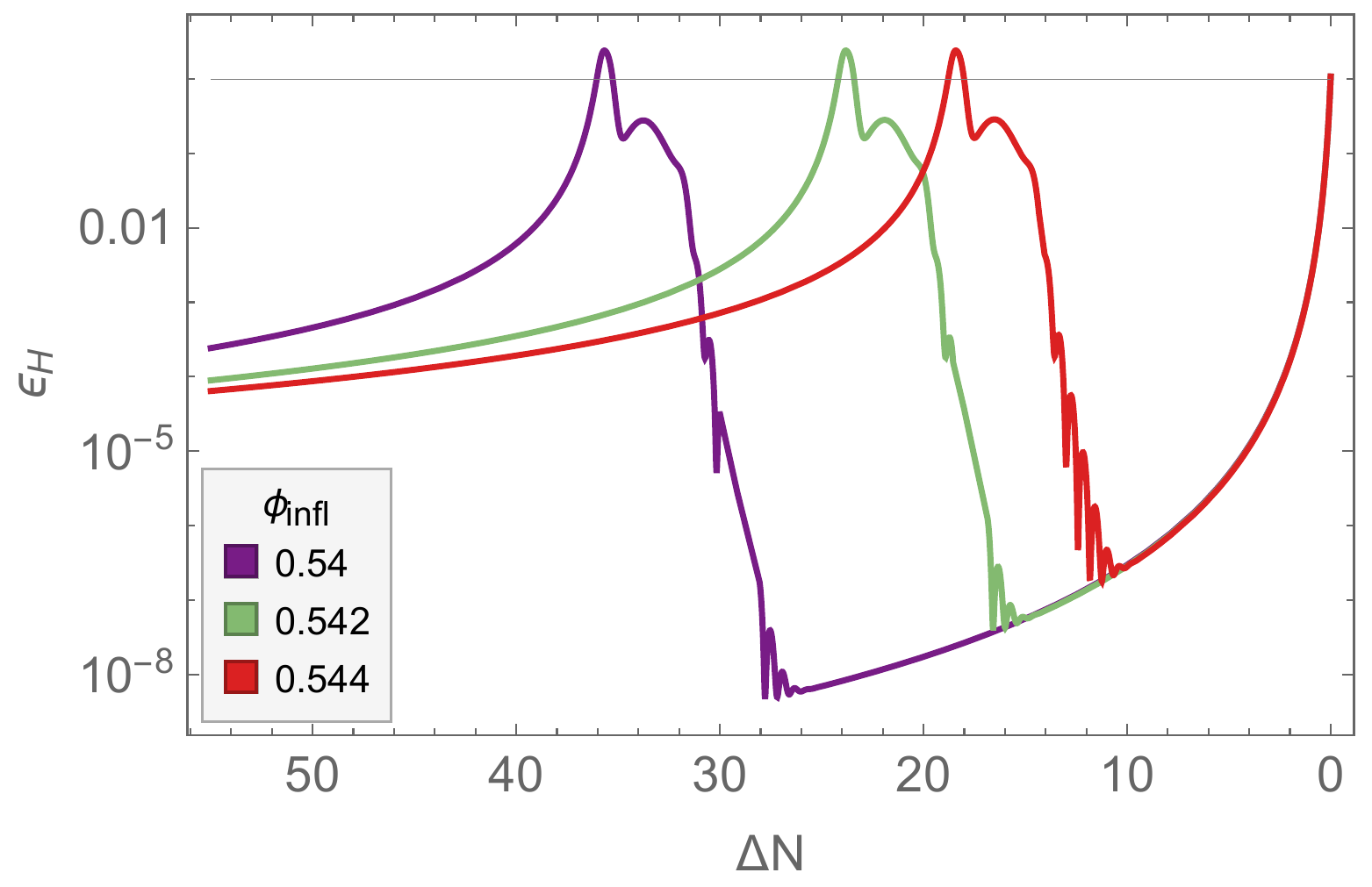}

  \end{subfigure}
  \begin{subfigure}[b]{0.48\textwidth}
    \includegraphics[width=\textwidth]{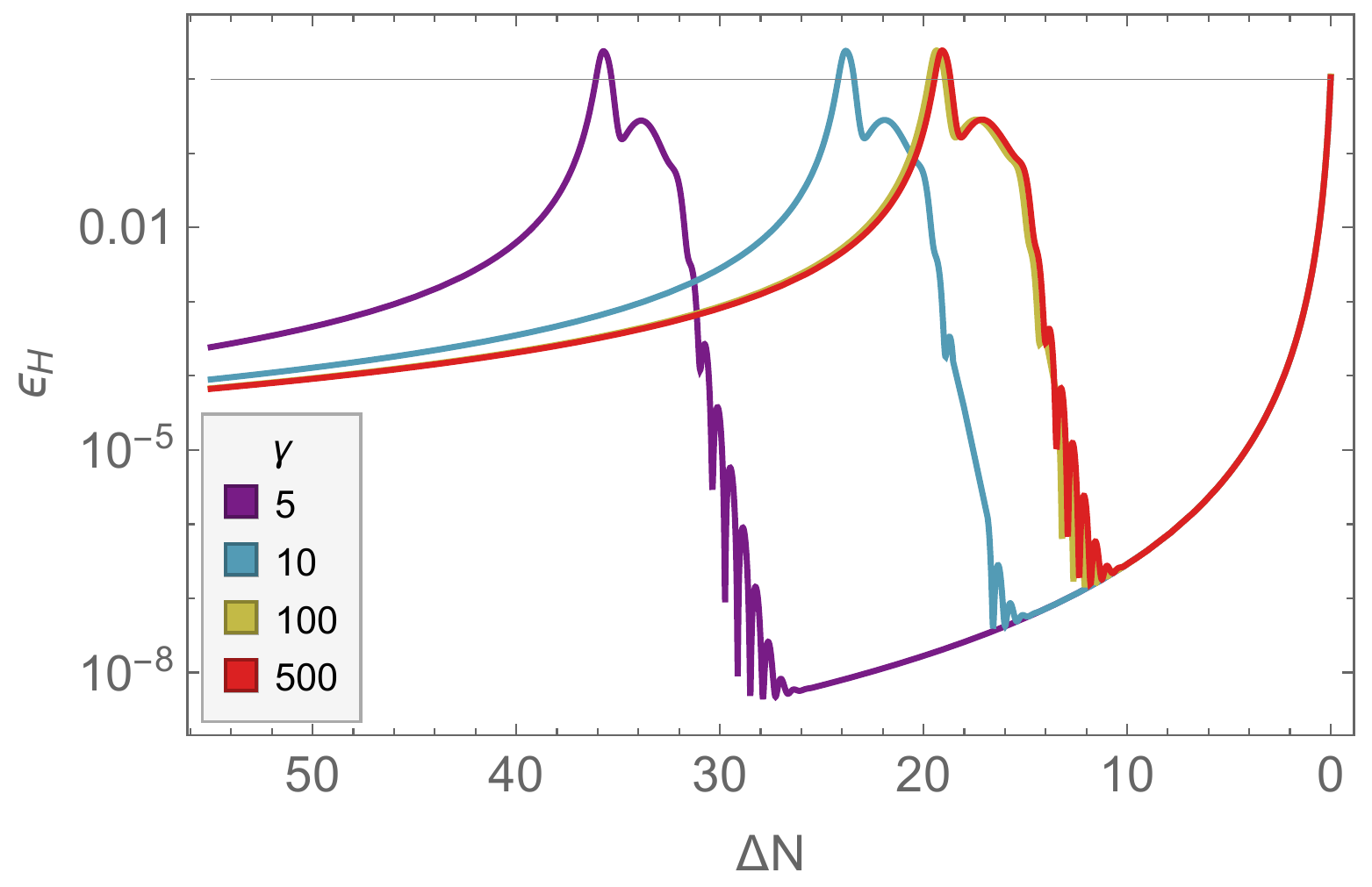}

  \end{subfigure}
\caption{\textit{Left panel}: effect of varying the position of the inflection point in the multi-field potential \eqref{potential multifield} with parameters $\{\alpha=0.1,\, \gamma=10 \}$ and initial conditions $\{\theta_\text{in}=7\pi/10,\,\theta'_\text{in}=0,\, \phi_\text{in}=3.1, \, \phi'_\text{in}=0\}$. \textit{Right panel}: effect of changing the parameter $\gamma$ in the potential \eqref{potential multifield} with parameters $\{\alpha=0.1,\, \phi_\text{infl}=0.542\}$ and same initial conditions as in the left panel.}
  \label{fig:multifield vary phi_infl and gamma}
\end{figure}

First we numerically solve the background equations \eqref{EoM multifield 1}--\eqref{EoM multifield 3} for a model with $\{\alpha=0.1,\, \gamma=10 \}$, testing $\phi_\text{infl}=\{0.54,\,0.542,\,0.544 \}$, and show the resulting $\epsilon_H(N)$ against $\Delta N\equiv N_\text{end}-N$ in the left panel of figure~\ref{fig:multifield vary phi_infl and gamma}. We see that the value of $\phi_\text{infl}$ affects the time the fields spend around the inflection point and therefore the duration of the second slow-roll phase (which is itself an attractor solution regardless of the value $\phi_\text{infl}$). In other words, $\phi_\text{infl}$ determines the time at which the transition between the two inflationary phases happens.

Next, we test $\gamma=\{5,\,10,\,100,\,500\}$ for the multi-field potential \eqref{potential multifield} with parameters $\{\alpha=0.1,\, \phi_\text{infl}=0.542\}$.
We consider $\gamma>3$ so that the turn in field space happens within the last $60$ e-folds of inflationary evolution, i.e., so that the second slow-roll phase lasts less than $60$ e-folds.
The resulting profile for $\epsilon_H(N)$ is displayed in the right panel of figure \ref{fig:multifield vary phi_infl and gamma} and demonstrates that the value of $\gamma$ affects the duration of the second phase of inflation, similar to the effect of varying $\phi_\text{infl}$. Also, increasing $\gamma$ beyond $\gamma\sim 100$ does not significantly change the background evolution. 

In order to understand the numerical results displayed in figure \ref{fig:multifield vary phi_infl and gamma} it is useful to calculate the effective squared-mass of the angular field, ${m_\theta}^2\equiv \partial^2 U(r,\,\theta)/\partial \theta^2$. Using the multi-field potential \eqref{potential multifield}, written in terms of the (non-canonical) radial field $r$, and assuming $H^2\simeq U(r,\,\theta)/3$, yields
\begin{equation}
\label{mass of theta field}
    \frac{{m_\theta}^2}{H^2}
    = 18\, r\, r_\text{infl}\, \frac{\gamma}{1+\gamma} \frac{ \left( r^2+3r_\text{infl}^2-\frac{3 \,r\, r_\text{infl}}{1+\gamma} \right) \cos{\theta} -\frac{3\, \gamma\,r\,r_\text{infl}}{1+\gamma}\, \cos{2\theta}}{ \left( r^2 +3 r_\text{infl}^2 - 3\, r\, r_\text{infl} \, \frac{1+\gamma \,\cos{\theta}}{1+\gamma} \right)^2} \;,
\end{equation}
where $r_\text{infl}\equiv \tanh{\left(\phi_\text{infl}/\sqrt{6\alpha}\right)}$. 
In the large $\gamma$ limit, the mass \eqref{mass of theta field} for a given $r$ is independent of $\gamma$, 
\begin{equation}
\begin{split}
\label{mass of theta field large gamma}
   \lim_{\gamma\rightarrow \infty} \frac{{m_\theta}^2}{H^2}= 18\, r\, r_\text{infl} \frac{ \left( r^2+3r_\text{infl}^2\right) \cos{\theta} -3\,r\,r_\text{infl}\, \cos{2\theta}}{ \left( r^2 +3 r_\text{infl}^2 - 3\, r\, r_\text{infl}\cos{\theta} \right)^2} \;,
\end{split}    
\end{equation}
which explains why the background evolution remains unchanged for sufficiently large values of $\gamma$, as observed in the right panel of figure~\ref{fig:multifield vary phi_infl and gamma}. 

During the first phase of inflation the angular field is frozen, $\theta\simeq \theta_\text{in}$, and the radial field satisfies $r\simeq1$, see eq.~\eqref{canonical field transformation}, therefore the effective squared-mass of the angular field \eqref{mass of theta field} is completely determined by $\gamma$ and $\phi_\text{infl}$. This explains why variations of the position of the inflection point or changes in $\gamma$ have the same effect on the background evolution, i.e., both change the time at which the transition between the first and second phases of inflation happens. In this sense, the potential parameters $\phi_\text{infl}$ and $\gamma$ are degenerate and, for fixed $\alpha$ and $\theta_\text{in}$, what really determines the duration of the second slow-roll phase overall is the effective mass of the angular field. 
\begin{figure}
\centering
\captionsetup[subfigure]{justification=centering}
   \begin{subfigure}[b]{0.48\textwidth}
    \includegraphics[width=\textwidth]{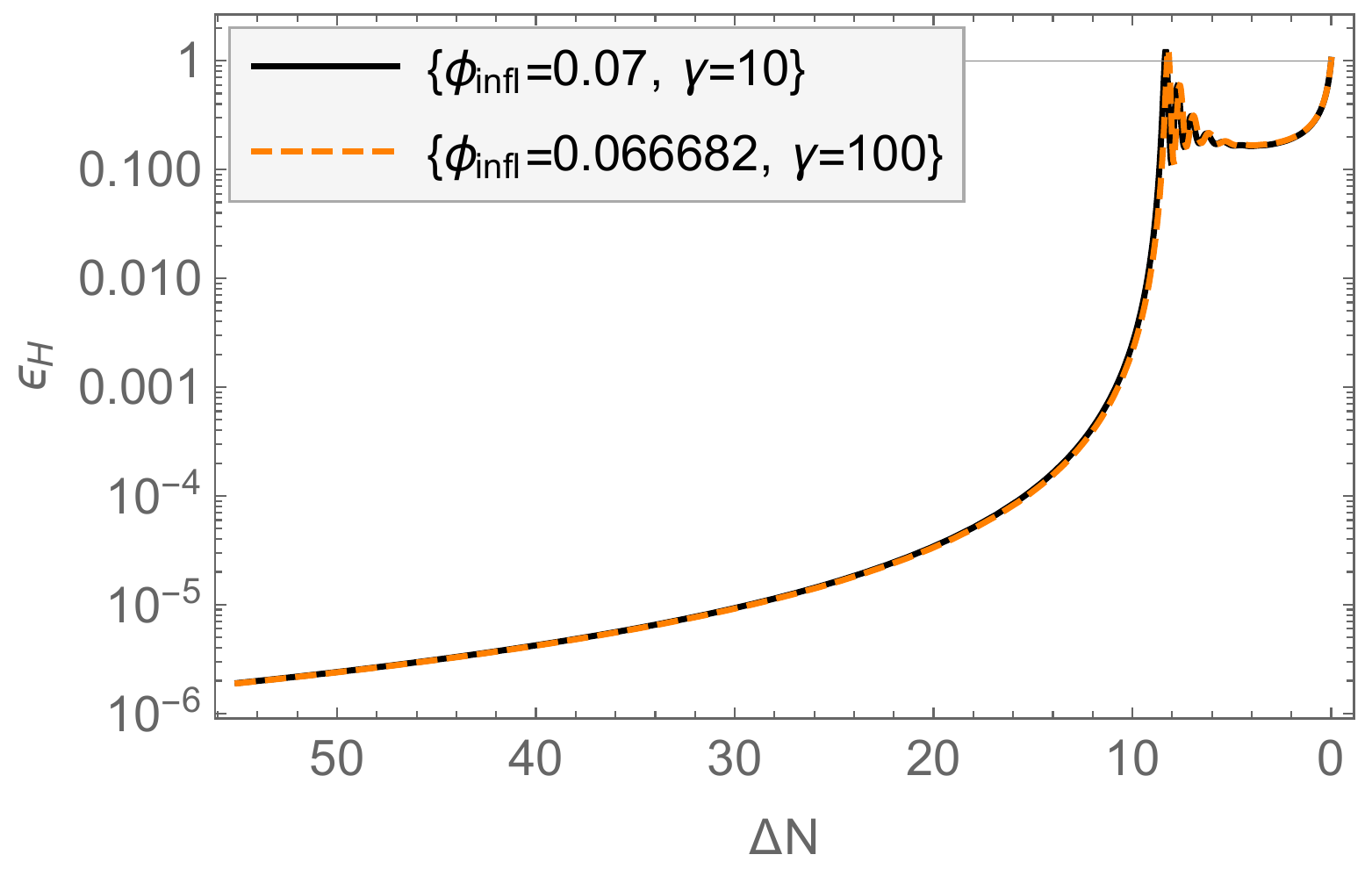}

  \end{subfigure}
  \begin{subfigure}[b]{0.47\textwidth}
    \includegraphics[width=\textwidth]{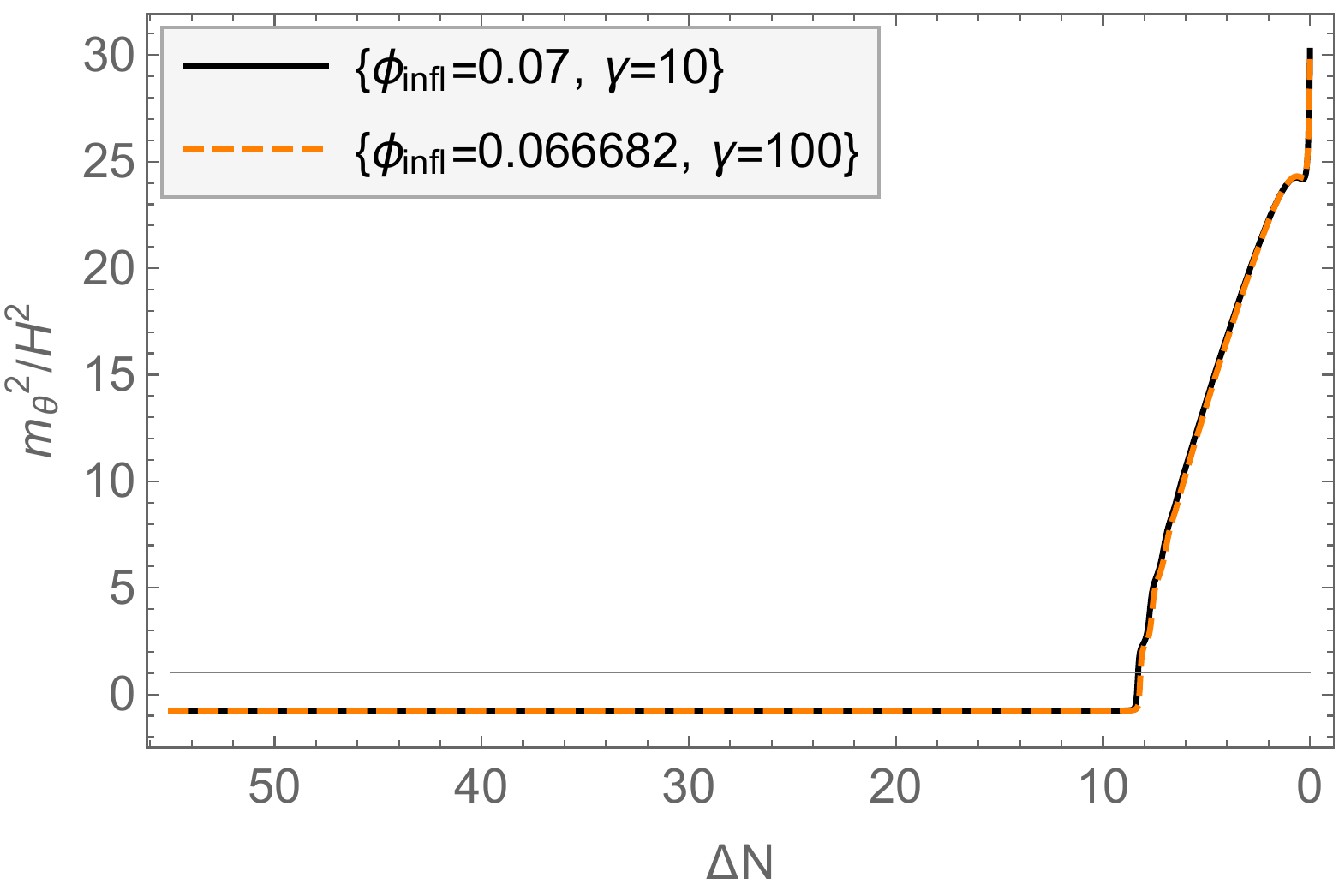}

  \end{subfigure}
\caption{Comparison between two models with same hyperbolic field-space curvature and the same initial value of the squared-mass of the angular field. The models parameters are listed in \eqref{models parameters mass matters 1} and~\eqref{models parameters mass matters 2} and the same initial conditions $\{\theta_\text{in}=7/10\,\pi ,\, \phi_\text{in}= 1.05\}$ are considered in both cases. We show the numerical evolution of the slow-roll parameter on the left and the squared-mass \eqref{mass of theta field} on the right. The two multi-field models display the same background evolution, which is due to selecting the values of $\gamma$ and $\phi_\text{infl}$ such that the angular mass is the same at the beginning of the evolution.}
  \label{fig:multifield angular mass matters}
\end{figure}

We demonstrate this by considering two different models formulated on the same hyperbolic field space, but whose potential parameters $\gamma$ and $\phi_\text{infl}$ are selected such that the initial mass of the angular field \eqref{mass of theta field} is the same, ${m_\theta}^2/H^2\simeq-0.762108$. The two models we consider are described by the parameters 
\begin{align}
\label{models parameters mass matters 1}
   \text{model}_1 &\rightarrow\{\alpha=0.005,\, \phi_\text{infl}=0.07,\, \gamma=10\},\\
\label{models parameters mass matters 2}
   \text{model}_2 &\rightarrow\{ \alpha=0.005,\, \phi_\text{infl}=0.066682,\, \gamma=100\} \;. 
\end{align}
The slow-roll parameter $\epsilon_H(N)$ and the angular field mass ${m_\theta}^2/H^2$ are represented respectively in the left and right panels of figure \ref{fig:multifield angular mass matters}. The comparison between the lines shows that the background evolution stemming from the same initial angular mass is the same in the two models and that $\gamma$ and $\phi_\text{infl}$ are not independent parameters. 

Given the degeneracy between $\gamma$ and $\phi_\text{infl}$, in section \ref{sec:robustness of single-field predictions} we choose to fix $\gamma$ and study the effect of changing the position of the inflection point and thus the initial mass of the angular field. The value of $\phi_\text{infl}$ affects the position of the transition between the first and second phases of evolution, and therefore it affects the CMB predictions of the model, so needs to be adjusted in order to produce a model which is not in tension with the CMB measurements, as we study in section~\ref{sec:robustness of single-field predictions}.

\section{2D hyperbolic field space: polar \textit{vs} planar coordinates}
\label{appendix:compare between polar and planar coordinates}
Different coordinate maps can be chosen to describe the (non-trivial) field space of multi-field models, where in this field space the coordinates are the fields themselves. The kinetic Lagrangian of $\alpha$--attractor models displayed in eq.~\eqref{alpha attractor metric} employs polar coordinates on the hyperbolic field space, with radial and angular fields, $(\phi, \,\theta)$, 
and the curvature of field space is $\mathcal{R}_\text{fs}=-4/(3\alpha)$.

Other coordinate maps have been used in the literature to describe hyperbolic field spaces with constant, negative curvature, see e.g.,~\cite{DiMarco:2002eb,DiMarco:2005nq,Lalak:2007vi} where planar coordinates have been selected. In this case, the kinetic Lagrangian reads
\begin{equation}
\label{planar coord L kin}
    \mathcal{L}_\text{planar}=-\frac{1}{2}(\partial u)^2-\frac{1}{2}\text{e}^{2 b u} (\partial v)^2 \;,
\end{equation}
where we label the set of planar coordinates as $(u,\,v)$ and the curvature of field space is $\mathcal{R}_\text{fs}=-2b^2$. Provided the curvature is the same, i.e., $b=\sqrt{2/(3\alpha)}$, the field-space geometry is the same as in eq.~\eqref{alpha attractor metric}, while the coordinate map selected is different. 

Planar coordinates were used in~\cite{Braglia:2020eai} to show how the hyperbolic geometry of field space could play a key role in enhancing the scalar power spectrum on small scales. In~\cite{Braglia:2020eai} the fields $(u, \,v)$ have a separable potential 
\begin{equation}
\label{Matteos potential planar coords}
    U(u,\,v)= U_0 \frac{u^2}{u_0^2+u^2}+\frac{1}{2} m_v^2v^2 \;,
\end{equation}
where $u$ has a plateau-type potential at large values of $u$, and the second field, $v$, has an apparently simple mass term. The authors of~\cite{Braglia:2020eai} demonstrate that the background evolution and the geometry of field space following from \eqref{planar coord L kin} and~\eqref{Matteos potential planar coords} could result in a transient tachyonic instability of the isocurvature perturbation, ${m_{s,\,\text{eff}}}^2/H^2<0$, which can lead to an enhancement of the scalar perturbation (see the discussion in section~\ref{sec: multi-field dynamics}).

Among all the possible combinations of model parameters discussed in~\cite{Braglia:2020eai}, we focus here on the inflationary potential described by $\{u_0=\sqrt{6},\, m_v^2=U_0/500\}$, with $b=7.84$ and initial conditions $\{u_\text{in}=7 ,\, v_\text{in}=7.31\}$, and refer the reader to the original work~\cite{Braglia:2020eai} for the equations describing the background evolution in planar coordinates. As demonstrated in~\cite{Braglia:2020eai}, this set of parameters and initial conditions produces a peak in the scalar power spectrum $P_\zeta=\mathcal{O}(10^{-2})$ located at the scales where the Laser Interferometer Space Antenna (LISA) operates. In this case the PBHs generated could potentially account for all of the dark matter in our Universe.
\begin{figure}
\centering
\includegraphics[scale=0.6]{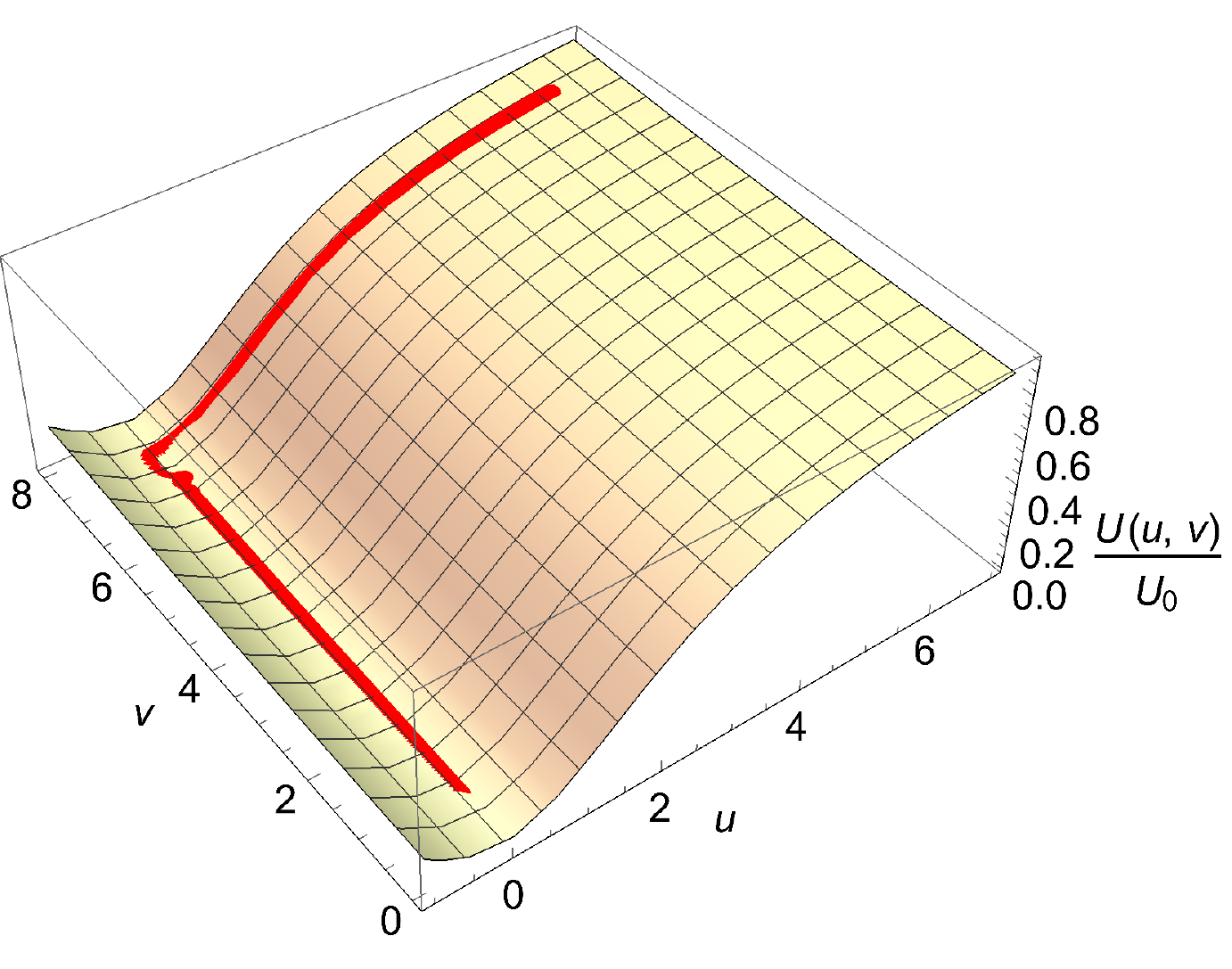}
\caption{Background evolution of the fields $u$ and $v$ represented on top of the potential profile, eq.~\eqref{Matteos potential planar coords}. We are here reproducing the background evolution of a model originally discussed in~\cite{Braglia:2020eai}. The model parameters and initial conditions considered are $\{u_0=\sqrt{6},\, m_v^2=U_0/500, \, b=7.84, \, u_\text{in}=7 ,\, v_\text{in}=7.31\}$.}
\label{fig:u and v evolution planar coords}
\end{figure}

In figure~\ref{fig:u and v evolution planar coords} the fields evolution is superimposed on top of the potential \eqref{Matteos potential planar coords}. The field $u$ drives a first stage of inflation, while $v$ is effectively frozen, with $v'$ suppressed by the geometrical factor $\text{e}^{-2b u}$ as long as $u\gg b^{-1}$, i.e., $u$ takes values larger than the curvature length of the field space. When $u\sim b^{-1}$ the suppression is lifted and $v$ starts evolving, driving a second stage of inflation as $u$ settles into its effective minimum. At the transition between the two inflationary stages, slow roll is violated and the effective squared-mass of the isocurvature perturbation briefly becomes negative, see the top-right panel of figure 1 in~\cite{Braglia:2020eai}.

We compare here the planar and polar coordinates description of an hyperbolic field space. Although different coordinate choices in field space are physically equivalent, once the form of the kinetic Lagrangian of the model is fixed, a specific choice for the potential, such as the one in eq.~\eqref{Matteos potential planar coords}, distinguishes between different physical models.
Our aim is to place  the inflationary model discussed in~\cite{Braglia:2020eai} (and therefore potentially other models formulated using planar coordinates, eq.~\eqref{planar coord L kin}) in the context of multi-field $\alpha$--attractors described using polar coordinates, and hence understand better the mechanism that allows for the enhancement of the curvature perturbation in that model.

In section~\ref{sec: mapping between polar and planar coordinates} we derive a coordinate transformation which allows us to transform from one coordinate map to the other. We re-analyse the model described in~\cite{Braglia:2020eai} employing polar coordinates in section \ref{sec:Braglias model in polar coordinates}. We find that the potential is singular in polar coordinates, sharing the same singularity as the kinetic Lagrangian, and initial conditions close to the singularity are necessary in order to enhance the scalar perturbation. This explains why the second field $v$ can lead to observable effects in this model, contrary to what was found in~\cite{Achucarro:2017ing} for $\alpha$--attractor models, i.e., for models with non-singular potentials. 

\subsection{Mapping between polar and planar coordinates}
\label{sec: mapping between polar and planar coordinates}
2D hyperbolic spaces with constant negative curvature (H$_2$) can be identified with spacelike hyperboloids embedded in a 3D Minkowski spacetime~\cite{Ball:2019atb}. This embedding procedure provides an intermediate step to map between polar and planar coordinates in the field space. Using the 3D Minkowski spacetime line element
\begin{equation}
\label{M3}
\mathrm{d}s^2 = -\mathrm{d}t^2 + \mathrm{d}x^2 + \mathrm{d}y^2 \;,
\end{equation}
surfaces with a fixed timelike displacement from the origin, are given by 
\begin{equation}
\label{H2constraint}
t^2 - x^2 - y^2 = R^2 \;.
\end{equation}
These surfaces have hyperbolic (H$_2$) geometry, while hyperboloids with a fixed spacelike displacement from the origin have dS$_2$ geometry~\cite{Ball:2019atb,Spradlin:2001pw}. 

Using polar coordinates, the line element of the hyperbolic field space is
\begin{equation}
\label{line element polar}
    \mathrm{d}s_\text{polar}^2 = \mathrm{d}\phi^2 + R^2 \sinh^2(\phi/R) \mathrm{d}\theta^2 \;\;\;\;\;\;(0\leq\phi<+\infty,\,0\leq\theta<2\pi )\;,
\end{equation}
where the curvature length of field space is $R\equiv \sqrt{3\alpha/2}$. Choosing coordinates $(\phi,\,\theta)$ on the hyperboloid \eqref{H2constraint} such that
\begin{align}
\label{polarM3}
t &= R \cosh (\phi/R) \;,\\
x &= R \sinh (\phi/R) \cos\theta \;,\\
\label{polarM3 last}
y &= R \sinh (\phi/R) \sin\theta \;,
\end{align}
the line element \eqref{M3} yields the H$_2$ line element in polar coordinates, eq.~\eqref{line element polar}. The polar coordinates cover the whole upper ($t>0$) hyperboloid for $0\leq\phi<+\infty$. 

The line element of the hyperbolic field space using planar coordinates is 
\begin{equation}
\label{line element planar}
    \mathrm{d}s_\text{planar}^2 = \mathrm{d}u^2 + e^{2bu} \mathrm{d}v^2 \;\;\;\;\;\;(-\infty< u<+\infty,\, -\infty< v<+\infty)\;, 
\end{equation}
where the curvature length of field space is $R=b^{-1}$. It is useful to first rewrite the line element \eqref{line element planar} in a conformally-flat form, 
\begin{equation}
\label{line element planar conformal}
    \mathrm{d}s_\text{planar, conf}^2=\Omega^2\left(\mathrm{d}w^2+\mathrm{d}v^2\right)\;. 
\end{equation}
This is achieved by means of the transformation
\begin{equation}
    w=-\frac{\text{e}^{-bu}}{b} \;,
\end{equation}
which leads to $\Omega^2\equiv 1/(-bw)^2$. Note that we have chosen integration constants such that for $-\infty < u<\infty$ we have $-\infty< w<0$. Choosing coordinates $(w,v)$ on the hyperboloid \eqref{H2constraint} such that
\begin{align}
\label{planarM3}
t &= - \frac{R^2}{2w} \left( 1+ \frac{v^2}{R^2} + \frac{w^2}{R^2} \right) \;,\\
x &= - \frac{R^2}{2w} \left( 1- \frac{v^2}{R^2}  - \frac{w^2}{R^2} \right) \;,\\
\label{planarM3 last}
y &= - \frac{Rv}{w} \;,
\end{align}
the line element \eqref{M3} yields the H$_2$ line element \eqref{line element planar conformal}, with $R=b^{-1}$. 

Using \eqref{planarM3}--\eqref{planarM3 last} to express the conformal planar coordinates $(w,\,v)$ in terms of those in Minkowski spacetime gives
\begin{align}
w &= - \frac{R^2}{t+x} \;, \\
v &= \frac{Ry}{t+x} \;.
\end{align}
Substituting \eqref{polarM3}--\eqref{polarM3 last} in the above expressions gives the conformal planar coordinates in terms of the polar coordinates. Finally expressing the conformal planar coordinates $(w,\,v)$ in terms of the planar coordinates $(u,\,v)$ yields 
\begin{align}
\label{u coord map}
    u &= R \ln \left[  \cosh (\phi/R) + \sinh (\phi/R) \cos\theta \right] \,,\\
\label{v coord map}
    v &= \frac{R \sinh (\phi/R) \sin\theta}{\cosh (\phi/R) + \sinh (\phi/R) \cos\theta} \,.
\end{align}

\subsection{A hyperbolic model with a singular potential}
\label{sec:Braglias model in polar coordinates}

In order to analyse the model of~\cite{Braglia:2020eai} using polar coordinates, we use the kinetic Lagrangian \eqref{alpha attractor metric} and we express the potential \eqref{Matteos potential planar coords} in terms of polar coordinates $(\phi,\,\theta)$ by means of the coordinate map \eqref{u coord map}--\eqref{v coord map}
\begin{multline}
\label{Matteos potential in polar coords}
    U(\phi,\,\theta)= U_0\, \Big\{\frac{\left(R \ln \left[  \cosh (\phi/R) + \sinh (\phi/R) \cos\theta \right]\right)^2}{6+\left(R \ln \left[  \cosh (\phi/R) 
    + \sinh (\phi/R) \cos\theta \right]\right)^2 }\\
    +\frac{1}{2\times 500} \left(\frac{R \sinh (\phi/R) \sin\theta}{\cosh (\phi/R) + \sinh (\phi/R) \cos\theta}\right)^2 \Big\}\;.
\end{multline}
In equation \eqref{Matteos potential in polar coords}, $R=\sqrt{3\alpha/2}$ and the model parameters have been substituted according to the parameters chosen in figure~\ref{fig:u and v evolution planar coords}. In particular, for the hyperbolic field space to be the same, $\mathcal{R}_\text{fs, polar}=\mathcal{R}_\text{fs, planar}\simeq -123$, we set $\alpha=2/(3b^2)\simeq0.01$.
\begin{figure}
\centering
\captionsetup[subfigure]{justification=centering}
   \begin{subfigure}[b]{0.455\textwidth}
    \includegraphics[width=\textwidth]{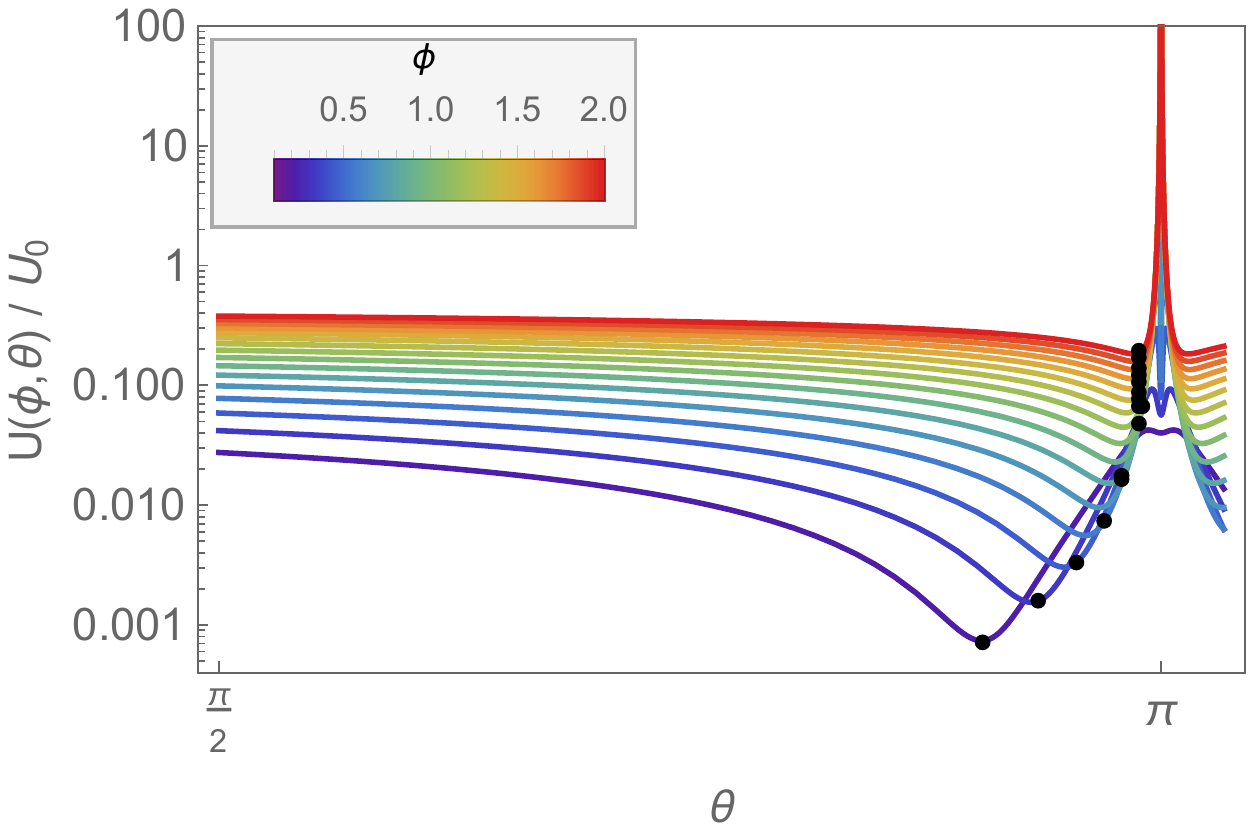}
  \end{subfigure}
  \begin{subfigure}[b]{0.47\textwidth}
    \includegraphics[width=\textwidth]{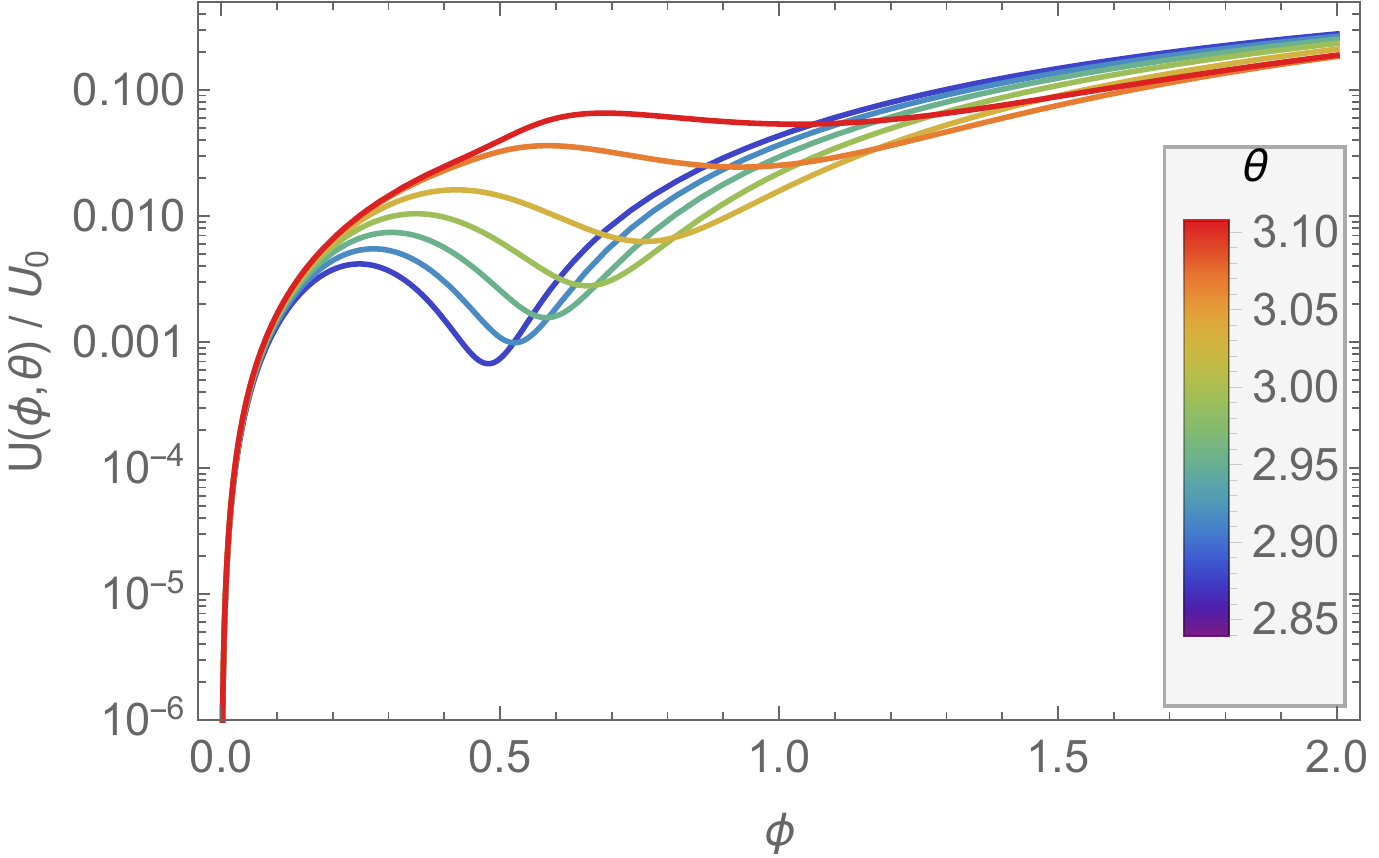}
  \end{subfigure}
 
 \caption{\textit{Left panel:} profile of the potential \eqref{Matteos potential in polar coords} as a function of the angular field $\theta$ for different fixed values of $\phi$. The selected range of $\phi$ values shows the potential's divergence at $\theta=\pi$. Black dots show the background evolution of the fields superimposed on the potential. Earlier stages of inflation correspond to larger values of $\phi$. \textit{Right panel:} profile of the potential \eqref{Matteos potential in polar coords} for fixed values of $\theta$ as a function of the radial field $\phi$. The range of $\theta$ corresponds to the values taken by the field during the background evolution displayed in figure \ref{fig:Matteos potential polar evo} and discussed in the main text.}
  \label{fig:Matteos potential profile polar coords}
\end{figure}

While being fairly simple in planar coordinates, the potential looks much more complicated when transformed to polar coordinates. The second term in eq.~\eqref{Matteos potential in polar coords} corresponds to the mass term for $v$ in the original potential \eqref{Matteos potential planar coords} and is singular at $\theta\to\pi$ for large values of $\phi$. We visualise the two-field potential as a function of $\phi$ and $\theta$ in figure~\ref{fig:Matteos potential profile polar coords}; from the left panel one can see that the potential diverges at $\theta=\pi$ for $\phi\gtrsim1$.
\begin{figure}
\centering
\captionsetup[subfigure]{justification=centering}
   \begin{subfigure}[b]{0.47\textwidth}
    \includegraphics[width=\textwidth]{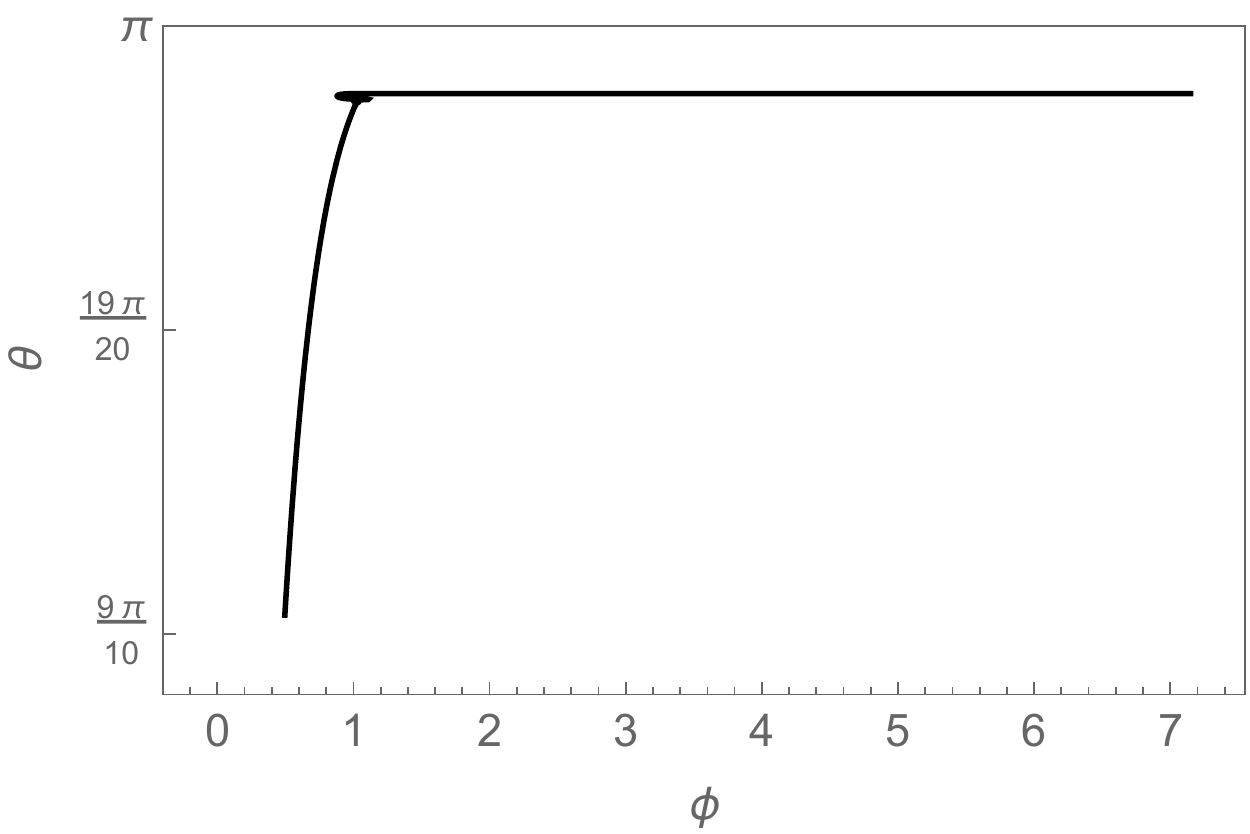}
  \end{subfigure}
  \begin{subfigure}[b]{0.48\textwidth}
    \includegraphics[width=\textwidth]{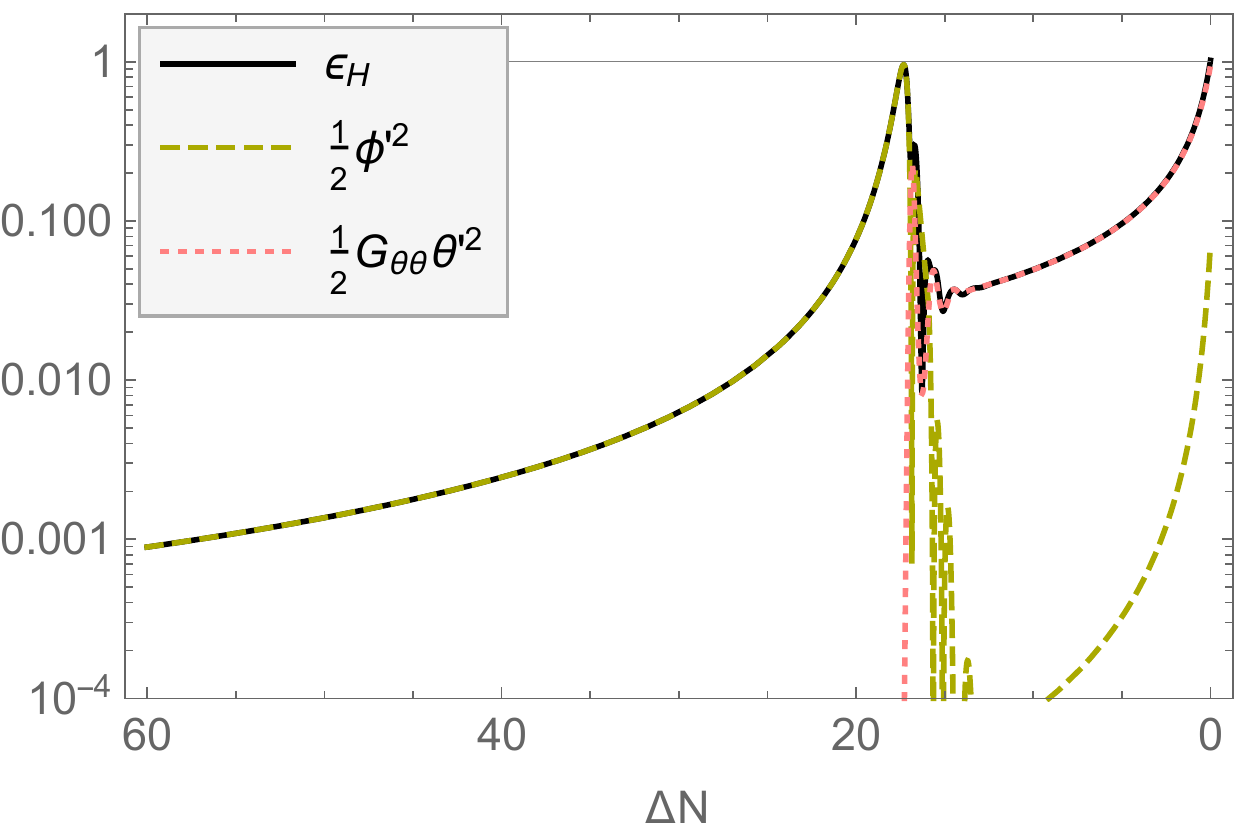}
  \end{subfigure}
 
 \caption{\textit{Left panel:} evolution of the background fields over the final 60 e-folds of inflation driven by the potential \eqref{Matteos potential in polar coords} in polar coordinates. \textit{Right panel:} evolution of $\epsilon_H$ over the last 60 e-folds of inflation for the same potential. The black line shows $\epsilon_H$, while the coloured lines show the contributions from the radial (green-dashed line) and angular (pink-dotted line) fields.}
  \label{fig:Matteos potential polar evo}
\end{figure}

We numerically solve eqs.~\eqref{EoM multifield 1}--\eqref{EoM multifield 3} to obtain the background evolution for $\phi$ and $\theta$. We select the initial conditions $\{ \phi_\text{in}= 7.1504 , \, \theta_\text{in}= 3.1067\}$ and slow-roll initial conditions for the velocities of the fields. We choose this set of initial conditions as they produce the same background evolution in terms of $u$ and $v$ shown in figure~\ref{fig:u and v evolution planar coords}.
The corresponding evolution of the fields in polar coordinates, $\phi$ and $\theta$, and the slow-roll parameter, $\epsilon_H$, is shown in figure \ref{fig:Matteos potential polar evo}. In particular, in the left panel we plot the trajectory in field space, showing how $\phi$ drives a first stage of inflation, after which there is a turn in field space and $\theta$, previously frozen, starts evolving. In the right panel of the same figure, the evolution of the slow-roll parameter $\epsilon_H$ and its components are shown against the number of e-folds to the end of inflation, $\Delta N\equiv N_\text{end}-N$. As expected, the major contribution to $\epsilon_H$ in the first phase of inflation comes from the kinetic energy of $\phi$, while the evolution of $\theta$ dominates a second stage of inflation.  Between the two phases, the slow-roll approximation is violated ($\epsilon_H\simeq1$).
\begin{figure}
\centering
\includegraphics[scale=0.6]{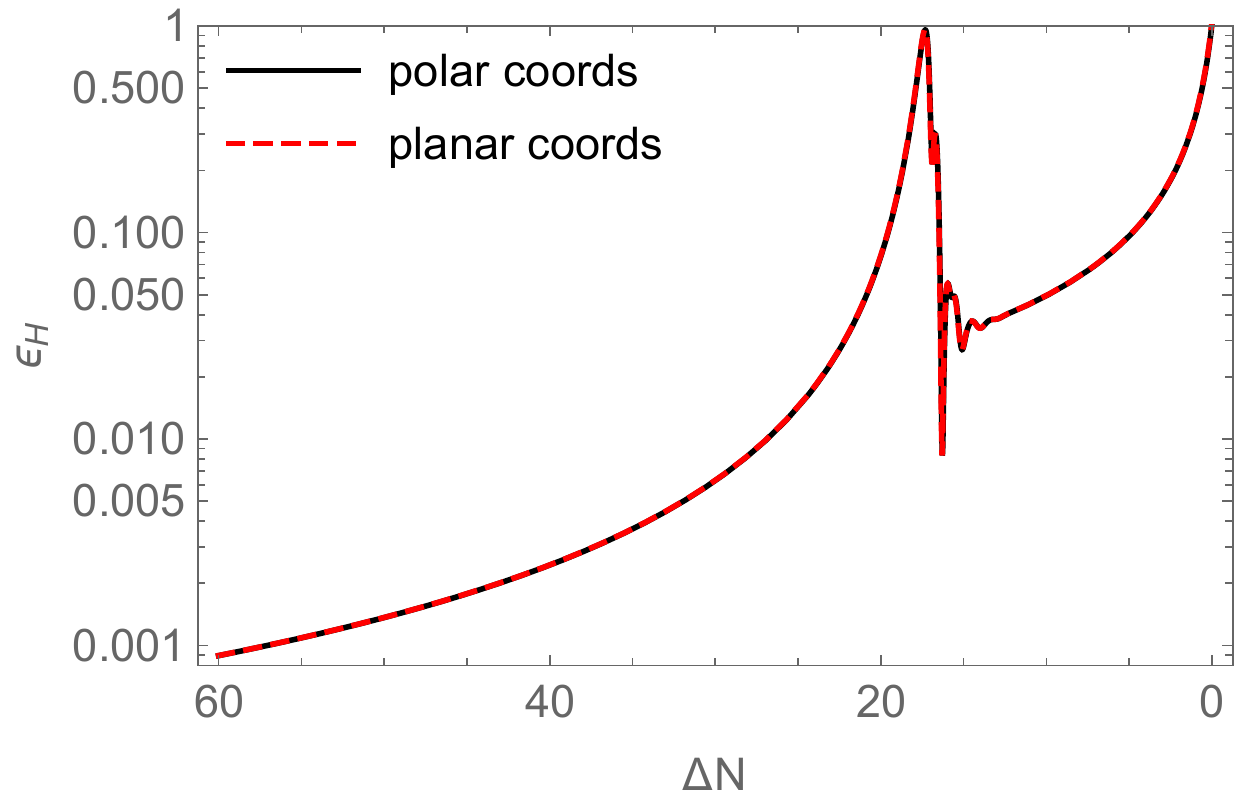}
\caption{Numerical solutions for the slow-roll parameter $\epsilon_H$ for the same model evolved using polar coordinates (black line) or planar coordinates (red-dashed line). The numerical solutions, using the corresponding potentials in polar \eqref{Matteos potential in polar coords} or planar \eqref{Matteos potential planar coords} coordinates, are identical.}
\label{fig:epsilon polar vs planar}
\end{figure}

The numerical solutions obtained with the polar coordinates description with the potential \eqref{Matteos potential in polar coords} is identical to that employing planar coordinates with the potential \eqref{Matteos potential planar coords}, as expected given the one-to-one correspondence between the two models. To show this, we compare the slow-roll parameter $\epsilon_H$. Using polar coordinates, $\epsilon_H(N)$ is 
\begin{equation}
\label{epsilon H polar}
   \epsilon_H(N)
   =\frac{1}{2}\left(\phi'^2+\frac{3\alpha}{2}\sinh^2{\left(\sqrt{\frac{2}{3\alpha}}\phi \right)}\theta'^2\right)    \;,
\end{equation}
see eq.~\eqref{EoM multifield 1}. When employing planar coordinates, we have instead
\begin{equation}
\label{epsilon H planar}
   \epsilon_H(N)
   =\frac{1}{2}\left(u'^2+\text{e}^{2b u}v'^2\right)  \;.
\end{equation}
Substituting in the corresponding numerical solutions for the fields, we show $\epsilon_H$ obtained from eq.~\eqref{epsilon H polar} and~\eqref{epsilon H planar} in figure~\ref{fig:epsilon polar vs planar}. As expected the two lines coincide exactly. 

While we have been focusing on a configuration which was chosen in~\cite{Braglia:2020eai} to produce a peak in the scalar power spectrum (and consequently in the induced second-order GWs) at LISA scales, a range of different initial conditions in field space are discussed in~\cite{Braglia:2020eai}  (see table~1 therein). In particular, varying the initial condition $v_\text{in}$ allows them to move the peak in the scalar power spectrum to scales where other future GW detectors could operate, e.g., SKA, BBO and ET. Inverting \eqref{u coord map} and~\eqref{v coord map} enables us to convert a set of initial conditions $(u_\text{in},\,v_\text{in})$ into the corresponding set in polar coordinates $(\phi_\text{in},\, \theta_\text{in})$. We have checked that the initial conditions listed in table~1 of~\cite{Braglia:2020eai} are all within $1.5\%$ of $\theta_\text{in}=\pi$. Thus we see that the configurations associated with enhanced scalar fluctuations on small scales stem from initial conditions very close to the singularity in the potential at $\theta=\pi$. As already pointed out in~\cite{Garcia-Saenz:2018ifx}, when the potential and the kinetic Lagrangian share the same singularity, large-scale observables are sensitive to the specific shape of the potential and to the initial conditions. 

It is straightforward to show that the kinetic Lagrangian \eqref{L kin r theta} and the potential \eqref{Matteos potential in polar coords} share the same pole in the conformal polar coordinates $(r,\,\theta)$, where $r$ is defined in \eqref{canonical field transformation}. By following a similar procedure to what was done in deriving eqs.~\eqref{u coord map} and~\eqref{v coord map}, we obtain the coordinate transformation
\begin{align}
\label{planar coordinate u in terms of r}
   u &= R\, \ln \left[ \frac{1+r^2+2r\cos\theta}{1-r^2}\right]\;,\\
\label{planar coordinate v in terms of r}
   v &= \frac{2R\,r\,\sin\theta}{1+r^2+2r\cos\theta}\;.
\end{align}
In order to assess the behaviour of the fields close to $\theta=\pi$, we define $\delta\equiv\pi-\theta$ and expand eqs.~\eqref{planar coordinate u in terms of r} and~\eqref{planar coordinate v in terms of r} to obtain
\begin{align}
\lim_{\delta\rightarrow0}u &= R\, \ln \left[ \frac{1-r}{1+r} + \frac{r\delta^2}{1-r^2} \right] \;,\\
    \lim_{\delta\rightarrow0}v &= \frac{2R\, r\, \delta}{(1-r)^2+r\delta^2} \;.
\end{align}
From the expression above it is clear that as $\delta\to0$ the term $m^2 v^2/2$ in the potential \eqref{Matteos potential planar coords} has a pole at $r=1$, as does the kinetic Lagrangian \eqref{L kin r theta}.


\bibliography{refs} 

\bibliographystyle{JHEP}

\end{document}